\documentclass[american,english,12pt]{article}
\usepackage{amsfonts}
\usepackage{amssymb}
\usepackage{graphicx}
\usepackage{amsmath}
\usepackage{latexsym}
\usepackage{rotating}
\usepackage{url}
\usepackage{natbib}
\usepackage{array,longtable}
\usepackage[colorlinks, linkcolor=blue, anchorcolor=blue, citecolor=blue, breaklinks]{hyperref}
\usepackage{float}
\usepackage{xr}
\usepackage{setspace}
\usepackage{hyperref}

\defcitealias{ahn2013}{AH}
\defcitealias{Jin_Lu_Su2024}{JLS}
\newtheorem{theorem}{Theorem}[section]

\newtheorem{lemma}{Lemma}

\newtheorem{remark}{Remark}

\newtheorem{assumption}{Assumption}
\numberwithin{assumption}{section}
\numberwithin{remark}{section}

\numberwithin{equation}{section}
\allowdisplaybreaks
\input{tcilatex}
\begin{document}

\title{Two-Way Mean Group Estimators for Heterogeneous Panel Models with
Fixed $T$\thanks{%
We thank Laura Serlenga for sharing her data. Lu gratefully acknowledges the
support from Chinese University of Hong Kong for the start-up fund. Su
gratefully acknowledges the NSFC for financial support under the grant
number 72133002. Address correspondence to: Liangjun Su, School of Economics
and Management, Tsinghua University, Beijing, 100084, China; Phone: +86 10
6278 9506; E-mail: sulj@sem.tsinghua.edu.cn.}}
\author{Xun Lu$^{a},$ Liangjun Su$^{b}$ \\
$^{a}$Department of Economics, Chinese University of Hong Kong\\
$^{b}$School of Economics and Management, Tsinghua University, China}
\date{}
\maketitle

\begin{abstract}
We consider a correlated random coefficient panel data model with two-way
fixed effects and interactive fixed effects in a fixed T framework. We
propose a two-way mean group (TW-MG) estimator for the expected value of the
slope coefficient and propose a leave-one-out jackknife method for valid
inference. We also consider a pooled estimator and provide a Hausman-type
test for poolability. Simulations demonstrate the excellent performance of
our estimators and inference methods in finite samples. We apply our new
methods to two datasets to examine the relationship between health-care
expenditure and income, and estimate a production function.\medskip

\medskip

\end{abstract}

\section{Introduction\label{Sec1}}

Heterogeneity is a fundamental characteristic of many economic data (see,
e.g., \cite{heckman2001micro}). The assumption\ of homogeneous slope
coefficients in traditional panel data models has been argued to be
unrealistic. Empirically, the homogeneity of slope coefficients is
frequently rejected; see, e.g., \cite{durlauf2001}, \cite{Browning2007}, 
\cite{phillips2007}, \cite{Su_Chen2013}, and \cite{lu2023}. In this paper,
we consider a heterogeneous panel data model with two-way fixed effects
(TWFEs) and interactive fixed effects (IFEs) in a fixed $T$ framework. This
type of general model usually requires a large $T,$ which is not applicable
to many panel data sets with only a modest number of time periods. With a
fixed $T,$ although we cannot consistently estimate individual slope
coefficients, we can still obtain a consistent estimator of the average
effects, which is usually of central interest in empirical work. Therefore,
we propose a mean-group (MG) type estimator for the expected value of the
slope coefficient based on the \textit{least squares dummy variable} (LSDV)
approach and a leave-one-out jackknife inference method.\footnote{%
Note that there is no simple within transformation we can apply to remove
the TWFEs when the slope coefficients vary with individuals.} One main
advantage of the new estimation and inference methods is their ease of
implementation. Despite their simplicity, they remain effective for a large
class of data-generating processes (DGPs), accommodating an arbitrarily
correlated slope coefficient with the regressors and allowing for the
presence of IFEs. This combination of simplicity and broad applicability can
make our new methods appealing to empirical researchers.

Specifically, we consider the following\ heterogeneous panel data model with
standard TWFEs and IFEs\footnote{%
Strictly speaking, the TWFEs can be absorbed into IFEs. Here we explicitly
allow the presence of both, as they play different roles in our estimation
procedure.}:%
\begin{equation}
y_{it}=\beta _{i}^{\prime }x_{it}+\alpha _{i\cdot }^{o}+\alpha _{\cdot
t}^{o}+\lambda _{i}^{o\prime }f_{t}^{o}+u_{it},\text{ }i=1,...,N\text{ and }%
t=1,...,T,  \label{y_eqn1}
\end{equation}%
where $x_{it}$ is a $K_{x}\times 1$ vector of observable regressors, $%
f_{t}^{o}$ and $\lambda _{i}^{o}$ are $R\times 1$ vectors of unobservable
factors and loadings, respectively, $\alpha _{i\cdot }^{o}$ and $\alpha
_{\cdot t}^{o}$ are the individual and time fixed effects, respectively, and 
$u_{it}$ is the error term. Here $\lambda _{i}^{\circ \prime }f_{t}^{\circ }$
in (\ref{y_eqn1}) is often referred to as IFEs in the model, which can be
correlated with $x_{it}.$ For the slope coefficient $\beta _{i},$ we assume
that 
\begin{equation}
\beta _{i}=\beta ^{0}+\eta _{i},  \label{beta_i}
\end{equation}%
with $E\left( \eta _{i}\right) =0$ so that $\beta ^{0}$ represents the
population average marginal effect of $x_{it}$ on $y_{it}.$ The key
parameter of interest is $\beta ^{0}.$ Even though we are not interested in
the fixed effect parameters, viz. $\alpha _{i\cdot }^{o},$ $\alpha _{\cdot
t}^{o},$ $\lambda _{i}^{o},$ and $f_{t}^{o},$ we find that it is convenient
to reformulate the model in (\ref{y_eqn1}) to ensure the unobserved factors
and loadings to have zero mean. For this purpose, we define 
\begin{equation*}
\lambda _{i}=\lambda _{i}^{o}-E(\lambda _{i}^{o})\text{ and }%
f_{t}=f_{t}^{o}-E(f_{t}^{o}),
\end{equation*}%
and rewrite the models in (\ref{y_eqn1}) equivalently as%
\begin{equation}
y_{it}=\beta _{i}^{\prime }x_{it}+\alpha _{i\cdot }+\alpha _{\cdot
t}+u_{it}^{\dag },  \label{y_eqn2}
\end{equation}%
where $\alpha _{i\cdot }=\alpha _{i\cdot }^{o}+\lambda _{i}^{o\prime
}E(f_{t}^{o}),$ $\alpha _{\cdot t}=\alpha _{\cdot t}^{o}+E\left( \lambda
_{i}^{o\prime }\right) f_{t},$ and $u_{it}^{\dag }=\lambda _{i}^{\prime
}f_{t}+u_{it}.$ We obtain the LSDV estimators of $\left\{ \alpha _{i\cdot
}\right\} _{i=1}^{N},\left\{ \alpha _{\cdot t}\right\} _{t=1}^{T}$ and $%
\left\{ \beta _{i}\right\} _{i=1}^{N}$ based on (\ref{y_eqn2})$.$ Let $\hat{%
\beta}_{i}$ denote the estimator of $\beta _{i},$ and our estimator of $%
\beta ^{0}$ is simply 
\begin{equation*}
\hat{\beta}^{MG}=\frac{1}{N}\sum_{i=1}^{N}\hat{\beta}_{i}.
\end{equation*}%
For simplicity, we name our estimator $\hat{\beta}^{MG}$ two-way-mean-group
(TW-MG) estimator.

There are three main features in our model. First, we consider a fixed $T$
framework and typically only require that $T>K_{x}+1$. Therefore, many
estimators, such as \cite{bai2009}'s principal component analysis (PCA)
estimator and \cite{pesaran2006}'s common correlated effects (CCE)
estimator, which typically require a large $T,$ are not applicable here. \ 

Second, we consider a heterogeneous panel data model where $\beta _{i}$
varies across individuals and can be correlated with $x_{it},$ so we have a 
\textit{correlated random coefficient} (CRC) panel in (\ref{y_eqn1}). For
our model, two popular estimators, namely, the pooled TWFE\ estimator and 
\cite{pesaran_smith1995}'s standard MG estimator, are not applicable for the
reasons discussed below. The pooled TWFE estimator of $\beta _{0}$ is based
on the following regression%
\begin{equation}
y_{it}=\beta ^{0\prime }x_{it}+\alpha _{i\cdot }+\alpha _{\cdot t}+\mathring{%
u}_{it},\text{ }i=1,...,N\text{ and }t=1,...,T,  \label{TWFE}
\end{equation}%
where $\mathring{u}_{it}=\eta _{i}x_{it}+u_{it}^{\dag }$ is treated as a
composite error term. The reason why the pooled estimator does not work here
is that $x_{it}$ can be correlated with $\eta _{i}x_{it}$ (and thus $%
\mathring{u}_{it}$), causing the typical endogeneity issue. For concrete
examples, see DGPs 3 and 4 in our simulations in Section \ref{Sim} where the
pooled TWFE\ estimator is inconsistent. See Remark \ref{Rmk3.1} for further
discussion on the correlated random coefficient. \cite{pesaran_smith1995}'s
MG estimator is based on the individual regression for each $i$ as follows,%
\begin{equation}
y_{it}=\beta _{i}^{\prime }x_{it}+\alpha _{i\cdot }+\ddot{u}_{it},\text{ }%
t=1,...,T,  \label{MG}
\end{equation}%
where $\ddot{u}_{it}=\alpha _{\cdot t}+u_{it}^{\dag }$ is treated as a
composite error term. The issue here is that $x_{it}$ can be correlated with 
$\alpha _{\cdot t}$ (and thus $\ddot{u}_{it}$)$,$ leading to an endogeneity
problem. For example, for all four DGPs in our simulations in Section \ref%
{Sim}, the standard MG estimator is inconsistent. For further discussion on
the relationship between our TW-MG and the standard MG estimators, see
Remark \ref{Rmk2.3}. Our TW-MG estimator can be thought of as a generalized
version of the pooled TWFE\ and standard MG estimators, as estimation
equations (\ref{TWFE}) and (\ref{MG}) are restricted versions of (\ref%
{y_eqn2}) with the restrictions of $\beta _{i}=\beta ^{0}$ and $\alpha
_{\cdot t}=0,$ respectively.

Third, we allow IFEs in our model. We argue that once we include TWFEs, the
demeaned IFEs can enter the error term under our key assumption that $%
\lambda _{i}$ and $x_{it}^{o}$ are uncorrelated, where $x_{it}^{o}$ denotes
the part of\ $x_{it}$ with its additive individual and time effects (if any)
removed. For a more precise statement of the key assumption, see Assumption %
\ref{Assmp3.1}(i) below. This point has already been made in the literature
(see, e.g., \cite{westerlund2019estimation} and \cite{kapetanios2023testing}
for the large $T$ case). Here we provide a rigorous asymptotic analysis for
a heterogenous model with a fixed $T.$ To illustrate this point, consider
the following simple example where 
\begin{equation*}
y_{it}=\beta _{i}x_{it}+\lambda _{i}^{o}f_{t}^{o}+u_{it},\text{ and }%
x_{it}=\gamma _{i}^{o}f_{t}^{o}+v_{it},
\end{equation*}%
with $K_{x}=1$ and $R=1.$ Suppose that $E\left( \lambda _{i}^{o}\right) =$ $%
E\left( \gamma _{i}^{o}\right) =1$ and $\lambda _{i}^{o},$ $\gamma _{i}^{o},$
$f_{t}^{o},$ $u_{it}$ and $v_{it}$ are mutually independent with $E\left(
u_{it}\right) =E\left( v_{it}\right) =0$. It appears that the IFEs, $\lambda
_{i}^{o}f_{t}^{o},$ cause much trouble, as 
\begin{equation*}
\text{Cov}\left( x_{it},\text{ }\lambda _{i}^{o}f_{t}^{o}\right)
=E[(f_{t}^{o})^{2}]\not=0,
\end{equation*}%
which can lead to an endogeneity issue. Furthermore, $\lambda
_{i}^{o}f_{t}^{o}$ can cause strong cross-sectional dependence in the $y$%
-equation. However, we show that our TW-MG estimator still works in this
case after controlling TWFEs, as the demeaned version of $\lambda
_{i}^{o}f_{t}^{o}$ does not cause the endogeneity issue and strong
cross-sectional dependence. Admittedly, we do rule out the case where $%
\lambda _{i}^{o}$ and $x_{it}^{o}$ are correlated. However, as we already
allow individual ($\alpha _{i\cdot })$ and time effects $(\alpha _{\cdot t})$
without any restrictions, the uncorrelated assumption between $\lambda
_{i}^{o}$ and $x_{it}^{o}$ may not be that strong. For the simple example
above, essentially, we require that the factor loadings $\lambda _{i}^{o}$
in the $y$-equation and $\gamma _{i}^{o}$ in the $x$-equation be
uncorrelated.\footnote{%
In fact, the classic paper of \cite{pesaran2006} for large $T$ maintains the
assumption that the factor loadings in the $y$-equation and the $x$-equation
are uncorrelated.} Recently, \cite{kapetanios2023testing} assume that both $%
x_{it}$ and $y_{it}$ follow a factor structure as in our simple example (or
more generally, the DGPs in Remark \ref{Rmk2.1}) and test the null
hypothesis that the loadings of $x$- and $y$-equations are uncorrelated.
They find that the null hypothesis is not rejected in thirteen out of
fourteen datasets considered.\footnote{%
Their test is based on the large $T$ framework and compares the TWFE
estimator and the PCA\ estimator.} Furthermore, to the best of our
knowledge, there is no well-accepted solution for estimating $\beta _{0}$ in
model (\ref{y_eqn1}) with a fixed $T$ when $\lambda _{i}^{o}$ and $%
x_{it}^{o} $ are correlated.

This paper makes several important methodological contributions. First, we
study the asymptotic properties of our TW-MG estimator for the general model
(\ref{y_eqn1}) and demonstrate its consistency and asymptotic mixture normal
distribution. In particular, we invoke the notion of stable convergence and
stable central limit theorem (CLT) (see, e.g., \cite{hausler2015stable} and 
\cite{van2023weak}) to take into account the randomness of the factors.
Second, we propose an inference method based on the leave-one-out jackknife
approach and provide rigorous justification. Although this method is easy to
implement, its theoretical justification is challenging because the
parameters involved ($\beta _{i}$'s) are not fixed when one cross-section
unit is deleted. We have to augment certain matrices and vectors in the
leave-one-out estimator to match those in the original estimator and show
the consistency of the variance estimator. See Remark \ref{Rmk3.2} below for
more details. Third, we examine the asymptotic properties of the pooled TWFE
estimator of $\beta _{0}$ for our model (\ref{y_eqn1}) under the additional
assumption that $\beta _{i}$ is random and uncorrelated with a demeaned
version of the regressor (Assumption \ref{Assmp4.1}(i) below). To the best
of our knowledge, this is also a new result for the pooled estimator, as we
allow for the presence of IFEs. Furthermore, we propose a Hausman-type test (%
\cite{hausman1978specification}) to assess the poolability, that is, whether
the TW-MG and pooled TWFE estimators converge to the same probability limit.
The test is also based on the leave-one-out jackknife method.

Our paper contributes to the extensive literature on panel data models with
IFEs and random coefficients (RCs). For a review of the RC panel data
models, see, e.g., \cite{hsiao2008random} and \cite{Hsiao2022}. For the
review of the panel data models with IFEs, see, e.g., \cite{baiwang2016}. In
the large $T$ framework, \cite{pesaran2006} introduces the CCE estimator for
both homogeneous and heterogeneous static panel data models with IFEs. For
the PCA\ approach, \cite{bai2009}, \cite{moon2015} and \cite{lu_su2016},
among others, propose\ estimators for homogeneous model. \cite%
{li2020efficient} and \cite{cao2024oracle} extend this by proposing a
PCA-type estimator for heterogeneous models. The literature for
heterogeneous panel data models with fixed $T$ is relatively sparse.
Although \cite{pesaran_smith1995}'s MG estimator is originally proposed for
the large $T$ case, it has been shown to be effective for fixed $T$ as well
(see, e.g., \cite{hsiao2019panel}). \cite{chamberlain1992efficiency}
proposes an estimator similar to ours but does not account for IFEs. More
recently, \cite{pesaran2024trimmed} introduce a trimmed version of
Chamberlain's estimator, which also excludes IFEs and employs a different
inference approach. \cite{westerlund2022cce} provide a pooled CCE estimator
for heterogeneous models.

The rest of the paper is structured as follows. In Section \ref{Sec2}, we
propose the estimation and inference methods. We aim to make Section \ref%
{Sec2} accessible to applied researchers who are interested in applying our
new methods. Section \ref{Sec3} examines the asymptotics of our TW-MG
estimator and the jackknife inference procedure. In Section \ref{Sec4}, we
study the pooled estimator and propose a Hausman-type test for poolability.
Section \ref{Sim} reports Monte Carlo simulation results. In Section \ref%
{App}, we apply our new methods to health-care expenditure data and
production data. Section \ref{Sec7} concludes. The appendix contains the
proofs of all main results. The online supplement contains the proofs of
some technical lemmas and additional simulation results.

\textit{Notation}. For a positive integer $n$, let $[n]\equiv \{1,2,\ldots
,n\},$ $n_{1}=n-1,$ $\mathbb{I}_{n}$ an $n\times n$ identity matrix, and $%
\mathbf{\iota }_{n}$ an $n\times 1$ vector of ones. For a real $m\times n$
matrix $A$, we use $\Vert A\Vert ,$ $\Vert A\Vert _{\text{sp }}$and $%
A^{\prime }$ to denote its Frobenius norm, spectral norm, and transpose,
respectively. When $A$ has full column rank, let $\mathbb{P}_{A}\equiv
A\left( A^{\prime }A\right) ^{-1}A^{\prime }$ and $\mathbb{M}_{A}\equiv 
\mathbb{I}_{m}-\mathbb{P}_{A}$, where $\equiv $ signifies a definitional
relationship. Let $\mu _{\max }(\cdot )$ and $\mu _{\min }(\cdot )$ denote
the largest and smallest eigenvalue of a real symmetric matrix,
respectively. Let $\otimes $ denote the Kronecker product and $\mathbf{1}%
\{\cdot \}$ the usual indicator function. Let $\mathbf{0}_{a\times b}$
denote an $a\times b$ matrix of zeros. Let bdiag$\left(
A_{1},...,A_{N}\right) $ denote the block diagonal matrix. The symbol $%
\overset{p}{\rightarrow }$ denotes convergence in probability, $\overset{d}{%
\rightarrow }$ convergence in distribution, and plim probability limit. We
use $a_{N}\lesssim b_{N}$ to denote that $a_{N}/b_{N}=O_{p}\left( 1\right) $
as $N\rightarrow \infty .$

\section{Model, Estimation and Inference Procedure \label{Sec2}}

In this section, we introduce the model and estimation and inference
procedure.

\subsection{Model \label{Sec2.1}}

Our original model is specified in (\ref{y_eqn1}), and it is reformulated in
(\ref{y_eqn2}). Note that $E(u_{it}^{\dag })=0$ in (\ref{y_eqn2}) under some
weak assumptions (e.g., Assumption \ref{Assmp3.1}(i) in Section \ref{Sec3.1}
or Assumption \ref{Assmp4.1}(i) in Section \ref{Sec4.1}). More importantly, $%
u_{it}^{\dag }$ is also uncorrelated with a demeaned version of $x_{it}$ (by
removing its individual and time effects) under Assumption \ref{Assmp3.1},
which ensures that one can estimate $\beta _{i}$'s and $\beta ^{0}$ by the
usual least squares approach. We emphasize that the above reformulation is
solely for theoretical analysis and is not required for the implementation
of our estimation and inference procedure.

Since we focus on the large $N$ and fixed $T$ framework, it is hard to
consider a PCA estimate for $\beta ^{0}.$ Due to the presence of the time
effects $\alpha _{\cdot t}$ in the $y$-equation, one cannot pool the $T$
observations for each $i$ to run a time series regression of $y_{it}$ on $%
\left( x_{it}^{\prime },1\right) $ to estimate $\beta _{i}$ as in standard
MG estimation (see, e.g., \cite{pesaran_smith1995}). Below we propose a
TW-MG estimator of $\beta ^{0}$ via the LSDV approach. Even though we do not
formally address the missing value problem in this paper, it is well known
that the use of the LSDV approach can easily account for the missing value
issues in panel regressions.

\begin{remark}
\label{Rmk2.1}\textbf{(An example of }$x_{it}$\textbf{) }In the above setup,
we only specify the DGP for $y_{it}.$ If one follows the literature on the
CCE estimation (e.g., \cite{pesaran2006}), one can also specify a DGP for $%
x_{it}$ as follows:%
\begin{equation}
x_{it}=\mu _{i\cdot }^{o}+\mu _{\cdot t}^{o}+\gamma _{i}^{o\prime
}f_{t}^{o}+v_{it},  \label{x_eqn2a}
\end{equation}%
where $\gamma _{i}^{o}$ is an $R\times K_{x}$ matrix of factor loadings in
the $x$-equation, $\mu _{i\cdot }^{o}$ and $\mu _{\cdot t}^{o}$ are the
individual and time fixed effects, respectively, and $v_{it}$ is the
zero-mean error term. As above, one can reformulate $\gamma _{i}^{o}$ and $%
f_{t}^{o}$ to have zero mean such that 
\begin{equation}
x_{it}=\mu _{i\cdot }+\mu _{\cdot t}+\gamma _{i}^{\prime }f_{t}+v_{it},
\label{x_eqn2}
\end{equation}%
where $\gamma _{i}=\gamma _{i}^{o}-E(\gamma _{i}^{o}),$ $%
f_{t}=f_{t}^{o}-E(f_{t}^{o}),$ $\mu _{i\cdot }=\mu _{i\cdot }^{o}+\gamma
_{i}^{o\prime }E(f_{t}^{o})$ and $\mu _{\cdot t}=\mu _{\cdot t}^{o}+E\left(
\gamma _{i}^{o\prime }\right) f_{t}.$ In this paper we do not need to
specify a DGP for $x_{it}.$ Even so, it is worth noting that if $x_{it}$ is
indeed generated via ($\ref{x_eqn2}$), then only its double-demeaned
version, $x_{it}^{o},$ will enter the asymptotics of our TW-MG estimator
below.
\end{remark}

\subsection{Estimation procedure \label{Sec2.2}}

To present the estimation procedure, we introduce some notations. Since $%
\alpha _{i\cdot }$ and $\alpha _{\cdot t}$ are not separately identified, we
need to impose a location restriction for the LSDV estimation. Below, we
impose the restriction that 
\begin{equation*}
\sum_{t=1}^{T}\alpha _{\cdot t}=0
\end{equation*}%
in the estimation procedure without requiring the true values $\left\{
\alpha _{i\cdot }\right\} $ to satisfy such a restriction. Let $\mathbf{%
\alpha }^{\left( 1\right) }\equiv \left( \alpha _{1\cdot },...,\alpha
_{N\cdot }\right) ^{\prime },$ $\mathbf{\alpha }^{\left( 2\right) }\equiv
\left( \alpha _{\cdot 1},...,\alpha _{\cdot T-1}\right) ^{\prime },$ and $%
\mathbf{\beta }\equiv (\beta _{1}^{\prime },...,\beta _{N}^{\prime
})^{\prime }.$ Define 
\begin{equation}
\mathbf{D}^{\left( 1\right) }=\mathbb{I}_{N}\otimes \mathbf{\iota }_{T}\text{
and }\mathbf{D}^{\left( 2\right) }=\mathbf{\iota }_{N}\otimes \left( 
\begin{array}{c}
\mathbb{I}_{T_{1}} \\ 
-\mathbf{\iota }_{T_{1}}^{\prime }%
\end{array}%
\right) ,  \label{DD1}
\end{equation}%
where $T_{1}\equiv T-1.$ Let $\mathbf{y}_{i}=\left( y_{i1},...,y_{iT}\right)
^{\prime },$ \ $\mathbf{Y}=\left( \mathbf{y}_{1}^{\prime },...,\mathbf{y}%
_{N}^{\prime }\right) ^{\prime },$ $\mathbf{x}_{i}=\left(
x_{i1},...,x_{iT}\right) ^{\prime }$ and $\mathbf{X}=$bdiag$(\mathbf{x}%
_{1},...,\mathbf{x}_{N}).$ Define $\mathbf{u}_{i}=\left(
u_{i1,}...,u_{iT}\right) ^{\prime },$\ $\mathbf{U}=\left( \mathbf{u}%
_{1}^{\prime },...,\mathbf{u}_{N}^{\prime }\right) ^{\prime },$ $F=\left(
f_{1},...,f_{T}\right) ^{\prime }$ and $\mathbf{\lambda =(}\lambda
_{1}^{\prime },...,\lambda _{N}^{\prime }\mathbf{)}^{\prime }.$ Define $%
\mathbf{u}_{i}^{\dag }$ and $\mathbf{U}^{\dag }$ analogously. Then we can
rewrite the model in (\ref{y_eqn2}) in matrix form:%
\begin{equation}
\mathbf{Y}=\mathbf{X\beta }+\mathbf{D}^{\left( 1\right) }\mathbf{\alpha }%
^{\left( 1\right) }+\mathbf{D}^{\left( 2\right) }\mathbf{\alpha }^{\left(
2\right) }+\mathbf{U}^{\dag }=\mathbf{X\beta }+\mathbf{D\alpha }+\mathbf{U}%
^{\dag },  \label{Y_eqn1}
\end{equation}%
where $\mathbf{D}=\left( \mathbf{D}^{\left( 1\right) },\mathbf{D}^{\left(
2\right) }\right) ,$ $\mathbf{\alpha }=(\mathbf{\alpha }^{\left( 1\right)
\prime },\mathbf{\alpha }^{\left( 2\right) \prime })^{\prime },$ and $%
\mathbf{U}^{\dag }=(\mathbb{I}_{N}\otimes F)\mathbf{\lambda }+\mathbf{U}.$

By partialling out $\mathbf{D}$ in (\ref{Y_eqn1}), we obtain the following
LSDV estimator of $\mathbf{\beta }$: 
\begin{equation}
\mathbf{\hat{\beta}}\equiv (\hat{\beta}_{1}^{\prime },...,\hat{\beta}%
_{N}^{\prime })^{\prime }\equiv \left( \mathbf{X}^{\prime }{\mathbb{M}}_{%
\mathbf{D}}\mathbf{X}\right) ^{-1}\mathbf{X}^{\prime }{\mathbb{M}}_{\mathbf{D%
}}\mathbf{Y.}  \label{betai_est}
\end{equation}%
Then our TW-MG estimator of $\beta ^{0}$ is given by 
\begin{equation*}
\hat{\beta}^{MG}=\frac{1}{N}\sum_{i=1}^{N}\hat{\beta}_{i}.
\end{equation*}%
Let $\mathbf{S}_{i,N}=(\mathbf{0}_{K_{x}\times K_{x}},...,\mathbf{0}%
_{K_{x}\times K_{x}},\mathbb{I}_{K_{x}},\mathbf{0}_{K_{x}\times K_{x}},...,%
\mathbf{0}_{K_{x}\times K_{x}})$ be a $K_{x}\times NK_{x}$ selection matrix
that has $\mathbb{I}_{K_{x}}$ in its $i$th block. Then $\hat{\beta}^{MG}$
can also be written as 
\begin{equation}
\hat{\beta}^{MG}=\frac{1}{N}\sum_{i=1}^{N}\mathbf{S}_{i,N}(\mathbf{X}%
^{\prime }{\mathbb{M}}_{\mathbf{D}}\mathbf{X)}^{-1}\mathbf{X}^{\prime }{%
\mathbb{M}}_{\mathbf{D}}\mathbf{Y}=\frac{1}{N}\mathbf{S}_{N}(\mathbf{X}%
^{\prime }{\mathbb{M}}_{\mathbf{D}}\mathbf{X)}^{-1}\mathbf{X}^{\prime }{%
\mathbb{M}}_{\mathbf{D}}\mathbf{Y,}  \label{betahat1}
\end{equation}%
where $\mathbf{S}_{N}=\mathbf{\iota }_{N}^{\prime }\otimes \mathbb{I}%
_{K_{x}}=(\mathbb{I}_{K_{x}},...,\mathbb{I}_{K_{x}}).$

\begin{remark}
\label{Rmk2.2}\textbf{(An alternative representation)} Note that $\mathbf{D}%
=(\mathbf{D}^{\left( 1\right) },\mathbf{D}^{\left( 2\right) }),$ where $%
\mathbf{D}^{\left( 1\right) }$ and $\mathbf{D}^{\left( 2\right) }$ are
defined in (\ref{DD1}). It is easy to verify that 
\begin{eqnarray*}
\mathbf{D}^{\prime }\mathbf{D} &\mathbf{=}&\left( 
\begin{array}{cc}
T\mathbb{I}_{N} & \mathbf{0}_{N\times T_{1}} \\ 
\mathbf{0}_{T_{1}\times N} & N\left( \mathbb{I}_{T_{1}}+\mathbf{\iota }%
_{T_{1}}\mathbf{\iota }_{T_{1}}^{\prime }\right)%
\end{array}%
\right) , \\
\mathbb{P}_{\mathbf{D}} &=&\mathbf{D}\left( \mathbf{D}^{\prime }\mathbf{D}%
\right) ^{-1}\mathbf{D}^{\prime }=\frac{\mathbf{\iota }_{N}\mathbf{\iota }%
_{N}^{\prime }}{N}\otimes \mathbb{M}_{\mathbf{\iota }_{T}}+\mathbb{I}%
_{N}\otimes \frac{\mathbf{\iota }_{T}\mathbf{\iota }_{T}^{\prime }}{T},\text{
and} \\
\mathbb{M}_{\mathbf{D}} &=&\mathbb{I}_{NT}-\mathbb{P}_{\mathbf{D}}=\mathbf{D}%
\left( \mathbf{D}^{\prime }\mathbf{D}\right) ^{-1}\mathbf{D}^{\prime }=%
\mathbb{M}_{\mathbf{\iota }_{N}}\otimes \mathbb{M}_{\mathbf{\iota }_{T}}.
\end{eqnarray*}%
As a result, we can also write the TW-MG estimator $\hat{\beta}^{MG}$ as
follows:%
\begin{equation}
\hat{\beta}^{MG}=\frac{1}{N}\mathbf{S}_{N}\left[ \mathbf{X}^{\prime }(%
\mathbb{M}_{\mathbf{\iota }_{N}}\otimes \mathbb{M}_{\mathbf{\iota }_{T}})%
\mathbf{X}\right] ^{-1}\mathbf{X}^{\prime }(\mathbb{M}_{\mathbf{\iota }%
_{N}}\otimes \mathbb{M}_{\mathbf{\iota }_{T}})\mathbf{Y.}  \label{betahat2}
\end{equation}%
Clearly, $\mathbb{M}_{\mathbf{\iota }_{N}}\otimes \mathbb{M}_{\mathbf{\iota }%
_{T}}$ is the double-demean (or two-way within group transformation)
operator that removes the individual and time fixed effects in both $y_{it}$
and $x_{it}.$
\end{remark}

\begin{remark}
\label{Rmk2.3}(\textbf{Comparison with standard MG\ estimator}) Since the
seminal work of \cite{pesaran_smith1995}, MG\ estimation has been widely
applied in panel data analyses to estimate the average marginal effect. To
the best of our knowledge, standard MG estimation is applied to heterogenous
panel data models with only individual effects. In this case, one can use
the $T$ observations for cross-section unit $i$ and estimate the
individual-specific slope coefficients $\beta _{i}$ via a time series
regression and then average the resulting estimates. Due to the presence of
time effects $\alpha _{\cdot t}$ in (\ref{y_eqn2}), this approach is not
applicable. In contrast, in our TW-MG estimation procedure above, we first
pool all $NT$ observations across $i$ and $t$ and estimate $\left\{ \beta
_{i}\right\} _{i\in \left[ N\right] }$ jointly and then average the
resulting estimates.
\end{remark}

\begin{remark}
\label{Rmk2.4}\textbf{(Ridge-type estimator)} When $T$ is small, to improve
the finite sample performance, we can modify the estimator in (\ref{betahat2}%
) as 
\begin{equation}
\hat{\beta}^{MG,Ridge}=\frac{1}{N}\mathbf{S}_{N}\left[ \frac{1}{T}\mathbf{X}%
^{\prime }(\mathbb{M}_{\mathbf{\iota }_{N}}\otimes \mathbb{M}_{\mathbf{\iota 
}_{T}})\mathbf{X}+\kappa _{N}\cdot \mathbb{I}_{NK_{x}}\right] ^{-1}\frac{1}{T%
}\mathbf{X}^{\prime }(\mathbb{M}_{\mathbf{\iota }_{N}}\otimes \mathbb{M}_{%
\mathbf{\iota }_{T}})\mathbf{Y,}  \label{ridge}
\end{equation}%
where $\kappa _{N}$ is a tuning parameter. This is a ridge-type estimator.
For the choice of $\kappa _{N},$ we can set $\kappa _{N}=c_{\kappa }\frac{1}{%
N},$ where $c_{\kappa }$ is a constant. We can show that all our theoretical
results below hold for $\hat{\beta}^{MG,Ridge}$ as well with this choice of $%
\kappa _{N}.$ In the simulations and applications below, to account for the
scale of the data, we let $c_{\kappa }$ be the median of $\left\{ \det
\left( \frac{1}{T}\sum_{t=1}^{T}\ddot{x}_{it}\ddot{x}_{it}^{\prime }\right)
\right\} _{i\in \lbrack N]},$ where $\det $ stands for determinant and $%
\ddot{x}_{it}$ is the double-demeaned version of $x_{it}$: $\ddot{x}%
_{it}=x_{it}-\frac{1}{N}\sum_{j=1}^{N}x_{jt}-\frac{1}{T}\sum_{s=1}^{T}x_{is}+%
\frac{1}{NT}\sum_{j=1}^{N}\sum_{s=1}^{T}x_{js}.$
\end{remark}

\subsection{Inference procedure \label{Sec2.3}}

Below we show that $\hat{\beta}^{MG}$ is asymptotically mixture normal under
some regularity conditions. We propose a jackknife method to estimate the
asymptotic variance of $\hat{\beta}^{MG}$ and to conduct inference for $%
\beta ^{0}.$ Following (\ref{betahat2}), we can obtain the leave-one-out
estimator of $\beta ^{0}$ in (\ref{y_eqn2}) by leaving out the observations
for the $i$th cross-sectional unit:%
\begin{equation}
\hat{\beta}^{MG\left( -i\right) }=\frac{1}{N_{1}}\mathbf{S}_{N_{1}}\left[ 
\mathbf{X}^{\left( -i\right) \prime }\left( \mathbb{M}_{\mathbf{\iota }%
_{N_{1}}}\otimes \mathbb{M}_{\mathbf{\iota }_{T}}\right) \mathbf{X}^{\left(
-i\right) }\right] ^{-1}\mathbf{X}^{\left( -i\right) \prime }\left( \mathbb{M%
}_{\mathbf{\iota }_{N_{1}}}\otimes \mathbb{M}_{\mathbf{\iota }_{T}}\right) 
\mathbf{Y}^{\left( -i\right) },  \label{beta_leave_one_out1}
\end{equation}%
where $N_{1}\equiv N-1,$ $\mathbf{X}^{\left( -i\right) \prime }$ is the
leave-one-out version\ of $\mathbf{X}^{\prime }$ by deleting its $i$th row
block: $\mathbf{X}^{\left( -i\right) \prime }=$bdiag$(\mathbf{x}_{1}^{\prime
},$ $...,\mathbf{x}_{i-1}^{\prime },\mathbf{x}_{i+1}^{\prime },...,\mathbf{x}%
_{N}^{\prime }),$ and $\mathbf{Y}^{\left( -i\right) }=\left( \mathbf{y}%
_{1}^{\prime },...,\mathbf{y}_{i-1}^{\prime },\mathbf{y}_{i+1}^{\prime },...,%
\mathbf{y}_{N}^{\prime }\right) ^{\prime }.$

Given the jackknife estimates $\{\hat{\beta}^{MG\left( -i\right) }\}_{i\in %
\left[ N\right] },$ we propose to estimate the asymptotic variance of $\sqrt{%
N}(\hat{\beta}^{MG}-\beta ^{0})$ by 
\begin{equation*}
\hat{\Omega}^{MG}=(N-1)\dsum\limits_{i=1}^{N}\left( \hat{\beta}^{MG\left(
-i\right) }-\overline{\hat{\beta}}^{_{MG}}\right) \left( \hat{\beta}%
^{MG\left( -i\right) }-\overline{\hat{\beta}}^{_{MG}}\right) ^{\prime },
\end{equation*}%
where $\overline{\hat{\beta}}^{_{MG}}\equiv \frac{1}{N}\sum_{i=1}^{N}\hat{%
\beta}^{MG\left( -i\right) }.$ Let $\mathbb{S}$ be a $K_{x}\times 1$
selection vector, which defines the scalar parameter of interest: $\mathbb{S}%
^{\prime }\beta ^{0}.$ Let $z_{\tau /2}$ denote the upper $\tau /2$-quantile
of $N(0,1).$ Then we construct the $100(1-\tau )\%$ confidence interval (CI)
of $\mathbb{S}^{\prime }\beta ^{0}$ as 
\begin{equation}
\left[ \mathbb{S}^{\prime }\hat{\beta}^{MG}-z_{\tau /2}\sqrt{\mathbb{S}%
^{\prime }\hat{\Omega}^{MG}\mathbb{S}/N},\text{ \ \ }\mathbb{S}^{\prime }%
\hat{\beta}^{MG}+z_{\tau /2}\sqrt{\mathbb{S}^{\prime }\hat{\Omega}^{MG}%
\mathbb{S}/N}\right] .  \label{CI1}
\end{equation}

\section{Asymptotic Properties of $\hat{\protect\beta}^{MG}$ \label{Sec3}}

In this section, we study the asymptotic properties of $\hat{\beta}^{MG}$.
We first state the assumptions and show that $\sqrt{N}(\hat{\beta}%
^{MG}-\beta ^{0})$ follows a version of stable CLT. Then we show that the
above jackknife method can estimate the asymptotic variance of $\sqrt{N}(%
\hat{\beta}^{MG}-\beta ^{0})$ consistently.

\subsection{Basic assumptions \label{Sec3.1}}

To proceed, we introduce some notations. Let $\mathbf{\hat{Q}}=\frac{1}{T}%
\mathbf{X}^{\prime }{\mathbb{M}}_{\mathbf{D}}\mathbf{X}\ $and $\mathbf{\ddot{%
X}}={\mathbb{M}}_{\mathbf{D}}\mathbf{X}.$ Noting that $\mathbb{M}_{\mathbf{D}%
}=\mathbb{M}_{\mathbf{\iota }_{N}}\otimes \mathbb{M}_{\mathbf{\iota }_{T}},$
we can write $\mathbf{\ddot{X}}=$bdiag$\left( \mathbf{\ddot{x}}_{1},...,%
\mathbf{\ddot{x}}_{N}\right) ,$ where $\mathbf{\ddot{x}}_{1}=(\ddot{x}%
_{i1},...,\ddot{x}_{iT})^{\prime }$ and $\ddot{x}_{it}=x_{it}-\frac{1}{N}%
\sum_{j=1}^{N}x_{jt}-\frac{1}{T}\sum_{s=1}^{T}x_{is}+\frac{1}{NT}%
\sum_{j=1}^{N}\sum_{s=1}^{T}x_{js}.$ In addition,%
\begin{equation*}
\mathbf{\hat{Q}}=\frac{1}{T}\mathbf{\ddot{X}}^{\prime }\mathbf{\ddot{X}}=%
\frac{1}{T}\mathbf{X}^{o\prime }{\mathbb{M}}_{\mathbf{D}}\mathbf{X}^{o},
\end{equation*}%
where $\mathbf{X}^{o}=$bdiag$\left( \mathbf{x}_{1}^{o},...,\mathbf{x}%
_{N}^{o}\right) ,$ $\mathbf{x}_{i}^{o}=(x_{i1}^{o},...,x_{iT}^{o})^{\prime
}, $ and $x_{it}^{o}$ denotes the part of\ $x_{it}$ with its additive
individual and time effects (if any) removed. For example, if $x_{it}$ is
generated from ($\ref{x_eqn2}$), then $x_{it}^{o}=\gamma _{i}^{\prime
}f_{t}+v_{it}$. In the case where $x_{it}$ does not contain any additive
term that is varying along either $i$ or $t$ but not both, we have that $%
x_{it}^{o}=x_{it}$ and $\mathbf{x}_{i}^{o}=\mathbf{x}_{i}.$ Define 
\begin{equation*}
\hat{\omega}_{j}=\sum_{i=1}^{N}[\mathbf{\hat{Q}}^{-1}]_{ij}\text{ and }\chi
_{j}^{\prime }=\frac{1}{T}\sum_{i=1}^{N}\hat{\omega}_{i}\mathbf{x}%
_{i}^{o\prime }\mathbb{M}_{\mathbf{\iota }_{T}}m_{ij},
\end{equation*}%
where $m_{ij}=\mathbf{1}\left\{ i=j\right\} -\frac{1}{N}$ denotes the $%
\left( i,j\right) $th element of $\mathbb{M}_{\mathbf{\iota }_{N}}.$

Let $\mathcal{F}_{N,0}^{MG}=\sigma \left( \{\mathbf{x}_{i}^{o}\}_{1\leq
i\leq N},F\right) ,$ the minimal sigma-field generated by $\{\mathbf{x}%
_{i}^{o}\}_{1\leq i\leq N}$ and $F.\footnote{%
If $x_{it}$ is generated from ($\ref{x_eqn2}$), we can equivalently define $%
\mathcal{F}_{N,0}^{MG}=\sigma \left( \{\left( \mathbf{\gamma }_{i},\mathbf{v}%
_{i}\right) \}_{1\leq i\leq N},F\right) .$}$ Let $\mathcal{F}%
_{N,j}^{MG}=\sigma (\mathcal{F}_{N,0}^{MG},\left\{ \mathbf{u}_{i},\lambda
_{i}\right\} _{i=1}^{j})$ for $j\in \left[ N\right] .$ Let $%
\max_{i}=\max_{1\leq i\leq N},$ and $\min_{i}=\min_{1\leq i\leq N}$. \ Let $%
\underline{c},$ $\overline{c}\in \left( 0,\infty \right) $ be some generic
constants that do not depend on $N$ and whose values may vary across places.
Let $E_{A}\left( \cdot \right) =E\left( \cdot |A\right) .$ Let $\tilde{q}%
_{ii}=\frac{1}{T}\mathbf{x}_{i}^{o\prime }{\mathbb{M}}_{\mathbf{\iota }_{T}}%
\mathbf{x}_{i}^{o}.$ To study the asymptotic properties of $\hat{\beta}%
^{MG}, $ we impose the following two assumptions.

\begin{assumption}
\label{Assmp3.1} (i) Conditional on $\mathcal{F}_{N,0}^{MG},$ $\left\{ (%
\mathbf{u}_{i},\lambda _{i})\right\} _{i\in \left[ N\right] }$ are
independent across $i$ with zero mean and $E_{\mathcal{F}_{N,0}^{MG}}[\left%
\Vert \mathbf{u}_{i}\right\Vert ^{4+2\delta }+\left\Vert \lambda
_{i}\right\Vert ^{4+2\delta }]\lesssim 1$ for some $\delta >0;$

(ii) $\left\{ \eta _{i}\right\} _{i\in \left[ N\right] }$ are independent
across $i$ with zero mean, $\frac{1}{N}\sum_{i=1}^{N}E(\eta _{i}\eta
_{i}^{\prime })\rightarrow \Omega _{\eta },$ and $E\left\Vert \eta
_{i}\right\Vert ^{2+\delta }\leq \overline{c};$

(iii) $E_{\mathcal{F}_{N,i-1}^{MG}}(\mathbf{u}_{i}\eta _{i}^{\prime })=0;$

(iv) $E_{F}\left\Vert \mathbf{x}_{i}^{o}\right\Vert ^{4+2\delta }\lesssim 1.$%
\ 
\end{assumption}

\begin{assumption}
\label{Assmp3.2}(i) $\tilde{q}_{ii}$ has full rank almost surely (a.s.) and $%
\frac{1}{N}\sum_{i=1}^{N}\left\Vert \tilde{q}_{ii}^{-1}\right\Vert
^{2}=O_{p}(1)$;

(ii) $\frac{1}{N}\sum_{i=1}^{N}\chi _{i}^{\prime }E_{\mathcal{F}_{N,0}^{MG}}[%
\mathbf{u}_{i}^{\dag }\mathbf{u}_{i}^{\dag \prime }]\chi _{i}\overset{p}{%
\rightarrow }\Omega _{1}^{MG}$ and $\Omega _{1}^{MG}$ is positive definite
almost surely (a.s.).
\end{assumption}

Assumption \ref{Assmp3.1}(i)--(iii) impose some dependence and moment
conditions on the unobserved components, i.e., $\mathbf{u}_{i},\lambda _{i},$
and $\eta _{i}.$ Noting that $\mathbf{X}^{o}$ is measurable with respect to $%
\mathcal{F}_{N,0}^{MG},$ Assumption \ref{Assmp3.1}(i) implies the
independence of $(\mathbf{u}_{i},\lambda _{i})$ across $i$ conditional on $%
\mathbf{X}^{o}.$ The zero conditional mean condition ensures that the
composite error term $u_{it}^{\dag }$ ($=\lambda _{i}^{\prime }f_{t}+u_{it}$%
) can be treated as a usual error term in least squares regression as the
individual and time effects can be partialled out via the usual partitioned
regressions. Clearly, Assumption \ref{Assmp3.1}(i) essentially requires that
the regressors $x_{it}$, after its individual and time effects are removed,
are strictly exogenous so that a dynamic panel is ruled out. Assumption \ref%
{Assmp3.1}(ii) imposes standard conditions on the heterogeneous parameters $%
\eta _{i}.$ Note that we allow $\Omega _{\eta }$ to be positive semidefinite
so that some elements in $\beta _{i}$ are allowed to be homogenous across $%
i. $ Assumption \ref{Assmp3.1}(iii) requires the conditional
uncorrelatedness between $\mathbf{u}_{i}$ and $\eta _{i}$ because $E(\mathbf{%
u}_{i}|\mathcal{F}_{N,i-1}^{MG})=0$ under Assumption \ref{Assmp3.1}(i).
Assumption \ref{Assmp3.1}(iv) imposes some moment conditions on $\mathbf{x}%
_{i}^{o}$ that will be used to verify certain Lyapunov conditions. In case
where $x_{it}$ is generated according to (\ref{x_eqn2}), this moment
condition can be replaced by\medskip

\noindent \textbf{Assumption 3.1(iv*)} $E\left\Vert \gamma _{i}\right\Vert
^{4+2\delta }+E\left\Vert \mathbf{v}_{i}\right\Vert ^{4+2\delta }\leq 
\overline{c},$ where $\mathbf{v}_{i}=(v_{i1},...,v_{iT})^{\prime }.$\medskip

It is worth mentioning that we do not impose any conditions on the
individual and time effects, viz. $\alpha _{i\cdot }$ and $\alpha _{\cdot t}$
for the $y$-equation and $\mu _{i\cdot }$ and $\mu _{\cdot t}$ in the $x$%
-equation if $x_{it}$ is generated according to (\ref{x_eqn2}) as they are
partialled out in the LSDV regression. Neither do we impose any condition on
the unobserved factor $f_{t}$ due to our focus on the fixed $T$ framework.
In other words, we allow various kinds of dependence, trending or
nonstationary behavior in $\alpha _{\cdot t},$ $\mu _{\cdot t}$ and $f_{t}.$
In addition, we do not impose any dependence condition on $\left\{
f_{t},u_{it},x_{it}\right\} $ along the time dimension.

Assumption \ref{Assmp3.2}(i) specifies some rank and moment conditions on
the use of TW-MG\ estimator when a time effect is present in the $y$%
-equation, which implicitly requires that $T>K_{x}+1.$ When $T\leq K_{x}$,
there is no way to consider the MG estimator even without the individual or
time effects in the $y$-equation. In this case, one can only consider the
pooled regression to estimate the population average marginal effect $\beta
^{0}$ as in the next section. When $T=K_{x}+1,$ we run into the irregular
case discussed in \cite{pesaran2024trimmed} even if $\tilde{q}_{ii}$ may
still be full rank a.s.. Note that in the fixed $T$-framework, $\tilde{q}%
_{ii}$ can be of full rank but with minimum eigenvalue that may take values
with positive probability in the neighborhood of 0. A sufficient condition
for the second part of Assumption \ref{Assmp3.2}(i) to hold is $E\left\Vert 
\tilde{q}_{ii}^{-1}\right\Vert ^{2}\leq \bar{c}.$ Assumption \ref{Assmp3.2}%
(ii)\ specifies a convergence condition on the conditional second moments of 
$\chi _{i}^{\prime }\mathbf{u}_{i}^{\dag },$ which appears in the Bahadur
representation of the TW-MG estimator $\hat{\beta}^{MG}.$ Due to the
presence of the unobserved factor component, $F,$ which enters both $\chi
_{i}$ and $\mathbf{u}_{i}^{\dag },$ the probability limit object $\Omega
_{1}^{MG}$ is random unless all factors inside $F$ are nonrandom. This
condition facilitates the application of a stable martingale CLT as
introduced in \cite{hausler2015stable}.

\begin{remark}
\label{Rmk3.1} \textbf{(Allowance of Correlated Random Coefficients)} We
allow correlation between $\eta _{i}$ and $x_{it}$ via various components in 
$x_{it}$ including $\mu _{i\cdot },$ $\gamma _{i}$ and $v_{it}$ if $x_{it}$
is generated according to (\ref{x_eqn2}). This largely broadens potential
applications of \textit{correlated random coefficient} (CRC) panel data
models. For a textbook treatment on random coefficient panels, see Chapter
13 in \cite{Hsiao2022}. For some recent applications and studies on CRC
panels, see \cite{arellano2012identifying}, \cite{graham2012identification},
and \cite{lu2023}, among others.
\end{remark}

\subsection{Asymptotic normality of $\hat{\protect\beta}^{MG}$ \label{Sec3.2}%
}

To state the first main result in this paper, let $\mathcal{F}%
_{0}^{MG}=\sigma \left( \{\mathbf{x}_{i}^{o}\}_{1\leq i\leq \infty
},F\right) .$ The following theorem reports the stable convergence of $\sqrt{%
N}(\hat{\beta}^{MG}-\beta ^{0}).$

\begin{theorem}
\label{Thm3.1} Suppose that Assumptions \ref{Assmp3.1}--\ref{Assmp3.2} hold.
Then 
\begin{equation}
\sqrt{N}(\hat{\beta}^{MG}-\beta ^{0})=\frac{1}{\sqrt{N}}\sum_{i=1}^{N}\left(
\chi _{i}^{\prime }\mathbf{u}_{i}^{\dagger }+\eta _{i}\right) \rightarrow
(\Omega ^{MG})^{1/2}\mathbb{Z}\mathcal{\ \ F}_{0}^{MG}\text{-stably as }%
N\rightarrow \infty ,  \label{SCLT1}
\end{equation}%
where $\Omega ^{MG}=\Omega _{1}^{MG}+\Omega _{\eta },\ \mathbb{Z}\backsim
N\left( 0,\mathbb{I}_{K_{x}}\right) ,$ and $\mathbb{Z}$ is independent of $%
\mathcal{F}_{0}^{MG}.$
\end{theorem}

Theorem \ref{Thm3.1} indicates that the TW-MG estimator $\hat{\beta}^{MG}$
is $\sqrt{N}$-consistent and follows a stable CLT under some regularity
conditions. The sequence $\{\sqrt{N}(\hat{\beta}^{MG}-\beta ^{0})\}_{N\geq
1} $ converges $\mathcal{F}$-stably as $N\rightarrow \infty $ to a mixture
normal distribution whose variance is random but measurable with respect to
the limit sigma-field $\mathcal{F}_{0}^{MG}$. We refer the readers directly
to \cite{hausler2015stable} and \cite{van2023weak} for the notion of stable
convergence and stable CLT. For recent applications of the concept of stable
convergence in econometrics and statistics, see \cite{lee2019stable}, \cite%
{cavaliere2020inference}, \cite{Jin_Miao_Su2021}, \cite{mikusheva2021many}
and \cite{hahn2024econometric}. It is worth mentioning that the usual weak
convergence concept, a standard tool for asymptotic analysis in
econometrics, is not adequate for our asymptotic analysis. Stable
convergence is a middle ground between convergence in probability and
convergence in distribution. It is weaker than convergence in probability
but stronger than weak convergence. Importantly, it allows for a stochastic
normalization in the central limit theorem. Intuitively, stable convergence
can be thought of as a weak convergence conditional on certain sigma-algebra
($\mathcal{F}_{0}^{MG}$ here) so that it can be recast as a convergence of
certain conditional distributions.

\subsection{Asymptotic inference on $\protect\beta ^{0}$ \label{Sec3.3}}

In this subsection, we address the problem of inference on $\mathbb{S}%
^{\prime }\beta ^{0}$ based on the jackknife method for the TW-MG
estimation. We add the following condition.

\begin{assumption}
\label{Assmp3.3}(i)$\frac{1}{N}\sum_{i=1}^{N}\left\Vert \tilde{q}%
_{ii}^{-1}\right\Vert ^{2}\left\Vert \mathbf{x}_{i}^{o}\right\Vert
^{2s}=O_{p}(1)$ for $s=0,1$.

(ii) $\delta >1.$
\end{assumption}

Assumption \ref{Assmp3.3}(i) strengthens Assumption \ref{Assmp3.2}(i). As
sufficient condition for it to hold is $\frac{1}{N}\sum_{i=1}^{N}E[\left%
\Vert \tilde{q}_{ii}^{-1}\right\Vert ^{2}\left\Vert \mathbf{x}%
_{i}^{o}\right\Vert ^{2s}]=O(1)$ for $s=0,1.$ Assumption \ref{Assmp3.3}(ii)
strengthens the moment condition in \ref{Assmp3.1} by requiring $\delta >1.$
Both parts of Assumption \ref{Assmp3.3} are reasonable because consistent
estimation of the variance-covariance matrix generally requires stronger
conditions than those needed for the asymptotic distribution result.

The following theorem establishes the consistency of $\hat{\Omega}^{MG}\ $%
and the distributional limit for a normalized version of $\sqrt{N}\mathbb{S}%
^{\prime }(\hat{\beta}^{MG}-\beta ^{0}).$

\begin{theorem}
\label{Thm3.2}Suppose Assumptions \ref{Assmp3.1}--\ref{Assmp3.3} hold. Then

(i) $\hat{\Omega}^{MG}=\frac{1}{N}\sum_{j=1}^{N}(\chi _{j}^{\prime }\mathbf{u%
}_{j}^{\dagger }+\eta _{j})(\chi _{j}^{\prime }\mathbf{u}_{j}^{\dagger
}+\eta _{j})^{\prime }+o_{p}\left( 1\right) =\Omega ^{MG}+o_{p}\left(
1\right) ;$

(ii) $(\mathbb{S}^{\prime }\hat{\Omega}^{MG}\mathbb{S})^{-1/2}\sqrt{N}%
\mathbb{S}^{\prime }(\hat{\beta}^{MG}-\beta ^{0})\overset{d}{\rightarrow }%
N\left( 0,1\right) \ $as $N\rightarrow \infty .$
\end{theorem}

Theorem \ref{Thm3.2}(i) indicates that we can estimate $\Omega ^{MG}$
consistently via the jackknife method and Theorem \ref{Thm3.2}(ii) justifies
the asymptotic validity of the jackknife confidence interval defined in (\ref%
{CI1}).

\begin{remark}
\label{Rmk3.2} \textbf{(High dimensional jackknife)} The jackknife method is
a popular method to estimate the asymptotic variance of an estimator whose
dimension does not grow with the sample size; see Chapter 10.3 in \cite%
{hansen2022econometrics} for a brief account. However, our TW-MG estimator $%
\hat{\beta}^{MG}$ is based on $\mathbf{\hat{\beta}},$ whose row dimension
diverges to infinity linearly in the sample size $N.$ To study the
asymptotic expansion of $\hat{\beta}^{MG\left( -i\right) },$ we have to
study that of $\mathbf{\hat{\beta}}^{\left( -i\right) },$ which is an $%
N_{1}K_{x}\times 1$ vector. The change of the row dimension of $\mathbf{\hat{%
\beta}}$ to that of $\mathbf{\hat{\beta}}^{\left( -i\right) }$ substantially
complicates the asymptotic analysis. Let $\mathbf{\hat{Q}}^{\left( -i\right)
}\equiv \frac{1}{T}\mathbf{X}^{\left( -i\right) \prime }(\mathbb{M}_{\mathbf{%
\iota }_{N_{1}}}\otimes \mathbb{M}_{\mathbf{\iota }_{T}})\mathbf{X}^{\left(
-i\right) }=\frac{1}{T}\mathbf{X}^{o\left( -i\right) \prime }(\mathbb{M}_{%
\mathbf{\iota }_{N_{1}}}\otimes \mathbb{M}_{\mathbf{\iota }_{T}})\mathbf{X}%
^{o\left( -i\right) },$ where $\mathbf{X}^{o\left( -i\right) \prime }$ is
the leave-one-out version of $\mathbf{X}^{o}.$ Noting that 
\begin{equation*}
\mathbf{\hat{\beta}}=\mathbf{\hat{Q}}^{-1}\frac{1}{T}\mathbf{X}^{\prime }(%
\mathbb{M}_{\mathbf{\iota }_{N}}\otimes \mathbb{M}_{\mathbf{\iota }_{T}})%
\mathbf{Y}\text{ and }\mathbf{\hat{\beta}}^{\left( -i\right) }=\left[ 
\mathbf{\hat{Q}}^{\left( -i\right) }\right] ^{-1}\frac{1}{T}\mathbf{X}%
^{\left( -i\right) \prime }(\mathbb{M}_{\mathbf{\iota }_{N_{1}}}\otimes 
\mathbb{M}_{\mathbf{\iota }_{T}})\mathbf{Y}^{\left( -i\right) },\text{ }
\end{equation*}%
we make the following decompositions: 
\begin{eqnarray*}
\mathbf{\hat{Q}} &\equiv &\frac{1}{T}\mathbf{X}^{o\prime }\left( {\mathbb{I}}%
_{N}\otimes {\mathbb{M}}_{\mathbf{\iota }_{T}}\right) \mathbf{X}^{o}-\frac{1%
}{T}\mathbf{X}^{o\prime }\left( {\mathbb{P}}_{\mathbf{\iota }_{N}}\otimes {%
\mathbb{M}}_{\mathbf{\iota }_{T}}\right) \mathbf{X}^{o}\equiv \mathbf{Q}%
^{\left( d\right) }-\mathbf{CC}^{\prime }\text{ and} \\
\mathbf{\hat{Q}}^{\left( -i\right) } &=&\frac{1}{T}\mathbf{X}^{o\left(
-i\right) \prime }\left( {\mathbb{I}}_{N_{1}}\otimes {\mathbb{M}}_{\mathbf{%
\iota }_{T}}\right) \mathbf{X}^{o\left( -i\right) }-\frac{1}{T}\mathbf{X}%
^{o\left( -i\right) \prime }({\mathbb{P}}_{\mathbf{\iota }_{N_{1}}}\otimes {%
\mathbb{M}}_{\mathbf{\iota }_{T}})\mathbf{X}^{o\left( -i\right) }\equiv 
\mathbf{Q}^{\left( -i,d\right) }-\mathbf{C}^{\left( -i\right) }\mathbf{C}%
^{\left( -i\right) \prime },
\end{eqnarray*}%
where $\mathbf{C}=(NT)^{-1/2}\mathbf{X}^{o\prime }[$\textbf{$\iota $}$%
_{N}\otimes {\mathbb{M}}_{\mathbf{\iota }_{T}}]$ and $\mathbf{C}^{\left(
-i\right) }=\left( N_{1}T\right) ^{-1/2}\mathbf{X}^{o\left( -i\right) \prime
}[$\textbf{$\iota $}$_{N_{1}}\otimes {\mathbb{M}}_{\mathbf{\iota }_{T}}].$
By the Sherman-Morrison Woodbury matrix identity (e.g., Fact 6.4.31 in \cite%
{bernstein2005}), we have%
\begin{equation*}
\left[ \mathbf{\hat{Q}}^{\left( -i\right) }\right] ^{-1}=\left[ \mathbf{Q}%
^{\left( -i,d\right) }\right] ^{-1}\left\{ \mathbf{Q}^{\left( -i,d\right) }+%
\mathbf{C}^{\left( -i\right) }[\mathbb{I}_{K_{x}T}+\mathbf{C}^{\left(
-i\right) \prime }\mathbf{C}^{\left( -i\right) }]^{-1}\mathbf{C}^{\left(
-i\right) \prime }\right\} \left[ \mathbf{Q}^{\left( -i,d\right) }\right]
^{-1}.
\end{equation*}%
A similar expression holds for $\mathbf{\hat{Q}}^{-1}.$ Due to the mismatch
of the dimensions of various matrix pairs such as $\mathbf{Q}^{\left(
-i,d\right) }$ and $\mathbf{Q}^{\left( d\right) },$ $\mathbf{C}^{\left(
-i\right) }$ and $\mathbf{C},$ $\mathbf{X}^{\left( -i\right) }$ and $\mathbf{%
X},$ and $\mathbf{Y}^{\left( -i\right) }$ and $\mathbf{Y},$ one key step in
our analysis is to augment the leave-one-out version of various
matrices/vectors to match the dimension of the original version. For
example, we augment the $N_{1}K_{x}\times N_{1}K_{x}$ block diagonal matrix $%
\mathbf{Q}^{\left( -i,d\right) }$ to $\mathbf{\breve{Q}}^{\left( -i,d\right)
}$ (an $NK_{x}\times NK_{x}$ block diagonal matrix) such that 
\begin{equation*}
\left[ \mathbf{\breve{Q}}^{\left( -i,d\right) }\right] _{jj}=\left\{ 
\begin{array}{ll}
\left[ \mathbf{Q}^{\left( -i,d\right) }\right] _{jj} & \text{if }j<i \\ 
\mathbf{0}_{K_{x}\times K_{x}} & \text{if }j=i \\ 
\left[ \mathbf{Q}^{\left( -i,d\right) }\right] _{j-1,j-1}\text{ \ \ } & 
\text{if }j>i%
\end{array}%
\right. ,
\end{equation*}%
where $\left[ \mathbf{Q}^{\left( -i,d\right) }\right] _{jj}$ denotes the $%
\left( j,j\right) $th block of $\mathbf{Q}^{\left( -i,d\right) }$ of size $%
K_{x}\times K_{x}.$ We do similar things for matrices $\mathbf{C}^{\left(
-i\right) },$ $\mathbf{X}^{\left( -i\right) },$ $\mathbf{Y}^{\left(
-i\right) },$ etc., to obtain their augmented versions $\mathbf{\breve{C}}%
^{\left( -i\right) },$ $\mathbf{\breve{X}}^{\left( -i\right) },$ $\mathbf{%
\breve{Y}}^{\left( -i\right) },$ etc. By using these augmented matrices
associated with some tedious matrix algebra, we can ultimately establish the
link between $\hat{\beta}^{MG\left( -i\right) }-\frac{1}{N_{1}}\sum_{j\neq
i}^{N}\beta _{j}$ and $\hat{\beta}^{MG}-\frac{1}{N}\sum_{j=1}^{N}\beta _{j}:$
\begin{equation}
\hat{\beta}^{MG\left( -i\right) }-\frac{1}{N_{1}}\sum_{j\neq i}^{N}\beta
_{j}=\frac{N}{N_{1}}\left( \hat{\beta}^{MG}-\frac{1}{N}\sum_{j=1}^{N}\beta
_{j}\right) +\Delta _{i2}^{MG}+\Delta _{i3}^{MG}+o_{p}(N^{-1}),
\label{jackknife0}
\end{equation}%
where $\Delta _{i2}^{MG}\equiv -\frac{1}{N_{1}T}\hat{\omega}_{i}\mathbf{x}%
_{i}^{o\prime }{\mathbb{M}}_{\mathbf{\iota }_{T}}\mathbf{u}_{i}^{\dag },$ $%
\Delta _{i3}^{MG}\equiv \frac{1}{NN_{1}}\sum_{j=1}^{N}\hat{\omega}_{j}\frac{1%
}{T}\mathbf{x}_{j}^{o\prime }{\mathbb{M}}_{\mathbf{\iota }_{T}}\mathbf{u}%
_{i}^{\dag },\ $and $o_{p}(N^{-1})$ holds uniformly in $i\in \left[ N\right]
.$ $\Delta _{i2}^{MG}+\Delta _{i3}^{MG}$ plays the role of a summand in the
influence function of $\hat{\beta}^{MG}-\frac{1}{N}\sum_{j=1}^{N}\beta _{j}$%
. The result in (\ref{jackknife0}) is a fundamental result that helps
deliver the main result in Theorem \ref{Thm3.2}(i).
\end{remark}

\begin{remark}
\label{Rmk3.3}\textbf{(Alternative approaches)} In the MG estimation
literature, one commonly uses the following formula to estimate the
asymptotic variance of $\hat{\beta}^{MG}$ 
\begin{equation*}
\frac{1}{N}\sum_{i=1}^{N}\left( \hat{\beta}_{i}-\hat{\beta}^{MG}\right)
\left( \hat{\beta}_{i}-\hat{\beta}^{MG}\right) ^{\prime }.
\end{equation*}%
However, it turns out that this variance estimator is hard to justify in our
framework. The main reason is that $\hat{\beta}_{i}$ is not a consistent
estimator of $\beta _{i}$ in the fixed $T$ framework. Alternatively, we may
consider cross-sectional bootstrap by resampling the data along the
cross-sectional dimension. We find that the cross-sectional bootstrap and
our leave-one-out jackknife method perform similarly in simulations.
However, the asymptotic validity of the cross-sectional bootstrap is much
harder to justify theoretically; therefore, we propose the jackknife method
here.
\end{remark}

\section{Pooled Estimation and Inference\label{Sec4}}

In this section, we consider the pooled estimation for the model in (\ref%
{y_eqn1}) and study the inference theory.

\subsection{Estimation and inference procedure \label{Sec4.1}}

We rewrite the model in (\ref{y_eqn1}) as follows:%
\begin{equation}
y_{it}=\beta ^{0\prime }x_{it}+\alpha _{i\cdot }+\alpha _{\cdot
t}+u_{it}^{\ddag },  \label{y_eqn3}
\end{equation}%
where $u_{it}^{\ddag }=u_{it}+\lambda _{i}^{\prime }f_{t}+\eta _{i}^{\prime
}x_{it}$. Let $X=\mathbf{(x}_{1}^{\prime },...,\mathbf{x}_{N}^{\prime
})^{\prime }$ and $\mathbf{\eta }=(\eta _{1}^{\prime },...,\eta _{N}^{\prime
})^{\prime }.$ Then we can rewrite the model in (\ref{y_eqn3}) in matrix
form:%
\begin{equation}
\mathbf{Y}=X\beta ^{0}+\mathbf{D\alpha }+\mathbf{U}^{\ddag },  \label{Y_eqn2}
\end{equation}%
where $\mathbf{U}^{\ddag }\equiv \mathbf{U}+(\mathbb{I}_{N}\otimes F)\mathbf{%
\lambda +X\eta }.$ By partialling out $\mathbf{D}$ in (\ref{Y_eqn2}), we
obtain the following pooled estimator of $\beta ^{0}$: 
\begin{equation}
\hat{\beta}^{P}\equiv \left( X^{\prime }{\mathbb{M}}_{\mathbf{D}}X\right)
^{-1}X^{\prime }{\mathbb{M}}_{\mathbf{D}}\mathbf{Y.}  \label{B_tilde1}
\end{equation}

Below we show that $\hat{\beta}^{P}$ is asymptotically mixture normal under
some regularity conditions. We propose a jackknife method to estimate its
asymptotic variance. As in (\ref{B_tilde1}), we can obtain the leave-one-out
pooled estimator of $\beta ^{0}$ by leaving out the observations for the $i$%
th cross-section unit:%
\begin{equation}
\hat{\beta}^{P\left( -i\right) }=\left[ X^{\left( -i\right) \prime }(\mathbb{%
M}_{\mathbf{\iota }_{N_{1}}}\otimes \mathbb{M}_{\mathbf{\iota }%
_{T}})X^{\left( -i\right) }\right] ^{-1}X^{\left( -i\right) \prime }(\mathbb{%
M}_{\mathbf{\iota }_{N_{1}}}\otimes \mathbb{M}_{\mathbf{\iota }_{T}})\mathbf{%
Y}^{\left( -i\right) },  \label{theta_leave_one_out3}
\end{equation}%
where $X^{\left( -i\right) }$ is the leave-one-out version\ of $X:X^{\left(
-i\right) }=(\mathbf{x}_{1}^{\prime },...,\mathbf{x}_{i-1}^{\prime },\mathbf{%
x}_{i+1}^{\prime },...,\mathbf{x}_{N}^{\prime })^{\prime }$. Given the
jackknife estimates $\{\hat{\beta}^{P\left( -i\right) }\}_{i\in \left[ N%
\right] },$ we propose to estimate the asymptotic variance of $\sqrt{N}(\hat{%
\beta}^{P}-\beta ^{0})$ by 
\begin{equation*}
\hat{\Omega}^{P}=(N-1)\dsum\limits_{i=1}^{N}\left( \hat{\beta}^{P\left(
-i\right) }-\overline{\hat{\beta}}^{_{P}}\right) \left( \hat{\beta}^{P\left(
-i\right) }-\overline{\hat{\beta}}^{_{P}}\right) ^{\prime },
\end{equation*}%
where $\overline{\hat{\beta}}^{_{P}}=\frac{1}{N}\sum_{i=1}^{N}\hat{\beta}%
^{P\left( -i\right) }.$ Then one can construct the confidence intervals for
element of $\mathbb{S}\beta ^{0}$ as in (\ref{CI1}) with $\hat{\Omega}^{MG}$
replaced by $\hat{\Omega}^{P}$.

\subsection{Asymptotic properties \label{Sec4.2}}

To study the asymptotic properties of $\hat{\beta}^{P},$ we add some
notations and assumptions. Let $\mathbf{\dot{x}}_{i}\equiv \mathbb{M}_{%
\mathbf{\iota }_{T}}\mathbf{x}_{i}=(\dot{x}_{i1},...,\dot{x}_{iT})^{\prime }$
and $\mathbf{\dot{u}}_{i}=\mathbb{M}_{\mathbf{\iota }_{T}}\mathbf{u}_{i}=(%
\dot{u}_{i1},...,\dot{u}_{iT})^{\prime }.$ Let $\mathbf{\dot{X}}\equiv {%
\mathbb{({\mathbb{I}}_{N}\otimes M}}_{\mathbf{\iota }_{T}})\mathbf{X}=$bdiag$%
(\mathbf{\dot{x}}_{1},...,\mathbf{\dot{x}}_{N})$ and $\mathbf{\dot{U}}%
^{\ddagger }=\mathbf{\dot{U}}+(\mathbb{I}_{N}\otimes \dot{F})\mathbf{\lambda
+\dot{X}\eta }=(\mathbf{\dot{u}}_{1}^{\ddagger \prime },...,\mathbf{\dot{u}}%
_{N}^{\ddagger \prime })^{\prime }$ where $\mathbf{\dot{u}}_{i}^{\ddagger }=(%
\dot{u}_{i1}^{\ddagger },...,\dot{u}_{iT}^{\ddagger })^{\prime }$ and $\dot{u%
}_{it}^{\ddagger }=\dot{u}_{it}+\dot{f}_{t}^{\prime }\lambda _{i}+\dot{x}%
_{it}^{\prime }\eta _{i}.$ Let 
\begin{equation*}
\hat{Q}\equiv \frac{1}{NT}X^{\prime }{\mathbb{M}}_{\mathbf{D}}X=\frac{1}{NT}%
\ddot{X}^{\prime }\ddot{X},
\end{equation*}%
where $\ddot{X}={\mathbb{M}}_{\mathbf{D}}X.$ Let $\mathcal{F}_{0}^{P}=\sigma
\left( \{\mathbf{\dot{x}}_{i}\}_{1\leq i\leq \infty },F\right) ,$ $\mathcal{F%
}_{N,0}^{P}=\sigma (\{\mathbf{\dot{x}}_{i}\}_{1\leq i\leq N},F)$ and $%
\mathcal{F}_{N,j}^{P}=\sigma (\mathcal{F}_{N,0}^{P},\left\{ (\mathbf{u}%
_{i},\lambda _{i},\eta _{i})\right\} _{i=1}^{j})$ for $j\in \left[ N\right]
. $ Note that $\mathcal{F}_{N,0}^{P}$ (resp. $\mathcal{F}_{0}^{P}$) slightly
differs from $\mathcal{F}_{N,0}^{MG}$ (resp. $\mathcal{F}_{0}^{MG}$) because 
$x_{it}^{\prime }\eta _{i}$ now enters the error term $u_{it}^{\ddag }$ in
the pooled regression and the time effects in $x_{it}$ (if any) cannot be
removed in the estimation procedure.

\begin{assumption}
\label{Assmp4.1} (i) Conditional on $\mathcal{F}_{N,0}^{P},$ $\left\{ (%
\mathbf{u}_{i},\lambda _{i},\eta _{i})\right\} _{i\in \left[ N\right] }$ are
independent across $i$ with zero mean and $E_{\mathcal{F}_{N,0}^{P}}[\left%
\Vert \mathbf{u}_{i}\right\Vert ^{4+2\delta }+\left\Vert \lambda
_{i}\right\Vert ^{4+2\delta }+\left\Vert \eta _{i}\right\Vert ^{4+2\delta
}]\lesssim 1$ for some $\delta >0;$

(ii) $E_{F}[\left\Vert \mathbf{\dot{x}}_{i}\right\Vert ^{4+2\delta
}]\lesssim 1.$
\end{assumption}

\begin{assumption}
\label{Assmp4.2}(i) $\hat{Q}\overset{p}{\rightarrow }Q$ and $Q$ is positive
definite a.s.;

(ii) $\frac{1}{NT^{2}}\sum_{i=1}^{N}\mathbf{\ddot{x}}_{i}^{\prime }E_{%
\mathcal{F}_{N,0}^{P}}\left[ \mathbf{\dot{u}}_{i}^{\ddag }\mathbf{\dot{u}}%
_{i}^{\ddag \prime }\right] \mathbf{\ddot{x}}_{i}=\Omega ^{P}+o_{p}\left(
1\right) \ $and $\Omega ^{P}$ is positive definite a.s..
\end{assumption}

Assumption \ref{Assmp4.1}(i) parallels Assumption \ref{Assmp3.1}(i)--(iii)
and it imposes some dependence and moment conditions on the unobserved
components, i.e., $\mathbf{u}_{i},\lambda _{i},$ and $\eta _{i}.$ Since $%
x_{it}^{\prime }\eta _{i}$ is a treated as part of the composite error term $%
u_{it}^{\ddagger },$ we now need to assume that $\eta _{i}$ has finite $%
\left( 4+2\delta \right) $th moment conditional on $\mathcal{F}_{N,0}^{P}.$
The condition that $E\left( \eta _{i}|\mathcal{F}_{N,0}^{P}\right) =\mathbf{0%
}_{K_{x}}$ rules out potential correlations between $\eta _{i}$ and $\mathbf{%
\dot{x}}_{i}$ via components in $x_{it}$ other than the individual effects.
That is, if there is any correlation between $\eta _{i}$ and $x_{it}$ as in
CRC\ panels, the correlation can only be caused by $\eta _{i}$ and a
time-invariant additive component in $x_{it}.$ When $x_{it}$ is generated
according to (\ref{x_eqn2}), $\eta _{i}$ can be correlated with $\mu
_{i\cdot }$ but not $\mu _{\cdot t},$ $\gamma _{i}$ and $v_{it}.$ Assumption %
\ref{Assmp4.1}(ii) parallels Assumption \ref{Assmp3.1}(iv). It is easy to
see that Assumption \ref{Assmp4.1}(ii) implies that $E_{F}[\left\Vert 
\mathbf{\dot{x}}_{i}^{o}\right\Vert ^{4+2\delta }]\lesssim 1$ and $%
E_{F}\left\Vert \mathbf{\ddot{x}}_{i}\right\Vert ^{4+2\delta }\lesssim 1.$

Assumption \ref{Assmp4.2}(i)-(ii) parallels Assumption \ref{Assmp3.2}%
(i)--(ii). In comparison with Assumption \ref{Assmp3.2}(i), Assumption \ref%
{Assmp4.2}(i) only requires the probability limit of $Q$ be positive
definite a.s., which may hold even when $T\leq K_{x}\ $so that the TW-MG
estimator $\hat{\beta}^{MG}$ is not well defined. Assumption \ref{Assmp4.2}%
(ii)\ specifies a convergence condition on the conditional second moments of 
$\mathbf{\ddot{x}}_{i}^{\prime }\mathbf{\dot{u}}_{i}^{\ddag },$ which
appears in the Bahadur representation of the pooled estimator $\hat{\beta}%
^{P}.$ Due to the presence of the unobserved factor component, $F,$ which
surely enters $\mathbf{\dot{u}}_{i}^{\ddag }$ and may also enter $\mathbf{%
\ddot{x}}_{i}$ if $x_{it}$ also contains a factor structure, the probability
limit object $\Omega ^{P}$ is random unless all factors inside $F$ are
nonrandom. This condition facilitates the application of a stable martingale
CLT as introduced in \cite{hausler2015stable}.

The following theorem reports the stable convergence of $\sqrt{N}(\hat{\beta}%
^{P}-\beta ^{0}).$

\begin{theorem}
\label{Thm4.1} Suppose that Assumptions \ref{Assmp4.1}--\ref{Assmp4.2} hold.
Then 
\begin{equation*}
\sqrt{N}(\hat{\beta}^{P}-\beta ^{0})=\hat{Q}^{-1}\frac{1}{\sqrt{N}}%
\sum_{i=1}^{N}\frac{1}{T}\mathbf{\ddot{x}}_{i}^{\prime }\mathbf{\dot{u}}%
_{i}^{\ddag }\rightarrow (Q^{-1}\Omega ^{P}Q^{-1})^{1/2}\mathbb{Z}\mathcal{\
\ F}_{0}^{P}\text{-stably as }N\rightarrow \infty ,
\end{equation*}%
where $\mathbb{Z}\backsim N\left( 0,\mathbb{I}_{K_{x}}\right) $ and is
independent of $\mathcal{F}_{0}^{P}.$
\end{theorem}

Theorem \ref{Thm4.1} shows that the sequence $\{\sqrt{N}(\hat{\beta}%
^{MG}-\beta ^{0})\}_{N\geq 1}$ converges $\mathcal{F}_{0}^{P}$-stably as $%
N\rightarrow \infty $ to a mixture normal distribution whose variance is
random but measurable with respect to the limit sigma-field $\mathcal{F}%
_{0}^{P}$.

The following theorem establishes the consistency of $\hat{\Omega}^{P}.$

\begin{theorem}
\label{Thm4.2}Suppose Assumptions \ref{Assmp4.1}--\ref{Assmp4.2} hold. Then

(i) $\hat{\Omega}^{P}=\frac{1}{N}\sum_{j=1}^{N}\mathbf{\ddot{x}}_{j}^{\prime
}\mathbf{\dot{u}}_{j}^{\ddag }\mathbf{\dot{u}}_{j}^{\ddag \prime }\mathbf{%
\ddot{x}}_{j}+o_{p}\left( 1\right) =\Omega ^{P}+o_{p}\left( 1\right) ;$

(ii) $(\mathbb{S}^{\prime }\hat{\Omega}^{P}\mathbb{S})^{-1/2}\sqrt{N}\mathbb{%
S}^{\prime }(\hat{\beta}^{P}-\beta ^{0})\overset{d}{\rightarrow }N\left(
0,1\right) .$
\end{theorem}

Theorem \ref{Thm4.2} indicates that we can estimate $\Omega ^{P}$
consistently via the jackknife method. It justifies the use of the normal
critical values for asymptotic inference based on the jackknife estimate of
the asymptotic variance.

\subsection{Specification tests for the poolability\label{Sec4.3}}

In this subsection, we propose a test for the poolability based on the
results in the early subsections. In general, $\hat{\beta}^{MG}$ is
consistent with $\beta ^{0}$ no matter whether one has a homogenous panel
with a slope coefficient $\beta ^{0}$ or a heterogenous panel with slope
coefficients whose mean is given by $\beta ^{0}$, and it also allows for
arbitrary correlations between $\beta _{i}$ and $x_{it}$ in a CRC panel. In
contrast, $\hat{\beta}^{P}$ is generally inconsistent when $\beta _{i}$ is
correlated with a time-varying component of $x_{it}$ so we allow $\hat{\beta}%
^{P}$ to converge to something different from $\beta ^{0}\ $in this case.
For this reason, we consider the following null hypothesis: 
\begin{equation}
\mathbb{H}_{0}:\text{plim}_{N\rightarrow \infty }\hat{\beta}^{MG}=\text{plim}%
_{N\rightarrow \infty }\hat{\beta}^{P}=\beta ^{0}.  \label{H0}
\end{equation}%
The alternative hypothesis is the negation of $\mathbb{H}_{0}:$ 
\begin{equation*}
\mathbb{H}_{1}:\text{plim}_{N\rightarrow \infty }\hat{\beta}^{MG}\neq \text{%
plim}_{N\rightarrow \infty }\hat{\beta}^{P}.
\end{equation*}%
Below we shall maintain the condition that plim$_{N\rightarrow \infty }\hat{%
\beta}^{MG}=\beta ^{0}\ $and assume that the probability limit of $\hat{\beta%
}^{P}$ is given by $\beta ^{P}$ under $\mathbb{H}_{1}.$ A failure to reject $%
\mathbb{H}_{0}$ justifies the common practice of pooling all cross-sectional
units to estimate the slope coefficients in a panel despite the potential
presence of slope heterogeneity. If the panel is indeed homogeneous, the
common slope $\beta ^{0}$ indicates the marginal effect; otherwise, it
indicates the population average marginal effect.

Under $\mathbb{H}_{0},$ we have%
\begin{eqnarray*}
\sqrt{N}(\hat{\beta}^{MG}-\hat{\beta}^{P}) &=&\sqrt{N}(\hat{\beta}%
^{MG}-\beta ^{0})-\sqrt{N}(\hat{\beta}^{P}-\beta ^{0}) \\
&=&\frac{1}{\sqrt{N}}\sum_{i=1}^{N}\left( \chi _{i}^{\prime }\mathbf{u}%
_{i}^{\dagger }+\eta _{i}\right) -\hat{Q}^{-1}\frac{1}{\sqrt{N}}%
\sum_{i=1}^{N}\mathbf{\ddot{x}}_{i}^{\prime }\mathbf{\dot{u}}_{i}^{\ddag },
\end{eqnarray*}%
which has a well behaved limit distribution if we assume that Assumptions %
\ref{Assmp3.1}--\ref{Assmp3.2} and \ref{Assmp4.1}--\ref{Assmp4.2} are all
satisfied. In this case, we can estimate the asymptotic variance of $\sqrt{N}%
(\hat{\beta}^{MG}-\hat{\beta}^{P})$ via the jackknife method:%
\begin{equation*}
\hat{\Omega}^{\Delta }=(N-1)\dsum\limits_{i=1}^{N}\left( \hat{\Delta}%
^{\left( -i\right) }-\overline{\hat{\Delta}}\right) \left( \hat{\Delta}%
^{\left( -i\right) }-\overline{\hat{\Delta}}\right) ^{\prime },
\end{equation*}%
where $\hat{\Delta}^{\left( -i\right) }=\hat{\beta}^{MG\left( -i\right) }-%
\hat{\beta}^{P\left( -i\right) }$ and $\overline{\hat{\Delta}}=\frac{1}{N}%
\sum_{i=1}^{N}\hat{\Delta}^{\left( -i\right) }.$ Following the lead of \cite%
{hausman1978specification}, we propose the following test statistic: 
\begin{equation}
J_{N}=N(\hat{\beta}^{MG}-\hat{\beta}^{P})^{\prime }\left[ \hat{\Omega}%
^{\Delta }\right] ^{-1}(\hat{\beta}^{MG}-\hat{\beta}^{P}).  \label{test}
\end{equation}

Below we only assume that Assumptions \ref{Assmp3.1}--\ref{Assmp3.3} are
always satisfied under $\mathbb{H}_{0}$ and $\mathbb{H}_{1}\ $so that both
estimators are well defined. Under $\mathbb{H}_{0},$ we also assume that
Assumptions \ref{Assmp4.1}--\ref{Assmp4.2} are satisfied so that the pooled
estimator $\hat{\beta}^{P}$ is consistent with $\beta ^{0}$. Under $\mathbb{H%
}_{1},$ we dispense with Assumptions \ref{Assmp4.1}--\ref{Assmp4.2} so that $%
\hat{\beta}^{P}$ does not converge to $\beta ^{0}$ in general.

Let $\mathcal{F}_{N}=\mathcal{\sigma (}\{\mathbf{\dot{x}}_{i}^{o}\}_{i\leq
N},F).$ To study the asymptotic properties of $J_{N}$ under $\mathbb{H}_{0}$
and $\mathbb{H}_{1},$ we add the following two assumptions.

\begin{assumption}
\label{Assmp4.3}$\frac{1}{N}\sum_{i=1}^{N}E_{\mathcal{F}_{N}}\left\{ \left[
\chi _{i}^{\prime }\mathbf{u}_{i}^{\dagger }+\eta _{i}-Q^{-1}\mathbf{\ddot{x}%
}_{i}^{\prime }\mathbf{\dot{u}}_{i}^{\ddag }\right] \left[ \chi _{i}^{\prime
}\mathbf{u}_{i}^{\dagger }+\eta _{i}-Q^{-1}\mathbf{\ddot{x}}_{i}^{\prime }%
\mathbf{\dot{u}}_{i}^{\ddag }\right] ^{\prime }\right\} \overset{p}{%
\rightarrow }\Omega ^{\Delta },$ where $\Omega ^{\Delta }$ is positive
definite a.s.
\end{assumption}

\begin{assumption}
\label{Assmp4.4}(i) plim$_{N\rightarrow \infty }\hat{\beta}^{P}=\beta ^{P}$
for some $\beta ^{P}\neq \beta ^{0};$

(ii) $\hat{\Omega}^{\Delta }\overset{p}{\rightarrow }\mathring{\Omega}%
^{\Delta },$ where $\mathring{\Omega}^{\Delta }$ is positive definite a.s.
\end{assumption}

Assumption \ref{Assmp4.3} specifies a condition to ensure that the
asymptotic conditional variance of $\sqrt{N}(\hat{\beta}^{MG}$ $-\hat{\beta}%
^{P})$ given $\mathcal{F}_{N}$ is well behaved under $\mathbb{H}_{0}.$
Assumption \ref{Assmp4.4}(i) restricts that $\hat{\beta}^{P}$ be a
consistent estimator of $\beta ^{P},$ which is different from $\beta ^{0}.$
Assumption \ref{Assmp4.4}(ii) imposes that $\hat{\Omega}^{\Delta }$ be
asymptotically nonsingular. We will use Assumptions \ref{Assmp4.3} and \ref%
{Assmp4.4} for the study of the asymptotic properties of $J_{N}$ under $%
\mathbb{H}_{0}\ $and $\mathbb{H}_{1},$ respectively.

The following theorem reports the asymptotic properties of $J_{N}$ under $%
\mathbb{H}_{0}\ $and $\mathbb{H}_{1}.$

\begin{theorem}
\label{Thm4.3} Suppose that Assumptions \ref{Assmp3.1}--\ref{Assmp3.3} hold.

(i) If Assumptions \ref{Assmp4.1}--\ref{Assmp4.3} hold, then under $\mathbb{H%
}_{0},$ $J_{N}\overset{d}{\rightarrow }\chi ^{2}\left( K_{x}\right) \mathcal{%
\ }$as $N\rightarrow \infty .$

(ii) If Assumption \ref{Assmp4.4} holds, then under $\mathbb{H}_{1},$ $%
P\left( J_{N}>c_{N}\right) \rightarrow 1$ for any $c_{N}=o(N)$.
\end{theorem}

Theorem \ref{Thm4.3} indicates that $J_{N}$ has the usual asymptotic $\chi
^{2}$-distribution under $\mathbb{H}_{0}$ and it diverges to infinity at
rate $N$ under $\mathbb{H}_{1}.\ $As a result, the test is consistent and
one can use the critical values from the $\chi ^{2}\left( K_{x}\right) $
distribution to make a decision: one rejects the null when $J_{N}>\chi
_{\tau }^{2}\left( K_{x}\right) ,$ the upper $\tau $-quantile of the $\chi
^{2}\left( K_{x}\right) $ distribution.

\section{Monte Carlo Simulations\label{Sim}}

In this section, we conduct some Monte Carlo simulations to evaluate the
proposed TW-MG\ estimator and jackknife method.

\subsection{DGPs\label{Sim_DGP} and implementations}

We consider four DGPs where $y_{it}$ and $x_{it}$ follow equations (\ref%
{y_eqn1}) and (\ref{x_eqn2}), respectively. DGPs 1--3 are simple DGPs with
one regressor $(K_{x}=1)$, covering three different cases of $\beta _{i}$, \
to confirm our main theoretical results. In DGP 1, $\beta _{i}$ is
homogeneous; in DGP 2, $\beta _{i}$ is random and independent of $x_{it};$
and in DGP 3, $\beta _{i}$ enters the DGP\ of $x_{it}.$ With this design,
the pooled-type estimator is consistent for DGPs 1 and 2 and inconsistent
for DGP\ 3. In contrast, the MG-type estimator is consistent for all three
DGPs. The conventional MG estimator (\cite{pesaran_smith1995}), without
accounting for time fixed effects, is inconsistent for all three DGPs due to
the presence of time fixed effects. DGP 4 is more general and realistic with
two regressors ($K_{x}=2$), and auto-correlated and heteroskedastic error
terms as described below. In DGP 4, only the MG-type estimator is
consistent. In all four DGPs, we include both TWFEs and IFEs and also ensure
that the key Assumption \ref{Assmp3.1}(i) holds. In Section \ref{SecS2} of
the online supplement, we report the simulation results for two more DGPs
with TWFEs only.

Specifically, in DGPs 1-3, $K_{x}=R=1,$ $\beta ^{0}=1.$ Let $(\lambda
_{i}^{o},f_{t}^{o},\gamma _{i}^{o},u_{it},\xi _{it},\eta _{i}^{\ast
},v_{it}^{\ast })$ be generated as $i.i.d.$ $N\left( (\mathbf{\iota }%
_{3}^{\prime },\mathbf{0}_{4\times 1}^{\prime })^{\prime },\mathbb{I}%
_{7}\right) $ random variables where $\mathbf{\iota }_{k}$ and $\mathbf{0}%
_{k\times 1}$ denote a $k\times 1$ vector of ones and zeros, respectively.
We let $\alpha _{i\cdot }^{o}=\mu _{i\cdot }^{o}=\lambda _{i}^{o},$ and $%
\alpha _{\cdot t}^{o}=\mu _{\cdot t}^{o}=f_{t}^{o}.$ Now we specify $\eta
_{i}$ and $v_{it}$ for DGPs 1-3. In DGP 1, $\eta _{i}=0,$ and in DGPs\ 2 and
3, $\eta _{i}=\eta _{i}^{\ast }$. \ In DGPs 1 and 2, $v_{it}=v_{it}^{\ast },$
and in DGP$\ $3, $v_{it}=\beta _{i}\xi _{it}+v_{it}^{\ast }.$

In DGP 4, $K_{x}=2$ and $R=1.$ Let $(\lambda _{i}^{o},f_{t}^{o},\gamma
_{1,i}^{o},\gamma _{2,i}^{o},u_{it}^{\ast \ast },\xi _{it},\eta _{1,i}^{\ast
},\eta _{2,i}^{\ast },v_{1,it}^{\ast },v_{2,it}^{\ast })$ be generated as $%
i.i.d.$ $N\left( (\mathbf{\iota }_{4}^{\prime },\mathbf{0}_{6\times
1}^{\prime })^{\prime },\mathbb{I}_{10}\right) $ random variables. We let $%
\alpha _{i\cdot }^{o}=\mu _{1,i\cdot }^{o}=\mu _{2,i\cdot }^{o}=\lambda
_{i}^{o},$ and $\alpha _{\cdot t}^{o}=\mu _{1,\cdot t}^{o}=\mu _{2,\cdot
t}^{o}=f_{t}^{o}.$ Let $x_{it}=\left( x_{1,it},x_{2,it}\right) $ which are
generated as $x_{1,it}=\mu _{1,i\cdot }^{o}+\mu _{1,\cdot t}^{o}+\gamma
_{1,i}^{o}f_{t}^{o}+v_{1,it}$ and $x_{2,it}=\mu _{2,i\cdot }^{o}+\mu
_{2,\cdot t}^{o}+\gamma _{2,i}^{o}f_{t}^{o}+v_{2,it}$ and let $\beta
_{i}=(\beta _{1,i},\beta _{2,i})^{\prime }=\beta ^{0}+(\eta _{1,i},\eta
_{2,i})^{\prime },$ where $\beta ^{0}=(1,1)^{\prime }.$ For $\eta _{1,i},$ $%
\eta _{2,i},$ $v_{1,it}$ and $v_{2,it},$ we let%
\begin{equation*}
v_{1,it}=\beta _{1,i}\xi _{it}+v_{1,it}^{\ast },\text{ }v_{2,it}=v_{2,it}^{%
\ast },\text{ }\eta _{1,i}=\eta _{1,i}^{\ast }\text{ and }\eta _{2,i}=\gamma
_{2,i}^{o}-1+\eta _{2,i}^{\ast }.
\end{equation*}%
$u_{it}$ is generated as 
\begin{equation*}
u_{it}=\sqrt{1+0.25x_{1,it}^{2}}u_{it}^{\ast }\text{ and }u_{it}^{\ast
}=0.25u_{i,t-1}^{\ast }+u_{it}^{\ast \ast }.
\end{equation*}%
Note that $\beta _{1,i}$ enters the DGP of $x_{1,it}$ via $v_{1,it},$ and $%
\eta _{2,i}$ (and $\beta _{2,i}$) is correlated with the factor loading $%
\gamma _{2,i}^{o}$ in the $x$-equation. Therefore, the pooled-type\
estimators of $\beta ^{0}$ are inconsistent.

For each DGP, we compare the six estimators: $(i)$ the TW-MG estimator in (%
\ref{betahat1}), $(ii)$ the TW-MG ridge estimator in (\ref{ridge}), $(iii)$
the TW-pooled estimator in (\ref{B_tilde1}), $(iv)$ \cite{pesaran2006}'s
CCE-pooled estimator, $(v)$ \cite{pesaran2006}'s CCE-MG estimator, and $(vi)$
\cite{pesaran_smith1995}'s standard MG estimator. Table \ref{consistency}
below summarizes the consistency of the estimators for the four DGPs. So,
only the TW-MG, TW-MG ridge, and CCE-MG estimators are consistent for all
four DGPs. For the CCE-MG estimator, the consistency results are merely our
conjecture, and we are not aware of any rigorous studies on the CCE-MG
estimator for heterogeneous models with fixed $T.$\footnote{%
We conjecture that theoretical analysis can be challenging. For example, we
will encounter the issue of the second moment matrix of the estimated
factors becoming asymptotically singular when the number of proxies exceeds
the number of factors, as pointed by \cite{karabiyik2017} for the large $T$
case (see, also \cite{andrews1987}). In addition, it is unclear how to
conduct valid inference for the CCE-MG estimators when $T$ is fixed.} Even
so, we observe below that our TW-MG and TW-MG ridge estimators substantially
outperform the CCE-MG estimator when $T$ is small. \cite{westerlund2022cce}
study the CCE-pooled estimator for heterogeneous models for fixed $T$, which
requires that the slope coefficients be uncorrelated with the regressors.
The choice of tuning parameter for the TW-MG ridge estimator is discussed in
Remark \ref{Rmk2.4}. For the CCE-pooled and CCE-MG estimators, we also
include the individual fixed effects.\footnote{%
If we do not include the individual effects, the performance is worse in
general for these four DGPs.} We consider different combinations of $\left(
N,T\right) $ as shown in the tables below. For DGPs 1-3 where $K_{x}=1,$ $T$
starts with $\,T=3.$ For DGP 4 where $K_{x}=2,$ $T$ starts with $T=4.$ The
number of replications is 250. 
\begin{table}[tph]
\caption{Consistency of the estimators}
\label{consistency}
\begin{center}
{\small $%
\begin{tabular}{ccccccc}
\hline\hline
& TW-MG & TW-MG ridge & TW-pooled & CCE-MG & CCE-pooled & MG \\ \hline
DGP 1 & consistent & consistent & consistent & consistent & consistent & 
inconsistent \\ \hline
DGP 2 & consistent & consistent & consistent & consistent & consistent & 
inconsistent \\ \hline
DGP 3 & consistent & consistent & inconsistent & consistent & inconsistent & 
inconsistent \\ \hline
DGP 4 & consistent & consistent & inconsistent & consistent & inconsistent & 
inconsistent \\ \hline
\end{tabular}%
$}
\end{center}
\par
{\small Notes: \textquotedblleft TW-MG\textquotedblright\ refers to the
two-way-mean-group estimator proposed in (\ref{betahat1}), \textquotedblleft
TW-MG ridge\textquotedblright\ refers to the ridge-type estimator proposed
in (\ref{ridge}), \textquotedblleft TW-pooled\textquotedblright\ refers to
the pooled two-way fixed effects estimator in (\ref{B_tilde1}),
\textquotedblleft CCE-pooled\textquotedblright\ refers to \cite{pesaran2006}%
's CCE\ pooled estimator, and \textquotedblleft CCE-MG\textquotedblright\
refers to \cite{pesaran2006}'s CCE\ mean group estimator, and
\textquotedblleft MG\textquotedblright\ refers to \cite{pesaran_smith1995}'s
standard mean-group estimator. }
\end{table}

\subsection{Simulation results\label{Sim_results}}

Table \ref{estimation1} presents the performance of the six estimators
discussed above for the four DGPs. For each estimator, we report its bias$%
\times 10$ and mean-squared-error\ (MSE)$\times 100$. To save space, the
variance is not reported but available upon request. For DGP 4 where there
are two regressors with $\beta ^{0}=(\beta _{1}^{0},\beta _{2}^{0})^{\prime
},$ we only report the results for $\beta _{1}^{0}$ to save space, as the
results for $\beta _{2}^{0}$ are similar. Among the four DGPs, the standard
MG estimator is inconsistent, so we mainly compare the other five
estimators. For DGP 1 where $\beta _{i}$ is homogeneous, all five estimators
are consistent. However, when $T$ is small ($T=3$), the TW-pooled estimator
performs best, as it is based on the most parsimonious model. The
second-best estimator is the TW-MG ridge estimator, followed by the TW-MG
estimator, CCE-pooled and CCE-MG estimators. The difference in the
performance is substantial. For example, when $N=50$ and $T=3,$ the MSEs$%
\times 100$ of the TW-pooled, TW-MG ridge, TW-MG, CCE-pooled and CCE-MG
estimators are 1.31, 3.02, 7.35, 16.60 and 34.33, respectively. However when 
$T$ is large ($T=10),$ the performances of the five estimators are all good
and similar. For DGP 2 where $\beta _{i}$ is heterogeneous but independent
of $x_{it},$ our TW-MG ridge estimator is the best when $T$ is small, by
achieving a smaller variance. When $T$ is large, the performances of the
TW-MG ridge, TW-MG and CCE-MG estimators are similar, and they outperform
the CCE-pooled and TW-pooled estimators. For DGPs 3 and 4, the pooled
estimators are inconsistent for the reason discussed above. It is clear that
its bias of the TW-pooled and CCE-pooled estimators does not converge to
zero when the sample size increases. In contrast, the bias of the TW-MG,
TW-MG ridge and CCE-MG estimators is all small. Again, when $T$ is small,
the TW-MG ridge estimator outperforms the TW-MG estimator, which in turn
outperforms the CCE-MG estimators. When $T$ is large, the three estimators
all perform well.

Table \ref{coverage1} shows the empirical coverage of the 95\% CI in (\ref%
{CI1}) for $\beta ^{0}$ in DGPs 1-3 and $\beta _{1}^{0}$ in DGP\ 4 based on
the leave-one-out jackknife inference method proposed in Section \ref{Sec2.3}%
. We consider both TW-MG and TW-MG ridge estimators. The actual coverages
are all close to the nominal 95\% for all sample sizes and DGPs. This
confirms the asymptotic validity of the leave-one-out jackknife method.
Table \ref{test_results} reports the rejection frequency of testing $\mathbb{%
H}_{0}$ in (\ref{H0}) with a nominal size of $5\%$. We consider two types of
tests, one based on the TW-MG estimator and the other based on the TW-MG
ridge estimators. As discussed above, in DGPs 1 and 2, both the TW-MG (TW-MG
ridge) and pooled estimators are consistent, so these two DGPs are under the
null. Under the null, the rejection frequency of both tests is close to $5\%$
for all $N$ and $T$ considered. For DGPs 3 and 4 which are under the
alternative, the rejection frequency of both tests increases with $N$ and $%
T, $ demonstrating the power of the tests. When $N$ is large, such as $N=$ $%
200, $ the rejection frequency of both tests is close to $100\%$. We also
observe that when the sample size is small (e.g., $N=50$ and $T=3$), the
test based on the TW-MG ridge estimator is more powerful than that based on
the TW-MG estimator, as expected. \renewcommand{\arraystretch}{0.9} %
\setlength{\tabcolsep}{2.0pt}

\begin{table}[tph]
\caption{Estimation Results}
\label{estimation1}
\begin{center}
{\small 
\begin{tabular}{ccccccccccccccccc}
\hline\hline
&  &  & \multicolumn{6}{c}{bias$\times 10$} &  &  & \multicolumn{6}{c}{MSE$%
\times 100$} \\ \cline{4-9}\cline{12-17}
&  &  & TW- & TW-MG & TW- & CCE- & CCE- &  & \ \ \ \  & \ \ \ \  & TW- & 
TW-MG & TW- & CCE- & CCE- &  \\ 
$N$ & $T$ &  & MG & ridge & pooled & MG & pooled & MG &  &  & MG & ridge & 
pooled & MG & pooled & MG \\ \cline{1-2}\cline{4-9}\cline{9-17}
\multicolumn{17}{c}{DGP 1} \\ \hline
50 & 3 &  & -0.12 & -0.39 & -0.07 & -0.04 & -0.24 & 7.42 &  &  & 7.35 & 3.02
& 1.31 & 34.33 & 16.60 & 72.78 \\ 
100 & 3 &  & -0.04 & -0.28 & 0.04 & 0.49 & -0.04 & 7.38 &  &  & 6.42 & 1.74
& 0.58 & 23.98 & 12.44 & 70.86 \\ 
200 & 3 &  & -0.25 & -0.28 & 0.01 & -0.27 & 0.03 & 7.07 &  &  & 2.08 & 0.99
& 0.31 & 13.10 & 11.35 & 63.13 \\ 
50 & 5 &  & -0.05 & -0.36 & 0.05 & 0.13 & 0.04 & 7.40 &  &  & 1.20 & 1.21 & 
0.75 & 7.69 & 1.06 & 58.53 \\ 
100 & 5 &  & 0.01 & -0.16 & -0.04 & -0.13 & 0.00 & 7.38 &  &  & 0.53 & 0.53
& 0.36 & 5.07 & 0.52 & 58.05 \\ 
200 & 5 &  & -0.01 & -0.09 & -0.07 & 0.09 & 0.02 & 7.39 &  &  & 0.30 & 0.29
& 0.23 & 2.58 & 0.25 & 57.74 \\ 
50 & 10 &  & -0.06 & -0.32 & -0.03 & -0.02 & -0.02 & 7.41 &  &  & 0.62 & 0.69
& 0.74 & 0.43 & 0.36 & 56.80 \\ 
100 & 10 &  & 0.01 & -0.12 & 0.03 & -0.01 & -0.02 & 7.40 &  &  & 0.26 & 0.27
& 0.32 & 0.22 & 0.14 & 56.12 \\ 
200 & 10 &  & -0.01 & -0.07 & 0.02 & -0.01 & 0.01 & 7.10 &  &  & 0.11 & 0.11
& 0.15 & 0.11 & 0.07 & 51.65 \\ \hline
\multicolumn{17}{c}{DGP 2} \\ \hline
50 & 3 &  & -0.02 & -0.30 & 0.19 & -0.02 & -1.10 & 7.53 &  &  & 9.76 & 4.84
& 8.46 & 29.86 & 118.60 & 76.74 \\ 
100 & 3 &  & -0.01 & -0.25 & 0.07 & 0.30 & -0.62 & 7.41 &  &  & 7.78 & 2.79
& 4.29 & 18.66 & 95.24 & 73.19 \\ 
200 & 3 &  & -0.22 & -0.26 & -0.01 & 0.08 & -0.08 & 7.10 &  &  & 2.46 & 1.44
& 1.77 & 18.78 & 95.47 & 63.83 \\ 
50 & 5 &  & 0.06 & -0.26 & 0.27 & 0.13 & 0.11 & 7.50 &  &  & 3.38 & 3.20 & 
7.19 & 9.18 & 4.77 & 61.78 \\ 
100 & 5 &  & 0.02 & -0.14 & -0.04 & -0.02 & -0.02 & 7.39 &  &  & 1.59 & 1.56
& 3.65 & 5.17 & 2.81 & 59.25 \\ 
200 & 5 &  & 0.03 & -0.06 & -0.04 & 0.17 & -0.03 & 7.42 &  &  & 0.76 & 0.75
& 1.29 & 6.17 & 1.30 & 58.81 \\ 
50 & 10 &  & 0.10 & -0.16 & 0.18 & 0.14 & 0.18 & 7.57 &  &  & 2.55 & 2.44 & 
5.99 & 2.28 & 2.60 & 61.19 \\ 
100 & 10 &  & 0.02 & -0.12 & -0.02 & 0.01 & -0.02 & 7.41 &  &  & 1.35 & 1.33
& 3.30 & 1.35 & 1.64 & 57.38 \\ 
200 & 10 &  & 0.01 & -0.05 & 0.02 & 0.01 & -0.01 & 7.12 &  &  & 0.64 & 0.63
& 1.18 & 0.61 & 0.77 & 52.69 \\ \hline
\multicolumn{17}{c}{DGP 3} \\ \hline
50 & 3 &  & 0.11 & -0.24 & 5.05 & 0.44 & 11.91 & 5.55 &  &  & 5.43 & 3.43 & 
37.61 & 35.99 & 264.24 & 43.69 \\ 
100 & 3 &  & 0.02 & -0.14 & 5.13 & 0.01 & 15.00 & 5.58 &  &  & 2.70 & 2.10 & 
31.97 & 10.68 & 303.12 & 41.62 \\ 
200 & 3 &  & -0.06 & -0.17 & 5.31 & -0.05 & 16.98 & 5.62 &  &  & 1.36 & 1.02
& 31.38 & 7.85 & 352.15 & 41.30 \\ 
50 & 5 &  & 0.14 & -0.10 & 5.15 & 0.37 & 5.96 & 5.68 &  &  & 2.66 & 2.47 & 
35.21 & 8.59 & 43.56 & 36.61 \\ 
100 & 5 &  & 0.04 & -0.08 & 4.97 & 0.03 & 6.25 & 5.59 &  &  & 1.36 & 1.33 & 
30.12 & 3.98 & 44.11 & 35.43 \\ 
200 & 5 &  & 0.00 & -0.06 & 5.09 & 0.16 & 6.61 & 5.51 &  &  & 0.65 & 0.64 & 
28.22 & 1.93 & 46.01 & 33.59 \\ 
50 & 10 &  & 0.11 & -0.08 & 5.00 & 0.30 & 6.35 & 5.71 &  &  & 2.20 & 2.10 & 
31.39 & 2.07 & 45.15 & 35.66 \\ 
100 & 10 &  & -0.04 & -0.13 & 4.99 & 0.06 & 6.35 & 5.55 &  &  & 1.22 & 1.21
& 29.01 & 1.21 & 43.52 & 32.93 \\ 
200 & 10 &  & 0.02 & -0.03 & 5.15 & 0.04 & 6.56 & 5.29 &  &  & 0.55 & 0.54 & 
28.00 & 0.54 & 44.42 & 29.81 \\ \hline
\multicolumn{17}{c}{DGP\ 4} \\ \hline
50 & 4 &  & -0.18 & -0.07 & 5.18 & -0.35 & 11.70 & 2.73 &  &  & 16.43 & 6.10
& 39.50 & 30.40 & 251.05 & 21.79 \\ 
100 & 4 &  & 0.09 & -0.10 & 5.00 & -0.01 & 12.20 & 2.80 &  &  & 7.86 & 3.71
& 32.63 & 14.16 & 286.55 & 16.15 \\ 
200 & 4 &  & 0.04 & 0.06 & 5.37 & -0.05 & 15.34 & 2.71 &  &  & 7.78 & 1.66 & 
32.65 & 9.55 & 346.44 & 16.06 \\ 
50 & 5 &  & 0.11 & -0.08 & 5.29 & -0.65 & 6.27 & 2.86 &  &  & 5.43 & 3.99 & 
39.86 & 42.57 & 55.73 & 13.41 \\ 
100 & 5 &  & 0.02 & -0.11 & 4.89 & -0.42 & 6.16 & 2.84 &  &  & 2.98 & 2.40 & 
30.62 & 23.46 & 46.97 & 10.90 \\ 
200 & 5 &  & 0.04 & 0.00 & 5.16 & -0.33 & 6.58 & 2.77 &  &  & 1.56 & 1.29 & 
29.69 & 25.54 & 48.73 & 9.39 \\ 
50 & 10 &  & 0.13 & -0.09 & 4.96 & 0.11 & 6.44 & 2.88 &  &  & 2.60 & 2.40 & 
33.03 & 3.67 & 47.97 & 10.68 \\ 
100 & 10 &  & -0.04 & -0.14 & 4.95 & -0.06 & 6.31 & 2.70 &  &  & 1.66 & 1.61
& 29.69 & 2.17 & 43.53 & 8.79 \\ 
200 & 10 &  & 0.00 & -0.05 & 5.19 & 0.01 & 6.52 & 2.65 &  &  & 0.75 & 0.74 & 
28.91 & 1.04 & 44.26 & 7.80 \\ \hline
\end{tabular}%
}
\end{center}
\par
{\small Notes: see the notes to Table \ref{consistency}. }
\end{table}

\begin{table}[tph]
\caption{Inference Results}
\label{coverage1}
\begin{center}
{\small \ \tabcolsep 2 mm 
\begin{tabular}{cccccccccccc}
\hline\hline
\multicolumn{12}{c}{Empirical coverage of 95\% CI for $\beta ^{0}$} \\ \hline
&  &  & \multicolumn{4}{c}{based on TW-MG estimator} & \ \ \ \ \  & 
\multicolumn{4}{c}{based on TW-MG ridge estimator} \\ 
\cline{4-7}\cline{9-10}\cline{11-12}
$N$ & $T$ &  & DGP\ 1 & DGP\ 2 & DGP\ 3 & DGP\ 4 &  & DGP\ 1 & DGP\ 2 & DGP\
3 & DGP\ 4 \\ \cline{1-2}\cline{4-7}\cline{9-10}\cline{11-12}
50 & 3 &  & 0.96 & 0.96 & 0.95 & 0.96 &  & 0.96 & 0.94 & 0.95 & 0.96 \\ 
100 & 3 &  & 0.98 & 0.95 & 0.94 & 0.95 &  & 0.94 & 0.94 & 0.95 & 0.95 \\ 
200 & 3 &  & 0.97 & 0.96 & 0.96 & 0.98 &  & 0.94 & 0.95 & 0.96 & 0.97 \\ 
50 & 5 &  & 0.95 & 0.96 & 0.96 & 0.95 &  & 0.93 & 0.95 & 0.96 & 0.96 \\ 
100 & 5 &  & 0.97 & 0.94 & 0.94 & 0.96 &  & 0.95 & 0.96 & 0.93 & 0.95 \\ 
200 & 5 &  & 0.94 & 0.96 & 0.96 & 0.95 &  & 0.93 & 0.95 & 0.95 & 0.94 \\ 
50 & 10 &  & 0.94 & 0.95 & 0.95 & 0.96 &  & 0.92 & 0.94 & 0.96 & 0.95 \\ 
100 & 10 &  & 0.95 & 0.95 & 0.96 & 0.94 &  & 0.93 & 0.95 & 0.95 & 0.92 \\ 
200 & 10 &  & 0.97 & 0.96 & 0.96 & 0.94 &  & 0.96 & 0.96 & 0.97 & 0.95 \\ 
\hline
\end{tabular}%
}
\end{center}
\par
{\small Notes: 95\% CI\ refers to 95\% confidence interval based on the
jackknife method. \textquotedblleft TW-MG\textquotedblright\ refers to the
two-way-mean-group estimator proposed in (\ref{betahat1}) and
\textquotedblleft TW-MG ridge\textquotedblright\ refers to the ridge-type
estimator proposed in (\ref{ridge}). For DGP\ 4 where $K_{x}=2$, the
shortest $T$ is 4 instead of $3$. }
\end{table}

\begin{table}[tph]
\caption{Testing Results}
\label{test_results}
\begin{center}
{\small \ \tabcolsep 1.8 mm 
\begin{tabular}{cccccccccccccccc}
\hline\hline
\multicolumn{16}{c}{Rejection frequency of testing $\mathbb{H}_{0}$} \\ 
\hline
&  &  & \multicolumn{6}{c}{based on TW-MG estimator} & \ \ \ \ \  & 
\multicolumn{6}{c}{based on TW-MG ridge estimator} \\ 
\cline{4-9}\cline{11-14}\cline{13-16}
&  &  & \multicolumn{2}{c}{size} &  &  & \multicolumn{2}{c}{power} &  & 
\multicolumn{2}{c}{size} &  &  & \multicolumn{2}{c}{power} \\ 
\cline{4-5}\cline{8-9}\cline{11-12}\cline{15-16}
$N$ & $T$ &  & DGP\ 1 & DGP\ 2 &  &  & DGP\ 3 & DGP\ 4 &  & DGP\ 1 & DGP\ 2
&  &  & DGP\ 3 & DGP\ 4 \\ 
\cline{1-2}\cline{4-5}\cline{8-9}\cline{11-12}\cline{15-16}
50 & 3 &  & 0.03 & 0.04 &  &  & 0.44 & 0.51 &  & 0.02 & 0.05 &  &  & 0.56 & 
0.68 \\ 
100 & 3 &  & 0.02 & 0.04 &  &  & 0.68 & 0.77 &  & 0.04 & 0.06 &  &  & 0.74 & 
0.86 \\ 
200 & 3 &  & 0.03 & 0.03 &  &  & 0.87 & 0.99 &  & 0.04 & 0.02 &  &  & 0.91 & 
1.00 \\ 
50 & 5 &  & 0.06 & 0.08 &  &  & 0.67 & 0.74 &  & 0.05 & 0.08 &  &  & 0.73 & 
0.80 \\ 
100 & 5 &  & 0.02 & 0.04 &  &  & 0.84 & 0.95 &  & 0.04 & 0.06 &  &  & 0.86 & 
0.97 \\ 
200 & 5 &  & 0.02 & 0.05 &  &  & 0.97 & 1.00 &  & 0.03 & 0.04 &  &  & 0.98 & 
1.00 \\ 
50 & 10 &  & 0.06 & 0.06 &  &  & 0.83 & 0.91 &  & 0.09 & 0.07 &  &  & 0.84 & 
0.94 \\ 
100 & 10 &  & 0.05 & 0.04 &  &  & 0.96 & 1.00 &  & 0.06 & 0.04 &  &  & 0.97
& 1.00 \\ 
200 & 10 &  & 0.04 & 0.05 &  &  & 1.00 & 1.00 &  & 0.06 & 0.06 &  &  & 1.00
& 1.00 \\ \hline
\end{tabular}%
}
\end{center}
\par
{\small Notes: The test of $\mathbb{H}_{0}$ (\ref{H0}) is based on the test
statistic (\ref{test}). The nominal size is $5\%$. \textquotedblleft
TW-MG\textquotedblright\ refers to the two-way-mean-group estimator proposed
in (\ref{betahat1}) and \textquotedblleft TW-MG ridge\textquotedblright\
refers to the ridge-type estimator proposed in (\ref{ridge}). For DGP\ 4
where $K_{x}=2$, the shortest $T$ is 4 instead of $3$. }
\end{table}

\section{Empirical Applications\label{App}}

In this section, we apply our new TW-MG estimator and jackknife method to
two datasets. The first application examines the relationship between
health-care expenditure and income, while the second estimates a production
function. In the first application, we reject the null of poolability,
whereas in the second, we fail to reject it.

\subsection{Application 1: Health-care expenditure and income\label{App1}}

In this application, we use the data from \cite{baltagi2017health} and \cite%
{kapetanios2023testing} to investigate the relationship between health-care
expenditure and income for a panel of 167 countries covering the period
1995--2012 $(N=167$ and $T=18).$ Here $y_{it}$ is the per-capita health-care
spending in the $i$th country at year $t.$ There are two regressors $\left(
K_{x}=2\right) $. The key regressor of interest is the income level measured
by per capita GDP. The second regressor is the rate of public expenditure
over total health expenditure as a control variable. For a more detailed
description of these two variables, see \cite{baltagi2017health}.

Let $\beta ^{0}=(\beta _{1}^{0},\beta _{2}^{0})^{\prime },$ where $\beta
_{1}^{0}$ and $\beta _{2}^{0}$ represent the expected values of slope
coefficients of income and public expenditure rate, respectively. Table \ref%
{health_table} reports the estimation results with four estimation methods,
namely the TW-MG, TW-MG ridge, TW-pooled, and MG estimators. We find that
TW-MG and TW-MG ridge estimates are almost identical, which are
substantially different from the TW-pooled and MG estimates. Therefore, our
discussion below focuses on the TW-MG, TW-pooled and MG estimates. For $%
\beta _{1}^{0},$ the three estimates are 0.94, 0.72, and 1.87, respectively.
For $\beta _{2}^{0},$ the three estimates are 0.30, 0.02 and 0.76,
respectively. If we treat our TW-MG estimator as a benchmark, the TW-pooled
estimator tends to under-estimate, while the standard MG estimator
over-estimates. For $\beta _{1}^{0},$ all three estimation methods indicate
a positive and statistically significant average effect of income, though
the magnitudes differ. For $\beta _{2}^{0},$ while all estimation methods
show a positive effect, only the TW-MG and MG estimates are statistically
significant at the 5\% level; the TW-pooled estimate is not. Figure \ref%
{health_hist} plots the histogram of the estimates of $\beta _{i}=(\beta
_{1i},\beta _{2i})^{\prime }$ based on (\ref{betai_est}). Although these
estimators are not consistent in the fixed $T$ framework, we observe
substantial variations, suggesting the heterogeneity. Notably, most
estimates of $\beta _{1i}$ are positive, demonstrating a robust positive
effect of income.

We conduct the test for the hypotheses of poolability introduced in Section %
\ref{Sec4.3}. In addition to testing poolability in estimating both
parameters in ($\beta _{1}^{0},\beta _{2}^{0}$) together, we also test the
poolability for each individual parameter. Specifically, let $\hat{\beta}%
^{P}=(\hat{\beta}_{1}^{P},\hat{\beta}_{2}^{P})^{\prime }$ and $\hat{\beta}%
^{MG}=(\hat{\beta}_{1}^{MG},\hat{\beta}_{2}^{MG})^{\prime },$ and test the
following three hypotheses: 
\begin{equation}
\mathbb{H}_{0\ell }:\text{plim}_{N\rightarrow \infty }\hat{\beta}_{\ell
}^{MG}=\text{plim}_{N\rightarrow \infty }\hat{\beta}_{\ell }^{P}\text{ for }%
\ell \in \left[ 2\right] ,\text{ and }\mathbb{H}_{0}:\text{plim}%
_{N\rightarrow \infty }\hat{\beta}^{MG}=\text{plim}_{N\rightarrow \infty }%
\hat{\beta}^{P}.  \label{H_app}
\end{equation}%
As shown in Table \ref{test_health}, we reject $\mathbb{H}_{01},$ $\mathbb{H}%
_{02}$ and $\mathbb{H}_{0}$ all with $p$-values of 0.027, 0.019 and 0.009,
respectively, based on the TW-MG estimator and the jackknife method. We also
report the step-down Holm adjusted $p$-values (\cite{holm1979simple})
accounting for multiple testing, which are all smaller than 0.05.\footnote{%
Suppose that we have conducted the tests for $K^{\ }$individual hypotheses.
We order the individual \textit{p}-values from the smallest to the largest
as $p_{\left( 1\right) }\leq p_{\left( 2\right) }\leq ...\leq p_{\left(
K\right) }$ with their corresponding null hypotheses labeled accordingly as $%
\mathbb{H}_{0\left( 1\right) },$ $\mathbb{H}_{0\left( 2\right) },...,$ $%
\mathbb{H}_{0\left( K\right) }.$ The step-down Holm adjusted $p$-values for
testing $\mathbb{H}_{0\left( k\right) }$ is calculated as: adjusted-$%
p_{\left( k\right) }=\min \left( \left( K-k+1\right) p_{\left( k\right)
},1\right) .$%
\par
{}} Therefore, we conclude that we cannot pool the data to estimate the
average effects, confirming the substantial differences in the estimation
results discussed above. 
\begin{table}[tph]
\caption{Estimation results for the health expenditure application $(N=167,$ 
$T=18)$}
\label{health_table}
\begin{center}
{\small 
\begin{tabular}{ccccccccccccc}
\hline\hline
&  & \multicolumn{2}{c}{TW-MG} &  & \multicolumn{2}{c}{TW-MG ridge} &  & 
\multicolumn{2}{c}{TW-pooled} &  & \multicolumn{2}{c}{MG} \\ 
\cline{3-7}\cline{9-13}
&  & estimates & 95\% CI &  & estimates & 95\% CI &  & estimates & 95\% CI & 
& estimates & 95\% CI \\ \hline
\multicolumn{1}{l}{$\beta _{1}^{0}$} &  & 0.94 & [0.68 1.20] &  & 0.94 & 
[0.68 1.20] &  & 0.72 & [0.55 0.98] &  & 1.87 & [1.64 2.10] \\ 
\multicolumn{1}{l}{$\beta _{2}^{0}$} &  & 0.30 & [0.06 0.54] &  & 0.30 & 
[0.06 0.54] &  & 0.02 & [-0.03 0.26] &  & 0.76 & [0.33 1.19] \\ \hline
\end{tabular}%
}
\end{center}
\par
{\small Notes: $\beta _{1}^{0}$ and $\beta _{2}^{0}$ are the expected values
of slope coefficients of income and public expenditure rate, respectively.
\textquotedblleft TW-MG\textquotedblright\ refers to the two-way-mean-group
estimator proposed in (\ref{betahat1}), \textquotedblleft TW-MG
ridge\textquotedblright\ refers to the ridge-type estimator proposed in (\ref%
{ridge}), \textquotedblleft TW-pooled\textquotedblright\ refers to the
pooled two-way fixed effects estimator in (\ref{B_tilde1}), and
\textquotedblleft MG\textquotedblright\ refers to \cite{pesaran_smith1995}'s
standard mean-group estimator. 95\% CI\ refers to 95\% confidence interval
based on the proposed jackknife method. }
\end{table}
\begin{table}[tph]
\caption{Testing results for the health expenditure application $(N=167,$ $%
T=18)$}
\label{test_health}
\begin{center}
{\small \ \tabcolsep 3 mm 
\begin{tabular}{ccccccc}
\hline\hline
& Null hypothesis & $\ \ \ \ \mathbb{H}_{01}$ $\ \ \ \ $ &  & $\ \ \ \ 
\mathbb{H}_{02}$ $\ \ \ \ $ &  & $\ \ \ \ \mathbb{H}_{0}$ $\ \ \ \ $ \\ 
\hline
& test statistics & 4.90 &  & 5.53 &  & 9.38 \\ 
\multicolumn{1}{l}{} & $p$-value & 0.027 &  & 0.019 &  & 0.009 \\ 
\multicolumn{1}{l}{} & Holm adjusted\ $p$-values & 0.027 &  & 0.038 &  & 
0.027 \\ \hline
\end{tabular}%
}
\end{center}
\par
{\small Notes: The hypotheses $\mathbb{H}_{01},$ $\mathbb{H}_{02}$ and $%
\mathbb{H}_{0}$ are defined in (\ref{H_app}). The test is based on the
two-way-mean-group (TW-MG) estimators and the jackknife method proposed in
Section \ref{Sec4.3}. }
\end{table}
{\normalsize 
\begin{figure}[tbp]
{\normalsize \centering
\includegraphics[keepaspectratio=true,scale=0.50,trim={3cm 0cm
				0	0cm},clip]{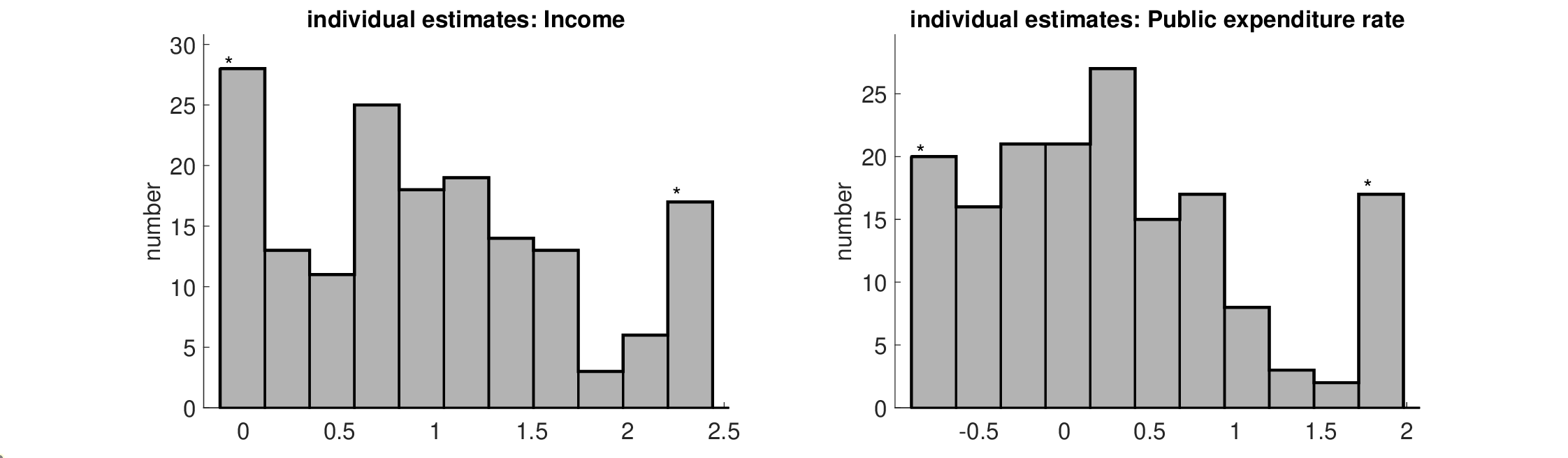}  }
\caption{Histogram of estimates of individual slope coefficients in the
health expenditure application}
\label{health_hist}
\end{figure}
}

\subsection{Application 2: Production function\label{App2}}

In this subsection, we apply our new methods to estimate a production
function using a balanced panel from \cite{blundell2000gmm}. The dataset
consists of 508 R\&D-performing U.S. manufacturing companies for 8 years,
1982--1988, so $N=508$ and $T=8.$\footnote{%
The original sample consists of 509 firms. However, we drop one firm because
its labor input (the first regressor) remains unchanged over the sample
period.} Here, the dependent variable $y_{it}$ is the logarithms of the
sales of firm $i$ in year $t$ and there are two regressors $(K_{x}=2)$, the\
logarithms of its labor (employment) and the logarithm of its capital stock,
which are inputs to the production function. For the detailed definitions of
the variables, see \cite{blundell2000gmm} and \cite{mairessehall}.

Table \ref{estimation_prod} reports the estimation results based on the four
estimation methods. Let $\beta ^{0}=(\beta _{1}^{0},\beta _{2}^{0})^{\prime
},$ where $\beta _{1}^{0}$ and $\beta _{2}^{0}$ represent the average return
to labor and capital, respectively. For $\beta _{1}^{0},$ the TW-MG, TW-MG
ridge, TW-pooled and MG estimates are similar, with the values of 0.67,
0.67, 0.65 and 0.63, respectively, all of which are statistically
significant. For $\beta _{2}^{0},$ the TW-MG, TW-MG ridge, and TW-pooled
estimates are similar with the values of 0.21, 0.21, and 0.23, respectively,
which are smaller than the MG estimate, 0.32. Again, all four estimates are
statistically significant. Figure \ref{prod} plots the histogram of the
estimates of $\beta _{i}=(\beta _{1i},\beta _{2i})^{\prime }$ based on (\ref%
{betai_est}), where we also observe some variations, though most of them are
positive

Similar to the first application in Section \ref{App1}, we conduct the test
for poolability. We test the three hypotheses as in (\ref{H_app}). As shown
in Table \ref{testing_prod}, we fail to reject $\mathbb{H}_{01},$ $\mathbb{H}%
_{02},$ and $\mathbb{H}_{0}$ all with $p$-values of 0.56, 0.67, and 0.84,
respectively. Therefore, for this application, we do not have strong
evidence against poolability. 
\begin{table}[tph]
\caption{Estimation results for the production function application $(N=508,$
$T=8)$}
\label{estimation_prod}
\begin{center}
{\small 
\begin{tabular}{ccccccccccccc}
\hline\hline
&  & \multicolumn{2}{c}{TW-MG} &  & \multicolumn{2}{c}{TW-MG ridge} &  & 
\multicolumn{2}{c}{TW-pooled} &  & \multicolumn{2}{c}{MG} \\ 
\cline{3-7}\cline{9-13}
&  & estimates & 95\% CI &  & estimates & 95\% CI &  & estimates & 95\% CI & 
& estimates & 95\% CI \\ \hline
\multicolumn{1}{l}{$\beta _{1}^{0}$} &  & 0.67 & [0.61 0.74] &  & 0.67 & 
[0.61 0.74] &  & 0.65 & [0.59 0.72] &  & 0.63 & [0.57 0.69] \\ 
\multicolumn{1}{l}{$\beta _{2}^{0}$} &  & 0.21 & [0.12 0.31] &  & 0.21 & 
[0.12 0.31] &  & 0.23 & [0.17 0.33] &  & 0.32 & [0.26 0.39] \\ \hline
\end{tabular}%
}
\end{center}
\par
{\small Notes: $\beta _{1}^{0}$ and $\beta _{2}^{0}$ are the expected values
of slope coefficients of labor and capital, respectively. \textquotedblleft
TW-MG\textquotedblright\ refers to the two-way-mean-group estimator proposed
in (\ref{betahat1}), \textquotedblleft TW-MG ridge\textquotedblright\ refers
to the ridge-type estimator proposed in (\ref{ridge}), \textquotedblleft
TW-pooled\textquotedblright\ refers to the pooled two-way fixed effects
estimator in (\ref{B_tilde1}), and \textquotedblleft MG\textquotedblright\
refers to \cite{pesaran_smith1995}'s standard mean-group estimator. 95\% CI\
refers to 95\% confidence interval based on the proposed jackknife method. }
\end{table}
\begin{table}[tph]
\caption{Testing results for the production function application $(N=508,$ $%
T=8)$}
\label{testing_prod}
\begin{center}
{\small \ \tabcolsep 3 mm 
\begin{tabular}{ccccccc}
\hline\hline
& Null hypothesis & $\ \ \ \ \mathbb{H}_{01}$ $\ \ \ \ $ &  & $\ \ \ \ 
\mathbb{H}_{02}$ $\ \ \ \ $ &  & $\ \ \ \ \mathbb{H}_{0}$ $\ \ \ \ $ \\ 
\hline
& test statistics & 0.26 &  & 0.18 &  & 0.36 \\ 
\multicolumn{1}{l}{} & $p$-value & 0.56 &  & 0.67 &  & 0.84 \\ 
\multicolumn{1}{l}{} & Holm adjusted\ $p$-values & 1 &  & 1 &  & 0.84 \\ 
\hline
\end{tabular}%
}
\end{center}
\par
{\small Notes: The hypotheses $\mathbb{H}_{01},$ $\mathbb{H}_{02}$ and $%
\mathbb{H}_{0}$ are defined in (\ref{H_app}). The test is based on the
two-way-mean-group (TW-MG) estimators and the jackknife method proposed in
Section \ref{Sec4.3}. }
\end{table}
{\normalsize 
\begin{figure}[tbp]
{\normalsize \centering
\includegraphics[keepaspectratio=true,scale=0.50,trim={3cm 0.5cm
				0	0.05cm},clip]{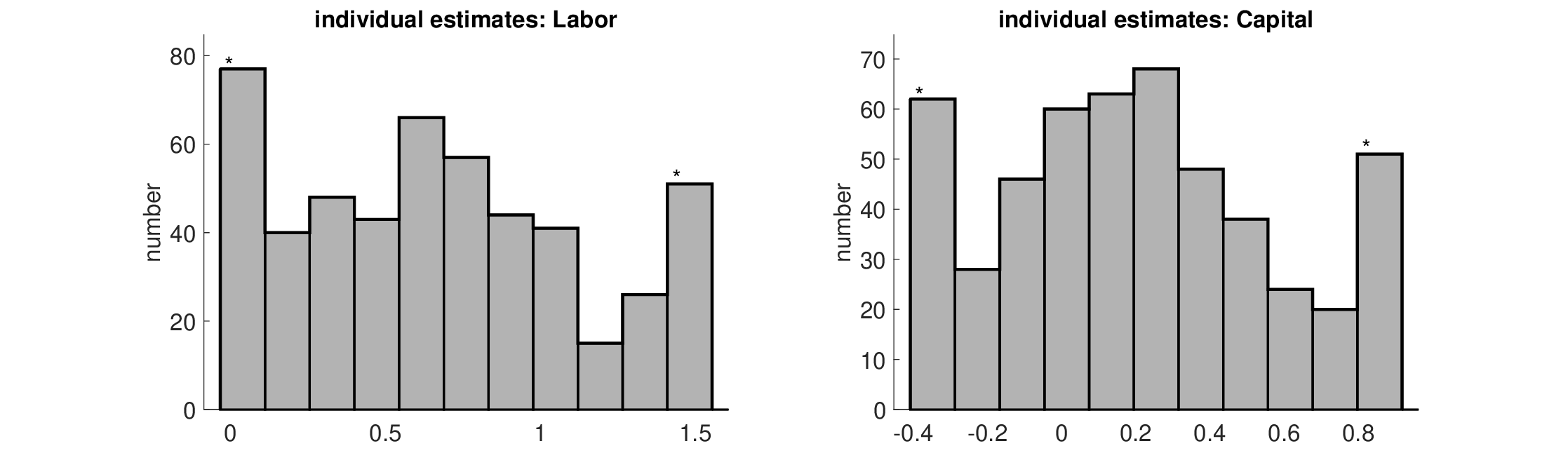}  }
\caption{Histogram of estimates of individual slope coefficients in the
production function application}
\label{prod}
\end{figure}
}

\section{Conclusion\label{Sec7}}

This paper considers estimation and inference for the population average of
slope coefficients in heterogeneous panels with two-way additive fixed
effects (TWFEs) and interactive fixed effects (IFEs). We propose two methods
to estimate the parameter of interest. The first method extends the mean
group (MG) approach to accommodate both individual and time effects, which
is valid for general correlated random coefficient (CRC) models but imposes
stronger rank conditions. The second method pools all observations to
estimate the parameter of interest, which may not be valid for some CRC
models but imposes weaker rank conditions. For both estimation methods, we
establish stable convergence for the estimators and propose a jackknife
method to estimate their asymptotic variance-covariance matrices. In
addition, we propose a Hausman-type test to assess poolability, which may
justify the commonly used pooled estimator in empirical research in the
presence of slope heterogeneity. Simulations demonstrate that our methods
perform well in finite samples.

There are many interesting topics for further research. First, we may extend
our methods to dynamic models and allow for weakly exogenous regressors.
Second, we may address the endogeneity issue in heterogeneous panels by
extending our methods. Third, we may consider nonlinear heterogeneous panels
with short $T$ and TWFEs and/or IFEs. These extensions are highly
challenging, but worth exploring.

{\footnotesize \setstretch{0.75} \setlength{\bibsep}{5pt} 
\bibliographystyle{apalike}
\bibliography{Hetero_panel2025}
}

\newpage 

\appendix%
\setcounter{page}{1}\setcounter{footnote}{0} \setcounter{section}{0} %
\setcounter{lemma}{0} \setcounter{table}{0} \setcounter{figure}{0} %
\renewcommand{\thesection}{S\arabic{section}} %
\renewcommand{\thefigure}{S\arabic{figure}} \renewcommand{\thetable}{S%
\arabic{table}} \renewcommand{\thelemma}{S.\arabic{lemma}}

\begin{center}
{\Large Online Supplement to \textquotedblleft Two-Way Mean Group Estimators
for Heterogeneous Panel Models with Fixed $T$}{\Large \textquotedblright }%
\medskip

Xun Lu$^{a},$ Liangjun Su$^{b}$ \medskip

$^{a}$Department of Economics, Chinese University of Hong Kong

$^{b}$School of Economics and Management, Tsinghua University, China
\end{center}

This online supplement is composed of three sections. \ref{SecA} contains
the proofs of the main results in the paper by calling upon some technical
lemmas. Section \ref{SecS1} contains the proofs of some technical lemmas.
Section \ref{SecS2} contains some additional simulation results.

\section{Proofs of the Main Results in Section \protect\ref{Sec3} \label%
{SecA}}

In this appendix, we prove the main results in Section \ref{Sec3}. For a
generic matrix $A,$ which is often written as a block partitioned matrix, we
will use $\left[ A\right] _{j},$ $\left[ A\right] _{\cdot k}$ and $\left[ A%
\right] _{jk}$ to denote its $j$th row block, its $k$th column block, and
its $\left( j,k\right) $th block, respectively, where the block size is
clear from the context but often related to $K_{x}$. For example, for $%
\mathbf{\hat{Q}}\equiv \frac{1}{T}\mathbf{X}^{o\prime }{\mathbb{M}}_{\mathbf{%
D}}\mathbf{X}^{o},$ an $NK_{x}\times NK_{x}$ matrix, we will use $[\mathbf{%
\hat{Q}}^{-1}]_{ij}$ to denote the $\left( i,j\right) $th block of $\mathbf{%
\hat{Q}}^{-1},$ each of which is of dimension $K_{x}\times K_{x}.$

\subsection{Proof of \textbf{Theorem \protect\ref{Thm3.1}} \label{SecA.1}}

\noindent \textbf{Proof of Theorem \ref{Thm3.1}}

Note that 
\begin{equation}
\mathbf{Y=X\beta }+\mathbf{D\alpha }+\mathbf{U}^{\dagger }=\mathbf{X\beta }+%
\mathbf{D\alpha }+(\mathbf{U}+(\mathbb{I}_{N}\otimes F)\mathbf{\lambda }).
\label{Y1}
\end{equation}%
Noting that $\mathbb{M}_{\mathbf{D}}=\mathbb{M}_{\mathbf{\iota }_{N}}\otimes 
\mathbb{M}_{\mathbf{\iota }_{T}},$ we have $\mathbb{M}_{\mathbf{D}}\mathbf{X}%
=\mathbb{M}_{\mathbf{D}}\mathbf{X}^{o}\ $where $\mathbf{X}^{o}\ $is a
version of $\mathbf{X}$ with its individual and time effects removed if any.
By the definition of $\mathbf{\hat{\beta},}$ we have 
\begin{equation}
\mathbf{\hat{\beta}}-\mathbf{\beta }=\left[ \mathbf{X}^{\prime }\mathbb{M}_{%
\mathbf{D}}\mathbf{X}\right] ^{-1}\mathbf{X}^{\prime }\mathbb{M}_{\mathbf{D}}%
\mathbf{Y}-\mathbf{\beta }=\left[ \frac{1}{T}\mathbf{X}^{o\prime }\mathbb{M}%
_{\mathbf{D}}\mathbf{X}^{o}\right] ^{-1}\frac{1}{T}\mathbf{X}^{o\prime }%
\mathbb{M}_{\mathbf{D}}\mathbf{U}^{\dagger }=\mathbf{\hat{Q}}^{-1}\frac{1}{T}%
\mathbf{X}^{o\prime }{\mathbb{M}}_{\mathbf{D}}\mathbf{U}^{\dagger },
\label{BB_hat1}
\end{equation}%
where $\mathbf{\hat{Q}}=\frac{1}{T}\mathbf{X}^{o\prime }{\mathbb{M}}_{%
\mathbf{D}}\mathbf{X}^{o}$. Then 
\begin{eqnarray*}
\hat{\beta}^{MG}-\beta ^{0} &=&\frac{1}{N}\mathbf{S}_{N}\mathbf{\hat{Q}}^{-1}%
\frac{1}{T}\mathbf{X}^{o\prime }{\mathbb{M}}_{\mathbf{D}}\mathbf{U}^{\dagger
}+\frac{1}{N}\sum_{i=1}^{N}\eta _{i} \\
&=&\frac{1}{N}\sum_{i=1}^{N}\sum_{j=1}^{N}\sum_{k=1}^{N}\left[ \mathbf{\hat{Q%
}}^{-1}\right] _{ij}\frac{1}{T}\mathbf{x}_{j}^{o\prime }\mathbb{M}_{\mathbf{%
\iota }_{T}}\mathbf{u}_{k}^{\dagger }m_{jk}+\frac{1}{N}\sum_{i=1}^{N}\eta
_{i} \\
&=&\frac{1}{N}\sum_{i=1}^{N}\sum_{j=1}^{N}\hat{\omega}_{i}\frac{1}{T}\mathbf{%
x}_{i}^{o\prime }\mathbb{M}_{\mathbf{\iota }_{T}}\mathbf{u}_{j}^{\dagger
}m_{ij}+\frac{1}{N}\sum_{j=1}^{N}\eta _{j}=\frac{1}{N}\sum_{j=1}^{N}\left(
\chi _{j}^{\prime }\mathbf{u}_{j}^{\dagger }+\eta _{j}\right) ,
\end{eqnarray*}%
where $m_{ij}=\mathbf{1}\left\{ i=j\right\} -\frac{1}{N}$ denotes the $%
\left( i,j\right) $th element of $\mathbb{M}_{\mathbf{\iota }_{N}},$ $\hat{%
\omega}_{j}=\sum_{i=1}^{N}[\mathbf{\hat{Q}}^{-1}]_{ij}$ and $\chi
_{j}^{\prime }=\frac{1}{T}\sum_{i=1}^{N}\hat{\omega}_{i}\mathbf{x}%
_{i}^{o\prime }\mathbb{M}_{\mathbf{\iota }_{T}}m_{ij}.$ It follows that $%
\sqrt{N}(\hat{\beta}^{MG}-\beta ^{0})=\sum_{j=1}^{N}\frac{1}{\sqrt{N}}\zeta
_{Nj},$ where $\zeta _{Nj}=\chi _{j}^{\prime }\mathbf{u}_{j}^{\dagger }+\eta
_{j}.$ Below, we consider stochastic convergence of $\frac{1}{\sqrt{N}}%
\sum_{j=1}^{N}\chi _{j}^{\prime }\mathbf{u}_{j}^{\dagger }$ and $\frac{1}{%
\sqrt{N}}\sum_{j=1}^{N}\eta _{j},$ respectively.\footnote{%
Let $\mathcal{F}_{N,0}^{o}=\mathcal{F}_{N,0}$ and $\mathcal{F}%
_{N,j}^{o}=\sigma (\mathcal{F}_{N,0}^{o},\left\{ (\mathbf{u}_{i},\lambda
_{i},\eta _{i})\right\} _{i=1}^{j})$ for $j\in \left[ N\right] .$ Note that $%
\chi _{j}$ and $F$ are measurable with respect to $\mathcal{F}_{N,j}^{o}$
for any $j.$ If we assume that conditional on $\mathcal{F}_{N,0}^{o},$ $%
\left\{ \mathbf{u}_{i},\lambda _{i},\eta _{i}\right\} _{i=1}^{N}$ are
independent with mean 0, we can verify $\left\{ \zeta _{Nj},\mathcal{F}%
_{N,j}^{o}\right\} $ forms an m.d.s. so that we can establish the stable
martingale CLT for $\sum_{j=1}^{N}Z_{Nj}$ directly. However, this assumption
rules out the potential dependence between $\eta _{i}$ and $\mathbf{x}%
_{i}^{o}$ allowed in some correlated random coefficient models. This
explains why we consider the convergence of $\frac{1}{\sqrt{N}}%
\sum_{j=1}^{N}\chi _{j}^{\prime }\mathbf{u}_{j}^{\dagger }$ and $\frac{1}{%
\sqrt{N}}\sum_{j=1}^{N}\eta _{j}$ separately.}

Recall that $\mathcal{F}_{0}^{MG}=\sigma \left( \{\mathbf{x}%
_{i}^{o}\}_{1\leq i\leq \infty },F\right) ,$ $\mathcal{F}_{N,0}^{MG}=\sigma
\left( \{\mathbf{x}_{i}^{o}\}_{1\leq i\leq N}\mathbf{,}F\right) $ and $%
\mathcal{F}_{N,j}^{MG}=\sigma (\mathcal{F}_{N,0}^{MG},\left\{ (\mathbf{u}%
_{i},\lambda _{i})\right\} _{i=1}^{j})$ for $j\in \left[ N\right] .$ Note
that $\chi _{j}$ and $F$ are measurable with respect to $\mathcal{F}%
_{N,j}^{MG}$ for any $j.$ By Assumption \ref{Assmp3.1}(i), 
\begin{equation*}
E\left( \chi _{j}^{\prime }\mathbf{u}_{j}^{\dagger }|\mathcal{F}%
_{N,j-1}^{MG}\right) =\chi _{j}^{\prime }\left[ E\left( \mathbf{u}_{j}|%
\mathcal{F}_{N,j-1}^{MG}\right) +F\cdot E\left( \lambda _{j}|\mathcal{F}%
_{N,j-1}^{MG}\right) \right] =\chi _{j}^{\prime }\left[ E\left( \mathbf{u}%
_{j}|\mathcal{F}_{N,0}^{MG}\right) +F\cdot E\left( \lambda _{j}|\mathcal{F}%
_{N,0}^{MG}\right) \right] =0.
\end{equation*}%
So $\{\chi _{j}^{\prime }\mathbf{u}_{j}^{\dagger },\mathcal{F}_{N,j}^{MG}\}$
forms a martingale difference sequence (m.d.s.). Let $\kappa $ be an
arbitrary $K_{x}\times 1$ nonrandom vector with $\left\Vert \kappa
\right\Vert =1.$ By the Cram\'{e}r-Wold device and the stable martingale CLT
(e.g., Corollary 3.19 and Theorem 6.1 in \cite{hausler2015stable}), it
suffices to show that%
\begin{eqnarray}
\sum_{j=1}^{N}E\left[ \left\Vert N^{-1/2}\chi _{j}^{\prime }\mathbf{u}%
_{j}^{\dagger }\right\Vert ^{2+\delta }|\mathcal{F}_{N,j-1}^{MG}\right]
&=&o_{p}\left( 1\right) \text{ and }  \label{mtg_CLT1a} \\
\sum_{j=1}^{N}E\left[ \left( \kappa ^{\prime }N^{-1/2}\chi _{j}^{\prime }%
\mathbf{u}_{j}^{\dagger }\right) ^{2}|\mathcal{F}_{N,j-1}^{MG}\right]
-\kappa ^{\prime }\Omega _{1}^{MG}\kappa &=&o_{p}\left( 1\right) ,
\label{mtg_CLT1b}
\end{eqnarray}%
where the Lyapunov condition in (\ref{mtg_CLT1a}) implies the Lindeberg
condition (CLB) in Theorem 6.1 of \cite{hausler2015stable}. Let $E_{A}\left(
\cdot \right) =E\left( \cdot |A\right) .$ By Lemma \ref{LemA.1} below,
Assumption \ref{Assmp3.1}(i, iii) and H\"{o}lder inequality, 
\begin{eqnarray*}
\max_{j}\left\Vert \chi _{j}\right\Vert ^{2+\delta } &=&\left\Vert \frac{1}{%
NT}\sum_{i=1}^{N}\hat{\omega}_{i}\mathbf{x}_{i}^{o\prime }\mathbb{M}_{%
\mathbf{\iota }_{T}}m_{ij}\right\Vert ^{2+\delta }\leq \left\{ \frac{1}{NT}%
\sum_{i=1}^{N}\left\Vert \hat{\omega}_{i}\right\Vert \left\Vert \mathbf{x}%
_{i}^{o}\right\Vert \right\} ^{2+\delta } \\
&\lesssim &\left\{ \frac{1}{NT}\sum_{i=1}^{N}\left\Vert \hat{\omega}%
_{i}\right\Vert ^{(4+2\delta )/(3+2\delta )}\right\} ^{(3+2\delta
)/2}\left\{ \frac{1}{NT}\sum_{i=1}^{N}\left\Vert \mathbf{x}%
_{i}^{o}\right\Vert ^{4+2\delta }\right\} ^{1/2}\lesssim 1\text{ and} \\
\max_{j}E_{\mathcal{F}_{N,j-1}^{MG}}\left[ \left\Vert \mathbf{u}%
_{j}^{\dagger }\right\Vert ^{2+\delta }\right] &=&\max_{j}E_{\mathcal{F}%
_{N,j-1}^{MG}}[\left\Vert \mathbf{u}_{j}+F\lambda \right\Vert ^{2+\delta }]
\\
&\lesssim &\max_{j}E_{\mathcal{F}_{N,j-1}^{MG}}[\left\Vert \mathbf{u}%
_{j}\right\Vert ^{2+\delta }]+\left\Vert F\right\Vert ^{2+\delta }\max_{j}E_{%
\mathcal{F}_{N,j-1}^{MG}}[\left\Vert \lambda \right\Vert ^{2+\delta }] \\
&=&\max_{j}E_{\mathcal{F}_{N,0}^{MG}}[\left\Vert \mathbf{u}_{j}\right\Vert
^{2+\delta }]+\left\Vert F\right\Vert ^{2+\delta }E_{\mathcal{F}%
_{N,0}^{MG}}[\left\Vert \lambda \right\Vert ^{2+\delta }]\lesssim 1.
\end{eqnarray*}%
It follows that%
\begin{equation*}
\sum_{j=1}^{N}E_{\mathcal{F}_{N,j-1}^{MG}}\left( \left\Vert N^{-1/2}\chi
_{j}^{\prime }\mathbf{u}_{j}^{\dagger }\right\Vert ^{2+\delta }\right)
\lesssim \frac{1}{N^{1+\delta /2}}\sum_{j=1}^{N}\left\Vert \chi
_{j}\right\Vert ^{2+\delta }E_{\mathcal{F}_{N,j-1}^{MG}}[||\mathbf{u}%
_{j}^{\dagger }||^{2+\delta }]\lesssim \frac{1}{N^{1+\delta /2}}%
\sum_{j=1}^{N}1=O_{p}(N^{-\delta /2}).
\end{equation*}%
For the condition in (\ref{mtg_CLT1b}), we have by Assumptions \ref{Assmp3.1}%
(i) and \ref{Assmp3.2}(ii)%
\begin{eqnarray*}
\sum_{j=1}^{N}E\left( \left( \kappa ^{\prime }N^{-1/2}\chi _{j}^{\prime }%
\mathbf{u}_{j}^{\dagger }\right) ^{2}|\mathcal{F}_{N,j-1}^{MG}\right)
&=&\kappa ^{\prime }\frac{1}{N}\sum_{j=1}^{N}\chi _{j}^{\prime }E\left[ 
\mathbf{u}_{j}^{\dagger }\mathbf{u}_{j}^{\dagger \prime }|\mathcal{F}%
_{N,j-1}^{MG}\right] \chi _{j}\kappa \\
&=&\kappa ^{\prime }\frac{1}{N}\sum_{j=1}^{N}\chi _{j}^{\prime }E\left[ 
\mathbf{u}_{j}^{\dagger }\mathbf{u}_{j}^{\dagger \prime }|\mathcal{F}%
_{N,0}^{MG}\right] \chi _{j}\kappa \overset{p}{\rightarrow }\kappa ^{\prime
}\Omega _{1}^{MG}\kappa .
\end{eqnarray*}%
Then 
\begin{equation*}
\frac{1}{\sqrt{N}}\sum_{j=1}^{N}\chi _{j}^{\prime }\mathbf{u}_{j}^{\dagger
}\rightarrow (\Omega _{1}^{MG})^{1/2}\mathbb{Z}_{1}\text{ \ }\mathcal{F}%
_{\infty }^{MG}\text{-stably,}
\end{equation*}%
where $\mathcal{F}_{\infty }^{MG}$ denotes the minimal sigma-field generated
by $\cup _{N=1}^{\infty }\mathcal{F}_{N,N}^{MG},$ $\mathbb{Z}_{1}\backsim
N(0,\mathbb{I}_{K_{x}})\ $and $Z_{1}$ is independent of $\mathcal{F}_{\infty
}^{MG}.$ Since $\mathcal{F}_{0}^{MG}\subset \mathcal{F}_{\infty }^{MG},$ the
above stable convergence implies that $\frac{1}{\sqrt{N}}\sum_{j=1}^{N}\chi
_{j}^{\prime }\mathbf{u}_{j}^{\dagger }\rightarrow (\Omega
_{1}^{MG})^{1/2}Z_{1}$ \ $\mathcal{F}_{0}^{MG}$-stably.

For $\frac{1}{\sqrt{N}}\sum_{j=1}^{N}\eta _{j},$ we observe that $\left(
\eta _{j},\mathcal{G}_{N,j}\right) $ is an m.d.s. with $E\left\Vert \eta
_{j}\right\Vert ^{2+\delta }\leq C<\infty $ under Assumption \ref{Assmp3.1}%
(ii), where $\mathcal{G}_{N,j}=\sigma (\left\{ \eta _{i}\right\}
_{i=1}^{j}). $ Then by the standard martingale CLT, $\frac{1}{\sqrt{N}}%
\sum_{j=1}^{N}\eta _{j}\overset{d}{\rightarrow }\Omega _{\eta }^{1/2}\mathbb{%
Z}_{2},$ where $\mathbb{Z}_{2}\backsim N(0,\mathbb{I}_{K_{x}})$ and $\Omega
_{\eta }\equiv \lim_{N\rightarrow \infty }\frac{1}{N}\sum_{j=1}^{N}E(\eta
_{j}\eta _{j}^{\prime })$ is nonrandom.

Next, noting that $E(\chi _{j}^{\prime }\mathbf{u}_{j}^{\dagger }|\mathcal{F}%
_{N,j-1})=\chi _{j}^{\prime }E(\mathbf{u}_{j}^{\dagger }|\mathcal{F}%
_{N,0})=0 $ under Assumption \ref{Assmp3.1}, 
\begin{equation*}
\text{Cov}\left( \chi _{j}^{\prime }\mathbf{u}_{j}^{\dagger },\eta _{j}|%
\mathcal{F}_{N,j-1}^{MG}\right) =E\left( \chi _{j}^{\prime }\mathbf{u}%
_{j}^{\dagger }\eta _{j}^{\prime }|\mathcal{F}_{N,j-1}^{MG}\right) =\chi
_{j}^{\prime }E\left( \mathbf{u}_{j}^{\dagger }\eta _{j}^{\prime }|\mathcal{F%
}_{N,j-1}^{MG}\right) =0,
\end{equation*}%
which holds irrespective of $E(\eta _{j}|\mathcal{F}_{N,j-1}^{MG})=0$ or
not. This ensures that $\frac{1}{\sqrt{N}}\sum_{j=1}^{N}\chi _{j}^{\prime }%
\mathbf{u}_{j}^{\dagger }$ and $\frac{1}{\sqrt{N}}\sum_{j=1}^{N}\eta _{j}$
are asymptotically independent, and 
\begin{equation*}
\sqrt{N}(\hat{\beta}^{MG}-\beta ^{0})=\frac{1}{\sqrt{N}}\sum_{j=1}^{N}\left(
\chi _{j}^{\prime }\mathbf{u}_{j}^{\dagger }+\eta _{j}\right) \rightarrow
\left( \Omega ^{MG}\right) ^{1/2}\mathbb{Z}\text{ \ }\mathcal{F}_{0}^{MG}%
\text{-stably,}
\end{equation*}%
where $\mathbb{Z}\backsim N(0,\mathbb{I}_{K_{x}})$ and $\Omega ^{MG}=\Omega
_{1}^{MG}+\Omega _{\eta }.$ ${\tiny \blacksquare }$

\begin{lemma}
\label{LemA.1} Suppose that Assumptions \ref{Assmp3.1}-\ref{Assmp3.2} hold.
Then $\frac{1}{N}\sum_{j=1}^{N}\left\Vert \hat{\omega}_{j}\right\Vert
^{(4+2\delta )/(3+2\delta )}=O_{p}(1).$
\end{lemma}

\subsection{\textbf{Proof of Theorem \protect\ref{Thm3.2} }\label{SecA.2}}

\noindent \textbf{Proof of Theorem \ref{Thm3.2}}

(i) Recall that $\hat{\beta}^{MG}=\frac{1}{N}\sum_{j=1}^{N}\mathbf{S}_{j,N}%
\left[ \mathbf{X}^{\prime }{\mathbb{M}}_{\mathbf{D}}\mathbf{X}\right] ^{-1}%
\mathbf{X}^{\prime }{\mathbb{M}}_{\mathbf{D}}\mathbf{Y.}$ The leave-one-out
MG estimator of $\beta ^{0}$ is given by 
\begin{equation*}
\hat{\beta}^{MG\left( -i\right) }=\frac{1}{N_{1}}\sum_{j=1}^{N_{1}}\mathbf{S}%
_{j,N_{1}}\left[ \mathbf{X}^{\left( -i\right) \prime }\left( {\mathbb{M}}_{%
\mathbf{\iota }_{N_{1}}}\otimes {\mathbb{M}}_{\mathbf{\iota }_{T}}\right) 
\mathbf{X}^{\left( -i\right) }\right] ^{-1}\mathbf{X}^{\left( -i\right)
\prime }\left( {\mathbb{M}}_{\mathbf{\iota }_{N_{1}}}\otimes {\mathbb{M}}_{%
\mathbf{\iota }_{T}}\right) \mathbf{Y}^{\left( -i\right) },
\end{equation*}%
where $\mathbf{X}^{\left( -i\right) \prime }$ and $\mathbf{Y}^{\left(
-i\right) }\ $are as defined in the main text. Define $\mathbf{U}^{\left(
-i\right) }$ and $\mathbf{U}^{\dag \left( -i\right) }$ analogously to $%
\mathbf{Y}^{\left( -i\right) }.$ Note that $\mathbf{U}^{\dag \left(
-i\right) }=\mathbf{U}^{\left( -i\right) }+(\mathbb{I}_{N_{1}}\otimes F)%
\mathbf{\lambda }^{\left( -i\right) },$ where $\mathbf{\lambda }^{\left(
-i\right) }=(\lambda _{1}^{\prime },...,\lambda _{i-1}^{\prime },\lambda
_{i+1}^{\prime },...,\lambda _{N}^{\prime })^{\prime }.$ Let $\mathbf{X}%
^{o\left( -i\right) }$ denote the leave-one-out version of $\mathbf{X}^{o}$
which removes the individual and time fixed effects from $\mathbf{X}^{\left(
-i\right) }$ if any. Note that 
\begin{equation*}
\hat{\beta}^{MG\left( -i\right) }=\frac{1}{N_{1}}\sum_{j=1}^{N_{1}}\mathbf{S}%
_{j,N_{1}}\left[ \mathbf{X}^{o\left( -i\right) \prime }\left( {\mathbb{M}}_{%
\mathbf{\iota }_{N_{1}}}\otimes {\mathbb{M}}_{\mathbf{\iota }_{T}}\right) 
\mathbf{X}^{o\left( -i\right) }\right] ^{-1}\mathbf{X}^{o\left( -i\right)
\prime }\left( {\mathbb{M}}_{\mathbf{\iota }_{N_{1}}}\otimes {\mathbb{M}}_{%
\mathbf{\iota }_{T}}\right) \mathbf{Y}^{\left( -i\right) }.
\end{equation*}

Define 
\begin{equation*}
\mathbf{\hat{Q}}^{\left( -i\right) }\equiv \frac{1}{T}\mathbf{X}^{o\left(
-i\right) \prime }\left( {\mathbb{M}}_{\mathbf{\iota }_{N_{1}}}\otimes {%
\mathbb{M}}_{\mathbf{\iota }_{T}}\right) \mathbf{X}^{o\left( -i\right) },
\end{equation*}%
which is the leave-one-out version of $\mathbf{\hat{Q}}$. We make the
following decompositions: 
\begin{eqnarray*}
\mathbf{\hat{Q}} &\equiv &\frac{1}{T}\mathbf{X}^{o\prime }\left( {\mathbb{I}}%
_{N}\otimes {\mathbb{M}}_{\mathbf{\iota }_{T}}\right) \mathbf{X}^{o}-\frac{1%
}{T}\mathbf{X}^{o\prime }\left( {\mathbb{P}}_{\mathbf{\iota }_{N}}\otimes {%
\mathbb{M}}_{\mathbf{\iota }_{T}}\right) \mathbf{X}^{o}\equiv \mathbf{Q}%
^{\left( d\right) }-\mathbf{CC}^{\prime }\text{ and} \\
\mathbf{\hat{Q}}^{\left( -i\right) } &=&\frac{1}{T}\mathbf{X}^{o\left(
-i\right) \prime }\left( {\mathbb{I}}_{N_{1}}\otimes {\mathbb{M}}_{\mathbf{%
\iota }_{T}}\right) \mathbf{X}^{o\left( -i\right) }-\frac{1}{T}\mathbf{X}%
^{o\left( -i\right) \prime }({\mathbb{P}}_{\mathbf{\iota }_{N_{1}}}\otimes {%
\mathbb{M}}_{\mathbf{\iota }_{T}})\mathbf{X}^{o\left( -i\right) }\equiv 
\mathbf{Q}^{\left( -i,d\right) }-\mathbf{C}^{\left( -i\right) }\mathbf{C}%
^{\left( -i\right) \prime },
\end{eqnarray*}%
where $\mathbf{C}=(NT)^{-1/2}\mathbf{X}^{o\prime }[$\textbf{$\iota $}$%
_{N}\otimes {\mathbb{M}}_{\mathbf{\iota }_{T}}],$ $\mathbf{C}^{\left(
-i\right) }=\left( N_{1}T\right) ^{-1/2}\mathbf{X}^{o\left( -i\right) \prime
}[$\textbf{$\iota $}$_{N_{1}}\otimes {\mathbb{M}}_{\mathbf{\iota }_{T}}]$
and the superscript $\left( d\right) $ signifies the corresponding matrix is
a block diagonal matrix. Let $\mathbb{Q}^{\left( -i,d\right) }=[\mathbf{Q}%
^{\left( -i,d\right) }]^{-1}.$ Let $\mathbf{\breve{Q}}^{\left( -i,d\right) }$
be the augmented version of $\mathbf{Q}^{\left( -i,d\right) }:$ 
\begin{equation*}
\left[ \mathbf{\breve{Q}}^{\left( -i,d\right) }\right] _{jj}=\left\{ 
\begin{array}{ll}
\left[ \mathbf{Q}^{\left( -i,d\right) }\right] _{jj} & \text{if }j<i \\ 
\mathbf{0}_{K_{x}\times K_{x}} & \text{if }j=i \\ 
\left[ \mathbf{Q}^{\left( -i,d\right) }\right] _{j-1,j-1}\text{ \ \ } & 
\text{if }j>i%
\end{array}%
\right. ,
\end{equation*}%
which is also a block diagonal matrix. Let $\mathbf{\breve{Q}}^{\left(
i,d\right) }=\mathbf{Q}^{\left( d\right) }-\mathbf{\breve{Q}}^{\left(
-i,d\right) }.$ Let $\mathbb{\breve{Q}}^{\left( -i,d\right) }$ be the
augmented version of $\mathbb{Q}^{\left( -i,d\right) }$ defined analogously,
which is also the Moore-Penrose generalized inverse of $\mathbf{\breve{Q}}%
^{\left( -i,d\right) }$. Let $\mathbf{\breve{C}}^{\left( -i\right) },$ $%
\mathbf{\breve{Y}}^{\left( -i\right) },$ and $\mathbf{\breve{U}}^{\dag
\left( -i\right) }$ be the augmented versions of $\mathbf{C}^{\left(
-i\right) },$ $\mathbf{Y}^{\left( -i\right) }$ and $\mathbf{U}^{\dag \left(
-i\right) },$ respectively: 
\begin{eqnarray*}
\left[ \mathbf{\breve{C}}^{\left( -i\right) }\right] _{j} &=&\left\{ 
\begin{array}{ll}
\left[ \mathbf{C}^{\left( -i\right) }\right] _{j} & \text{if }j<i \\ 
\mathbf{0}_{K_{x}\times T} & \text{if }j=i \\ 
\left[ \mathbf{C}^{\left( -i\right) }\right] _{j-1}\text{ \ } & \text{if }j>i%
\end{array}%
\right. ,\text{ }\left[ \mathbf{\breve{Y}}^{\left( -i\right) }\right]
_{j}=\left\{ 
\begin{array}{ll}
\left[ \mathbf{Y}^{\left( -i\right) }\right] _{j} & \text{if }j<i \\ 
\mathbf{0}_{N\times 1} & \text{if }j=i \\ 
\left[ \mathbf{Y}^{\left( -i\right) }\right] _{j-1}\text{ \ } & \text{if }j>i%
\end{array}%
\right. ,\text{ and } \\
\left[ \mathbf{\breve{U}}^{\dag \left( -i\right) }\right] _{j} &=&\left\{ 
\begin{array}{ll}
\left[ \mathbf{U}^{\dag \left( -i\right) }\right] _{j} & \text{if }j<i \\ 
\mathbf{0}_{N\times 1} & \text{if }j=i \\ 
\left[ \mathbf{U}^{\dag \left( -i\right) }\right] _{j-1}\text{ \ } & \text{%
if }j>i%
\end{array}%
\right. .
\end{eqnarray*}%
Next, we augment the $N_{1}K_{x}\times N_{1}T$ block diagonal matrix $%
\mathbf{X}^{o\left( -i\right) \prime }$ to an $NK_{x}\times NT$ block
diagonal matrix $\mathbf{\breve{X}}^{o\left( -i\right) \prime }$ whose $%
\left( j,j\right) $th block of size $K_{x}\times T$ is defined as follows 
\begin{equation*}
\left[ \mathbf{\breve{X}}^{o\left( -i\right) \prime }\right] _{jj}=\left\{ 
\begin{array}{ll}
\left[ \mathbf{X}^{o\left( -i\right) \prime }\right] _{jj} & \text{if }j<i
\\ 
\mathbf{0}_{K_{x}\times T} & \text{if }j=i \\ 
\left[ \mathbf{X}^{o\left( -i\right) \prime }\right] _{j-1,j-1}\text{ \ } & 
\text{if }j>i%
\end{array}%
\right. .
\end{equation*}%
It is easy to verify that $\mathbf{\breve{C}}^{\left( -i\right) }=\left(
N_{1}T\right) ^{-1/2}\mathbf{X}^{o\left( -i\right) \prime }[$\textbf{$\iota $%
}$_{N_{1}}\otimes {\mathbb{M}}_{\mathbf{\iota }_{T}}]=\left( N_{1}T\right)
^{-1/2}\mathbf{\breve{X}}^{o\left( -i\right) \prime }[$\textbf{$\iota $}$%
_{N}\otimes {\mathbb{M}}_{\mathbf{\iota }_{T}}].$ Define 
\begin{equation*}
\mathbf{\bar{C}}\equiv \left( N_{1}T\right) ^{-1/2}\mathbf{X}^{o\prime }[%
\mathbf{\iota }_{N}\otimes {\mathbb{M}}_{\mathbf{\iota }%
_{T}}]=(N/N_{1})^{1/2}\mathbf{C}.
\end{equation*}

Let $\mathbf{\breve{C}}^{\left( i\right) }\equiv \mathbf{\bar{C}}-\mathbf{%
\breve{C}}^{\left( -i\right) },$ $\mathbf{\breve{Y}}^{\left( i\right)
}\equiv \mathbf{Y}-\mathbf{\breve{Y}}^{\left( -i\right) },$ $\mathbf{\breve{U%
}}^{\left( i\right) }\equiv \mathbf{U}^{\dag }-\mathbf{\breve{U}}^{\dag
\left( -i\right) },$ and $\mathbf{\breve{X}}^{o\left( i\right) \prime
}\equiv \mathbf{X}^{o\prime }-\mathbf{\breve{X}}^{o\left( -i\right) \prime
}. $ Obviously, $\mathbf{\breve{C}}^{\left( i\right) },$ $\mathbf{\breve{Y}}%
^{\left( i\right) },$ $\mathbf{\breve{U}}^{\left( i\right) },$ and $\mathbf{%
\breve{X}}^{o\left( i\right) \prime }$ are sparse matrices with the
corresponding $i$th block of submatrix (or subvector) being nonzero and
other blocks being zero. Let $\mathbf{A}^{\left( -i\right) }=\mathbb{I}%
_{K_{x}T}+\mathbf{C}^{\left( -i\right) \prime }\mathbf{C}^{\left( -i\right)
} $. Define 
\begin{equation*}
\mathbb{A}^{\left( -i\right) }=(\mathbf{A}^{\left( -i\right) })^{-1},\text{ }%
\mathbb{Q}^{\left( d\right) }=(\mathbf{Q}^{\left( d\right) })^{-1},\text{ }%
\mathbb{Q}^{\left( -i,d\right) }=(\mathbf{Q}^{\left( -i,d\right) })^{-1}.
\end{equation*}%
Then for any $K_{x}\times K_{x}N_{1}$ generic matrix $A^{\left( -i\right) },$
we can augment it to a $K_{x}\times K_{x}N$ matrix $\breve{A}^{\left(
-i\right) }$ such that 
\begin{equation*}
\left[ \breve{A}^{\left( -i\right) }\right] _{\cdot j}=\left\{ 
\begin{array}{ll}
\left[ A^{\left( -i\right) }\right] _{\cdot j} & \text{if }j<i \\ 
\mathbf{0}_{K_{x}\times K_{x}} & \text{if }j=i \\ 
\left[ A^{\left( -i\right) }\right] _{\cdot j-1}\text{ \ \ } & \text{if }j>i%
\end{array}%
\right. .
\end{equation*}%
Let \textbf{$\iota $}$_{N}^{(-i)}$ denote \textbf{$\iota $}$_{N}$ with its $%
i $th element replaced by 0. A key observation is 
\begin{eqnarray}
&&A^{\left( -i\right) }\frac{1}{T}\mathbf{X}^{o\left( -i\right) \prime
}\left( {\mathbb{M}}_{\mathbf{\iota }_{N_{1}}}\otimes {\mathbb{M}}_{\mathbf{%
\iota }_{T}}\right) \mathbf{U}^{\dag \left( -i\right) }  \notag \\
&=&\breve{A}^{\left( -i\right) }\frac{1}{T}\mathbf{\breve{X}}^{o\left(
-i\right) \prime }\left( {\mathbb{M}}_{\mathbf{\iota }_{N}^{(-i)}}\otimes {%
\mathbb{M}}_{\mathbf{\iota }_{T}}\right) \mathbf{\breve{U}}^{\dag \left(
-i\right) }  \notag \\
&=&\breve{A}^{\left( -i\right) }\frac{1}{T}\mathbf{\breve{X}}^{o\left(
-i\right) \prime }\left( \left[ {\mathbb{M}}_{\mathbf{\iota }_{N}}+({\mathbb{%
P}}_{\mathbf{\iota }_{N}}-{\mathbb{P}}_{\mathbf{\iota }_{N}^{(-i)}})\right]
\otimes {\mathbb{M}}_{\mathbf{\iota }_{T}}\right) (\mathbf{U}^{\dag }-%
\mathbf{\breve{U}}^{\dag \left( i\right) })  \notag \\
&=&\breve{A}^{\left( -i\right) }\frac{1}{T}\left\{ \mathbf{X}^{o\prime
}\left( {\mathbb{M}}_{\mathbf{\iota }_{N}}\otimes {\mathbb{M}}_{\mathbf{%
\iota }_{T}}\right) \mathbf{U}^{\dag }-\mathbf{\breve{X}}^{o\left( i\right)
\prime }({\mathbb{M}}_{\mathbf{\iota }_{N}}\otimes {\mathbb{M}}_{\mathbf{%
\iota }_{T}})\mathbf{U}^{\dag }-\mathbf{\breve{X}}^{o\left( -i\right) \prime
}\left( {\mathbb{M}}_{\mathbf{\iota }_{N}}\otimes {\mathbb{M}}_{\mathbf{%
\iota }_{T}}\right) \mathbf{\breve{U}}^{\dag \left( i\right) }\right.  \notag
\\
&&\left. +\mathbf{\breve{X}}^{o\left( -i\right) \prime }[({\mathbb{P}}_{%
\mathbf{\iota }_{N}}-{\mathbb{P}}_{\mathbf{\iota }_{N}^{(-i)}})\otimes {%
\mathbb{M}}_{\mathbf{\iota }_{T}}]\mathbf{U}^{\dag \left( -i\right)
}\right\} \equiv \breve{A}^{\left( -i\right) }\sum_{s=1}^{4}L_{s}^{MG\left(
-i\right) }.  \label{L_decomp}
\end{eqnarray}%
Let $\mathbf{\beta }^{\left( -i\right) }=(\beta _{1}^{\prime },...,\beta
_{i-1}^{\prime },\beta _{i+1}^{\prime },...,\beta _{N}^{\prime })^{\prime },$
the leave-one-out version of $\mathbf{\beta }$. Let $\mathbf{D}^{\left(
1,-i\right) },$ $\mathbf{D}^{\left( 2,-i\right) }$ and $\mathbf{D}^{\left(
-i\right) }$ be the leave-one-out version of $\mathbf{D}^{\left( 1\right) },$
$\mathbf{D}^{\left( 2\right) }$ and $\mathbf{D}:$ 
\begin{equation}
\mathbf{D}^{\left( 1,-i\right) }=\mathbb{I}_{N_{1}}\otimes \mathbf{\iota }%
_{T},\text{ }\mathbf{D}^{\left( 2,-i\right) }=\mathbf{\iota }_{N_{1}}\otimes
\left( 
\begin{array}{c}
\mathbb{I}_{T_{1}} \\ 
-\mathbf{\iota }_{T_{1}}^{\prime }%
\end{array}%
\right) ,\text{ and }\mathbf{D}^{\left( -i\right) }=(\mathbf{D}^{\left(
1,-i\right) },\mathbf{D}^{\left( 2,-i\right) }),
\end{equation}%
all of which are actually not $i$-dependent. Let $\mathbf{\alpha }^{\left(
-i\right) }=(\mathbf{\alpha }^{\left( 1,-i\right) \prime },\mathbf{\alpha }%
^{\left( 2\right) \prime })^{\prime },$ where $\mathbf{\alpha }^{\left(
1,-i\right) }=(\alpha _{1\cdot },...,\alpha _{i-1\cdot },\alpha _{i+1\cdot
}, $ $...,\alpha _{N\cdot })^{\prime }.$ Noting that $\mathbf{Y}^{\left(
-i\right) }=\mathbf{X}^{\left( -i\right) }\mathbf{\beta }^{\left( -i\right)
}+\mathbf{D}^{\left( -i\right) }\mathbf{\alpha }^{\left( -i\right) }+\mathbf{%
U}^{\dag \left( -i\right) },$ we have by Sherman-Morrison Woodbury matrix
identity (e.g., Fact 6.4.31 in \cite{bernstein2005})\ and (\ref{L_decomp})%
\begin{eqnarray}
&&\hat{\beta}^{MG\left( -i\right) }-\frac{1}{N_{1}}\sum_{j\neq i}^{N}\beta
_{j}  \notag \\
&=&\frac{1}{N_{1}}\sum_{j=1}^{N_{1}}\mathbf{S}_{j,N_{1}}\left\{ \left[ \frac{%
1}{T}\mathbf{X}^{o\left( -i\right) \prime }\left( {\mathbb{M}}_{\mathbf{%
\iota }_{N_{1}}}\otimes {\mathbb{M}}_{\mathbf{\iota }_{T}}\right) \mathbf{X}%
^{o\left( -i\right) }\right] ^{-1}\frac{1}{T}\mathbf{X}^{o\left( -i\right)
\prime }({\mathbb{M}}_{\mathbf{\iota }_{N_{1}}}\otimes {\mathbb{M}}_{\mathbf{%
\iota }_{T}})\mathbf{Y}^{\left( -i\right) }-\mathbf{\beta }^{\left(
-i\right) }\right\}  \notag \\
&=&\frac{1}{N_{1}}\mathbf{S}_{N_{1}}\mathbb{Q}^{\left( -i,d\right) }\left\{ 
\mathbf{Q}^{\left( -i,d\right) }+\mathbf{C}^{\left( -i\right) }\mathbb{A}%
^{\left( -i\right) }\mathbf{C}^{\left( -i\right) \prime }\right\} \mathbb{Q}%
^{\left( -i,d\right) }\frac{1}{T}\mathbf{X}^{o\left( -i\right) \prime }({%
\mathbb{M}}_{\mathbf{\iota }_{N_{1}}}\otimes {\mathbb{M}}_{\mathbf{\iota }%
_{T}})\mathbf{U}^{\dag \left( -i\right) }  \notag \\
&=&\frac{1}{N_{1}}\mathbf{S}_{N}\mathbb{\breve{Q}}^{\left( -i,d\right)
}\left\{ \mathbf{\breve{Q}}^{\left( -i,d\right) }+\mathbf{\breve{C}}^{\left(
-i\right) }\mathbb{A}^{\left( -i\right) }\mathbf{\breve{C}}^{\left(
-i\right) \prime }\right\} \mathbb{\breve{Q}}^{\left( -i,d\right)
}\sum_{s=1}^{4}L_{s}^{MG\left( -i\right) }  \notag \\
&=&\frac{1}{N_{1}}\mathbf{S}_{N}\mathbb{Q}^{\left( d\right) }\left\{ \mathbf{%
\breve{Q}}^{\left( -i,d\right) }+\mathbf{\breve{C}}^{\left( -i\right) }%
\mathbb{A}^{\left( -i\right) }\mathbf{\breve{C}}^{\left( -i\right) \prime
}\right\} \mathbb{Q}^{\left( d\right) }\sum_{s=1}^{4}L_{s}^{MG\left(
-i\right) }  \notag \\
&=&\sum_{s=1}^{4}\frac{1}{N_{1}}\mathbf{S}_{N}\mathbb{Q}^{\left( d\right)
}\left\{ [\mathbf{Q}^{\left( d\right) }+\mathbf{\bar{C}}\mathbb{A}\mathbf{%
\bar{C}}^{\prime }]+\mathbf{\bar{C}}(\mathbb{A}^{\left( -i\right) }-\mathbb{A%
})\mathbf{\bar{C}}^{\prime }-\mathbf{\breve{Q}}^{\left( i,d\right) }-\mathbf{%
\breve{C}}^{\left( i\right) }\mathbb{A}^{\left( -i\right) }\mathbf{\bar{C}}%
^{\prime }-\mathbf{\bar{C}}\mathbb{A}^{\left( -i\right) }\mathbf{\breve{C}}%
^{\left( i\right) \prime }\right.  \notag \\
&&\left. +\mathbf{\breve{C}}^{\left( i\right) }\mathbb{A}^{\left( -i\right) }%
\mathbf{\breve{C}}^{\left( i\right) \prime }\right\} \cdot \mathbb{Q}%
^{(d)}L_{s}^{MG\left( -i\right) }  \notag \\
&\equiv &\sum_{s=1}^{4}B_{s}^{MG\left( -i\right) }.  \label{beta_mg_decomp}
\end{eqnarray}%
By Lemma \ref{LemA.2}(i)-(iv), we have that uniformly in $i\in \left[ N%
\right] ,$ 
\begin{equation}
\hat{\beta}^{MG\left( -i\right) }-\frac{1}{N_{1}}\sum_{j\neq i}^{N}\beta
_{j}=\frac{N}{N_{1}}\left( \hat{\beta}^{MG}-\frac{1}{N}\sum_{j=1}^{N}\beta
_{j}\right) +\Delta _{i2}^{MG}+\Delta _{i3}^{MG}+R_{i}o_{p}(N^{-1}),
\label{jackknife1}
\end{equation}%
where $\Delta _{i2}^{MG}\equiv -\frac{1}{N_{1}T}\hat{\omega}_{i}\mathbf{x}%
_{i}^{o\prime }{\mathbb{M}}_{\mathbf{\iota }_{T}}\mathbf{u}_{i}^{\dag },$ $%
\Delta _{i3}^{MG}\equiv \frac{1}{NN_{1}}\sum_{j=1}^{N}\hat{\omega}_{j}\frac{1%
}{T}\mathbf{x}_{j}^{o\prime }{\mathbb{M}}_{\mathbf{\iota }_{T}}\mathbf{u}%
_{i}^{\dag },$ and $\left\Vert R_{i}\right\Vert \leq \left\Vert \hat{\omega}%
_{i}\right\Vert +\left\Vert \tilde{q}_{ii}^{-1}\right\Vert +1.$ Then $\hat{%
\beta}^{MG\left( -i\right) }=\frac{N}{N_{1}}\hat{\beta}^{MG}-\frac{1}{N_{1}}%
\beta _{i}+\Delta _{i2}^{MG}+\Delta _{i3}^{MG}+o_{p}(N^{-1}),$ and%
\begin{eqnarray*}
\hat{\beta}^{MG\left( -i\right) }-\frac{1}{N}\sum_{j=1}^{N}\hat{\beta}%
^{MG\left( -j\right) } &=&-\frac{1}{N_{1}}\beta _{i}+\frac{1}{NN_{1}}%
\sum_{j=1}^{N}\beta _{j}+\left( \Delta _{i2}^{MG}-\frac{1}{N}%
\sum_{j=1}^{N}\Delta _{j2}^{MG}\right) +\left( \Delta _{i3}^{MG}-\frac{1}{N}%
\sum_{j=1}^{N}\Delta _{j3}^{MG}\right) \\
&&+R_{i}o_{p}(N^{-1}) \\
&=&-\frac{1}{N_{1}}\eta _{i}+\Delta _{i2}^{MG}+\Delta
_{i3}^{MG}+R_{i}o_{p}(N^{-1}),
\end{eqnarray*}%
where the second equality holds by the fact that $\beta _{i}=\beta ^{0}+\eta
_{i},$ $\frac{1}{NN_{1}}\sum_{j=1}^{N}\eta _{j}=o_{p}(N^{-1})$ under
Assumption \ref{Assmp3.1}(ii), $\frac{1}{N}\sum_{j=1}^{N}\Delta
_{j2}^{MG}=o_{p}(N^{-1})$ by Lemma \ref{LemA.3}(i),$\ $and $\frac{1}{N}%
\sum_{j=1}^{N}\Delta _{j3}^{MG}=o_{p}(N^{-1})$ by Lemma \ref{LemA.3}(ii).
Consequently, we have%
\begin{eqnarray}
\hat{\beta}^{MG\left( -i\right) }-\frac{1}{N}\sum_{j=1}^{N}\hat{\beta}%
^{MG\left( -j\right) } &=&-\frac{1}{N_{1}}\eta _{i}-\frac{1}{N_{1}T}\hat{%
\omega}_{i}\mathbf{x}_{i}^{o\prime }{\mathbb{M}}_{\mathbf{\iota }_{T}}%
\mathbf{u}_{i}^{\dag }+\frac{1}{NN_{1}}\sum_{j=1}^{N}\hat{\omega}_{j}\frac{1%
}{T}\mathbf{x}_{j}^{o\prime }{\mathbb{M}}_{\mathbf{\iota }_{T}}\mathbf{u}%
_{i}^{\dag }+R_{i}o_{p}(N^{-1})  \notag \\
&=&-\frac{1}{N_{1}}\eta _{i}-\frac{1}{N_{1}}\sum_{j=1}^{N}\hat{\omega}_{j}%
\frac{1}{T}\mathbf{x}_{j}^{o\prime }{\mathbb{M}}_{\mathbf{\iota }_{T}}%
\mathbf{u}_{i}^{\dag }m_{ji}+R_{i}o_{p}(N^{-1})  \notag \\
&=&-\frac{1}{N_{1}}(\chi _{i}^{\prime }\mathbf{u}_{i}^{\dag }+\eta
_{i})+R_{i}o_{p}(N^{-1})\text{ uniformly in }i\in \left[ N\right] ,
\label{beta_demean1}
\end{eqnarray}%
where recall that $\chi _{i}^{\prime }=\sum_{j=1}^{N}\hat{\omega}_{j}\frac{1%
}{T}\mathbf{x}_{j}^{o\prime }{\mathbb{M}}_{\mathbf{\iota }_{T}}m_{ji}$. Then 
\begin{eqnarray*}
\hat{\Omega}^{MG} &=&N_{1}\sum_{i=1}^{N}\left[ \hat{\beta}^{MG\left(
-i\right) }-\overline{\hat{\beta}}^{_{MG}}\right] \left[ \hat{\beta}%
^{MG\left( -i\right) }-\overline{\hat{\beta}}^{_{MG}}\right] ^{\prime }=%
\frac{1}{N_{1}}\sum_{i=1}^{N}(\chi _{i}^{\prime }\mathbf{u}_{i}^{\dag }+\eta
_{i})(\chi _{i}^{\prime }\mathbf{u}_{i}^{\dag }+\eta _{i})^{\prime }+o_{p}(1)
\\
&=&\frac{1}{N}\sum_{i=1}^{N}(\chi _{i}^{\prime }\mathbf{u}_{i}^{\dag }+\eta
_{i})(\chi _{i}^{\prime }\mathbf{u}_{i}^{\dag }+\eta _{i})^{\prime
}+o_{p}(1)\equiv \bar{\Omega}^{MG}+o_{p}(1),
\end{eqnarray*}%
where we use the fact that $\frac{1}{N}\sum_{i=1}^{N}\left\Vert \hat{\omega}%
_{i}\right\Vert ^{2}=O_{p}(1)$ by strengthening the last part of the proof
of Lemma \ref{LemA.1} under Assumption \ref{Assmp3.3}, and that $\frac{1}{N}%
\sum_{i=1}^{N}\left\Vert \tilde{q}_{ii}^{-1}\right\Vert ^{2}=O_{p}\left(
1\right) $ by Assumption \ref{Assmp3.3} and Markov inequality.

Now, 
\begin{equation*}
\bar{\Omega}^{MG}=\frac{1}{N}\sum_{i=1}^{N}\chi _{i}^{\prime }\mathbf{u}%
_{i}^{\dag }\mathbf{u}_{i}^{\dag \prime }\chi _{i}+\frac{1}{N}%
\sum_{i=1}^{N}\eta _{i}\eta _{i}{}^{\prime }+\frac{1}{N}\sum_{i=1}^{N}(\chi
_{i}^{\prime }\mathbf{u}_{i}^{\dag }\eta _{i}{}^{\prime }+\eta _{i}\mathbf{u}%
_{i}^{\dag \prime }\chi _{i})\equiv w_{1}+w_{2}+w_{3}.
\end{equation*}%
By the LLN for conditionally independent observations $\left\{ \chi
_{i}^{\prime }\mathbf{u}_{i}^{\dag }\right\} ,$ independent observations $%
\left\{ \eta _{i}\right\} ,$ and Assumption \ref{Assmp3.1} and \ref{Assmp3.2}%
(ii), 
\begin{eqnarray*}
w_{1} &=&\frac{1}{N}\sum_{i=1}^{N}\chi _{i}^{\prime }E\left( \mathbf{u}%
_{i}^{\dag }\mathbf{u}_{i}^{\dag \prime }|\mathcal{F}_{N,0}^{MG}\right) \chi
_{i}+o_{p}\left( 1\right) =\Omega _{1}^{MG}+o_{p}\left( 1\right) , \\
w_{2} &=&\frac{1}{N}\sum_{i=1}^{N}E(\eta _{i}\eta _{i}^{\prime
})+o_{p}\left( 1\right) =\Omega _{\eta }+o_{p}\left( 1\right) ,\text{ and} \\
w_{3} &=&\frac{1}{N}\sum_{i=1}^{N}E(\chi _{i}^{\prime }\mathbf{u}_{i}^{\dag
}\eta _{i}{}^{\prime }+\eta _{i}\mathbf{u}_{i}^{\dag \prime }\chi _{i}|%
\mathcal{F}_{N,0}^{MG})+o_{p}\left( 1\right) =o_{p}\left( 1\right) .
\end{eqnarray*}%
The above first result also follows from Proposition 6.16 in \cite%
{hausler2015stable} under Assumptions \ref{Assmp3.1}--\ref{Assmp3.2}. It
follows that $\bar{\Omega}^{MG}=\Omega _{1}^{MG}+\Omega _{\eta }+o_{p}\left(
1\right) =\Omega ^{MG}+o_{p}\left( 1\right) .$

(ii) By part (i), $\mathbb{S}^{\prime }\hat{\Omega}^{MG}\mathbb{S=\tau }%
^{2}+o_{p}\left( 1\right) ,$ where $\tau =\sqrt{\mathbb{S}^{\prime }\Omega
^{MG}\mathbb{S}}$ and $\Omega ^{MG}$ is $\mathcal{F}_{0}^{MG}$-measurable.
Then by the stable convergence analogue of Cram\'{e}r-Slutsky lemma (e.g.,
Theorem 3.18(b) in \cite{hausler2015stable}), we have%
\begin{equation*}
\left( \sqrt{N}\mathbb{S}^{\prime }(\hat{\beta}^{MG}-\beta ^{0}),\text{ }%
\mathbb{S}^{\prime }\hat{\Omega}^{MG}\mathbb{S}\right) \rightarrow \left(
\tau \mathbb{Z},\tau ^{2}\right) \text{ \ }\mathcal{F}_{0}^{MG}\text{-stably
as }N\rightarrow \infty .
\end{equation*}%
where $\mathbb{Z}$ $\backsim N\left( 0,1\right) $ and is independent of $%
\mathcal{F}_{0}^{MG}.$ The function $g:\mathbb{R}\times \mathbb{R}%
_{+}\rightarrow \mathbb{R}$ with $g\left( x,y\right) =x/\sqrt{y}$ with $y>0$
is Borel measurable and continuous. Then by the continuous mapping theorem
(e.g., Theorem 3.18(c) in \cite{hausler2015stable}), 
\begin{equation*}
(\mathbb{S}^{\prime }\hat{\Omega}^{MG}\mathbb{S})^{-1/2}\sqrt{N}\mathbb{S}%
^{\prime }(\hat{\beta}^{MG}-\beta ^{0})\overset{}{\rightarrow }\mathbb{Z}%
\text{ \ }\mathcal{F}_{0}^{MG}\text{-stably as }N\rightarrow \infty
\end{equation*}%
By the independence of $\mathbb{Z}$ and $\mathcal{F}_{0}^{MG},$ the above $%
\mathcal{F}_{0}^{MG}$-stable convergence is also $\mathcal{F}_{0}^{MG}$%
-mixing convergence (see Definition 3.15 in \cite{hausler2015stable} and the
remark after it). Then we can also write $(\mathbb{S}^{\prime }\hat{\Omega}%
^{MG}\mathbb{S})^{-1/2}\sqrt{N}\mathbb{S}^{\prime }(\hat{\beta}^{MG}-\beta
^{0})\overset{d}{\rightarrow }N\left( 0,1\right) .$ ${\tiny \blacksquare }$

\begin{lemma}
\label{LemA.2} Suppose that Assumptions \ref{Assmp3.1}--\ref{Assmp3.3} hold.
Then uniformly in $i\in \left[ N\right] ,$

(i) $B_{1}^{MG\left( -i\right) }=\frac{N}{N_{1}}(\hat{\beta}^{MG}-\frac{1}{N}%
\sum_{j=1}^{N}\beta _{j})-\frac{1}{N_{1}}\tilde{q}_{ii}^{-1}\frac{1}{T}%
\sum_{j=1}^{N}\mathbf{x}_{i}^{o\prime }{\mathbb{M}}_{\mathbf{\iota }_{T}}%
\mathbf{u}_{j}^{\dag }m_{ij}-\frac{1}{N_{1}}\mathbf{S}_{N}\mathbb{Q}^{\left(
d\right) }\mathbf{\bar{C}}\mathbb{A}^{\left( -i\right) }\mathbf{\breve{C}}%
^{\left( i\right) \prime }\mathbb{Q}^{\left( d\right) }\frac{1}{T}\mathbf{%
\breve{X}}^{o\left( i\right) }$ $({\mathbb{M}}_{\mathbf{\iota }_{N}}\otimes {%
\mathbb{M}}_{\mathbf{\iota }_{T}})\mathbf{U}^{\dag }+R_{1i}o_{p}(N^{-1})$
with $R_{1i}\leq \left\Vert \tilde{q}_{ii}^{-1}\right\Vert +1;$

(ii) $B_{2}^{MG\left( -i\right) }=-\frac{1}{N_{1}T}\hat{\omega}_{i}\mathbf{x}%
_{i}^{o\prime }{\mathbb{M}}_{\mathbf{\iota }_{T}}\mathbf{u}_{i}^{\dag }+%
\frac{1}{N_{1}}\tilde{q}_{ii}^{-1}\frac{1}{T}\sum_{j=1}^{N}\mathbf{x}%
_{i}^{o\prime }{\mathbb{M}}_{\mathbf{\iota }_{T}}\mathbf{u}_{j}^{\dag
}m_{ij}+\frac{1}{N_{1}}\mathbf{S}_{N}\mathbb{Q}^{\left( d\right) }\mathbf{%
\bar{C}}\mathbb{A}^{\left( -i\right) }\mathbf{\breve{C}}^{\left( i\right)
\prime }\mathbb{Q}^{\left( d\right) }\frac{1}{T}\mathbf{\breve{X}}^{o\left(
i\right) }$ $({\mathbb{M}}_{\mathbf{\iota }_{N}}$ $\otimes {\mathbb{M}}_{%
\mathbf{\iota }_{T}})\mathbf{U}^{\dag }+R_{2i}o_{p}(N^{-1})$ with $%
R_{2i}\leq \left\Vert \hat{\omega}_{i}\right\Vert +\left\Vert \tilde{q}%
_{ii}^{-1}\right\Vert ;$

(iii) $B_{3}^{MG\left( -i\right) }=\frac{1}{NN_{1}}\sum_{j=1}^{N}\hat{\omega}%
_{j}\frac{1}{T}\mathbf{x}_{j}^{o\prime }{\mathbb{M}}_{\mathbf{\iota }_{T}}%
\mathbf{u}_{i}^{\dag }+R_{3i}o_{p}\left( N^{-1}\right) \ $with $\left\Vert
R_{3i}\right\Vert \leq \left\Vert \hat{\omega}_{i}\right\Vert +\left\Vert 
\tilde{q}_{ii}^{-1}\right\Vert ;$

(iv) $B_{4}^{MG\left( -i\right) }=R_{4i}o_{p}\left( N^{-1}\right) \ $with $%
R_{1i}\leq \left\Vert \hat{\omega}_{i}\right\Vert +\left\Vert \tilde{q}%
_{ii}^{-1}\right\Vert +1.$\textbf{\ }
\end{lemma}

\begin{lemma}
\label{LemA.3} Suppose that Assumptions \ref{Assmp3.1}--\ref{Assmp3.3} hold.
Then

(i) $\frac{1}{N_{1}T}\sum_{i=1}^{N}\hat{\omega}_{i}\mathbf{x}_{i}^{o\prime }{%
\mathbb{M}}_{\mathbf{\iota }_{T}}\mathbf{u}_{i}^{\dag }=o_{p}(1);$

(ii) $\frac{1}{NN_{1}}\sum_{l=1}^{N}\sum_{j=1}^{N}\hat{\omega}_{j}\frac{1}{T}%
\mathbf{x}_{j}^{o\prime }{\mathbb{M}}_{\mathbf{\iota }_{T}}\mathbf{u}%
_{l}^{\dag }=o_{p}(1).$
\end{lemma}

\subsection{Proof of \textbf{Theorem \protect\ref{Thm4.1}} \label{SecA.3}}

\noindent \textbf{Proof of Theorem \ref{Thm4.1}}

Recall that $\ddot{X}\equiv {\mathbb{M}}_{\mathbf{D}}X=(\mathbf{\ddot{x}}%
_{1}^{\prime },...,\mathbf{\ddot{x}}_{N}^{\prime })^{\prime }$ and $\hat{Q}%
\equiv \frac{1}{NT}X^{\prime }{\mathbb{M}}_{\mathbf{D}}X=\frac{1}{NT}\ddot{X}%
^{\prime }\ddot{X}.$ Define 
\begin{eqnarray*}
\text{\ }\dot{X} &\equiv &{\mathbb{({\mathbb{I}}_{N}\otimes M}}_{\mathbf{%
\iota }_{T}})X=(\mathbf{\dot{x}}_{1}^{\prime },...,\mathbf{\dot{x}}%
_{N}^{\prime })^{\prime },\text{ } \\
\mathbf{\dot{U}} &\mathbf{\equiv }&{\mathbb{({\mathbb{I}}_{N}\otimes M}}_{%
\mathbf{\iota }_{T}})\mathbf{U}=(\mathbf{\dot{u}}_{1}^{\prime },...,\mathbf{%
\dot{u}}_{N}^{\prime })^{\prime },\text{ and} \\
\mathbf{\dot{X}} &\equiv &{\mathbb{({\mathbb{I}}_{N}\otimes M}}_{\mathbf{%
\iota }_{T}})\mathbf{X}=\text{bdiag}(\mathbf{\dot{x}}_{1},...,\mathbf{\dot{x}%
}_{N}),
\end{eqnarray*}%
where $\mathbf{\dot{x}}_{i}=(\dot{x}_{i1},...,\dot{x}_{iT})^{\prime }$ and $%
\mathbf{\dot{u}}_{t}=(\dot{u}_{i1},...,\dot{u}_{iT})^{\prime }.$ Let $\dot{F}%
={\mathbb{M}}_{\mathbf{\iota }_{T}}F=(\dot{f}_{1},...,\dot{f}_{T})^{\prime }$
and $\mathbf{\dot{U}}^{\ddagger }=\mathbf{\dot{U}}+(\mathbb{I}_{N}\otimes 
\dot{F})\mathbf{\lambda +\dot{X}\eta }=(\mathbf{\dot{u}}_{1}^{\ddagger
\prime },...,\mathbf{\dot{u}}_{N}^{\ddagger \prime })^{\prime },$ where $%
\mathbf{\dot{u}}_{i}^{\ddagger }=(\dot{u}_{i1}^{\ddagger },...,\dot{u}%
_{iT}^{\ddagger })^{\prime }$ and $\dot{u}_{it}^{\ddagger }=\dot{u}_{it}+%
\dot{f}_{t}^{\prime }\lambda _{i}+\dot{x}_{it}^{\prime }\eta _{i}.$ Noting
that $\mathbf{Y}=X\beta ^{0}+\mathbf{D\alpha }+\mathbf{U}^{\ddagger }\ $with 
$\mathbf{U}^{\ddagger }=\mathbf{U}+(\mathbb{I}_{N}\otimes F)\mathbf{\lambda
+X\eta ,}\ $we have 
\begin{equation*}
X^{\prime }{\mathbb{M}}_{\mathbf{D}}\mathbf{Y=}X^{\prime }{\mathbb{M}}_{%
\mathbf{D}}\left[ X\beta ^{0}+\mathbf{D\alpha }+\mathbf{U}^{\ddagger }\right]
=X^{\prime }{\mathbb{M}}_{\mathbf{D}}X\beta ^{0}+X^{\prime }{\mathbb{M}}_{%
\mathbf{D}}{\mathbb{({\mathbb{I}}}}_{N}{\mathbb{\otimes M}}_{\mathbf{\iota }%
_{T}})\mathbf{U}^{\ddagger }=\ddot{X}^{\prime }\ddot{X}\beta ^{0}+\ddot{X}%
^{\prime }\mathbf{\dot{U}}^{\ddagger }.
\end{equation*}%
It follows that 
\begin{equation*}
\sqrt{N}(\hat{\beta}^{P}-\beta ^{0})=\left( \frac{1}{NT}X^{\prime }{\mathbb{M%
}}_{\mathbf{D}}X\right) ^{-1}\frac{1}{NT}X^{\prime }{\mathbb{M}}_{\mathbf{D}}%
\mathbf{Y}-\beta ^{0}=\frac{1}{\sqrt{N}T}\hat{Q}^{-1}\ddot{X}^{\prime }%
\mathbf{\dot{U}}^{\ddagger }=\frac{1}{\sqrt{N}T}\hat{Q}^{-1}\sum_{i=1}^{N}%
\mathbf{\ddot{x}}_{i}^{\prime }\mathbf{\dot{u}}_{i}^{\ddagger }.
\end{equation*}%
Recall that $\mathcal{F}_{0}^{P}=\sigma \left( \{\mathbf{\dot{x}}%
_{i}\}_{1\leq i\leq \infty },F\right) ,$ $\mathcal{F}_{N,0}^{P}=\sigma (\{%
\mathbf{\dot{x}}_{i}\}_{1\leq i\leq N},F)$ and $\mathcal{F}_{N,j}^{P}=\sigma
(\mathcal{F}_{N,0}^{P},\left\{ (\mathbf{u}_{i},\lambda _{i},\eta
_{i})\right\} _{i=1}^{j})$ for $j\in \left[ N\right] .$ Note that $\mathbf{%
\ddot{x}}_{j},$ $\mathbf{\dot{x}}_{j},$ and $F$ are measurable with respect
to $\mathcal{F}_{N,j}^{P}$ for any $j.$ Then under Assumption \ref{Assmp4.1}%
(i), 
\begin{equation*}
E\left( \mathbf{\ddot{x}}_{j}^{\prime }\mathbf{\dot{u}}_{j}^{\ddag }|%
\mathcal{F}_{N,j-1}\right) =\mathbf{\ddot{x}}_{j}^{\prime }\left[ E\left( 
\mathbf{\dot{u}}_{j}|\mathcal{F}_{N,j-1}\right) +\dot{F}\cdot E\left(
\lambda _{j}|\mathcal{F}_{N,j-1}\right) +\mathbf{\dot{x}}_{j}E\left( \eta
_{j}|\mathcal{F}_{N,j-1}\right) \right] =0.
\end{equation*}%
So $\{\frac{1}{T}\mathbf{\ddot{x}}_{j}^{\prime }\mathbf{\dot{u}}%
_{j}^{\ddagger },\mathcal{F}_{N,j}^{P}\}$ forms a m.d.s. Let $\kappa $ be an
arbitrary $K_{x}\times 1$ nonrandom vector with $\left\Vert \kappa
\right\Vert =1.$ By the Cram\'{e}r-Wold device and the stable martingale CLT
(e.g., Corollary 3.19(iii) and Theorem 6.1 in \cite{hausler2015stable}), it
suffices to show that%
\begin{eqnarray}
\sum_{j=1}^{N}E\left[ \left\Vert N^{-1/2}T^{-1}\mathbf{\ddot{x}}_{j}^{\prime
}\mathbf{\dot{u}}_{j}^{\ddagger }\right\Vert ^{2+\delta }|\mathcal{F}%
_{N,j-1}^{P}\right] &=&o_{p}\left( 1\right) \text{ and }  \label{mtg_CLT2a}
\\
\sum_{j=1}^{N}E\left[ \left( \kappa ^{\prime }N^{-1/2}T^{-1}\mathbf{\ddot{x}}%
_{j}^{\prime }\mathbf{\dot{u}}_{j}^{\ddag }\right) ^{2}|\mathcal{F}%
_{N,j-1}^{P}\right] -\kappa ^{\prime }\Omega ^{P}\kappa &=&o_{p}\left(
1\right) ,  \label{mtg_CLT2b}
\end{eqnarray}%
where (\ref{mtg_CLT2a}) implies the Lindeberg condition (CLB) in Theorem 6.1
of \cite{hausler2015stable}. Noting that under Assumption \ref{Assmp4.1} 
\begin{eqnarray*}
E_{\mathcal{F}_{N,j-1}^{P}}\left[ \left\Vert \mathbf{\dot{u}}_{j}^{\ddagger
}\right\Vert ^{2+\delta }\right] &\lesssim &E_{\mathcal{F}%
_{N,j-1}^{P}}[\left\Vert \mathbf{\dot{u}}_{j}\right\Vert ^{2+\delta }]+E_{%
\mathcal{F}_{N,j-1}^{P}}[\left\Vert \dot{F}\lambda \right\Vert ^{2+\delta
}]+E_{\mathcal{F}_{N,j-1}^{P}}[\left\Vert \mathbf{\dot{x}}_{j}\eta
_{j}\right\Vert ^{2+\delta }] \\
&\lesssim &E_{\mathcal{F}_{N,j-1}^{P}}[\left\Vert \mathbf{\dot{u}}%
_{j}\right\Vert ^{2+\delta }]+\left\Vert \dot{F}\right\Vert ^{2+\delta }E_{%
\mathcal{F}_{N,j-1}^{P}}[\left\Vert \lambda \right\Vert ^{2+\delta
}]+\left\Vert \mathbf{\dot{x}}_{j}\right\Vert ^{2+\delta }E_{\mathcal{F}%
_{N,j-1}^{P}}[\left\Vert \eta _{j}\right\Vert ^{2+\delta }] \\
&=&E_{\mathcal{F}_{N,0}^{P}}[\left\Vert \mathbf{\dot{u}}_{j}\right\Vert
^{2+\delta }]+\left\Vert \dot{F}\right\Vert ^{2+\delta }E_{\mathcal{F}%
_{N,0}^{P}}[\left\Vert \lambda \right\Vert ^{2+\delta }]+\left\Vert \mathbf{%
\dot{x}}_{j}\right\Vert ^{2+\delta }E_{\mathcal{F}_{N,0}^{P}}[\left\Vert
\eta _{j}\right\Vert ^{2+\delta }]\lesssim 1+\left\Vert \mathbf{\dot{x}}%
_{j}\right\Vert ^{2+\delta },
\end{eqnarray*}%
we have%
\begin{eqnarray*}
\sum_{j=1}^{N}E_{\mathcal{F}_{N,j-1}^{P}}\left[ \left\Vert N^{-1/2}T^{-1}%
\mathbf{\ddot{x}}_{j}^{\prime }\mathbf{\dot{u}}_{j}^{\ddagger }\right\Vert
^{2+\delta }\right] &\lesssim &\frac{1}{N^{1+\delta /2}}\sum_{j=1}^{N}\left%
\Vert \mathbf{\ddot{x}}_{j}\right\Vert ^{2+\delta }E_{\mathcal{F}%
_{N,j-1}^{P}}\left[ \left\Vert \mathbf{\dot{u}}_{j}^{\ddagger }\right\Vert
^{2+\delta }\right] \\
&\lesssim &\frac{1}{N^{1+\delta /2}}\sum_{j=1}^{N}\left\Vert \mathbf{\ddot{x}%
}_{j}\right\Vert ^{2+\delta }\left( 1+\left\Vert \mathbf{\dot{x}}%
_{j}\right\Vert ^{2+\delta }\right) =O_{p}(N^{-\delta /2}).
\end{eqnarray*}%
Then (\ref{mtg_CLT2a}) holds. For the condition in (\ref{mtg_CLT2b}), we
have by Assumption \ref{Assmp4.1}(i) and Assumption \ref{Assmp4.2}(i) 
\begin{eqnarray*}
\sum_{j=1}^{N}E\left( \left( \kappa ^{\prime }N^{-1/2}T^{-1}\mathbf{\ddot{x}}%
_{j}^{\prime }\mathbf{\dot{u}}_{j}^{\ddagger }\right) ^{2}|\mathcal{F}%
_{N,j-1}^{P}\right) &=&\kappa ^{\prime }\frac{1}{NT^{2}}\sum_{j=1}^{N}%
\mathbf{\ddot{x}}_{j}^{\prime }E\left[ \mathbf{\dot{u}}_{j}^{\ddagger }%
\mathbf{\dot{u}}_{j}^{\ddagger \prime }|\mathcal{F}_{N,j-1}^{P}\right] 
\mathbf{\ddot{x}}_{j}\kappa \\
&=&\kappa ^{\prime }\frac{1}{NT^{2}}\sum_{j=1}^{N}\mathbf{\ddot{x}}%
_{j}^{\prime }E\left[ \mathbf{\dot{u}}_{j}^{\ddagger }\mathbf{\dot{u}}%
_{j}^{\ddagger \prime }|\mathcal{F}_{N,0}^{P}\right] \mathbf{\ddot{x}}%
_{j}\kappa \overset{p}{\rightarrow }\kappa ^{\prime }\Omega ^{P}\kappa .
\end{eqnarray*}%
Then by Theorem 6.1 in \cite{hausler2015stable}, 
\begin{equation*}
\frac{1}{\sqrt{N}}\sum_{j=1}^{N}\frac{1}{T}\mathbf{\ddot{x}}_{j}^{\prime }%
\mathbf{\dot{u}}_{j}^{\ddagger }\rightarrow (\Omega ^{P})^{1/2}\mathbb{Z}%
\text{ \ }\mathcal{F}_{\infty }^{P}\text{-stably,}
\end{equation*}%
where $\mathcal{F}_{\infty }^{P}$ denote the minimal sigma-field generated
by $\cup _{N=1}^{\infty }\mathcal{F}_{N,N}^{P},\ \mathbb{Z}\backsim N(0,%
\mathbb{I}_{K_{x}}),$ and $\mathbb{Z}$ is independent of\ $\mathcal{F}%
_{\infty }^{P}$. Since $\mathcal{F}_{0}^{P}\subset \mathcal{F}_{\infty
}^{P}, $ the above stable convergence implies that $\frac{1}{\sqrt{N}}%
\sum_{j=1}^{N}\frac{1}{T}\mathbf{\ddot{x}}_{j}^{\prime }\mathbf{\dot{u}}%
_{j}^{\ddagger }\rightarrow (\Omega ^{P})^{1/2}\mathbb{Z}$ $\mathcal{F}%
_{0}^{P}$-stably. By Assumption \ref{Assmp4.2}(i), $\hat{Q}=\frac{1}{NT}%
\sum_{i=1}^{N}\mathbf{\ddot{x}}_{i}^{\prime }\mathbf{\ddot{x}}_{i}\overset{p}%
{\rightarrow }Q$ which is positive definite a.s.. By the fact that $Q$ is
measurable with respect to $\mathcal{F}_{0}^{P}$ and Theorem 3.18(b)-(c) in 
\cite{hausler2015stable}, we have 
\begin{equation*}
\sqrt{N}(\hat{\beta}^{P}-\beta ^{0})=\frac{1}{\sqrt{N}}\hat{Q}%
^{-1}\sum_{i=1}^{N}\frac{1}{T}\mathbf{\ddot{x}}_{i}^{\prime }\mathbf{\dot{u}}%
_{i}^{\ddagger }\rightarrow (Q^{-1}\Omega ^{P}Q^{-1})^{1/2}\mathbb{Z}\text{
\ }\mathcal{F}^{P}\text{-stably.}
\end{equation*}%
This completes the proof of the theorem. ${\tiny \blacksquare }$

\subsection{Proof of \textbf{Theorem \protect\ref{Thm4.2}} \label{SecA.4}}

(i) Let $\mathbf{U}^{\ddagger \left( -i\right) }$ denote the leave-one-out
version of $\mathbf{U}^{\ddagger }$ by leaving the observation for the $i$th
cross-section unit out. Let $\breve{X}^{\left( -i\right) }$ be the augmented
version of $X^{\left( -i\right) }:$%
\begin{equation*}
\left[ \breve{X}^{\left( -i\right) }\right] _{j}=\left\{ 
\begin{array}{ll}
\left[ X^{\left( -i\right) }\right] _{j} & \text{if }j<i \\ 
\mathbf{0}_{T\times K_{x}} & \text{if }j=i \\ 
\left[ X^{\left( -i\right) }\right] _{j-1}\text{ \ \ } & \text{if }j>i%
\end{array}%
\right. ,
\end{equation*}%
where $\left[ X^{\left( -i\right) }\right] _{j}$ denotes the $j$th row block
of $X^{\left( -i\right) }$ of size $T\times K_{x}.$ Define $\mathbf{\breve{U}%
}^{\ddagger \left( -i\right) }$ analogously. Let $\mathbf{\iota }%
_{N}^{\left( -i\right) }$ be defined as before. Let $\breve{X}^{\left(
i\right) }=X-\breve{X}^{\left( -i\right) }\ $and $\mathbf{\breve{U}}%
^{\ddagger \left( i\right) }=\mathbf{U}^{\ddagger }-\mathbf{\breve{U}}%
^{\ddagger \left( -i\right) }.$ Let $\hat{Q}^{\left( -i\right) }=\frac{1}{%
N_{1}T}X^{\left( -i\right) \prime }(\mathbb{M}_{\mathbf{\iota }%
_{N_{1}}}\otimes \mathbb{M}_{\mathbf{\iota }_{T}})X^{\left( -i\right) }.$ We
make the following decomposition: 
\begin{eqnarray}
&&\frac{1}{N_{1}T}X^{\left( -i\right) \prime }(\mathbb{M}_{\mathbf{\iota }%
_{N_{1}}}\otimes \mathbb{M}_{\mathbf{\iota }_{T}})\mathbf{U}^{\ddagger
\left( -i\right) }  \notag \\
&=&\frac{1}{N_{1}T}\breve{X}^{\left( -i\right) \prime }(\mathbb{M}_{\mathbf{%
\iota }_{N}^{\left( -i\right) }}\otimes \mathbb{M}_{\mathbf{\iota }_{T}})%
\mathbf{\breve{U}}^{\ddagger \left( -i\right) }  \notag \\
&=&\frac{1}{N_{1}T}\breve{X}^{\left( -i\right) \prime }(\mathbb{M}_{\mathbf{%
\iota }_{N}}\otimes \mathbb{M}_{\mathbf{\iota }_{T}})\mathbf{\breve{U}}%
^{\ddagger \left( -i\right) }-\frac{1}{N_{1}T}\breve{X}^{\left( -i\right)
\prime }\left[ (\mathbb{P}_{\mathbf{\iota }_{N}^{\left( -i\right) }}-\mathbb{%
P}_{\mathbf{\iota }_{N}})\otimes \mathbb{M}_{\mathbf{\iota }_{T}}\right] 
\mathbf{\breve{U}}^{\ddagger \left( -i\right) }  \notag \\
&=&\frac{1}{N_{1}T}X^{\prime }(\mathbb{M}_{\mathbf{\iota }_{N}}\otimes 
\mathbb{M}_{\mathbf{\iota }_{T}})\mathbf{U}^{\ddagger }-\frac{1}{N_{1}T}%
\breve{X}^{\left( i\right) }(\mathbb{M}_{\mathbf{\iota }_{N}}\otimes \mathbb{%
M}_{\mathbf{\iota }_{T}})\mathbf{U}^{\ddagger }  \notag \\
&&-\frac{1}{N_{1}T}\breve{X}^{\left( -i\right) \prime }(\mathbb{M}_{\mathbf{%
\iota }_{N}}\otimes \mathbb{M}_{\mathbf{\iota }_{T}})\mathbf{\breve{U}}%
^{\ddagger \left( i\right) }-\frac{1}{N_{1}T}\breve{X}^{\left( -i\right) }%
\left[ (\mathbb{P}_{\mathbf{\iota }_{N}^{\left( -i\right) }}-\mathbb{P}_{%
\mathbf{\iota }_{N}})\otimes \mathbb{M}_{\mathbf{\iota }_{T}}\right] \mathbf{%
\breve{U}}^{\ddagger \left( -i\right) }  \notag \\
&\equiv &\sum_{s=1}^{4}L_{s}^{P(-i)}.  \label{Lp_decomp}
\end{eqnarray}%
Then%
\begin{equation}
\hat{\beta}^{P\left( -i\right) }-\beta ^{0}=\left[ \hat{Q}^{(-i)}\right]
^{-1}\frac{1}{N_{1}T}X^{\left( -i\right) \prime }(\mathbb{M}_{\mathbf{\iota }%
_{N_{1}}}\otimes \mathbb{M}_{\mathbf{\iota }_{T}})\mathbf{U}^{\ddag \left(
-i\right) }=\sum_{s=1}^{4}\left[ \hat{Q}^{(-i)}\right] ^{-1}L_{s}^{P(-i)}%
\equiv \sum_{s=1}^{4}B_{s}^{P\left( -i\right) }.  \label{beta_p_decomp}
\end{equation}%
By Lemma \ref{LemA.4} below, we have that uniformly in $i\in \left[ N\right]
,$ 
\begin{equation*}
\hat{\beta}^{P\left( -i\right) }-\beta ^{0}=\frac{N}{N_{1}}\left( \hat{\beta}%
^{P}-\beta ^{0}\right) -\frac{1}{N_{1}T}\hat{Q}^{-1}\mathbf{\ddot{x}}%
_{i}^{\prime }\mathbf{u}_{i}^{\ddagger }+o_{p}(N^{-1})=\frac{N}{N_{1}}\left( 
\hat{\beta}^{P}-\beta ^{0}\right) +\Delta _{i2}^{P}+o_{p}(N^{-1}),
\end{equation*}%
where $\Delta _{i2}^{P}\equiv -\frac{1}{N_{1}T}\hat{Q}^{-1}\mathbf{\ddot{x}}%
_{i}^{\prime }\mathbf{\dot{u}}_{i}^{\ddagger }.$ Then 
\begin{equation*}
\hat{\beta}^{P\left( -i\right) }-\frac{1}{N}\sum_{j=1}^{N}\hat{\beta}%
^{P\left( -j\right) }=\Delta _{i2}^{P}-\frac{1}{N}\sum_{j=1}^{N}\Delta
_{j2}^{P}+o_{p}(N^{-1})=\Delta _{i2}^{P}+o_{p}(N^{-1}),
\end{equation*}%
where the second equality holds by the fact that $\frac{1}{N}%
\sum_{j=1}^{N}\Delta _{j2}^{P}=\frac{1}{NN_{1}T}\sum_{i=1}^{N}\mathbf{\ddot{x%
}}_{i}^{\prime }\mathbf{\dot{u}}_{i}^{\ddagger }=o_{p}(N^{-1})$ under
Assumption \ref{Assmp4.1}. Consequently, we have%
\begin{equation}
\hat{\beta}^{P\left( -i\right) }-\frac{1}{N}\sum_{j=1}^{N}\hat{\beta}%
^{P\left( -j\right) }=-\frac{1}{N_{1}T}\hat{Q}^{-1}\mathbf{\ddot{x}}%
_{i}^{\prime }\mathbf{\dot{u}}_{i}^{\ddagger }+o_{p}(N^{-1})\text{ uniformly
in }i\in \left[ N\right] .  \label{beta_demean2}
\end{equation}%
Then by arguments as used in the proof of Theorem \ref{Thm3.1} 
\begin{eqnarray*}
\hat{\Omega}^{P} &=&N_{1}\sum_{i=1}^{N}\left[ \hat{\beta}^{P\left( -i\right)
}-\overline{\hat{\beta}}^{_{P}}\right] \left[ \hat{\beta}^{P\left( -i\right)
}-\overline{\hat{\beta}}^{_{P}}\right] ^{\prime }=\frac{1}{N_{1}T^{2}}\hat{Q}%
^{-1}\sum_{i=1}^{N}\mathbf{\ddot{x}}_{i}^{\prime }\mathbf{\dot{u}}%
_{i}^{\ddagger }\mathbf{\dot{u}}_{i}^{\ddagger \prime }\mathbf{\ddot{x}}_{i}%
\hat{Q}^{-1}+o_{p}(1) \\
&=&\frac{1}{N_{1}T^{2}}\hat{Q}^{-1}\sum_{i=1}^{N}\mathbf{\ddot{x}}%
_{i}^{\prime }E\left( \mathbf{\dot{u}}_{i}^{\ddagger }\mathbf{\dot{u}}%
_{i}^{\ddagger \prime }|\mathcal{F}_{N,0}^{P}\right) \mathbf{\ddot{x}}_{i}%
\hat{Q}^{-1}+o_{p}(1)=\Omega ^{P}+o_{p}(1).
\end{eqnarray*}

(ii) This follows from the result in part (i) and similar argument as used
in the proof of Theorem \ref{Thm3.2}. ${\tiny \blacksquare }\medskip $

\begin{lemma}
\label{LemA.4} Suppose that Assumptions \ref{Assmp4.1}--\ref{Assmp4.2} hold.
Then uniformly in $i\in \left[ N\right] ,$

(i) $B_{1}^{P\left( -i\right) }=\frac{N}{N_{1}}(\hat{\beta}^{P}-\beta
^{0})+o_{p}(N^{-1});$

(ii) $B_{2}^{P\left( -i\right) }=-\hat{Q}^{-1}\frac{1}{N_{1}T}\mathbf{\dot{x}%
}_{i}^{\prime }\mathbf{\dot{u}}_{i}^{\ddagger }+o_{p}(N^{-1});$

(iii) $B_{3}^{P\left( -i\right) }=-\frac{1}{N_{1}T}\hat{Q}^{-1}\mathbf{\ddot{%
x}}_{i}^{\prime }\mathbf{\dot{u}}_{i}^{\ddagger }+\hat{Q}^{-1}\frac{1}{N_{1}T%
}\mathbf{\dot{x}}_{i}^{\prime }\mathbf{\dot{u}}_{i}^{\ddagger
}+o_{p}(N^{-1});$

(iv) $B_{4}^{P\left( -i\right) }=o_{p}(N^{-1}).$
\end{lemma}

\subsection{Proof of \textbf{Theorem \protect\ref{Thm4.3}} \label{SecA.5}}

(i) Under Assumptions \ref{Assmp3.1}--\ref{Assmp3.3} and \ref{Assmp4.1}--\ref%
{Assmp4.2} and $\mathbb{H}_{0},$ we can follow the proof of Theorem \ref%
{Thm3.1} and show that 
\begin{eqnarray*}
\sqrt{N}(\hat{\beta}^{MG}-\hat{\beta}^{P}) &=&\sqrt{N}(\hat{\beta}%
^{MG}-\beta ^{0})-\sqrt{N}(\hat{\beta}^{P}-\beta ^{0}) \\
&=&\frac{1}{\sqrt{N}}\sum_{i=1}^{N}\left( \chi _{i}^{\prime }\mathbf{u}%
_{i}^{\dagger }+\eta _{i}\right) -\hat{Q}^{-1}\frac{1}{\sqrt{N}}%
\sum_{i=1}^{N}\mathbf{\ddot{x}}_{i}^{\prime }\mathbf{\dot{u}}_{i}^{\ddag
}+o_{p}\left( 1\right) \\
&=&\frac{1}{\sqrt{N}}\sum_{i=1}^{N}\left[ \chi _{i}^{\prime }\mathbf{u}%
_{i}^{\dagger }+\eta _{i}-Q^{-1}\mathbf{\ddot{x}}_{i}^{\prime }\mathbf{\dot{u%
}}_{i}^{\ddag }\right] +o_{p}\left( 1\right) \\
&\rightarrow &\left( \Omega ^{\Delta }\right) ^{1/2}\mathbb{Z}\text{ \ }%
\mathcal{F}_{0}\text{-stably,}
\end{eqnarray*}%
where $\mathcal{F}_{0}\mathcal{=\sigma (}\{\mathbf{\dot{x}}_{i}^{o}\}_{i\geq
1},F),$ $\Omega ^{\Delta }=$plim$\frac{1}{N}\sum_{i=1}^{N}E_{F_{N}}\left\{ %
\left[ \chi _{i}^{\prime }\mathbf{u}_{i}^{\dagger }+\eta _{i}-Q^{-1}\mathbf{%
\ddot{x}}_{i}^{\prime }\mathbf{\dot{u}}_{i}^{\ddag }\right] \left[ \chi
_{i}^{\prime }\mathbf{u}_{i}^{\dagger }+\eta _{i}-Q^{-1}\mathbf{\ddot{x}}%
_{i}^{\prime }\mathbf{\dot{u}}_{i}^{\ddag }\right] ^{\prime }\right\} $ with 
$\mathcal{F}_{N}\mathcal{=\sigma (}\{\mathbf{\dot{x}}_{i}^{o}\}_{i\geq
N},F), $ $\mathbb{Z\backsim N}\left( 0,\mathbb{I}_{K_{x}}\right) ,$ and $%
\mathbb{Z}$ is independent of $\Omega ^{\Delta }.$ By (\ref{beta_demean1})
and (\ref{beta_demean2}), we have 
\begin{equation*}
\hat{\Delta}^{\left( -i\right) }-\overline{\hat{\Delta}}=-\frac{1}{N_{1}}%
\left[ \chi _{i}^{\prime }\mathbf{u}_{i}^{\dag }+\eta _{i}-\hat{Q}^{-1}\frac{%
1}{T}\mathbf{\ddot{x}}_{i}^{\prime }\mathbf{\dot{u}}_{i}^{\ddagger }\right]
+o_{p}(N^{-1}).
\end{equation*}%
Then 
\begin{eqnarray*}
\hat{\Omega}^{\Delta } &=&N_{1}\dsum\limits_{i=1}^{N}\left( \hat{\Delta}%
^{\left( -i\right) }-\overline{\hat{\Delta}}\right) \left( \hat{\Delta}%
^{\left( -i\right) }-\overline{\hat{\Delta}}\right) ^{\prime } \\
&=&\frac{1}{N_{1}}\dsum\limits_{i=1}^{N}\left( \chi _{i}^{\prime }\mathbf{u}%
_{i}^{\dag }+\eta _{i}-\hat{Q}^{-1}\frac{1}{T}\mathbf{\ddot{x}}_{i}^{\prime }%
\mathbf{\dot{u}}_{i}^{\ddagger }\right) \left( \chi _{i}^{\prime }\mathbf{u}%
_{i}^{\dag }+\eta _{i}-\hat{Q}^{-1}\frac{1}{T}\mathbf{\ddot{x}}_{i}^{\prime }%
\mathbf{\dot{u}}_{i}^{\ddagger }\right) ^{\prime }+o_{p}(1) \\
&=&\frac{1}{N_{1}}\dsum\limits_{i=1}^{N}\left( \chi _{i}^{\prime }\mathbf{u}%
_{i}^{\dag }+\eta _{i}-Q^{-1}\frac{1}{T}\mathbf{\ddot{x}}_{i}^{\prime }%
\mathbf{\dot{u}}_{i}^{\ddagger }\right) \left( \chi _{i}^{\prime }\mathbf{u}%
_{i}^{\dag }+\eta _{i}-Q^{-1}\frac{1}{T}\mathbf{\ddot{x}}_{i}^{\prime }%
\mathbf{\dot{u}}_{i}^{\ddagger }\right) ^{\prime }+o_{p}(1)\overset{p}{%
\rightarrow }\Omega ^{\Delta }.
\end{eqnarray*}%
Using arguments as used in the proof of Theorem \ref{Thm3.2}(ii), we can
show that under $\mathbb{H}_{0},$ 
\begin{equation*}
\left( \hat{\Omega}^{\Delta }\right) ^{-1/2}\sqrt{N}(\hat{\beta}^{MG}-\hat{%
\beta}^{P})\rightarrow \mathbb{Z}\text{ }\mathcal{F}_{0}\text{-mixing as }%
N\rightarrow \infty .
\end{equation*}%
It follows that $J_{N}=N(\hat{\beta}^{MG}-\hat{\beta}^{P})^{\prime }\left[ 
\hat{\Omega}^{\Delta }\right] ^{-1}(\hat{\beta}^{MG}-\hat{\beta}^{P})\overset%
{d}{\rightarrow }\chi ^{2}\left( K_{x}\right) $ under $\mathbb{H}_{0}$ by
the continuous mapping theorem.

(ii) Under $\mathbb{H}_{1},$ we have%
\begin{equation*}
\sqrt{N}(\hat{\beta}^{MG}-\hat{\beta}^{P})=\sqrt{N}\left( \beta ^{0}-\beta
^{P}\right) +\sqrt{N}(\hat{\beta}^{MG}-\beta ^{0})-\sqrt{N}(\hat{\beta}%
^{P}-\beta ^{P})=\sqrt{N}\left( \beta ^{0}-\beta ^{P}\right) +o_{p}(\sqrt{N})
\end{equation*}%
and $\hat{\Omega}^{\Delta }\overset{p}{\rightarrow }\mathring{\Omega}%
^{\Delta }>0.$ Then%
\begin{eqnarray*}
J_{N} &=&N(\hat{\beta}^{MG}-\hat{\beta}^{P})^{\prime }\left[ \hat{\Omega}%
^{\Delta }\right] ^{-1}(\hat{\beta}^{MG}-\hat{\beta}^{P})^{\prime } \\
&=&\left[ \sqrt{N}\left( \beta ^{0}-\beta ^{P}\right) +o_{p}(\sqrt{N})\right]
\left[ \left( \mathring{\Omega}^{\Delta }\right) ^{-1}+o_{p}\left( 1\right) %
\right] \left[ \sqrt{N}\left( \beta ^{0}-\beta ^{P}\right) ^{\prime }+o_{p}(%
\sqrt{N})\right] \\
&=&N\left( \beta ^{0}-\beta ^{P}\right) \left( \mathring{\Omega}^{\Delta
}\right) ^{-1}\left( \beta ^{0}-\beta ^{P}\right) ^{\prime }+o_{p}\left(
N\right) \gtrsim \left[ \mu _{\max }\left( \mathring{\Omega}^{\Delta
}\right) \right] ^{-1}N\left\Vert \beta ^{0}-\beta ^{P}\right\Vert ^{2}.
\end{eqnarray*}%
Then $P\left( J_{N}>c_{N}\right) \rightarrow 1$ for any $c_{N}=o\left(
N\right) .$ ${\tiny \blacksquare }$

\section{Proofs of the Technical Lemmas \label{SecS1}}

In this section, we prove the technical lemmas in Section \ref{SecA}. During
the proof we also state and prove some supplementary lemmas.

\subsection{Proof of Lemma \protect\ref{LemA.1} \label{SecS1.1}}

\noindent \textbf{Proof of Lemma \ref{LemA.1}. \ }

Recall that $\hat{\omega}_{j}=\sum_{i=1}^{N}\left[ \mathbf{\hat{Q}}^{-1}%
\right] _{ij},$ where $\mathbf{\hat{Q}}=\frac{1}{T}\mathbf{X}^{\prime }{%
\mathbb{M}}_{\mathbf{D}}\mathbf{X}=\frac{1}{T}\mathbf{X}^{o\prime }{\mathbb{M%
}}_{\mathbf{D}}\mathbf{X}^{o}.$ Let $\mathbf{C}=\left( NT\right) ^{-1/2}%
\mathbf{X}^{o\prime }[\mathbf{\iota }_{N}\otimes {\mathbb{M}}_{\mathbf{\iota 
}_{T}}].$ Then 
\begin{equation*}
\mathbf{\hat{Q}}=\frac{1}{T}\mathbf{X}^{o\prime }\left( {\mathbb{I}}%
_{N}\otimes {\mathbb{M}}_{\mathbf{\iota }_{T}}\right) \mathbf{X}^{o}-\frac{1%
}{T}\mathbf{X}^{o\prime }\left( {\mathbb{P}}_{\mathbf{\iota }_{N}}\otimes {%
\mathbb{M}}_{\mathbf{\iota }_{T}}\right) \mathbf{X}^{o}=\mathbf{Q}^{\left(
d\right) }-\mathbf{CC}^{\prime },
\end{equation*}%
where $\mathbf{Q}^{\left( d\right) }=\frac{1}{T}\mathbf{X}^{o\prime }\left( {%
\mathbb{I}}_{N}\otimes {\mathbb{M}}_{\mathbf{\iota }_{T}}\right) \mathbf{X}%
^{o}$ is a block diagonal matrix with\ $\left( i,i\right) $th block given by 
$\tilde{q}_{ii}\equiv \left[ \mathbf{Q}^{\left( d\right) }\right] _{ii}=%
\frac{1}{T}\mathbf{x}_{i}^{o\prime }{\mathbb{M}}_{\mathbf{\iota }_{T}}%
\mathbf{x}_{i}^{o}.$ By Sherman-Morrison Woodbury matrix identity (e.g.,
Fact 6.4.31 in \cite{bernstein2005}), 
\begin{equation}
\mathbf{\hat{Q}}^{-1}=(\mathbf{Q}^{d})^{-1}+(\mathbf{Q}^{d})^{-1}\mathbf{C}%
\left( \mathbb{I}_{T}+\mathbf{C}^{\prime }\mathbf{C}\right) ^{-1}\mathbf{C}%
^{\prime }(\mathbf{Q}^{d})^{-1}.  \label{Qa_inverse}
\end{equation}%
Then 
\begin{eqnarray}
\hat{\omega}_{j} &=&\sum_{i=1}^{N}\left[ \mathbf{\hat{Q}}^{-1}\right]
_{ij}=\sum_{i=1}^{N}\left[ (\mathbf{Q}^{d})^{-1}\right] _{ij}+\sum_{i=1}^{N}%
\left[ (\mathbf{Q}^{d})^{-1}\mathbf{C}\left( \mathbb{I}_{T}+\mathbf{C}%
^{\prime }\mathbf{C}\right) ^{-1}\mathbf{C}^{\prime }(\mathbf{Q}^{d})^{-1}%
\right] _{ij}  \notag \\
&=&\tilde{q}_{jj}^{-1}+\sum_{i=1}^{N}\tilde{q}_{ii}^{-1}\left[ \mathbf{C}%
\left( \mathbb{I}_{T}+\mathbf{C}^{\prime }\mathbf{C}\right) ^{-1}\mathbf{C}%
^{\prime }\right] _{ij}\tilde{q}_{jj}^{-1}.  \label{Qa_summation}
\end{eqnarray}%
Write the $\left( NK_{x}\times T\right) $ matrix $\mathbf{C}$ as $\mathbf{C}%
=\left( \mathbf{c}_{1}^{\prime },...,\mathbf{c}_{N}^{\prime }\right)
^{\prime }$ where $\mathbf{c}_{i}=\left( NT\right) ^{-1/2}\mathbf{x}%
_{i}^{o\prime }{\mathbb{M}}_{\mathbf{\iota }_{T}}$ is a $K_{x}\times T$
matrix with rank $K_{x}$ uniformly in $i\in \left[ N\right] $ w.p.a.1 under
Assumption \ref{Assmp3.2}(i). Then by H\"{o}lder inequality and Assumptions %
\ref{Assmp3.1}(iv) and \ref{Assmp3.2}(i), 
\begin{eqnarray}
&&\sum_{i=1}^{N}\left\Vert \tilde{q}_{ii}^{-1}\left[ \mathbf{C}\left( 
\mathbb{I}_{T}+\mathbf{C}^{\prime }\mathbf{C}\right) ^{-1}\mathbf{C}^{\prime
}\right] _{ij}\right\Vert  \notag \\
&=&\sum_{i=1}^{N}\left\Vert \tilde{q}_{ii}^{-1}\right\Vert \left\Vert 
\mathbf{c}_{i}\left( \mathbb{I}_{T}+\mathbf{C}^{\prime }\mathbf{C}\right)
^{-1}\mathbf{c}_{j}^{\prime }\right\Vert \leq
K_{x}^{1/2}\sum_{i=1}^{N}\left\Vert \tilde{q}_{ii}^{-1}\right\Vert
\left\Vert \mathbf{c}_{i}\left( \mathbb{I}_{T}+\mathbf{C}^{\prime }\mathbf{C}%
\right) ^{-1}\mathbf{c}_{j}^{\prime }\right\Vert _{\text{sp}}  \notag \\
&\lesssim &\left\Vert \mathbf{c}_{j}\right\Vert \sum_{i=1}^{N}\left\Vert 
\tilde{q}_{ii}^{-1}\right\Vert \left\Vert \mathbf{c}_{i}\right\Vert \leq
\left( NT\right) ^{-1}\left\Vert \mathbf{x}_{j}^{o}\right\Vert
\sum_{i=1}^{N}\left\Vert \tilde{q}_{ii}^{-1}\right\Vert \left\Vert \mathbf{x}%
_{i}^{o}\right\Vert  \notag \\
&\lesssim &T^{-1/2}\left\Vert \mathbf{x}_{j}^{o}\right\Vert \left\{ \frac{1}{%
N}\sum_{i=1}^{N}\left\Vert \tilde{q}_{ii}^{-1}\right\Vert ^{(4+2\delta
)/(3+2\delta )}\right\} ^{(3+2\delta )/(4+2\delta )}\left\{ \frac{1}{N}%
\sum_{i=1}^{N}\left\Vert T^{-1/2}\mathbf{x}_{i}^{o}\right\Vert ^{4+2\delta
}\right\} ^{1/(4+2\delta )}  \notag \\
&\lesssim &T^{-1/2}\left\Vert \mathbf{x}_{j}^{o}\right\Vert ,  \label{CC1}
\end{eqnarray}%
where the first inequality holds by the fact that $\left\Vert A\right\Vert
\leq \sqrt{\text{rank}\left( A\right) }\left\Vert A\right\Vert _{\text{sp}}$%
. Then by Assumptions \ref{Assmp3.1}(iii) and \ref{Assmp3.2}(i), 
\begin{eqnarray}
\left\Vert \hat{\omega}_{j}\right\Vert &\leq &\left\Vert \tilde{q}%
_{jj}^{-1}\right\Vert +\left\Vert \sum_{i=1}^{N}\tilde{q}_{ii}^{-1}\left[ 
\mathbf{C}\left( \mathbb{I}_{T}+\mathbf{C}^{\prime }\mathbf{C}\right) ^{-1}%
\mathbf{C}^{\prime }\right] _{ij}\tilde{q}_{jj}^{-1}\right\Vert  \notag \\
&\leq &\left\Vert \tilde{q}_{jj}^{-1}\right\Vert \left\{
1+\sum_{i=1}^{N}\left\Vert \tilde{q}_{ii}^{-1}\left[ \mathbf{C}\left( 
\mathbb{I}_{T}+\mathbf{C}^{\prime }\mathbf{C}\right) ^{-1}\mathbf{C}^{\prime
}\right] _{ij}\right\Vert \right\}  \notag \\
&\leq &\left\Vert \tilde{q}_{jj}^{-1}\right\Vert \left\{
1+T^{-1/2}\left\Vert \mathbf{x}_{j}^{o}\right\Vert \right\}
\label{omega_hat1}
\end{eqnarray}%
and 
\begin{eqnarray*}
\frac{1}{N}\sum_{j=1}^{N}\left\Vert \hat{\omega}_{j}\right\Vert ^{(4+2\delta
)/(3+2\delta )} &\leq &\frac{1}{N}\sum_{j=1}^{N}\left\Vert \tilde{q}%
_{jj}^{-1}\right\Vert ^{(4+2\delta )/(3+2\delta )}+\frac{1}{N}\sum_{j=1}^{N}%
\left[ \left\Vert \tilde{q}_{jj}^{-1}\right\Vert \left\Vert \mathbf{x}%
_{j}^{o}\right\Vert \right] ^{(4+2\delta )/(3+2\delta )} \\
&\lesssim &1+\left\{ \frac{1}{N}\sum_{j=1}^{N}\left\Vert \tilde{q}%
_{jj}^{-1}\right\Vert ^{2}\right\} ^{(2+\delta )/(3+2\delta )}\left\{ \frac{1%
}{N}\sum_{j=1}^{N}\left[ \left\Vert \mathbf{x}_{j}^{o}\right\Vert \right]
^{(4+2\delta )/(1+\delta )}\right\} ^{(1+\delta )/(3+2\delta )} \\
&=&O_{p}\left( 1\right) .
\end{eqnarray*}
This completes the proof of the lemma. ${\tiny \blacksquare }$

\subsection{\textbf{Proof of Lemma \protect\ref{LemA.2}}}

To prove Lemma \ref{LemA.2}\textbf{, }we state and prove a technical lemma.

\begin{lemma}
\label{LemS.1} Suppose that Assumptions \ref{Assmp3.1}--\ref{Assmp3.2} hold.
Then

(i) $\max_{j}\left\Vert \left[ \mathbf{C}\right] _{j}\right\Vert \lesssim
\max_{j}N^{-1/2}||\mathbf{x}_{j}^{o}||=O_{p}(N^{-1/2+1/(4+2\delta )}),$ $%
\max_{j}\sum_{t}\left\Vert [\mathbf{C}]_{jt}\right\Vert
=O_{p}(N^{-1/2+1/(4+2\delta )}),$ and $\sum_{j}\left\Vert [\mathbf{C}%
]_{j}\right\Vert =O_{p}(N^{1/2});$

(ii) $\max_{s}\sum_{t}\left\Vert \left[ \mathbf{A}^{-1}\right]
_{ts}\right\Vert =O_{p}(1),$ $\max_{i,s}\sum_{t}\left\Vert \left[ (\mathbf{A}%
^{(-i)})^{-1}\right] _{ts}\right\Vert =O_{p}(1),$ $\max_{i,t,s}\left\Vert %
\left[ \mathbf{A}^{(-i)}-\mathbf{A}\right] _{t,s}\right\Vert
=O_{p}(N^{-1+1/(2+\delta )}),$ $\max_{i,t}\sum_{s}\left\Vert \left[ \mathbf{A%
}^{(-i)}-\mathbf{A}\right] _{t,s}\right\Vert =O_{p}(N^{-1+1/(2+\delta )}),$
and $\max_{i}\left\Vert \mathbf{A}^{(-i)}-\mathbf{A}\right\Vert _{\text{sp}%
}= $ $O_{p}(N^{-1+1/(2+\delta )});$

(iii) $\max_{i}\left\Vert \mathbb{A}^{\left( -i\right) }-\mathbb{A}%
\right\Vert _{\text{sp}}=O_{p}(N^{-1+1/(2+\delta )}),$ $\max_{i,t,s}\left%
\Vert [\mathbb{A}^{\left( -i\right) }-\mathbb{A]}_{ts}\right\Vert
=O_{p}(N^{-1+1/(2+\delta )});$ and $\max_{t}$ $\sum_{s}||[\mathbb{A}^{\left(
-i\right) }$ $-\mathbb{A]}_{t,s}||=O_{p}(N^{-1+1/(2+\delta )});$

(iv) $\frac{1}{T}\sum_{j=1}^{N}\mathbf{x}_{i}^{o\prime }{\mathbb{M}}_{%
\mathbf{\iota }_{T}}\mathbf{u}_{j}^{\dag }m_{ij}=\frac{1}{T}\mathbf{x}%
_{i}^{o\prime }{\mathbb{M}}_{\mathbf{\iota }_{T}}\mathbf{u}_{i}^{\dag
}+O_{p}(N^{-1/2+1/(4+2\delta )})=O_{p}(N^{1/(2+\delta )})\ $uniformly in $%
i\in \left[ N\right] ;$

(v) $\max_{s}||\frac{1}{T}\sum_{k}[\mathbf{C}^{\prime }]_{sk}\tilde{q}%
_{kk}^{-1}\sum_{l}\mathbf{x}_{k}^{o\prime }{\mathbb{M}}_{\mathbf{\iota }_{T}}%
\mathbf{u}_{l}^{\dag }m_{kl}||=O_{p}\left( N^{1/(4+2\delta )}\right) ;$

(vi) $\max_{i}\left\Vert {\mathbb{P}}_{\mathbf{\iota }_{N}}-{\mathbb{P}}_{%
\mathbf{\iota }_{N}^{\left( -i\right) }}\right\Vert _{\text{sp}%
}=O_{p}(N^{-1/2});$

(vii) $\max_{i,s}\left\Vert \frac{1}{T}\sum_{j=1,j\neq i}^{N}[\mathbf{C}%
^{\prime }]_{sj}\tilde{q}_{jj}^{-1}\mathbf{x}_{j}^{o\prime }{\mathbb{M}}_{%
\mathbf{\iota }_{T}}\mathbf{u}_{i}^{\dag }[{\mathbb{P}}_{\mathbf{\iota }%
_{N}}-{\mathbb{P}}_{\mathbf{\iota }_{N}^{\left( -i\right)
}}]_{ji}\right\Vert =O_{p}(N^{-1/2+1/(4+2\delta )});$

(viii) $\max_{i}\frac{1}{N_{1}T}\left\Vert \frac{1}{N}\sum_{k=1}^{N}\mathbf{x%
}_{i}^{o\prime }{\mathbb{M}}_{\mathbf{\iota }_{T}}\mathbf{u}_{k}^{\dag
}\right\Vert =O_{p}(N^{-3/2+1/(4+\delta )})=o_{p}\left( N^{-1}\right) \ $and 
$\max_{i}\frac{1}{NN_{1}}\left\Vert \frac{1}{T}\mathbf{x}_{i}^{o\prime }{%
\mathbb{M}}_{\mathbf{\iota }_{T}}\mathbf{u}_{i}^{\dag }\right\Vert
=O_{p}\left( N^{-2+1/(2+\delta )}\right) =o_{p}\left( N^{-1}\right) .$
\end{lemma}

\noindent \textbf{Proof of Lemma \ref{LemS.1}}

(i) Recall that $\mathbf{C}=\left( NT\right) ^{-1/2}\mathbf{X}^{o\prime }[%
\mathbf{\iota }_{N}\otimes {\mathbb{M}}_{\mathbf{\iota }_{T}}].$ Under
Assumption \ref{Assmp3.1}(iii)%
\begin{equation*}
\max_{j}\left\Vert \left[ \mathbf{C}\right] _{j}\right\Vert =\frac{1}{%
(NT)^{1/2}}\max_{j}\left\Vert \mathbf{x}_{j}^{o\prime }{\mathbb{M}}_{\mathbf{%
\iota }_{T}}\right\Vert \lesssim \frac{1}{(NT)^{1/2}}\max_{j}\left\Vert 
\mathbf{x}_{j}^{o}\right\Vert =O_{p}(N^{-1/2+1/(4+2\delta )}).
\end{equation*}%
Noting that $\left\Vert \left[ \mathbf{C}\right] _{jt}\right\Vert \lesssim
(NT)^{-1/2}\sum_{s=1}^{T}\left\Vert x_{js}^{o}\right\Vert \left\vert \mathbf{%
1}\left\{ t=s\right\} -\frac{1}{T}\right\vert ,$ we have 
\begin{eqnarray*}
\max_{j}\sum_{t}\left\Vert \left[ \mathbf{C}\right] _{jt}\right\Vert
&\lesssim &\frac{1}{(NT)^{1/2}}\max_{j}\sum_{s,t=1}^{T}\left\Vert
x_{js}^{o}\right\Vert \left\vert \mathbf{1}\left\{ t=s\right\} -\frac{1}{T}%
\right\vert \lesssim \frac{1}{(NT)^{1/2}}\max_{j}\sum_{s=1}^{T}\left\Vert
x_{js}^{o}\right\Vert \\
&\lesssim &N^{-1/2}\max_{j}\left\Vert \mathbf{x}_{j}^{o}\right\Vert
=O_{p}(N^{-1/2+1/(4+2\delta )}), \\
\max_{t}\sum_{j}\left\Vert \left[ \mathbf{C}\right] _{jt}\right\Vert
&\lesssim &\frac{1}{(NT)^{1/2}}\max_{t}\sum_{j}\sum_{s=1}^{T}\left\Vert
x_{js}^{o}\right\Vert \left\vert \mathbf{1}\left\{ t=s\right\} -\frac{1}{T}%
\right\vert \\
&\lesssim &N^{-1/2}\sum_{j}\left\Vert \mathbf{x}_{j}^{o}\right\Vert
=O_{p}(N^{1/2}).
\end{eqnarray*}%
The last result also implies that $\sum_{j}\left\Vert [\mathbf{C]}%
_{j}\right\Vert =O_{p}(N^{1/2})$ as $T$ is fixed.

(ii) Recall that $\mathbf{A}=\mathbb{I}_{K_{x}T}+\mathbf{C}^{\prime }\mathbf{%
C}$ and $\mathbf{CC}^{\prime }=T^{-1}\mathbf{X}^{o\prime }[\mathbf{\iota }%
_{N}(\mathbf{\iota }_{N}^{\prime }\mathbf{\iota }_{N})^{-1/2}\otimes {%
\mathbb{M}}_{\mathbf{\iota }_{T}}][(\mathbf{\iota }_{N}^{\prime }\mathbf{%
\iota }_{N})^{-1/2}\mathbf{\iota }_{N}^{\prime }\otimes {\mathbb{M}}_{%
\mathbf{\iota }_{T}}]\mathbf{X}^{o}=T^{-1}\mathbf{X}^{o\prime }(\mathbb{P}_{%
\mathbf{\iota }_{N}}\otimes {\mathbb{M}}_{\mathbf{\iota }_{T}})\mathbf{X}^{o}%
\mathbf{.}$ Since $T$ is fixed, we have%
\begin{equation*}
\max_{s}\sum_{t}\left\Vert \left[ \mathbf{A}^{-1}\right] _{ts}\right\Vert
\lesssim \left\Vert \mathbf{A}^{-1}\right\Vert \lesssim \left\Vert \mathbf{A}%
^{-1}\right\Vert _{\text{sp}}\leq 1.
\end{equation*}%
Then $\max_{s}\sum_{t}\left\Vert \left[ \mathbf{A}^{-1}\right]
_{ts}\right\Vert =O_{p}\left( 1\right) .$ Analogously, $\sum_{t}\left\Vert %
\left[ (\mathbf{A}^{(-i)})^{-1}\right] _{ts}\right\Vert =O_{p}(1).$ Let $%
\mathbf{\iota }_{N}^{(i)}=\mathbf{\iota }_{N}-\mathbf{\iota }_{N}^{(-i)}.$
Noting that $\mathbf{C}^{\left( -i\right) }=\left( N_{1}T\right) ^{-1/2}%
\mathbf{X}^{o\left( -i\right) \prime }[\mathbf{\iota }_{N_{1}}\otimes {%
\mathbb{M}}_{\mathbf{\iota }_{T}}]$ and $\mathbf{C}=\left( NT\right) ^{-1/2}%
\mathbf{X}^{o\prime }[\mathbf{\iota }_{N}\otimes {\mathbb{M}}_{\mathbf{\iota 
}_{T}}],$ we can use $\mathbf{\iota }_{N}^{(-i)}=\mathbf{\iota }_{N}^{\prime
}-\mathbf{\iota }_{N}^{(i)}$ to make the following decomposition%
\begin{eqnarray*}
&&\mathbf{A}^{(-i)}-\mathbf{A} \\
&=&\mathbf{C}^{\left( -i\right) \prime }\mathbf{C}^{\left( -i\right) }-%
\mathbf{C}^{\prime }\mathbf{C} \\
&=&\frac{1}{N_{1}T}[\mathbf{\iota }_{N_{1}}^{\prime }\otimes {\mathbb{M}}_{%
\mathbf{\iota }_{T}}]\mathbf{X}^{o\left( -i\right) }\mathbf{X}^{o\left(
-i\right) \prime }[\mathbf{\iota }_{N_{1}}\otimes {\mathbb{M}}_{\mathbf{%
\iota }_{T}}]-\mathbf{C}^{\prime }\mathbf{C} \\
&=&\frac{1}{N_{1}T}[\mathbf{\iota }_{N}^{\prime }\otimes {\mathbb{M}}_{%
\mathbf{\iota }_{T}}]\mathbf{\breve{X}}^{o\left( -i\right) }\mathbf{\breve{X}%
}^{o\left( -i\right) \prime }[\mathbf{\iota }_{N}\otimes {\mathbb{M}}_{%
\mathbf{\iota }_{T}}]-\mathbf{C}^{\prime }\mathbf{C} \\
&=&\frac{1}{N_{1}T}(\mathbf{\iota }_{N}^{(-i)\prime }\otimes {\mathbb{M}}_{%
\mathbf{\iota }_{T}})\mathbf{X}^{o}\mathbf{X}^{o\prime }(\mathbf{\iota }%
_{N}^{(-i)}\otimes {\mathbb{M}}_{\mathbf{\iota }_{T}})-\mathbf{C}^{\prime }%
\mathbf{C} \\
&=&\left[ \frac{1}{N_{1}T}(\mathbf{\iota }_{N}^{\prime }\otimes {\mathbb{M}}%
_{\mathbf{\iota }_{T}})\mathbf{X}^{o}\mathbf{X}^{o\prime }(\mathbf{\iota }%
_{N}\otimes {\mathbb{M}}_{\mathbf{\iota }_{T}})-\mathbf{C}^{\prime }\mathbf{C%
}\right] -\frac{1}{N_{1}T}(\mathbf{\iota }_{N}^{\prime }\otimes {\mathbb{M}}%
_{\mathbf{\iota }_{T}})\mathbf{X}^{o}\mathbf{X}^{o\prime }(\mathbf{\iota }%
_{N}^{(i)}\otimes {\mathbb{M}}_{\mathbf{\iota }_{T}}) \\
&&-\frac{1}{N_{1}T}(\mathbf{\iota }_{N}^{(i)\prime }\otimes {\mathbb{M}}_{%
\mathbf{\iota }_{T}})\mathbf{X}^{o}\mathbf{X}^{o\prime }(\mathbf{\iota }%
_{N}\otimes {\mathbb{M}}_{\mathbf{\iota }_{T}})+\frac{1}{N_{1}T}(\mathbf{%
\iota }_{N}^{(i)\prime }\otimes {\mathbb{M}}_{\mathbf{\iota }_{T}})\mathbf{X}%
^{o}\mathbf{X}^{o\prime }(\mathbf{\iota }_{N}^{(i)}\otimes {\mathbb{M}}_{%
\mathbf{\iota }_{T}})\equiv \sum_{l=1}^{4}d_{l}^{\left( -i\right) }.
\end{eqnarray*}%
Clearly, $\frac{1}{N_{1}T}[\mathbf{\iota }_{N}^{\prime }\otimes {\mathbb{M}}%
_{\mathbf{\iota }_{T}}]\mathbf{XX}^{\prime }[\mathbf{\iota }_{N}\otimes {%
\mathbb{M}}_{\mathbf{\iota }_{T}}]-\mathbf{C}^{\prime }\mathbf{C}=\left( 
\frac{1}{N_{1}}-\frac{1}{N}\right) \frac{1}{T}[\mathbf{\iota }_{N}^{\prime
}\otimes {\mathbb{M}}_{\mathbf{\iota }_{T}}]\mathbf{XX}^{\prime }[\mathbf{%
\iota }_{N}\otimes {\mathbb{M}}_{\mathbf{\iota }_{T}}]=\frac{1}{N_{1}}%
\mathbf{C}^{\prime }\mathbf{C.}$ Noting that the elements of ${\mathbb{M}}_{%
\mathbf{\iota }_{T}}$ are uniformly bounded, ${\mathbb{M}}_{\mathbf{\iota }%
_{T}}$ plays a role similar to an identity matrix. In addition, 
\begin{equation*}
\left[ \mathbf{C}^{\prime }\mathbf{C}\right] _{ts}=\left( NT\right) ^{-1}[(%
\mathbf{\iota }_{N}^{\prime }\otimes {\mathbb{M}}_{\mathbf{\iota }_{T}})%
\mathbf{X}^{o}\mathbf{X}^{o\prime }(\mathbf{\iota }_{N}\otimes {\mathbb{M}}_{%
\mathbf{\iota }_{T}})]_{ts}=\left( NT\right) ^{-1}\sum_{i=1}^{N}[{\mathbb{M}}%
_{\mathbf{\iota }_{T}}\mathbf{x}_{i}^{o}\mathbf{x}_{i}^{o\prime }{\mathbb{M}}%
_{\mathbf{\iota }_{T}}]_{ts}=\left( NT\right) ^{-1}\sum_{i=1}^{N}\dot{x}%
_{it}^{o}\dot{x}_{is}^{o\prime },
\end{equation*}%
where $\dot{x}_{it}^{o}=x_{it}^{o}-\frac{1}{T}\sum_{r=1}^{T}x_{ir}^{o}.$
Then by Assumption \ref{Assmp3.1}(iii) 
\begin{equation*}
\max_{i,t,s}\left\Vert \left[ d_{l}^{\left( -i\right) }\right]
_{ts}\right\Vert =\frac{1}{N_{1}}\max_{t,s}\left\Vert \left[ \mathbf{C}%
^{\prime }\mathbf{C}\right] _{ts}\right\Vert \lesssim \frac{1}{NN_{1}T}%
\max_{t}\sum_{i=1}^{N}\left\Vert \dot{x}_{it}^{o}\right\Vert \left\Vert \dot{%
x}_{is}^{o}\right\Vert \lesssim \frac{1}{NN_{1}T}\sum_{i=1}^{N}\left\Vert 
\mathbf{\dot{x}}_{i}^{o}\right\Vert ^{2}=O_{p}(N^{-1}),
\end{equation*}%
where $\mathbf{\dot{x}}_{i}^{o}=(\dot{x}_{i1}^{o\prime },...,\dot{x}%
_{iT}^{o\prime })^{\prime }.$ Similarly, noting that $\left[ (\mathbf{\iota }%
_{N}^{\prime }\otimes {\mathbb{M}}_{\mathbf{\iota }_{T}})\mathbf{X}^{o}%
\mathbf{X}^{o\prime }(\mathbf{\iota }_{N}^{(i)}\otimes {\mathbb{M}}_{\mathbf{%
\iota }_{T}})\right] _{ts}=[(\mathbf{\iota }_{N}^{(i)\prime }\otimes {%
\mathbb{M}}_{\mathbf{\iota }_{T}})\mathbf{X}^{o}\mathbf{X}^{o\prime }$ $(%
\mathbf{\iota }_{N}^{(i)}\otimes {\mathbb{M}}_{\mathbf{\iota }_{T}})]_{ts}$ $%
=\dot{x}_{it}^{o}\dot{x}_{is}^{o\prime },$ we have%
\begin{equation*}
\max_{i,t,s}\left\Vert \left[ d_{l}^{\left( -i\right) }\right]
_{ts}\right\Vert \lesssim \frac{1}{N_{1}T}\left( \max_{i,t}\left\Vert \dot{x}%
_{it}^{o}\right\Vert \right) ^{2}\lesssim \frac{1}{N_{1}T}\left(
\max_{i}\left\Vert \mathbf{\dot{x}}_{i}^{o}\right\Vert \right)
^{2}=O_{p}(N^{-1+1/(2+\delta )})\text{ for }l=2,3,4\text{.}
\end{equation*}%
It follows that 
\begin{eqnarray*}
\max_{i,t,s}\left\Vert \left[ \mathbf{A}^{(-i)}-\mathbf{A}\right]
_{ts}\right\Vert &=&O_{p}(N^{-1+1/(2+\delta )})\text{ and} \\
\max_{i,t}\sum_{s}\left\Vert \left[ \mathbf{A}^{(-i)}-\mathbf{A}\right]
_{t,s}\right\Vert &\lesssim &\frac{1}{N_{1}NT}\sum_{s=1}^{T}\sum_{i=1}^{N}%
\left\Vert \dot{x}_{it}^{o}\right\Vert \left\Vert \dot{x}_{is}^{o}\right%
\Vert +\frac{1}{NT}\max_{i,t}\left\Vert \dot{x}_{it}^{o}\right\Vert
\sum_{s}\left\Vert \dot{x}_{is}^{o}\right\Vert \\
&\lesssim &\frac{1}{N_{1}NT}\sum_{i=1}^{N}\left\Vert \mathbf{\dot{x}}%
_{i}^{o}\right\Vert ^{2}+\frac{1}{NT}\max_{i}\left\Vert \mathbf{\dot{x}}%
_{i}^{o}\right\Vert ^{2}=O_{p}(N^{-1+1/(2+\delta )}).
\end{eqnarray*}%
By the relationship between spectral norm and absolute column/row sum norms,
we have%
\begin{equation*}
\max_{i}\left\Vert \mathbf{A}^{(-i)}-\mathbf{A}\right\Vert _{\text{sp}}\leq
\left\{ \max_{i,t}\sum_{s}\left\Vert \left[ \mathbf{A}^{(-i)}-\mathbf{A}%
\right] _{t,s}\right\Vert \right\} ^{1/2}\left\{
\max_{i,s}\sum_{t}\left\Vert \left[ \mathbf{A}^{(-i)}-\mathbf{A}\right]
_{t,s}\right\Vert \right\} ^{1/2}=O_{p}(N^{-1+1/(2+\delta )}).
\end{equation*}

(iii) Noting that $\mathbb{A}^{\left( -i\right) }-\mathbb{A}=[\mathbf{A}%
^{(-i)}]^{-1}-\mathbf{A}^{-1}=-[\mathbf{A}^{(-i)}]^{-1}(\mathbf{A}^{(-i)}-%
\mathbf{A})\mathbf{A}^{-1}$ and both $\mathbf{A}^{-1}$ and $[\mathbf{A}%
^{(-i)}]^{-1}$ have bounded spectral norm, we have by part (ii), $%
\max_{i}\left\Vert \mathbb{A}^{\left( -i\right) }-\mathbb{A}\right\Vert _{%
\text{sp}}\lesssim \max_{i}\left\Vert \mathbf{A}^{(-i)}-\mathbf{A}%
\right\Vert _{\text{sp}}=O_{p}(N^{-1+1/(2+\delta )}).$ Next, 
\begin{eqnarray*}
\max_{i,t,s}\left\Vert [\mathbb{A}^{\left( -i\right) }-\mathbb{A]}%
_{ts}\right\Vert &=&\left\Vert \sum_{r,q}\left[ [\mathbf{A}^{(-i)}]^{-1}%
\right] _{tr}\left[ \mathbf{A}^{(-i)}-\mathbf{A}\right] _{rq}\left[ \mathbf{A%
}^{-1}\right] _{qs}\right\Vert \\
&\lesssim &\max_{i,t,s}\left\Vert \left[ \mathbf{A}^{(-i)}-\mathbf{A}\right]
_{ts}\right\Vert \max_{i,t}\sum_{r}\left\Vert \left[ (\mathbf{A}^{(-i)})^{-1}%
\right] _{tr}\right\Vert \max_{s}\sum_{q}\left\Vert \left[ \mathbf{A}^{-1}%
\right] _{qs}\right\Vert \\
&=&O_{p}(N^{-1+1/(2+\delta )})O_{p}(1)O_{p}(1)=O_{p}(N^{-1+1/(2+\delta )}),%
\text{ and} \\
\max_{t}\sum_{s}\left\Vert [\mathbb{A}^{\left( -i\right) }-\mathbb{A}%
]_{t,s}\right\Vert &\lesssim &\max_{i,t}\sum_{s}\left\Vert \left[ \mathbf{A}%
^{(-i)}-\mathbf{A}\right] _{ts}\right\Vert \max_{i,t}\sum_{r}\left\Vert %
\left[ (\mathbf{A}^{(-i)})^{-1}\right] _{tr}\right\Vert
\max_{s}\sum_{q}\left\Vert \left[ \mathbf{A}^{-1}\right] _{qs}\right\Vert \\
&=&O_{p}(N^{-1+1/(2+\delta )})O_{p}(1)O_{p}(1)=O_{p}(N^{-1+1/(2+\delta )}).
\end{eqnarray*}

(iv) Noting that $m_{ij}=\mathbf{1}\left\{ i=j\right\} -\frac{1}{N},$ $\frac{%
1}{T}\sum_{j=1}^{N}\mathbf{x}_{i}^{o\prime }{\mathbb{M}}_{\mathbf{\iota }%
_{T}}\mathbf{u}_{j}^{\dag }m_{ij}=\frac{1}{T}\mathbf{x}_{i}^{o\prime }{%
\mathbb{M}}_{\mathbf{\iota }_{T}}\mathbf{u}_{i}^{\dag }-\frac{1}{NT}%
\sum_{j=1}^{N}\mathbf{x}_{i}^{o\prime }{\mathbb{M}}_{\mathbf{\iota }_{T}}%
\mathbf{u}_{j}^{\dag }.$ Under Assumption \ref{Assmp3.1}(i) and (iii), we
have 
\begin{eqnarray*}
\frac{1}{T}\left\Vert \mathbf{x}_{i}^{o\prime }{\mathbb{M}}_{\mathbf{\iota }%
_{T}}\mathbf{u}_{i}^{\dag }\right\Vert &\lesssim &\max_{i}\left\Vert \mathbf{%
x}_{i}^{o}\right\Vert \max_{i}\left\Vert \mathbf{u}_{i}^{\dag }\right\Vert
=O_{p}(N^{1/(4+2\delta )})O_{p}(N^{1/(4+2\delta )})=O_{p}(N^{1/(2+\delta )})%
\text{, and} \\
\max_{i}\left\Vert \frac{1}{NT}\sum_{j=1}^{N}\mathbf{x}_{i}^{o\prime }{%
\mathbb{M}}_{\mathbf{\iota }_{T}}\mathbf{u}_{j}^{\dag }\right\Vert &\lesssim
&\max_{i}\left\Vert \mathbf{x}_{i}^{o}\right\Vert \frac{1}{NT}\left\Vert
\sum_{j=1}^{N}\mathbf{u}_{j}^{\dag }\right\Vert =O_{p}(N^{-1/2+1/(4+2\delta
)}).
\end{eqnarray*}%
Then $\frac{1}{T}\sum_{j=1}^{N}\mathbf{x}_{i}^{o\prime }{\mathbb{M}}_{%
\mathbf{\iota }_{T}}\mathbf{u}_{j}^{\dag }m_{ij}=\frac{1}{T}\mathbf{x}%
_{i}^{o\prime }{\mathbb{M}}_{\mathbf{\iota }_{T}}\mathbf{u}_{i}^{\dag
}+O_{p}(N^{-1/2+1/(4+2\delta )})=O_{p}(N^{1/(2+\delta )})$ uniformly in $%
i\in \left[ N\right] $. $\medskip $

(v) We make the following decomposition:%
\begin{eqnarray*}
&&\frac{1}{T}\sum_{k}[\mathbf{C}^{\prime }]_{sk}\tilde{q}_{kk}^{-1}\sum_{l}%
\mathbf{x}_{k}^{o\prime }{\mathbb{M}}_{\mathbf{\iota }_{T}}\mathbf{u}%
_{l}^{\dag }m_{kl} \\
&=&\frac{1}{T}\sum_{k}[\mathbf{C}^{\prime }]_{sk}\tilde{q}_{kk}^{-1}\mathbf{x%
}_{k}^{o\prime }{\mathbb{M}}_{\mathbf{\iota }_{T}}\mathbf{u}_{k}+\frac{1}{T}%
\sum_{k}[\mathbf{C}^{\prime }]_{sk}\tilde{q}_{kk}^{-1}\mathbf{x}%
_{k}^{o\prime }{\mathbb{M}}_{\mathbf{\iota }_{T}}F\lambda _{k} \\
&&-\frac{1}{NT}\sum_{k}[\mathbf{C}^{\prime }]_{sk}\tilde{q}_{kk}^{-1}\sum_{l}%
\mathbf{x}_{k}^{o\prime }{\mathbb{M}}_{\mathbf{\iota }_{T}}\mathbf{u}_{l}-%
\frac{1}{NT}\sum_{k}[\mathbf{C}^{\prime }]_{sk}\tilde{q}_{kk}^{-1}\sum_{l}%
\mathbf{x}_{k}^{o\prime }{\mathbb{M}}_{\mathbf{\iota }_{T}}F\lambda _{l} \\
&\equiv &\tilde{d}_{1s}+\tilde{d}_{2s}-\tilde{d}_{3s}-\tilde{d}_{4s}.
\end{eqnarray*}%
Noting that $\mathbf{C}=\left( NT\right) ^{-1/2}\mathbf{X}^{o\prime }[%
\mathbf{\iota }_{N}\otimes {\mathbb{M}}_{\mathbf{\iota }_{T}}],$ we have $%
\tilde{d}_{1s}=\frac{1}{\left( NT\right) ^{1/2}T}\sum_{k}[{\mathbb{M}}_{%
\mathbf{\iota }_{T}}\mathbf{x}_{k}^{o}]_{s}\tilde{q}_{kk}^{-1}\mathbf{x}%
_{k}^{o\prime }{\mathbb{M}}_{\mathbf{\iota }_{T}}\mathbf{u}_{k}.$ Under
Assumptions \ref{Assmp3.1} and \ref{Assmp3.3}(i), 
\begin{eqnarray*}
E_{\mathcal{F}_{N,0}}\left( \tilde{d}_{1s}\right) &=&\frac{1}{\left(
NT\right) ^{1/2}T}\sum_{k}[{\mathbb{M}}_{\mathbf{\iota }_{T}}\mathbf{x}%
_{k}^{o}]_{s}\tilde{q}_{kk}^{-1}\mathbf{x}_{k}^{o\prime }{\mathbb{M}}_{%
\mathbf{\iota }_{T}}E\left( \mathbf{u}_{k}\right) =0,\text{ and} \\
\left\Vert \text{Var}_{\mathcal{F}_{N,0}}\left( \tilde{d}_{1s}\right)
\right\Vert &=&\frac{1}{NT^{2}}\left\Vert \sum_{k}[{\mathbb{M}}_{\mathbf{%
\iota }_{T}}\mathbf{x}_{k}^{o}]_{s}\tilde{q}_{kk}^{-1}\mathbf{x}%
_{k}^{o\prime }{\mathbb{M}}_{\mathbf{\iota }_{T}}E\left( \mathbf{u}_{k}%
\mathbf{u}_{k}^{\prime }\right) {\mathbb{M}}_{\mathbf{\iota }_{T}}\mathbf{x}%
_{k}^{o}\tilde{q}_{kk}^{-1}([{\mathbb{M}}_{\mathbf{\iota }_{T}}\mathbf{x}%
_{k}^{o}]_{s})^{\prime }\right\Vert \\
&\lesssim &\left\{ \frac{1}{NT^{2}}\sum_{k}\left\Vert \mathbf{x}%
_{k}^{o}\right\Vert ^{2}\left\Vert \tilde{q}_{kk}^{-1}\right\Vert
^{2}\right\} \max_{i}\left\Vert \mathbf{x}_{i}^{o}\right\Vert
^{2}=O_{p}(N^{1/(2+\delta )}).
\end{eqnarray*}%
Then $\tilde{d}_{1s}=O_{p}\left( N^{1/(4+2\delta )}\right) $ by the
conditional Chebyshev inequality. Similarly, we can show that 
\begin{eqnarray*}
\tilde{d}_{2s} &=&\frac{1}{\left( NT\right) ^{1/2}T}\sum_{k}[{\mathbb{M}}_{%
\mathbf{\iota }_{T}}\mathbf{x}_{k}^{o}]_{s}\tilde{q}_{kk}^{-1}\mathbf{x}%
_{k}^{o\prime }{\mathbb{M}}_{\mathbf{\iota }_{T}}F\lambda _{k}=O_{p}\left(
N^{1/(4+2\delta )}\right) , \\
\tilde{d}_{3s} &=&\frac{1}{\left( NT\right) ^{3/2}}\sum_{k}[{\mathbb{M}}_{%
\mathbf{\iota }_{T}}\mathbf{x}_{k}^{o}]_{sk}\tilde{q}_{kk}^{-1}\sum_{l}%
\mathbf{x}_{k}^{o\prime }{\mathbb{M}}_{\mathbf{\iota }_{T}}\mathbf{u}%
_{l}=O_{p}\left( N^{1/(4+2\delta )}\right) ,\text{ and} \\
\tilde{d}_{4s} &=&\frac{1}{\left( NT\right) ^{3/2}}\sum_{k}[{\mathbb{M}}_{%
\mathbf{\iota }_{T}}\mathbf{x}_{k}^{o}]_{sk}\tilde{q}_{kk}^{-1}\sum_{l}%
\mathbf{x}_{k}^{o\prime }{\mathbb{M}}_{\mathbf{\iota }_{T}}F\lambda
_{l}=O_{p}\left( N^{1/(4+2\delta )}\right) .
\end{eqnarray*}%
Then $\max_{s}||\frac{1}{T}\sum_{k}[\mathbf{C}^{\prime }]_{sk}\tilde{q}%
_{kk}^{-1}\sum_{l}\mathbf{x}_{k}^{o\prime }{\mathbb{M}}_{\mathbf{\iota }_{T}}%
\mathbf{u}_{l}^{\dag }m_{kl}||=O_{p}\left( N^{1/(4+2\delta )}\right) $ by
the fact that $T$ is fixed. Similarly, we can show that $||\frac{1}{T}[%
\mathbf{C}^{\prime }]_{\cdot k}\tilde{q}_{kk}^{-1}\sum_{l}\mathbf{x}%
_{k}^{o\prime }{\mathbb{M}}_{\mathbf{\iota }_{T}}\mathbf{u}_{l}^{\dag
}m_{kl}||=O_{p}\left( N^{1/(4+2\delta )}\right) .\medskip $

(vi) Note that%
\begin{eqnarray}
{\mathbb{P}}_{\mathbf{\iota }_{N}}-{\mathbb{P}}_{\mathbf{\iota }_{N}^{\left(
-i\right) }} &=&\frac{1}{N}\mathbf{\iota }_{N}\mathbf{\iota }_{N}^{\prime }-%
\frac{1}{N_{1}}\mathbf{\iota }_{N}^{\left( -i\right) }\mathbf{\iota }%
_{N}^{\left( -i\right) \prime }=\frac{1}{N_{1}}\left( \mathbf{\iota }_{N}%
\mathbf{\iota }_{N}^{\prime }-\mathbf{\iota }_{N}^{\left( -i\right) }\mathbf{%
\iota }_{N}^{\left( -i\right) \prime }\right) +\left( \frac{1}{N}-\frac{1}{%
N_{1}}\right) \mathbf{\iota }_{N}\mathbf{\iota }_{N}^{\prime }  \notag \\
&=&\frac{1}{N_{1}}\mathbf{\iota }_{N}^{\left( i\right) }\mathbf{\iota }%
_{N}^{\left( i\right) \prime }+\frac{1}{N_{1}}\mathbf{\iota }_{N}^{\left(
-i\right) }\mathbf{\iota }_{N}^{\left( i\right) \prime }+\frac{1}{N_{1}}%
\mathbf{\iota }_{N}^{\left( i\right) }\mathbf{\iota }_{N}^{\left( -i\right)
\prime }-\frac{1}{NN_{1}}\mathbf{\iota }_{N}\mathbf{\iota }_{N}^{\prime
}\equiv \sum_{l=1}^{4}\Theta _{il},  \label{P_decomp}
\end{eqnarray}%
where $\mathbf{\iota }_{N}^{\left( i\right) }\equiv \mathbf{\iota }_{N}-%
\mathbf{\iota }_{N}^{\left( -i\right) }.$ Then 
\begin{eqnarray*}
\left\Vert {\mathbb{P}}_{\mathbf{\iota }_{N}}-{\mathbb{P}}_{\mathbf{\iota }%
_{N}^{\left( -i\right) }}\right\Vert _{\text{sp}} &\lesssim &\left\Vert 
\frac{1}{N_{1}}\mathbf{\iota }_{N}^{\left( i\right) }\mathbf{\iota }%
_{N}^{\left( i\right) \prime }\right\Vert +\frac{1}{N_{1}}\left\Vert \mathbf{%
\iota }_{N}^{\left( -i\right) }\right\Vert \left\Vert \mathbf{\iota }%
_{N}^{\left( i\right) }\right\Vert +\frac{1}{NN_{1}}\left\Vert \mathbf{\iota 
}_{N}\mathbf{\iota }_{N}^{\prime }\right\Vert
=O(N^{-1})+O(N^{-1/2})+O_{p}(N^{-1}) \\
&=&O_{p}(N^{-1/2}).
\end{eqnarray*}

(vii) Using the decomposition in (\ref{P_decomp}), we define $\vartheta
_{si,l}\equiv \frac{1}{T}\sum_{j\neq i}^{N}[\mathbf{C}^{\prime }]_{sj}\tilde{%
q}_{jj}^{-1}\mathbf{x}_{j}^{o\prime }{\mathbb{M}}_{\mathbf{\iota }_{T}}%
\mathbf{u}_{i}^{\dag }\left[ \Theta _{il}\right] _{ji}$ for $l\in \left[ 4%
\right] .$ Noting that $\left[ \Theta _{i1}\right] _{ji}=0$ and $\left[
\Theta _{i3}\right] _{ji}=0$ for any $j\neq i,$ we have $\vartheta
_{si,1}=\vartheta _{si,3}=0.$ Noting that $\left[ \mathbf{\iota }%
_{N}^{\left( -i\right) }\mathbf{\iota }_{N}^{\left( i\right) \prime }\right]
_{ji}=1$ for any $j\neq i$ and $\left\Vert \left[ \mathbf{C}\right]
_{jt}\right\Vert \lesssim (NT)^{-1/2}\sum_{s=1}^{T}\left\Vert
x_{js}^{o}\right\Vert \left\vert \mathbf{1}\left\{ t=s\right\} -\frac{1}{T}%
\right\vert \lesssim N^{-1/2}\left\Vert \mathbf{x}_{j}^{o}\right\Vert ,$ we
have 
\begin{eqnarray*}
\max_{s,i}\left\Vert \vartheta _{si,2}\right\Vert &=&\max_{s,i}\frac{1}{%
N_{1}T}\left\Vert \sum_{j\neq i}^{N}[\mathbf{C}^{\prime }]_{sj}\tilde{q}%
_{jj}^{-1}\mathbf{x}_{j}^{o\prime }{\mathbb{M}}_{\mathbf{\iota }_{T}}\mathbf{%
u}_{i}^{\dag }\right\Vert \lesssim \frac{1}{N_{1}}\max_{s,i}\sum_{j\neq
i}^{N}\frac{1}{N^{1/2}}\left\Vert \mathbf{x}_{j}^{o}\right\Vert \left\Vert 
\tilde{q}_{jj}^{-1}\right\Vert \frac{1}{T}\left\Vert \mathbf{x}_{j}^{o\prime
}{\mathbb{M}}_{\mathbf{\iota }_{T}}\mathbf{u}_{i}^{\dag }\right\Vert \\
&\lesssim &\frac{1}{N^{1/2}}\left( N^{-1}\sum_{j}\left\Vert \mathbf{x}%
_{j}^{o}\right\Vert ^{2}\left\Vert \tilde{q}_{jj}^{-1}\right\Vert \right)
\max_{i}\left\Vert \mathbf{u}_{i}^{\dag }\right\Vert
=O_{p}(N^{-1/2+1/(4+2\delta )}), \\
\max_{s,i}\left\Vert \vartheta _{si,4}\right\Vert &=&\frac{1}{NN_{1}T}%
\max_{s,i}\left\Vert \sum_{j\neq i}^{N}[\mathbf{C}^{\prime }]_{sj}\tilde{q}%
_{jj}^{-1}\mathbf{x}_{j}^{o\prime }{\mathbb{M}}_{\mathbf{\iota }_{T}}\mathbf{%
u}_{i}^{\dag }\right\Vert \lesssim \frac{1}{NN_{1}T}\max_{s,i}\sum_{j\neq
i}^{N}\frac{1}{N^{1/2}}\left\Vert \mathbf{x}_{j}^{o}\right\Vert \left\Vert 
\tilde{q}_{jj}^{-1}\right\Vert \left\Vert \mathbf{x}_{j}^{o\prime }{\mathbb{M%
}}_{\mathbf{\iota }_{T}}\mathbf{u}_{i}^{\dag }\right\Vert \\
&\lesssim &\frac{1}{N^{3/2}}\left( \frac{1}{N}\sum_{j}\left\Vert \mathbf{x}%
_{j}^{o}\right\Vert ^{2}\left\Vert \tilde{q}_{jj}^{-1}\right\Vert \right)
\max_{i}\left\Vert \mathbf{u}_{i}^{\dag }\right\Vert
=O_{p}(N^{-3/2+1/(4+2\delta )}).
\end{eqnarray*}%
In sum, we have $\max_{s,i}\left\Vert \frac{1}{T}\sum_{j=1,j\neq i}^{N}[%
\mathbf{C}^{\prime }]_{sj}\tilde{q}_{jj}^{-1}\mathbf{x}_{j}^{o\prime }{%
\mathbb{M}}_{\mathbf{\iota }_{T}}\mathbf{u}_{i}^{\dag }[{\mathbb{P}}_{%
\mathbf{\iota }_{N}}-{\mathbb{P}}_{\mathbf{\iota }_{N}^{\left( -i\right)
}}]_{ji}\right\Vert =O_{p}(N^{-1/2+1/(4+2\delta )}).$

(viii) By (\ref{omega_hat1}) and Assumptions \ref{Assmp3.1}(i, iii) and \ref%
{Assmp3.2}(i) 
\begin{eqnarray*}
\max_{i}\frac{1}{N_{1}T}\left\Vert \frac{1}{N}\sum_{k=1}^{N}\mathbf{x}%
_{i}^{o\prime }{\mathbb{M}}_{\mathbf{\iota }_{T}}\mathbf{u}_{k}^{\dag
}\right\Vert &\leq &\frac{1}{N_{1}T}\max_{i}\left\Vert \mathbf{x}%
_{i}^{o}\right\Vert \frac{1}{N}\left\Vert \sum_{k=1}^{N}\mathbf{u}_{k}^{\dag
}\right\Vert \\
&=&N^{-1}O_{p}(N^{1/(4+2\delta )})O_{p}(N^{-1/2})=O_{p}(N^{-3/2+1/(4+2\delta
)})=o_{p}(N^{-1}),\text{ }
\end{eqnarray*}%
and 
\begin{eqnarray*}
\frac{1}{NN_{1}}\left\Vert \frac{1}{T}\mathbf{x}_{i}^{o\prime }{\mathbb{M}}_{%
\mathbf{\iota }_{T}}\mathbf{u}_{i}^{\dag }\right\Vert &\leq &\frac{1}{NN_{1}}%
\max_{i}\frac{1}{T}\left\Vert \mathbf{x}_{i}^{o\prime }{\mathbb{M}}_{\mathbf{%
\iota }_{T}}\mathbf{u}_{i}^{\dag }\right\Vert \lesssim \frac{1}{NN_{1}}%
\left\Vert \mathbf{x}_{i}^{o}\right\Vert \left\Vert \mathbf{u}_{i}^{\dag
}\right\Vert \\
&\lesssim &N^{-2}\max_{i}\left\Vert \mathbf{x}_{i}^{o}\right\Vert
\max_{i}\left\Vert \mathbf{u}_{i}^{\dag }\right\Vert
=O_{p}(N^{-2+1/(2+\delta )})=o_{p}\left( N^{-1}\right) .\text{ }{\tiny %
\blacksquare }
\end{eqnarray*}

\noindent \textbf{Proof of Lemma \ref{LemA.2}}

To prove Lemma \ref{LemA.2}, we make the following decompositions: 
\begin{eqnarray}
&&B_{h}^{MG\left( -i\right) }  \notag \\
&=&\frac{\mathbf{S}_{N}}{N_{1}}\mathbb{Q}^{(d)}\left\{ [\mathbf{Q}^{\left(
d\right) }\text{+}\mathbf{\bar{C}}\mathbb{A}\mathbf{\bar{C}}^{\prime }]+%
\mathbf{\bar{C}}(\mathbb{A}^{\left( -i\right) }-\mathbb{A})\mathbf{\bar{C}}%
^{\prime }-\mathbf{\breve{Q}}^{\left( i,d\right) }-\mathbf{\breve{C}}%
^{\left( i\right) }\mathbb{A}^{\left( -i\right) }\mathbf{\bar{C}}^{\prime }-%
\mathbf{\bar{C}}\mathbb{A}^{\left( -i\right) }\mathbf{\breve{C}}^{\left(
i\right) \prime }+\mathbf{\breve{C}}^{\left( i\right) }\mathbb{A}^{\left(
-i\right) }\mathbf{\breve{C}}^{\left( i\right) \prime }\right\}  \notag \\
&&\cdot \mathbb{Q}^{(d)}L_{h}^{MG\left( -i\right) }  \notag \\
&\equiv &\sum_{l=1}^{6}B_{h,l}^{\left( -i\right) }\text{ for }h\in \left[ 4%
\right] ,\text{ }  \label{B_decomp1}
\end{eqnarray}%
where, e.g., $B_{h,1}^{\left( -i\right) }=\frac{1}{N_{1}}\mathbf{S}_{N}%
\mathbb{Q}^{\left( d\right) }[\mathbf{Q}^{\left( d\right) }+\mathbf{\bar{C}}%
\mathbb{A}\mathbf{\bar{C}}^{\prime }]\mathbb{Q}^{\left( d\right)
}L_{h}^{MG\left( -i\right) }$ and $B_{h,2}^{\left( -i\right) }=\frac{1}{N_{1}%
}\mathbf{S}_{N}\mathbb{Q}^{\left( d\right) }[\mathbf{\bar{C}}(\mathbb{A}%
^{\left( -i\right) }-\mathbb{A})\mathbf{\bar{C}}^{\prime }]\mathbb{Q}%
^{\left( d\right) }L_{h}^{MG\left( -i\right) }.$ Below, we study $%
B_{h,l}^{\left( -i\right) }$ for $l\in \left[ 6\right] $ in parts (i)--(iv)
for $h\in \left[ 4\right] $ respectively.

(i) We study $B_{1,l}^{\left( -i\right) },$ $l\in \left[ 6\right] $, in
turn. For $B_{1,1}^{\left( -i\right) },$ we have%
\begin{equation*}
B_{1,1}^{\left( -i\right) }=\frac{1}{N_{1}}\mathbf{S}_{N}\mathbb{Q}^{\left(
d\right) }[\mathbf{Q}^{d}+\mathbf{C}\mathbb{A}\mathbf{C}^{\prime }]\mathbb{Q}%
^{\left( d\right) }L_{1}^{MG\left( -i\right) }+\frac{1}{N_{1}}\mathbf{S}_{N}%
\mathbb{Q}^{\left( d\right) }[\mathbf{\bar{C}}\mathbb{A}\mathbf{\bar{C}}%
^{\prime }-\mathbf{C}\mathbb{A}\mathbf{C}^{\prime }]\mathbb{Q}^{\left(
d\right) }L_{1}^{MG\left( -i\right) }\equiv B_{1,11}^{\left( -i\right)
}+B_{1,12}^{\left( -i\right) }.
\end{equation*}%
For $B_{1,11}^{\left( -i\right) },$ we have%
\begin{eqnarray*}
B_{1,11}^{\left( -i\right) } &=&\frac{N}{N_{1}}\frac{1}{N}\mathbf{S}_{N}%
\left[ \frac{1}{T}\mathbf{X}^{\prime }\left( {\mathbb{M}}_{\mathbf{\iota }%
_{N}}\otimes {\mathbb{M}}_{\mathbf{\iota }_{T}}\right) \mathbf{X}\right]
^{-1}\frac{1}{T}\mathbf{X}^{\prime }\left( {\mathbb{M}}_{\mathbf{\iota }%
_{N}}\otimes {\mathbb{M}}_{\mathbf{\iota }_{T}}\right) \mathbf{U}^{\dag }=%
\frac{N}{N_{1}}\frac{1}{N}\mathbf{S}_{N}\left( \mathbf{\hat{\beta}}-\mathbf{%
\beta }\right) \\
&=&\frac{N}{N_{1}}\left( \hat{\beta}^{MG}-\frac{1}{N}\sum_{j=1}^{N}\beta
_{j}\right) ,\text{ }
\end{eqnarray*}%
where we use the fact that $\mathbf{\hat{Q}}^{-1}=\left[ \frac{1}{T}\mathbf{X%
}^{\prime }\left( {\mathbb{M}}_{\mathbf{\iota }_{N}}\otimes {\mathbb{M}}_{%
\mathbf{\iota }_{T}}\right) \mathbf{X}\right] ^{-1}=\mathbb{Q}^{\left(
d\right) }[\mathbf{Q}^{d}+\mathbf{C}\mathbb{A}\mathbf{C}^{\prime }]\mathbb{Q}%
^{\left( d\right) }$.$\ $Note that 
\begin{eqnarray*}
B_{1,12}^{\left( -i\right) } &=&\frac{1}{N_{1}}\mathbf{S}_{N}\mathbb{Q}%
^{\left( d\right) }[\mathbf{\bar{C}}\mathbb{A}\mathbf{\bar{C}}^{\prime }-%
\mathbf{C}\mathbb{A}\mathbf{C}^{\prime }]\mathbb{Q}^{\left( d\right)
}L_{1}^{MG\left( -i\right) }\frac{1}{T}\mathbf{X}^{o\prime }\left( {\mathbb{M%
}}_{\mathbf{\iota }_{N}}\otimes {\mathbb{M}}_{\mathbf{\iota }_{T}}\right) 
\mathbf{U}^{\dag } \\
&=&\frac{1}{N_{1}}\mathbf{S}_{N}\mathbb{Q}^{\left( d\right) }(\mathbf{\bar{C}%
}-\mathbf{C})\mathbb{A}\mathbf{\bar{C}}^{\prime }\mathbb{Q}^{\left( d\right)
}\frac{1}{T}\mathbf{X}^{o\prime }\left( {\mathbb{M}}_{\mathbf{\iota }%
_{N}}\otimes {\mathbb{M}}_{\mathbf{\iota }_{T}}\right) \mathbf{U}^{\dag } \\
&&+\frac{1}{N_{1}}\mathbf{S}_{N}\mathbb{Q}^{\left( d\right) }\mathbf{C}%
\mathbb{A(}\mathbf{\bar{C}}-\mathbf{C})^{\prime }\mathbb{Q}^{\left( d\right)
}\frac{1}{T}\mathbf{X}^{o\prime }\left( {\mathbb{M}}_{\mathbf{\iota }%
_{N}}\otimes {\mathbb{M}}_{\mathbf{\iota }_{T}}\right) \mathbf{U}^{\dag
}\equiv B_{1,12,1}^{\left( -i\right) }+B_{1,12,2}^{\left( -i\right) }.
\end{eqnarray*}%
Note that $\mathbf{\bar{C}}=c_{1N}\mathbf{C}$ and $\mathbf{\bar{C}}-\mathbf{C%
}=c_{2N}\mathbf{C,}$ where $c_{1N}\equiv \left( N/N_{1}\right) ^{1/2}$ and $%
c_{2N}\equiv c_{1N}-1\lesssim N^{-1}.$ Let $\psi _{1s}\equiv \sum_{k}\left[ 
\mathbf{C}^{\prime }\right] _{sk}\tilde{q}_{kk}^{-1}\frac{1}{T}\sum_{l}%
\mathbf{x}_{k}^{o\prime }{\mathbb{M}}_{\mathbf{\iota }_{T}}\mathbf{u}%
_{l}^{\dag }m_{kl}.$ By Lemma \ref{LemS.1}(i)-(ii) and (v) and the fact that 
$\frac{1}{N}\sum_{j}\left\Vert \tilde{q}_{jj}^{-1}\right\Vert \lesssim 1,$ 
\begin{eqnarray*}
\left\Vert B_{1,12,1}^{\left( -i\right) }\right\Vert
&=&c_{1N}c_{2N}\left\Vert \frac{1}{N_{1}}\sum_{j}\tilde{q}%
_{jj}^{-1}\sum_{t,s}[\mathbf{C}]_{jt}\left[ \mathbb{A}\right] _{ts}\sum_{k}%
\left[ \mathbf{C}^{\prime }\right] _{sk}\tilde{q}_{kk}^{-1}\frac{1}{T}%
\sum_{l}\mathbf{x}_{k}^{o\prime }{\mathbb{M}}_{\mathbf{\iota }_{T}}\mathbf{u}%
_{l}^{\dag }m_{kl}\right\Vert \\
&\lesssim &N^{-1}\max_{j}\sum_{t}\left\Vert [\mathbf{C}]_{jt}\right\Vert
\max_{t}\sum_{s}\left\Vert \left[ \mathbb{A}\right] _{ts}\right\Vert
\max_{s}\left\Vert \psi _{1s}\right\Vert =N^{-1}O_{p}(N^{-1/2+1/(4+2\delta
)})O_{p}(1)O_{p}(N^{1/(4+2\delta )}) \\
&=&o_{p}\left( N^{-1}\right) ,\text{ and} \\
\left\Vert B_{1,12,2}^{\left( -i\right) }\right\Vert &=&\left\Vert \frac{1}{%
N_{1}}\sum_{j}\tilde{q}_{jj}^{-1}\sum_{t,s}[\mathbf{C}]_{jt}\left[ \mathbb{A}%
\right] _{ts}\sum_{k}\left[ (\mathbf{\bar{C}}-\mathbf{C})^{\prime }\right]
_{sk}\tilde{q}_{kk}^{-1}\frac{1}{T}\sum_{l}\mathbf{x}_{k}^{o\prime }{\mathbb{%
M}}_{\mathbf{\iota }_{T}}\mathbf{u}_{l}^{\dag }m_{kl}\right\Vert \\
&=&c_{2N}\left\Vert \frac{1}{N_{1}}\sum_{j}\tilde{q}_{jj}^{-1}\sum_{t,s}[%
\mathbf{C}]_{jt}\left[ \mathbb{A}\right] _{ts}\sum_{k}\left[ \mathbf{C}%
^{\prime }\right] _{sk}\tilde{q}_{kk}^{-1}\frac{1}{T}\sum_{l}\mathbf{x}%
_{k}^{o\prime }{\mathbb{M}}_{\mathbf{\iota }_{T}}\mathbf{u}_{l}^{\dag
}m_{kl}\right\Vert \\
&\lesssim &N^{-1}\max_{j}\sum_{t}\left\Vert [\mathbf{C}]_{jt}\right\Vert
\max_{t}\sum_{s}\left\Vert \left[ \mathbb{A}\right] _{ts}\right\Vert
\max_{s}\left\Vert \psi _{1s}\right\Vert =N^{-1}O_{p}(N^{-1/2+1/(4+2\delta
)})O_{p}(1)O_{p}(N^{1/(4+2\delta )}) \\
&=&o_{p}\left( N^{-1}\right) .
\end{eqnarray*}%
Then $\max_{i}\left\Vert B_{1,12}^{\left( -i\right) }\right\Vert
=o_{p}\left( N^{-1}\right) \ $and $B_{1,1}^{\left( -i\right) }=\frac{N}{N_{1}%
}(\hat{\beta}^{MG}-\frac{1}{N}\sum_{j=1}^{N}\beta _{j})+o_{p}\left(
N^{-1}\right) .$

By Lemma \ref{LemS.1}(iii) and (v) and the fact that $\mathbf{\bar{C}}=c_{1N}%
\mathbf{C}$ and $\frac{1}{N}\sum_{j}\left\Vert \tilde{q}_{jj}^{-1}\right%
\Vert \left\Vert \mathbf{x}_{j}^{o}\right\Vert \lesssim 1\mathbf{,}$%
\begin{eqnarray*}
\left\Vert B_{1,2}^{\left( -i\right) }\right\Vert &=&\left\Vert \frac{1}{%
N_{1}}\mathbf{S}_{N}\mathbb{Q}^{\left( d\right) }\mathbf{\bar{C}}(\mathbb{A}%
^{\left( -i\right) }-\mathbb{A})\mathbf{\bar{C}}^{\prime }\mathbb{Q}^{\left(
d\right) }\frac{1}{T}\mathbf{X}^{o\prime }\left( {\mathbb{M}}_{\mathbf{\iota 
}_{N}}\otimes {\mathbb{M}}_{\mathbf{\iota }_{T}}\right) \mathbf{U}^{\dag
}\right\Vert \\
&=&c_{1N}^{2}\left\Vert \frac{-1}{N_{1}}\sum_{j}\sum_{t,s}\tilde{q}_{jj}^{-1}%
\left[ \mathbf{C}\right] _{jt}[\mathbb{A}^{\left( -i\right) }-\mathbb{A}%
]_{t,s}\text{ }\sum_{k}[\mathbf{C}^{\prime }]_{sk}\tilde{q}_{kk}^{-1}\frac{1%
}{T}\sum_{l}\mathbf{x}_{k}^{o\prime }{\mathbb{M}}_{\mathbf{\iota }_{T}}%
\mathbf{u}_{l}^{\dag }m_{kl}\right\Vert \\
&\lesssim &\frac{1}{N^{3/2}}\sum_{j}\left\Vert \tilde{q}_{jj}^{-1}\right%
\Vert \left\Vert \mathbf{x}_{j}^{o}\right\Vert \max_{t}\sum_{s}\left\Vert [%
\mathbb{A}^{\left( -i\right) }-\mathbb{A}]_{t,s}\right\Vert
\max_{s}\left\Vert \psi _{1s}\right\Vert \  \\
&=&O_{p}(N^{-1/2})O_{p}(N^{-1+1/(2+\delta )})O_{p}(N^{1/(4+2\delta
)})=o_{p}(N^{-1})
\end{eqnarray*}%
where we use the condition $\delta >1$ by Assumption \ref{Assmp3.3}. Noting
that $\mathbf{\breve{Q}}^{\left( i,d\right) }$ is a block diagonal matrix
with a nonzero diagonal block only in its $i$th position and $\mathbf{\breve{%
Q}}^{\left( i,d\right) }\mathbb{Q}^{\left( d\right) }\mathbf{\breve{X}}%
^{\left( -i\right) }=\mathbf{0}_{NK_{x}\times NT}$ for each $i,$ we have%
\begin{eqnarray*}
B_{1,3}^{\left( -i\right) } &=&\frac{-1}{N_{1}}\mathbf{S}_{N}\mathbb{Q}%
^{\left( d\right) }\mathbf{\breve{Q}}^{\left( i,d\right) }\mathbb{Q}^{\left(
d\right) }\frac{1}{T}(\mathbf{\breve{X}}^{o\left( -i\right) }+\mathbf{\breve{%
X}}^{o\left( i\right) })^{\prime }\left( {\mathbb{M}}_{\mathbf{\iota }%
_{N}}\otimes {\mathbb{M}}_{\mathbf{\iota }_{T}}\right) \mathbf{U}^{\dag } \\
&=&\frac{-1}{N_{1}}\mathbf{S}_{N}\mathbb{Q}^{\left( d\right) }\mathbf{\breve{%
Q}}^{\left( i,d\right) }\mathbb{Q}^{\left( d\right) }\frac{1}{T}\mathbf{%
\breve{X}}^{o\left( i\right) \prime }\left( {\mathbb{M}}_{\mathbf{\iota }%
_{N}}\otimes {\mathbb{M}}_{\mathbf{\iota }_{T}}\right) \mathbf{U}^{\dag }=%
\frac{-1}{N_{1}}[\mathbb{Q}^{\left( d\right) }]_{ii}[\mathbf{\breve{Q}}%
^{\left( i,d\right) }]_{ii}[\mathbb{Q}^{\left( d\right) }]_{ii}\frac{1}{T}%
\sum_{j=1}^{N}\mathbf{x}_{i}^{o\prime }{\mathbb{M}}_{\mathbf{\iota }_{T}}%
\mathbf{u}_{j}^{\dag }m_{ij} \\
&=&\frac{-1}{N_{1}}[\mathbb{Q}^{\left( d\right) }]_{ii}\frac{1}{T}%
\sum_{j=1}^{N}\mathbf{x}_{i}^{o\prime }{\mathbb{M}}_{\mathbf{\iota }_{T}}%
\mathbf{u}_{j}^{\dag }m_{ij},\text{ }
\end{eqnarray*}%
where we use the fact that $[\mathbf{\breve{Q}}^{\left( i,d\right) }]_{ii}=[%
\mathbf{Q}^{\left( d\right) }]_{ii}=\{[\mathbb{Q}^{\left( d\right)
}]_{ii}\}^{-1}=\tilde{q}_{ii}.$ We keep the last expression for $%
B_{1,3}^{\left( -i\right) }$ as it will be cancelled with $B_{2,3}^{\left(
-i\right) }$ below. For $B_{1,4}^{\left( -i\right) },$ we have by Lemma \ref%
{LemS.1}(i), (ii) and (v)%
\begin{eqnarray*}
B_{1,4}^{\left( -i\right) } &=&\frac{-1}{N_{1}}\mathbf{S}_{N}\mathbb{Q}%
^{\left( d\right) }\mathbf{\breve{C}}^{\left( i\right) }\mathbb{A}^{\left(
-i\right) }\mathbf{\bar{C}}^{\prime }\mathbb{Q}^{\left( d\right) }\frac{1}{T}%
\mathbf{X}^{o\prime }\left( {\mathbb{M}}_{\mathbf{\iota }_{N}}\otimes {%
\mathbb{M}}_{\mathbf{\iota }_{T}}\right) \mathbf{U}^{\dag } \\
&=&-\tilde{q}_{ii}^{-1}c_{1N}^{2}\frac{1}{N_{1}}\sum_{s,t}\left[ \mathbf{C}%
\right] _{is}\left[ \mathbb{A}^{\left( -i\right) }\right] _{st}%
\sum_{j=1}^{N}[\mathbf{C}^{\prime }]_{tj}\tilde{q}_{jj}^{-1}\frac{1}{T}%
\sum_{k=1}^{N}\mathbf{x}_{j}^{o\prime }{\mathbb{M}}_{\mathbf{\iota }_{T}}%
\mathbf{u}_{k}^{\dag }m_{jk}\equiv -\tilde{q}_{ii}^{-1}R_{1,i}
\end{eqnarray*}%
where 
\begin{eqnarray*}
\max_{i}\left\Vert R_{1,i}\right\Vert &\lesssim &\max_{i}\frac{1}{N_{1}}%
\left\Vert \sum_{s,t}\left[ \mathbf{C}\right] _{is}\left[ \mathbb{A}^{\left(
-i\right) }\right] _{st}\psi _{1t}\right\Vert \lesssim
N^{-1}\max_{i}\sum_{s}\left\Vert \left[ \mathbf{C}\right] _{is}\right\Vert
\max_{i,s}\sum_{t}\left\Vert \left[ \mathbb{A}^{\left( -i\right) }\right]
_{st}\right\Vert \max_{t}\left\Vert \psi _{1t}\right\Vert \\
&=&N^{-1}O_{p}(N^{-1/2+1/(4+2\delta )})O_{p}(1)O_{p}(N^{1/(4+2\delta
)})=o_{p}\left( N^{-1}\right) .
\end{eqnarray*}
So $B_{1,4}^{\left( -i\right) }=\tilde{q}_{ii}^{-1}o_{p}(N^{-1})$ uniformly
in $i\in \left[ N\right] .$ Next, by the fact $\mathbf{\breve{C}}^{\left(
i\right) \prime }\mathbb{Q}^{\left( d\right) }\mathbf{\breve{X}}^{o\left(
-i\right) \prime }=\mathbf{0}_{K_{x}T\times NT},$ we have%
\begin{eqnarray*}
B_{1,5}^{\left( -i\right) } &=&\frac{-1}{N_{1}}\mathbf{S}_{N}\mathbb{Q}%
^{\left( d\right) }\mathbf{\bar{C}}\mathbb{A}^{\left( -i\right) }\mathbf{%
\breve{C}}^{\left( i\right) \prime }\mathbb{Q}^{\left( d\right) }\frac{1}{T}%
\left( \mathbf{\breve{X}}^{o\left( -i\right) }+\mathbf{\breve{X}}^{o\left(
i\right) }\right) ^{\prime }\left( {\mathbb{M}}_{\mathbf{\iota }_{N}}\otimes 
{\mathbb{M}}_{\mathbf{\iota }_{T}}\right) \mathbf{U}^{\dag } \\
&=&\frac{-1}{N_{1}}\mathbf{S}_{N}\mathbb{Q}^{\left( d\right) }\mathbf{\bar{C}%
}\mathbb{A}^{\left( -i\right) }\mathbf{\breve{C}}^{\left( i\right) \prime }%
\mathbb{Q}^{\left( d\right) }\frac{1}{T}\mathbf{\breve{X}}^{o\left( i\right)
\prime }\left( {\mathbb{M}}_{\mathbf{\iota }_{N}}\otimes {\mathbb{M}}_{%
\mathbf{\iota }_{T}}\right) \mathbf{U}^{\dag },
\end{eqnarray*}%
which is kept because it will be cancelled with $B_{2,5}^{\left( -i\right) }$
below. Let $\psi _{2i}\equiv \lbrack \mathbf{C}^{\prime }]_{\cdot i}\tilde{q}%
_{ii}^{-1}\frac{1}{T}\sum_{j=1}^{N}\mathbf{x}_{i}^{o\prime }{\mathbb{M}}_{%
\mathbf{\iota }_{T}}\mathbf{u}_{j}^{\dag }m_{ij}.$ For $B_{1,6}^{\left(
-i\right) },$ we have Lemma \ref{LemS.1}(i)-(ii) and Assumption \ref%
{Assmp3.3} 
\begin{eqnarray*}
B_{1,6}^{\left( -i\right) } &=&\frac{1}{N_{1}}\mathbf{S}_{N}\mathbb{Q}%
^{\left( d\right) }\mathbf{\breve{C}}^{\left( i\right) }\mathbb{A}^{\left(
-i\right) }\mathbf{\breve{C}}^{\left( i\right) \prime }\mathbb{Q}^{\left(
d\right) }\frac{1}{T}\mathbf{X}^{o\prime }\left( {\mathbb{M}}_{\mathbf{\iota 
}_{N}}\otimes {\mathbb{M}}_{\mathbf{\iota }_{T}}\right) \mathbf{U}^{\dag } \\
&=&\tilde{q}_{ii}^{-1}\frac{c_{1N}^{2}}{N_{1}}\left[ \mathbf{C}\right] _{i}%
\mathbb{A}^{\left( -i\right) }[\mathbf{C}^{\prime }]_{\cdot i}\tilde{q}%
_{ii}^{-1}\frac{1}{T}\sum_{j=1}^{N}\mathbf{x}_{i}^{o\prime }{\mathbb{M}}_{%
\mathbf{\iota }_{T}}\mathbf{u}_{j}^{\dag }m_{ij}\equiv \tilde{q}%
_{ii}^{-1}R_{1,2i},\text{ }
\end{eqnarray*}%
where%
\begin{equation*}
\max_{i}\left\Vert R_{1,2i}\right\Vert \lesssim N^{-1}\max_{i}\left\Vert 
\left[ \mathbf{C}\right] _{i}\right\Vert \left\Vert \psi _{2i}\right\Vert
=N^{-1}O_{p}(N^{-1/2+1/(4+2\delta )})O_{p}(N^{1/(4+2\delta )})=o_{p}\left(
N^{-1}\right) .
\end{equation*}
In sum, $B_{1}^{\left( -i\right) }=\frac{N}{N_{1}}(\hat{\beta}-\frac{1}{N}%
\sum_{j=1}^{N}\beta _{j})-\frac{1}{N_{1}}[\mathbb{Q}^{\left( d\right) }]_{ii}%
\frac{1}{T}\sum_{j=1}^{N}\mathbf{x}_{i}^{o\prime }{\mathbb{M}}_{\mathbf{%
\iota }_{T}}\mathbf{u}_{j}^{\dag }m_{ij}-\frac{1}{N_{1}}\mathbf{S}_{N}%
\mathbb{Q}^{\left( d\right) }\mathbf{\bar{C}}\mathbb{A}^{\left( -i\right) }%
\mathbf{\breve{C}}^{\left( i\right) \prime }\mathbb{Q}^{\left( d\right) }$ $%
\frac{1}{T}\mathbf{\breve{X}}^{o\left( i\right) \prime }({\mathbb{M}}_{%
\mathbf{\iota }_{N}}\otimes {\mathbb{M}}_{\mathbf{\iota }_{T}})\mathbf{U}%
^{\dag }+(\tilde{q}_{ii}^{-1}+\mathbb{I}_{K_{x}})o_{p}(N^{-1})$ uniformly in%
\textbf{\ }$i\in \left[ N\right] .\medskip $

(ii) We study $B_{2,l}^{\left( -i\right) },$ $l\in \left[ 6\right] $, in
turn. Noting that $\mathbf{\hat{Q}}^{-1}=\mathbb{Q}^{\left( d\right) }[%
\mathbf{Q}^{\left( d\right) }+\mathbf{C}\mathbb{A}\mathbf{C}^{\prime }]%
\mathbb{Q}^{\left( d\right) },$ we can following the analysis of $%
B_{1,12}^{\left( -i\right) }$ to show that uniformly in $i\in \left[ N\right]
,$ 
\begin{eqnarray*}
B_{2,1}^{\left( -i\right) } &=&\frac{1}{N_{1}}\mathbf{S}_{N}\mathbb{Q}%
^{\left( d\right) }[\mathbf{Q}^{\left( d\right) }+\mathbf{\bar{C}}\mathbb{A}%
\mathbf{\bar{C}}^{\prime }]\mathbb{Q}^{\left( d\right) }L_{2}^{MG\left(
-i\right) } \\
&=&\frac{1}{N_{1}}\mathbf{S}_{N}\mathbb{Q}^{\left( d\right) }[\mathbf{Q}%
^{\left( d\right) }+\mathbf{C}\mathbb{A}\mathbf{C}^{\prime }]\mathbb{Q}%
^{\left( d\right) }L_{2}^{MG\left( -i\right) }+\frac{1}{N_{1}}\mathbf{S}_{N}%
\mathbb{Q}^{\left( d\right) }[\mathbf{C}\mathbb{A}\mathbf{C}^{\prime }-%
\mathbf{\bar{C}}\mathbb{A}\mathbf{\bar{C}}^{\prime }]\mathbb{Q}^{\left(
d\right) }L_{2}^{MG\left( -i\right) } \\
&=&\frac{-1}{N_{1}}\mathbf{S}_{N}\mathbf{\hat{Q}}^{-1}\frac{1}{T}\mathbf{%
\breve{X}}^{o\left( i\right) \prime }({\mathbb{M}}_{\mathbf{\iota }%
_{N}}\otimes {\mathbb{M}}_{\mathbf{\iota }_{T}})\mathbf{U}^{\dag
}+\left\Vert \tilde{q}_{ii}^{-1}\right\Vert o_{p}(N^{-1}) \\
&=&\frac{-1}{N_{1}}\hat{\omega}_{i}\frac{1}{T}\sum_{k=1}^{N}\mathbf{x}%
_{i}^{o\prime }{\mathbb{M}}_{\mathbf{\iota }_{T}}\mathbf{u}_{k}^{\dag
}m_{ik}+\left\Vert \tilde{q}_{ii}^{-1}\right\Vert o_{p}(N^{-1})=\frac{-1}{%
N_{1}T}\hat{\omega}_{i}\mathbf{x}_{i}^{o\prime }{\mathbb{M}}_{\mathbf{\iota }%
_{T}}\mathbf{u}_{i}^{\dag }+\left( \left\Vert \tilde{q}_{ii}^{-1}\right\Vert
+\left\Vert \hat{\omega}_{i}\right\Vert \right) o_{p}(N^{-1}),\text{ }
\end{eqnarray*}%
where the first $o_{p}(N^{-1})$ term follows from similar arguments as used
in the analysis of $B_{1,12}^{\left( -i\right) }\ $and the last equality
holds because $\frac{1}{N_{1}T}\frac{1}{N}\sum_{k=1}^{N}\mathbf{x}%
_{i}^{o\prime }{\mathbb{M}}_{\mathbf{\iota }_{T}}\mathbf{u}_{k}^{\dag
}=o_{p}\left( N^{-1}\right) $ by Lemma \ref{LemS.1}(viii). For $%
B_{2,2}^{\left( -i\right) },$ $B_{2,4}^{\left( -i\right) }$ and $%
B_{2,6}^{\left( -i\right) },$ by Lemma \ref{LemS.1}(iii)-(iv), the fact that 
$\mathbf{\bar{C}}=\mathbf{\breve{C}}^{\left( i\right) }+\mathbf{\breve{C}}%
^{\left( -i\right) }$ and $\mathbf{\breve{C}}^{\left( -i\right) \prime }%
\mathbb{Q}^{\left( d\right) }\mathbf{\breve{X}}^{o\left( i\right) \prime }=%
\mathbf{0}_{K_{1}T\times NT}$ and Assumption \ref{Assmp3.3}, we have that
uniformly in $i\in \left[ N\right] ,$%
\begin{eqnarray*}
\max_{i}\left\Vert B_{2,2}^{\left( -i\right) }\right\Vert
&=&\max_{i}\left\Vert \frac{-1}{N_{1}}\mathbf{S}_{N}\mathbb{Q}^{\left(
d\right) }\mathbf{\bar{C}}(\mathbb{A}^{\left( -i\right) }-\mathbb{A})\mathbf{%
\bar{C}}^{\prime }\mathbb{Q}^{\left( d\right) }\frac{1}{T}\mathbf{\breve{X}}%
^{o\left( i\right) \prime }({\mathbb{M}}_{\mathbf{\iota }_{N}}\otimes {%
\mathbb{M}}_{\mathbf{\iota }_{T}})\mathbf{U}^{\dag }\right\Vert \text{ } \\
&=&\max_{i}\left\Vert \frac{-c_{1N}^{2}}{N_{1}}\sum_{j}\tilde{q}_{jj}^{-1}%
\left[ \mathbf{C}\right] _{j}[\mathbb{A}^{\left( -i\right) }-\mathbb{A}]%
\left[ \mathbf{C}^{\prime }\right] _{\cdot i}\tilde{q}_{ii}^{-1}\frac{1}{T}%
\sum_{k=1}^{N}\mathbf{x}_{i}^{o\prime }{\mathbb{M}}_{\mathbf{\iota }_{T}}%
\mathbf{u}_{k}^{\dag }m_{ik}\right\Vert \\
&=&\max_{i}\left\Vert \frac{-c_{1N}^{2}}{N_{1}}\sum_{j}\tilde{q}_{jj}^{-1}%
\left[ \mathbf{C}\right] _{j}[\mathbb{A}^{\left( -i\right) }-\mathbb{A}]\psi
_{2i}\right\Vert \\
&\lesssim &\frac{1}{N_{1}N^{1/2}}\sum_{j}\left\Vert \tilde{q}%
_{jj}^{-1}\right\Vert \left\Vert \mathbf{x}_{j}^{o}\right\Vert
\max_{i}\left\Vert \mathbb{A}^{\left( -i\right) }-\mathbb{A}\right\Vert
\max_{i}\left\Vert \psi _{2i}\right\Vert \\
&=&O_{p}\left( N^{-1/2}\right) O_{p}\left( N^{-1+1/(2+\delta )}\right)
O_{p}\left( N^{1/(4+2\delta )}\right) =o_{p}\left( N^{-1}\right)
\end{eqnarray*}%
and%
\begin{eqnarray*}
B_{2,4}^{\left( -i\right) }+B_{2,6}^{\left( -i\right) } &=&\frac{1}{N_{1}}%
\mathbf{S}_{N}\mathbb{Q}^{\left( d\right) }\mathbf{\breve{C}}^{\left(
i\right) }\mathbb{A}^{\left( -i\right) }\mathbf{\breve{C}}^{\left( i\right)
\prime }\mathbb{Q}^{\left( d\right) }\frac{1}{T}\mathbf{\breve{X}}^{o\left(
i\right) \prime }({\mathbb{M}}_{\mathbf{\iota }_{N}}\otimes {\mathbb{M}}_{%
\mathbf{\iota }_{T}})\mathbf{U}^{\dag } \\
&&+\frac{-1}{N_{1}}\mathbf{S}_{N}\mathbb{Q}^{\left( d\right) }\mathbf{\breve{%
C}}^{\left( i\right) }\mathbb{A}^{\left( -i\right) }\mathbf{\breve{C}}%
^{\left( i\right) \prime }\mathbb{Q}^{\left( d\right) }\frac{1}{T}\mathbf{%
\breve{X}}^{o\left( i\right) \prime }({\mathbb{M}}_{\mathbf{\iota }%
_{N}}\otimes {\mathbb{M}}_{\mathbf{\iota }_{T}})\mathbf{U}^{\dag }=0.\text{ }
\end{eqnarray*}%
Next, 
\begin{eqnarray*}
B_{2,3}^{\left( -i\right) } &=&\frac{1}{N_{1}}\mathbf{S}_{N}\mathbb{Q}%
^{\left( d\right) }\mathbf{\breve{Q}}^{\left( i,d\right) }\mathbb{Q}^{\left(
d\right) }\frac{1}{T}\mathbf{\breve{X}}^{o\left( i\right) \prime }({\mathbb{M%
}}_{\mathbf{\iota }_{N}}\otimes {\mathbb{M}}_{\mathbf{\iota }_{T}})\mathbf{U}%
^{\dag }=\frac{1}{N_{1}}[\mathbb{Q}^{\left( d\right) }]_{ii}[\mathbf{\breve{Q%
}}^{\left( i,d\right) }]_{ii}[\mathbb{Q}^{\left( d\right) }]_{ii}\frac{1}{T}%
\sum_{j=1}^{N}\mathbf{x}_{i}^{o\prime }{\mathbb{M}}_{\mathbf{\iota }_{T}}%
\mathbf{u}_{j}^{\dag }m_{ij} \\
&=&\frac{1}{N_{1}}[\mathbb{Q}^{\left( d\right) }]_{ii}\frac{1}{T}%
\sum_{j=1}^{N}\mathbf{x}_{i}^{o\prime }{\mathbb{M}}_{\mathbf{\iota }_{T}}%
\mathbf{u}_{j}^{\dag }m_{ij}=-B_{1,3}^{\left( -i\right) }\ \text{and } \\
B_{2,5}^{\left( -i\right) } &=&\frac{1}{N_{1}}\mathbf{S}_{N}\mathbb{Q}%
^{\left( d\right) }\mathbf{\bar{C}}\mathbb{A}^{\left( -i\right) }\mathbf{%
\breve{C}}^{\left( i\right) \prime }\mathbb{Q}^{\left( d\right) }\frac{1}{T}%
\mathbf{\breve{X}}^{o\left( i\right) \prime }({\mathbb{M}}_{\mathbf{\iota }%
_{N}}\otimes {\mathbb{M}}_{\mathbf{\iota }_{T}})\mathbf{U}^{\dag
}=-B_{1,5}^{\left( -i\right) }.
\end{eqnarray*}%
In sum, we have $B_{2}^{\left( -i\right) }=-\frac{1}{N_{1}}\hat{\omega}_{i}%
\frac{1}{T}\sum_{k=1}^{N}\mathbf{x}_{i}^{o\prime }{\mathbb{M}}_{\mathbf{%
\iota }_{T}}\mathbf{u}_{k}^{\dag }m_{ik}+\frac{1}{N_{1}}[\mathbb{Q}^{\left(
d\right) }]_{ii}\frac{1}{T}\sum_{j=1}^{N}\mathbf{x}_{i}^{o\prime }{\mathbb{M}%
}_{\mathbf{\iota }_{T}}\mathbf{u}_{j}^{\dag }m_{ij}$ $+\frac{1}{N_{1}}%
\mathbf{S}_{N}\mathbb{Q}^{\left( d\right) }\mathbf{\bar{C}}\mathbb{A}%
^{\left( -i\right) }\mathbf{\breve{C}}^{\left( i\right) \prime }$ $\mathbb{Q}%
^{\left( d\right) }\frac{1}{T}\mathbf{\breve{X}}^{o\left( i\right) \prime }({%
\mathbb{M}}_{\mathbf{\iota }_{N}}\otimes {\mathbb{M}}_{\mathbf{\iota }_{T}})%
\mathbf{U}^{\dag }+R_{2i}\ $uniformly in $i,$ where $R_{2i}=\left(
\left\Vert \hat{\omega}_{i}\right\Vert +\left\Vert \tilde{q}%
_{ii}^{-1}\right\Vert \right) o_{p}(N^{-1}).$ \medskip

(iii) We study $B_{3,l}^{\left( -i\right) },$ $l\in \left[ 6\right] $, in
turn. For $B_{3,1}^{\left( -i\right) },$ we have 
\begin{eqnarray*}
B_{3,1}^{\left( -i\right) } &=&\frac{1}{N_{1}}\mathbf{S}_{N}\mathbb{Q}%
^{\left( d\right) }[\mathbf{Q}^{\left( d\right) }+\mathbf{C}\mathbb{A}%
\mathbf{C}^{\prime }]\mathbb{Q}^{\left( d\right) }L_{3}^{MG\left( -i\right)
}+\frac{1}{N_{1}}\mathbf{S}_{N}\mathbb{Q}^{\left( d\right) }[\mathbf{\bar{C}}%
^{(-i)}\mathbb{A}\mathbf{\bar{C}}^{\prime }-\mathbf{C}\mathbb{A}\mathbf{C}%
^{\prime }]\mathbb{Q}^{\left( d\right) }L_{3}^{MG\left( -i\right) } \\
&=&\frac{-1}{N_{1}}\mathbf{S}_{N}\left[ \frac{1}{T}\mathbf{X}^{\prime
}\left( {\mathbb{M}}_{\mathbf{\iota }_{N}}\otimes {\mathbb{M}}_{\mathbf{%
\iota }_{T}}\right) \mathbf{X}\right] ^{-1}\frac{1}{T}\mathbf{\breve{X}}%
^{o\left( -i\right) \prime }\left( {\mathbb{M}}_{\mathbf{\iota }_{N}}\otimes 
{\mathbb{M}}_{\mathbf{\iota }_{T}}\right) \mathbf{\breve{U}}^{\dag \left(
i\right) }+\left\Vert \tilde{q}_{ii}^{-1}\right\Vert o_{p}(N^{-1}) \\
&=&\frac{-1}{N_{1}}\mathbf{S}_{N}\mathbf{\hat{Q}}^{-1}\frac{1}{T}\mathbf{%
\breve{X}}^{o\left( -i\right) \prime }\left( {\mathbb{M}}_{\mathbf{\iota }%
_{N}}\otimes {\mathbb{M}}_{\mathbf{\iota }_{T}}\right) \mathbf{\breve{U}}%
^{\dag \left( i\right) }+\left\Vert \tilde{q}_{ii}^{-1}\right\Vert
o_{p}(N^{-1})\text{ } \\
&=&\frac{-1}{N_{1}}\mathbf{S}_{N}\mathbf{\hat{Q}}^{-1}\frac{1}{T}\mathbf{%
\breve{X}}^{o\left( -i\right) \prime }\left( {\mathbb{M}}_{\mathbf{\iota }%
_{N}}\otimes {\mathbb{M}}_{\mathbf{\iota }_{T}}\right) \mathbf{\breve{U}}%
^{\dag \left( i\right) }+\left\Vert \tilde{q}_{ii}^{-1}\right\Vert
o_{p}\left( N^{-1}\right) \\
&=&\frac{-1}{N_{1}}\sum_{j=1,j\neq i}^{N}\hat{\omega}_{j}\frac{1}{T}\mathbf{x%
}_{j}^{o\prime }{\mathbb{M}}_{\mathbf{\iota }_{T}}\mathbf{u}_{i}^{\dag
}m_{ji}+\left\Vert \tilde{q}_{ii}^{-1}\right\Vert o_{p}\left( N^{-1}\right)
\\
&=&\frac{1}{NN_{1}}\sum_{j=1}^{N}\hat{\omega}_{j}\frac{1}{T}\mathbf{x}%
_{j}^{o\prime }{\mathbb{M}}_{\mathbf{\iota }_{T}}\mathbf{u}_{i}^{\dag
}+\left( \left\Vert \hat{\omega}_{i}\right\Vert +\left\Vert \tilde{q}%
_{ii}^{-1}\right\Vert \right) o_{p}\left( N^{-1}\right) ,\text{\textbf{\ }}
\end{eqnarray*}%
where the first $o_{p}(N^{-1})$ term follows from similar arguments as used
in the analysis of $B_{1,12}^{\left( -i\right) }\ $and the last equality
holds by\ the fact that $\frac{1}{NN_{1}}\left\Vert \frac{1}{T}\mathbf{x}%
_{i}^{o\prime }{\mathbb{M}}_{\mathbf{\iota }_{T}}\mathbf{u}_{i}^{\dag
}\right\Vert =o_{p}\left( N^{-1}\right) $ by Lemma \ref{LemS.1}(viii). For $%
B_{3,2}^{\left( -i\right) }$ and $B_{3,4}^{\left( -i\right) },$ we have by
Lemma \ref{LemS.1}(iii), (v) and (vii), 
\begin{eqnarray*}
B_{3,2}^{\left( -i\right) } &=&\frac{-1}{N_{1}}\mathbf{S}_{N}\mathbb{Q}%
^{\left( d\right) }\mathbf{\bar{C}}(\mathbb{A}^{\left( -i\right) }-\mathbb{A}%
)\mathbf{\bar{C}}^{\prime }\mathbb{Q}^{\left( d\right) }\frac{1}{T}\mathbf{%
\breve{X}}^{o\left( -i\right) \prime }\left( {\mathbb{M}}_{\mathbf{\iota }%
_{N}}\otimes {\mathbb{M}}_{\mathbf{\iota }_{T}}\right) \mathbf{\breve{U}}%
^{\dag \left( i\right) } \\
&=&\tilde{q}_{ii}^{-1}\frac{-1}{N_{1}}\sum_{t,s=1}^{T}\left[ \mathbf{\bar{C}}%
\right] _{it}[\mathbb{A}^{\left( -i\right) }-\mathbb{A}]_{ts}\sum_{j=1,j\neq
i}^{N}\left[ \mathbf{\bar{C}}^{\prime }\right] _{sj}\tilde{q}_{jj}^{-1}\frac{%
1}{T}\mathbf{x}_{j}^{o\prime }{\mathbb{M}}_{\mathbf{\iota }_{T}}\mathbf{u}%
_{i}^{\dag }m_{ji} \\
&=&\tilde{q}_{ii}^{-1}\frac{-1}{N_{1}}\sum_{t,s=1}^{T}\left[ \mathbf{\bar{C}}%
\right] _{it}[\mathbb{A}^{\left( -i\right) }-\mathbb{A}]_{ts}\sum_{j=1}^{N}%
\left[ \mathbf{\bar{C}}^{\prime }\right] _{sj}\tilde{q}_{jj}^{-1}\frac{1}{T}%
\mathbf{x}_{j}^{o\prime }{\mathbb{M}}_{\mathbf{\iota }_{T}}\mathbf{u}%
_{i}^{\dag }m_{ji} \\
&&+\tilde{q}_{ii}^{-1}\frac{1}{N_{1}}\sum_{t,s=1}^{T}\left[ \mathbf{\bar{C}}%
\right] _{it}[\mathbb{A}^{\left( -i\right) }-\mathbb{A}]_{ts}\left[ \mathbf{%
\bar{C}}^{\prime }\right] _{si}\tilde{q}_{ii}^{-1}\frac{1}{T}\mathbf{x}%
_{i}^{o\prime }{\mathbb{M}}_{\mathbf{\iota }_{T}}\mathbf{u}_{i}^{\dag }m_{ii}
\\
&\equiv &\tilde{q}_{ii}^{-1}R_{3,1i}+\tilde{q}_{ii}^{-1}R_{3,2i},\text{ and}
\\
B_{3,4}^{\left( -i\right) } &=&\frac{1}{N_{1}}\mathbf{S}_{N}\mathbb{Q}%
^{\left( d\right) }\mathbf{\breve{C}}^{\left( i\right) }\mathbb{A}^{\left(
-i\right) }\mathbf{\bar{C}}^{\prime }\mathbb{Q}^{\left( d\right) }\frac{1}{T}%
\mathbf{\breve{X}}^{o\left( -i\right) \prime }\left( {\mathbb{M}}_{\mathbf{%
\iota }_{N}}\otimes {\mathbb{M}}_{\mathbf{\iota }_{T}}\right) \mathbf{\breve{%
U}}^{\dag \left( i\right) } \\
&=&\tilde{q}_{ii}^{-1}\frac{1}{N_{1}}\sum_{t,s=1}^{T}\left[ \mathbf{\bar{C}}%
\right] _{it}[\mathbb{A}^{\left( -i\right) }]_{ts}\sum_{j=1,j\neq i}^{N}[%
\mathbf{\bar{C}}^{\prime }]_{sj}\tilde{q}_{jj}^{-1}\frac{1}{T}\mathbf{x}%
_{j}^{o\prime }{\mathbb{M}}_{\mathbf{\iota }_{T}}\mathbf{u}_{i}^{\dag }[{%
\mathbb{P}}_{\mathbf{\iota }_{N}}-{\mathbb{P}}_{\mathbf{\iota }_{N}^{\left(
-i\right) }}]_{ji}\equiv \tilde{q}_{ii}^{-1}R_{3,3i}
\end{eqnarray*}%
where 
\begin{eqnarray*}
\max_{i}\left\Vert R_{3,1i}\right\Vert &\lesssim
&N^{-1}\max_{i}\sum_{t=1}^{T}\left\Vert \left[ \mathbf{C}\right]
_{it}\right\Vert \max_{t,s}\left\Vert [\mathbb{A}^{\left( -i\right) }-%
\mathbb{A}]_{ts}\right\Vert \max_{i,s}\left\Vert \sum_{j=1}^{N}\left[ 
\mathbf{C}^{\prime }\right] _{sj}\tilde{q}_{jj}^{-1}\frac{1}{T}\mathbf{x}%
_{j}^{o\prime }{\mathbb{M}}_{\mathbf{\iota }_{T}}\mathbf{u}_{i}^{\dag
}m_{ji}\right\Vert \\
&=&N^{-1}O_{p}(N^{-1/2+1/(4+2\delta )})O_{p}(N^{-1+1/(2+\delta
)})O_{p}(1)=o_{p}\left( N^{-1}\right) ,\text{ } \\
\max_{i}\left\Vert R_{3,2i}\right\Vert &\lesssim
&N^{-1}\max_{i}\sum_{t=1}^{T}\left\Vert \left[ \mathbf{C}\right]
_{it}\right\Vert \max_{t,s}\left\Vert [\mathbb{A}^{\left( -i\right) }-%
\mathbb{A}]_{ts}\right\Vert \max_{i,s}\left\Vert \left[ \mathbf{C}^{\prime }%
\right] _{si}\tilde{q}_{ii}^{-1}\frac{1}{T}\mathbf{x}_{i}^{o\prime }{\mathbb{%
M}}_{\mathbf{\iota }_{T}}\mathbf{u}_{i}^{\dag }m_{ii}\right\Vert \\
&\lesssim &N^{-1}\max_{i}\sum_{t=1}^{T}\left\Vert \left[ \mathbf{C}\right]
_{it}\right\Vert \max_{t,s}\left\Vert [\mathbb{A}^{\left( -i\right) }-%
\mathbb{A}]_{ts}\right\Vert \left( \frac{1}{N^{1/2}}\sum_{i}\left\Vert 
\tilde{q}_{ii}^{-1}\right\Vert \left\Vert \mathbf{x}_{i}^{o}\right\Vert
^{2}\right) \max_{i}\left\Vert \mathbf{u}_{i}^{\dag }\right\Vert \\
&=&N^{-1}O_{p}(N^{-1/2+1/(4+2\delta )})O_{p}(N^{-1+1/(2+\delta
)})O_{p}(N^{1/2})O_{p}(N^{1/(4+2\delta )})=o_{p}\left( N^{-1}\right) ,\text{
and} \\
\max_{i}\left\Vert R_{3,3i}\right\Vert &\lesssim &N^{-1}\max_{i}\left(
\sum_{t=1}^{T}\left\Vert [\mathbf{C}]_{it}\right\Vert \right)
\max_{i,t}\left( \sum_{s=1}^{T}\left\Vert [\mathbb{A}^{\left( -i\right)
}]_{ts}\right\Vert \right) \\
&&\times \max_{i}\left\Vert \frac{1}{T}\sum_{j=1,j\neq i}^{N}[\mathbf{C}%
^{\prime }]_{sj}\tilde{q}_{jj}^{-1}\mathbf{x}_{j}^{o\prime }{\mathbb{M}}_{%
\mathbf{\iota }_{T}}\mathbf{u}_{i}^{\dag }[{\mathbb{P}}_{\mathbf{\iota }%
_{N}}-{\mathbb{P}}_{\mathbf{\iota }_{N}^{\left( -i\right) }}]_{ji}\right\Vert
\\
&=&N^{-1}O_{p}(N^{-1/2+1/(4+2\delta )})O_{p}\left( 1\right)
O_{p}(N^{-1/2+1/(4+2\delta )})=o_{p}\left( N^{-1}\right) .
\end{eqnarray*}%
So $B_{3,l}^{\left( -i\right) }=\tilde{q}_{ii}^{-1}o_{p}(N^{-1})\ $uniformly
in $i\in \left[ N\right] .$

Next, noting that $\mathbf{\breve{Q}}^{\left( i,d\right) }\mathbb{Q}^{\left(
d\right) }\mathbf{\breve{X}}^{o\left( -i\right) \prime }=\mathbf{0}%
_{NK_{x}\times NT}$ and $\mathbf{\breve{C}}^{\left( i\right) \prime }\mathbb{%
Q}^{\left( d\right) }\mathbf{\breve{X}}^{o\left( -i\right) \prime }=\mathbf{0%
}_{K_{1}T\times NT},$ we have%
\begin{eqnarray*}
B_{3,3}^{\left( -i\right) } &=&\frac{1}{N_{1}}\mathbf{S}_{N}\mathbb{Q}%
^{\left( d\right) }\mathbf{\breve{Q}}^{\left( i,d\right) }\mathbb{Q}^{\left(
d\right) }\frac{1}{T}\mathbf{\breve{X}}^{o\left( -i\right) \prime }\left( {%
\mathbb{M}}_{\mathbf{\iota }_{N}}\otimes {\mathbb{M}}_{\mathbf{\iota }%
_{T}}\right) \mathbf{\breve{U}}^{\dag \left( i\right) }=\mathbf{0}%
_{K_{x}\times 1}, \\
B_{3,5}^{\left( -i\right) } &=&\frac{1}{N_{1}}\mathbf{S}_{N}\mathbb{Q}%
^{\left( d\right) }\mathbf{\bar{C}}\mathbb{A}^{\left( -i\right) }\mathbf{%
\breve{C}}^{\left( i\right) \prime }\mathbb{Q}^{\left( d\right) }\frac{1}{T}%
\mathbf{\breve{X}}^{o\left( -i\right) \prime }\left( {\mathbb{M}}_{\mathbf{%
\iota }_{N}}\otimes {\mathbb{M}}_{\mathbf{\iota }_{T}}\right) \mathbf{\breve{%
U}}^{\dag \left( i\right) }=\mathbf{0}_{K_{x}\times 1}\text{ and } \\
B_{3,6}^{\left( -i\right) } &=&\frac{-1}{N_{1}}\mathbf{S}_{N}\mathbb{Q}%
^{\left( d\right) }\mathbf{\breve{C}}^{\left( i\right) }\mathbb{A}^{\left(
-i\right) }\mathbf{\breve{C}}^{\left( i\right) \prime }\mathbb{Q}^{\left(
d\right) }\frac{1}{T}\mathbf{\breve{X}}^{o\left( -i\right) \prime }\left( {%
\mathbb{M}}_{\mathbf{\iota }_{N}}\otimes {\mathbb{M}}_{\mathbf{\iota }%
_{T}}\right) \mathbf{\breve{U}}^{\dag \left( i\right) }=\mathbf{0}%
_{K_{x}\times 1}.
\end{eqnarray*}%
In sum, we have $B_{3}^{\left( -i\right) }=\frac{1}{NN_{1}}\sum_{j=1}^{N}%
\hat{\omega}_{j}\frac{1}{T}\mathbf{x}_{j}^{o\prime }{\mathbb{M}}_{\mathbf{%
\iota }_{T}}\mathbf{u}_{i}^{\dag }+R_{3i}$\textbf{\ }uniformly in\textbf{\ }$%
i\in \left[ N\right] ,\ $where $R_{3i}=\left( \left\Vert \hat{\omega}%
_{i}\right\Vert +\left\Vert \tilde{q}_{ii}^{-1}\right\Vert \right)
o_{p}(N^{-1}).\medskip $

(iv) We study $B_{4,l}^{\left( -i\right) },$ $l\in \left[ 6\right] $, in
turn. For $B_{4,1}^{\left( -i\right) },$ we have%
\begin{eqnarray*}
B_{4,1}^{\left( -i\right) } &=&\frac{1}{N_{1}}\mathbf{S}_{N}\mathbb{Q}%
^{\left( d\right) }[\mathbf{Q}^{\left( d\right) }+\mathbf{\bar{C}}\mathbb{A}%
\mathbf{\bar{C}}^{\prime }]\mathbb{Q}^{\left( d\right) }L_{4}^{MG\left(
-i\right) } \\
&=&\frac{1}{N_{1}}\mathbf{S}_{N}\mathbb{Q}^{\left( d\right) }[\mathbf{Q}%
^{\left( d\right) }+\mathbf{C}\mathbb{A}\mathbf{C}^{\prime }]\mathbb{Q}%
^{\left( d\right) }L_{4}^{MG\left( -i\right) }+\frac{1}{N_{1}}\mathbf{S}_{N}%
\mathbb{Q}^{\left( d\right) }[\mathbf{\bar{C}}\mathbb{A}\mathbf{\bar{C}}%
^{\prime }-\mathbf{C}\mathbb{A}\mathbf{C}^{\prime }]\mathbb{Q}^{\left(
d\right) }L_{4}^{MG\left( -i\right) } \\
&=&\frac{1}{N_{1}}\mathbf{S}_{N}\left[ \frac{1}{T}\mathbf{X}^{\prime }\left( 
{\mathbb{M}}_{\mathbf{\iota }_{N}}\otimes {\mathbb{M}}_{\mathbf{\iota }%
_{T}}\right) \mathbf{X}\right] ^{-1}\frac{1}{T}\mathbf{\breve{X}}^{o\left(
-i\right) \prime }[({\mathbb{P}}_{\mathbf{\iota }_{N}}-{\mathbb{P}}_{\mathbf{%
\iota }_{N}^{\left( -i\right) }})\otimes {\mathbb{M}}_{\mathbf{\iota }_{T}}]%
\mathbf{U}^{\dag \left( -i\right) }+\left\Vert \tilde{q}_{ii}^{-1}\right%
\Vert o_{p}(N^{-1})\text{ } \\
&=&\frac{1}{N_{1}}\mathbf{S}_{N}\mathbf{\hat{Q}}^{-1}\frac{1}{T}\mathbf{%
\breve{X}}^{o\left( -i\right) \prime }[({\mathbb{P}}_{\mathbf{\iota }_{N}}-{%
\mathbb{P}}_{\mathbf{\iota }_{N}^{\left( -i\right) }})\otimes {\mathbb{M}}_{%
\mathbf{\iota }_{T}}]\mathbf{U}^{\dag \left( -i\right) }+\left\Vert \tilde{q}%
_{ii}^{-1}\right\Vert o_{p}(N^{-1}) \\
&=&\frac{1}{N_{1}}\sum_{j=1,j\neq i}^{N}\hat{\omega}_{j}\sum_{k=1,k\neq
i}^{N}\frac{1}{T}\mathbf{x}_{j}^{o\prime }{\mathbb{M}}_{\mathbf{\iota }_{T}}%
\mathbf{u}_{k}^{\dag }[{\mathbb{P}}_{\mathbf{\iota }_{N}}-{\mathbb{P}}_{%
\mathbf{\iota }_{N}^{\left( -i\right) }}]_{jk}+\left\Vert \tilde{q}%
_{ii}^{-1}\right\Vert o_{p}(N^{-1}),
\end{eqnarray*}%
where the first $o_{p}(N^{-1})$ term follows from similar arguments as used
in the analysis of $B_{1,12}^{\left( -i\right) }.$ Let $\Theta _{il}$'s be
as defined in the decomposition of ${\mathbb{P}}_{\mathbf{\iota }_{N}}-{%
\mathbb{P}}_{\mathbf{\iota }_{N}^{\left( -i\right) }}$ in (\ref{P_decomp}).
Note that when $j\neq i$ and $k\neq i,$ we have $\left[ \Theta _{il}\right]
_{jk}=0$ for $l\in \lbrack 3].$ This implies that 
\begin{equation}
\lbrack {\mathbb{P}}_{\mathbf{\iota }_{N}}-{\mathbb{P}}_{\mathbf{\iota }%
_{N}^{\left( -i\right) }}]_{jk}=\left[ \Theta _{i4}\right] _{jk}=\frac{-1}{%
NN_{1}}\text{ when }j\neq i\text{ and }k\neq i.  \label{PP1}
\end{equation}%
Then 
\begin{eqnarray*}
B_{4,1}^{\left( -i\right) } &=&\frac{-1}{NN_{1}^{2}}\sum_{j=1,j\neq
i}^{N}\sum_{k=1,k\neq i}^{N}\hat{\omega}_{j}\frac{1}{T}\mathbf{x}%
_{j}^{o\prime }{\mathbb{M}}_{\mathbf{\iota }_{T}}\mathbf{u}_{k}^{\dag
}+o_{p}(N^{-1}) \\
&=&\frac{-1}{NN_{1}^{2}}\sum_{j,k=1}^{N}\hat{\omega}_{j}\frac{1}{T}\mathbf{x}%
_{j}^{o\prime }{\mathbb{M}}_{\mathbf{\iota }_{T}}\mathbf{u}_{k}^{\dag }+%
\frac{1}{NN_{1}^{2}}\sum_{j=1}^{N}\hat{\omega}_{j}\frac{1}{T}\mathbf{x}%
_{j}^{o\prime }{\mathbb{M}}_{\mathbf{\iota }_{T}}\mathbf{u}_{i}^{\dag }+%
\frac{1}{NN_{1}^{2}}\sum_{k=1}^{N}\hat{\omega}_{i}\frac{1}{T}\mathbf{x}%
_{i}^{o\prime }{\mathbb{M}}_{\mathbf{\iota }_{T}}\mathbf{u}_{k}^{\dag } \\
&&-\frac{1}{NN_{1}^{2}}\hat{\omega}_{i}\frac{1}{T}\mathbf{x}_{i}^{o\prime }{%
\mathbb{M}}_{\mathbf{\iota }_{T}}\mathbf{u}_{i}^{\dag }+o_{p}(N^{-1})\equiv
-I_{1}+I_{2i}+I_{3i}-I_{4i}.
\end{eqnarray*}%
Note that 
\begin{equation}
\frac{1}{N}\sum_{j=1}^{N}\left\Vert \frac{1}{N}\sum_{k=1}^{N}\frac{1}{T}%
\mathbf{x}_{j}^{o\prime }{\mathbb{M}}_{\mathbf{\iota }_{T}}\mathbf{u}%
_{k}^{\dag }\right\Vert ^{2}\lesssim \frac{1}{N}\sum_{j=1}^{N}\left\{ \frac{1%
}{T}\left\Vert \mathbf{x}_{j}^{o}\right\Vert \frac{1}{N}\left\Vert
\sum_{k=1}^{N}\mathbf{u}_{k}^{\dag }\right\Vert \right\} ^{2}=O_{p}(N^{-1})
\label{inter0}
\end{equation}%
by Markov equality. Then by Lemma \ref{LemA.1} and Cauchy-Schwarz inequality,%
\begin{eqnarray}
\left\Vert I_{1}\right\Vert &=&\frac{1}{N_{1}^{2}}\left\Vert \sum_{j=1}^{N}%
\hat{\omega}_{j}\frac{1}{N}\sum_{k=1}^{N}\frac{1}{T}\mathbf{x}_{j}^{o\prime }%
{\mathbb{M}}_{\mathbf{\iota }_{T}}\mathbf{u}_{k}^{\dag }\right\Vert \lesssim 
\frac{1}{N}\left\{ \frac{1}{N_{1}}\sum_{l=1}^{N}\left\Vert \hat{\omega}%
_{j}\right\Vert ^{2}\right\} ^{1/2}\left\{ \frac{1}{N}\sum_{j=1}^{N}\left%
\Vert \frac{1}{N}\sum_{k=1}^{N}\frac{1}{T}\mathbf{x}_{j}^{o\prime }{\mathbb{M%
}}_{\mathbf{\iota }_{T}}\mathbf{u}_{k}^{\dag }\right\Vert ^{2}\right\} ^{1/2}
\notag \\
&=&N^{-1}O_{p}(1)O_{p}(N^{-1/2})=o_{p}(N^{-1}).  \label{inter2}
\end{eqnarray}%
In addition, it is trivial to show that $\max_{i}\left\Vert
I_{2i}\right\Vert =o_{p}(N^{-1})$ and $I_{li}=\hat{\omega}_{i}o_{p}(N^{-1})$
for $l=3,4.$ Then $B_{4,1}^{\left( -i\right) }=\left( \left\Vert \hat{\omega}%
_{i}\right\Vert +1\right) o_{p}(N^{-1}).$

For $B_{4,2}^{\left( -i\right) },$ we have by (\ref{PP1}), and Lemma \ref%
{LemS.1}(iii) and (vi), 
\begin{eqnarray*}
\max_{i}\left\Vert B_{4,2}^{\left( -i\right) }\right\Vert
&=&\max_{i}\left\Vert \frac{1}{N_{1}}\mathbf{S}_{N}\mathbb{Q}^{\left(
d\right) }\mathbf{\bar{C}}(\mathbb{A}^{\left( -i\right) }-\mathbb{A})\mathbf{%
\bar{C}}^{\prime }\mathbb{Q}^{\left( d\right) }\frac{1}{T}\mathbf{\breve{X}}%
^{o\left( -i\right) \prime }[({\mathbb{P}}_{\mathbf{\iota }_{N}}-{\mathbb{P}}%
_{\mathbf{\iota }_{N}^{\left( -i\right) }})\otimes {\mathbb{M}}_{\mathbf{%
\iota }_{T}}]\mathbf{U}^{\dag \left( -i\right) }\right\Vert \\
&=&c_{1N}^{2}\max_{i}\left\Vert \frac{1}{N_{1}}\sum_{j}\sum_{k\neq i,l\neq
i}^{N}\tilde{q}_{jj}^{-1}[\mathbf{C]}_{j}[\mathbb{A}^{\left( -i\right) }-%
\mathbb{A]}[\mathbf{C}^{\prime }]_{\cdot k}\tilde{q}_{kk}^{-1}\frac{1}{T}%
\mathbf{x}_{k}^{o\prime }{\mathbb{M}}_{\mathbf{\iota }_{T}}\mathbf{u}%
_{l}^{\dag }[{\mathbb{P}}_{\mathbf{\iota }_{N}}-{\mathbb{P}}_{\mathbf{\iota }%
_{N}^{\left( -i\right) }}]_{kl}\right\Vert \\
&\lesssim &c_{1N}^{2}\max_{i}\left\Vert \frac{1}{NN_{1}^{2}}%
\sum_{j}\sum_{k\neq i,l\neq i}^{N}\tilde{q}_{jj}^{-1}[\mathbf{C]}_{j}[%
\mathbb{A}^{\left( -i\right) }-\mathbb{A]}[\mathbf{C}^{\prime }]_{\cdot k}%
\tilde{q}_{kk}^{-1}\frac{1}{T}\mathbf{x}_{k}^{o\prime }{\mathbb{M}}_{\mathbf{%
\iota }_{T}}\mathbf{u}_{l}^{\dag }\right\Vert \\
&\lesssim &\frac{1}{N}\left( \frac{1}{N^{1/2}}\sum_{j}\left\Vert \tilde{q}%
_{jj}^{-1}\right\Vert \left\Vert [\mathbf{C]}_{j}\right\Vert \right)
^{2}\max_{i}\left\Vert \mathbb{A}^{\left( -i\right) }-\mathbb{A}\right\Vert
_{\text{sp}}\max_{k}\left\Vert \mathbf{x}_{k}^{o}\right\Vert
\max_{i}\left\Vert \frac{1}{N}\sum_{l\neq i}^{N}\mathbf{u}_{l}^{\dag
}\right\Vert \\
&\lesssim &\frac{1}{N}\left( \frac{1}{N}\sum_{j}\left\Vert \tilde{q}%
_{jj}^{-1}\right\Vert \left\Vert \mathbf{x}_{j}^{o}\right\Vert \right)
^{2}\max_{i}\left\Vert \mathbb{A}^{\left( -i\right) }-\mathbb{A}\right\Vert
_{\text{sp}}\max_{k}\left\Vert \mathbf{x}_{k}^{o}\right\Vert
\max_{i}\left\Vert \frac{1}{N}\sum_{l\neq i}^{N}\mathbf{u}_{l}^{\dag
}\right\Vert \\
&=&N^{-1}O_{p}(1)O_{p}(N^{-1+1/(2+\delta )})O_{p}(N^{1/(2+\delta
)})O_{p}(N^{-1/2})=o_{p}(N^{-1}).
\end{eqnarray*}%
Similarly, for $B_{4,4}^{\left( -i\right) }$ we have by (\ref{PP1}), and
Lemma \ref{LemS.1}(i) and (iii)%
\begin{eqnarray*}
B_{4,4}^{\left( -i\right) } &=&\frac{1}{N_{1}}\mathbf{S}_{N}\mathbb{Q}%
^{\left( d\right) }\mathbf{\breve{C}}^{\left( i\right) }\mathbb{A}^{\left(
-i\right) }\mathbf{\bar{C}}^{\prime }\mathbb{Q}^{\left( d\right) }\frac{1}{T}%
\mathbf{\breve{X}}^{o\left( -i\right) \prime }[({\mathbb{P}}_{\mathbf{\iota }%
_{N}}-{\mathbb{P}}_{\mathbf{\iota }_{N}^{\left( -i\right) }})\otimes {%
\mathbb{M}}_{\mathbf{\iota }_{T}}]\mathbf{\breve{U}}^{\dag \left( -i\right) }%
\text{ \ \ } \\
&=&\tilde{q}_{ii}^{-1}\frac{1}{N_{1}}\left[ \mathbf{\bar{C}}\right] _{i\cdot
}\mathbb{A}^{\left( -i\right) }\sum_{j\neq i}\sum_{k\neq i}[\mathbf{\bar{C}}%
^{\prime }]_{\cdot j}\tilde{q}_{jj}^{-1}\text{ }\frac{1}{T}\mathbf{x}%
_{j}^{\prime }{\mathbb{M}}_{\mathbf{\iota }_{T}}\mathbf{u}_{k}^{\dag }[{%
\mathbb{P}}_{\mathbf{\iota }_{N}}-{\mathbb{P}}_{\mathbf{\iota }_{N}^{\left(
-i\right) }}]_{jk} \\
&=&\tilde{q}_{ii}^{-1}c_{1N}^{2}\frac{-1}{NN_{1}^{2}}\left[ \mathbf{C}\right]
_{i\cdot }\mathbb{A}^{\left( -i\right) }\sum_{j\neq i}[\mathbf{C}^{\prime
}]_{\cdot j}\tilde{q}_{jj}^{-1}\text{ }\sum_{k\neq i}\frac{1}{T}\mathbf{x}%
_{j}^{\prime }{\mathbb{M}}_{\mathbf{\iota }_{T}}\mathbf{u}_{k}^{\dag }\equiv 
\tilde{q}_{ii}^{-1}R_{4i},
\end{eqnarray*}%
where 
\begin{eqnarray*}
\max_{i}\left\Vert R_{4i}\right\Vert &\lesssim &\max_{i}\left( \left\Vert [%
\mathbf{C}]_{i\cdot }\right\Vert \right) \max_{i}\left\Vert \mathbb{A}%
^{\left( -i\right) }\right\Vert \left\{ \frac{1}{N^{1/2}}\sum_{j=1}\left%
\Vert \mathbf{x}_{j}^{o}\right\Vert \left\Vert \tilde{q}_{jj}^{-1}\right%
\Vert \right\} \max_{i,j}\frac{1}{NN_{1}^{2}}\left\Vert \sum_{k\neq i}\frac{1%
}{T}\mathbf{x}_{j}^{o\prime }{\mathbb{M}}_{\mathbf{\iota }_{T}}\mathbf{u}%
_{k}^{\dag }\right\Vert \\
&=&O_{p}(N^{-1/2+1/(4+2\delta )})O_{p}\left( 1\right)
O_{p}(N^{1/2})O_{p}(N^{-5/2+1/(4+2\delta )})=o_{p}\left( N^{-1}\right) \text{%
.}
\end{eqnarray*}%
As in the analyses of $B_{3,3}^{\left( -i\right) },$ $B_{3,5}^{\left(
-i\right) },$ and $B_{3,6}^{\left( -i\right) },$ we use the fact that $%
\mathbf{\breve{Q}}^{\left( i,d\right) }\mathbb{Q}^{\left( d\right) }\frac{1}{%
T}\mathbf{\breve{X}}^{o\left( -i\right) \prime }=\mathbf{0}_{NK_{x}\times
NT} $ and $\mathbf{\breve{C}}^{\left( i\right) \prime }\mathbb{Q}^{\left(
d\right) }\frac{1}{T}\mathbf{\breve{X}}^{o\left( -i\right) \prime }=\mathbf{0%
}_{K_{x}T\times NT}$ to obtain that 
\begin{eqnarray*}
B_{4,3}^{\left( -i\right) } &=&\frac{-1}{N_{1}}\mathbf{S}_{N}\mathbb{Q}%
^{\left( d\right) }\mathbf{\breve{Q}}^{\left( i,d\right) }\mathbb{Q}^{\left(
d\right) }\frac{1}{T}\mathbf{\breve{X}}^{o\left( -i\right) \prime }[({%
\mathbb{P}}_{\mathbf{\iota }_{N}}-{\mathbb{P}}_{\mathbf{\iota }_{N}^{\left(
-i\right) }})\otimes {\mathbb{M}}_{\mathbf{\iota }_{T}}]\mathbf{U}^{\dag
(-i)}=\mathbf{0}_{K_{x}\times 1}, \\
B_{4,5}^{\left( -i\right) } &=&\frac{-1}{N_{1}}\mathbf{S}_{N}\mathbb{Q}%
^{\left( d\right) }\mathbf{\bar{C}}\mathbb{A}^{\left( -i\right) }\mathbf{%
\breve{C}}^{\left( i\right) \prime }\mathbb{Q}^{\left( d\right) }\frac{1}{T}%
\mathbf{\breve{X}}^{o\left( -i\right) \prime }[({\mathbb{P}}_{\mathbf{\iota }%
_{N}}-{\mathbb{P}}_{\mathbf{\iota }_{N}^{\left( -i\right) }})\otimes {%
\mathbb{M}}_{\mathbf{\iota }_{T}}]\mathbf{U}^{\dag (-i)}=\mathbf{0}%
_{K_{x}\times 1},\text{ and} \\
B_{4,6}^{\left( -i\right) } &=&\frac{1}{N_{1}}\mathbf{S}_{N}\mathbb{Q}%
^{\left( d\right) }\mathbf{\breve{C}}^{\left( i\right) }\mathbb{A}^{\left(
-i\right) }\mathbf{\breve{C}}^{\left( i\right) \prime }\mathbb{Q}^{\left(
d\right) }\frac{1}{T}\mathbf{\breve{X}}^{o\left( -i\right) \prime }[({%
\mathbb{P}}_{\mathbf{\iota }_{N}}-{\mathbb{P}}_{\mathbf{\iota }_{N}^{\left(
-i\right) }})\otimes {\mathbb{M}}_{\mathbf{\iota }_{T}}]\mathbf{U}^{\dag
(-i)}=\mathbf{0}_{K_{x}\times 1}.
\end{eqnarray*}%
In sum, we have $B_{4}^{\left( -i\right) }=\left( \left\Vert \tilde{q}%
_{ii}^{-1}\right\Vert +\left\Vert \hat{\omega}_{i}\right\Vert +1\right)
o_{p}\left( N^{-1}\right) \ $uniformly in\textbf{\ }$i\in \left[ N\right] .$
This completes the proof of Lemma \ref{LemA.2}. ${\tiny \blacksquare }$

\subsection{Proof of Lemma \protect\ref{LemA.3} \label{SecS1.3}}

(i) We make the following decomposition 
\begin{eqnarray*}
\frac{1}{N_{1}T}\sum_{i=1}^{N}\hat{\omega}_{i}\mathbf{x}_{i}^{o\prime }{%
\mathbb{M}}_{\mathbf{\iota }_{T}}\mathbf{u}_{i}^{\dag } &=&\frac{1}{N_{1}T}%
\sum_{i=1}^{N}\hat{\omega}_{i}\mathbf{x}_{i}^{o\prime }\mathbf{u}_{i}+\frac{1%
}{N_{1}T}\sum_{i=1}^{N}\hat{\omega}_{i}\mathbf{x}_{i}^{o\prime }F\lambda
_{i}-\frac{1}{N_{1}T^{2}}\sum_{i=1}^{N}\hat{\omega}_{i}\mathbf{x}%
_{i}^{o\prime }\mathbf{\iota }_{T}\mathbf{\iota }_{T}^{\prime }\mathbf{u}_{i}
\\
&&-\frac{1}{N_{1}T^{2}}\sum_{i=1}^{N}\hat{\omega}_{i}\mathbf{x}_{i}^{o\prime
}\mathbf{\iota }_{T}\mathbf{\iota }_{T}^{\prime }F\lambda _{i}\equiv
II_{1}+II_{2}-II_{3}-II_{4}.
\end{eqnarray*}%
By Assumption \ref{Assmp3.1}, Lemma \ref{LemA.1} and Markov inequalities%
\begin{eqnarray*}
E_{\mathcal{F}_{N,0}^{MG}}\left( II_{1}\right) &=&\frac{1}{N_{1}T}%
\sum_{i=1}^{N}\hat{\omega}_{i}\mathbf{x}_{i}^{o\prime }E_{\mathcal{F}%
_{N,0}^{MG}}(\mathbf{u}_{i})=0,\text{ and} \\
\left\Vert \text{Var}_{\mathcal{F}_{N,0}^{MG}}\left( II_{1}\right)
\right\Vert &=&\frac{1}{N_{1}^{2}T^{2}}\left\Vert \sum_{i=1}^{N}\hat{\omega}%
_{i}\mathbf{x}_{i}^{o\prime }E_{\mathcal{F}_{N,0}^{MG}}(\mathbf{u}_{i}%
\mathbf{u}_{i}^{\prime })\mathbf{x}_{i}^{o}\hat{\omega}_{i}^{\prime
}\right\Vert \lesssim \frac{1}{N_{1}^{2}T^{2}}\sum_{i=1}^{N}\left\Vert \hat{%
\omega}_{i}\right\Vert ^{2}\left\Vert \mathbf{x}_{i}^{o}\right\Vert ^{2} \\
&\lesssim &\frac{1}{N}\left( \frac{1}{N}\sum_{i=1}^{N}\left\Vert \hat{\omega}%
_{i}\right\Vert ^{2}\right) \max_{i}\left\Vert \mathbf{x}_{i}^{o}\right\Vert
^{2}=N^{-1}O_{p}(1)O_{p}(N^{1/(2+\delta )})=o_{p}(1)
\end{eqnarray*}%
where we use the fact that 
\begin{equation*}
\frac{1}{N}\sum_{i=1}^{N}\left\Vert \hat{\omega}_{i}\right\Vert ^{2}\lesssim 
\frac{1}{N}\sum_{i=1}^{N}\left\Vert \tilde{q}_{jj}^{-1}\right\Vert
^{2}\{1+T^{-1}\left\Vert \mathbf{x}_{i}^{o}\right\Vert ^{2}\}=O_{p}\left(
1\right)
\end{equation*}%
by following the proof of Lemma \ref{LemA.1} and Assumption \ref{Assmp3.3}.
Then $II_{1}=o_{p}(1)$ by the conditional Chebyshev inequality. Analogously,
we can show that $II_{l}=o_{p}(1)$ for $l=2,3,4.$ Then $\frac{1}{N_{1}T}%
\sum_{i=1}^{N}\hat{\omega}_{i}\mathbf{x}_{i}^{o\prime }{\mathbb{M}}_{\mathbf{%
\iota }_{T}}\mathbf{u}_{i}^{\dag }=o_{p}(1).$

(ii) This is shown in the derivation of (\ref{inter2}). ${\tiny \blacksquare 
}$

\subsection{Proof of Lemma \protect\ref{LemA.4} \label{SecS1.4}}

Note that 
\begin{equation*}
\hat{Q}^{\left( -i\right) }=\frac{1}{N_{1}T}X^{\left( -i\right) \prime }(%
\mathbb{M}_{\mathbf{\iota }_{N_{1}}}\otimes \mathbb{M}_{\mathbf{\iota }%
_{T}})X^{\left( -i\right) }=\frac{1}{N_{1}T}X^{o\left( -i\right) \prime }(%
\mathbb{M}_{\mathbf{\iota }_{N_{1}}}\otimes \mathbb{M}_{\mathbf{\iota }%
_{T}})X^{o\left( -i\right) },
\end{equation*}%
where $X^{o\left( -i\right) }$ is a version of $X^{\left( -i\right) }$ with
both individual and time effects removed if they are contained in $x_{it}$: $%
X^{o\left( -i\right) }=(\mathbf{x}_{1}^{o\prime },...,\mathbf{x}%
_{i-1}^{o\prime },\mathbf{x}_{i+1}^{o\prime },...,\mathbf{x}_{N}^{o\prime
})^{\prime }.$ To prove Lemma \ref{LemA.4}, we add some notations. Let $%
C=N^{-1}T^{-1/2}X^{o\prime }(\mathbf{\iota }_{N}\otimes \mathbb{M}_{\mathbf{%
\iota }_{T}}),$ $C^{\left( -i\right) }=N_{1}^{-1}T^{-1/2}X^{o\left(
-i\right) \prime }(\mathbf{\iota }_{N_{1}}\otimes \mathbb{M}_{\mathbf{\iota }%
_{T}}),$ $Q^{\left( d\right) }=\frac{1}{NT}X^{o\prime }(\mathbb{I}%
_{N}\otimes \mathbb{M}_{\mathbf{\iota }_{T}})X^{o},\ $and $Q^{\left(
-i,d\right) }=\frac{1}{N_{1}T}X^{o\left( -i\right) \prime }(\mathbb{I}%
_{N_{1}}\otimes \mathbb{M}_{\mathbf{\iota }_{T}})X^{o\left( -i\right) }.$
Then%
\begin{eqnarray*}
\hat{Q} &=&\frac{1}{NT}X^{o\prime }(\mathbb{I}_{N}\otimes \mathbb{M}_{%
\mathbf{\iota }_{T}})X^{o}-\frac{1}{NT}X^{o\prime }(\mathbb{P}_{\mathbf{%
\iota }_{N}}\otimes \mathbb{M}_{\mathbf{\iota }_{T}})X^{o}=Q^{\left(
d\right) }-CC^{\prime }\text{ and} \\
\hat{Q}^{\left( -i\right) } &=&\frac{1}{N_{1}T}X^{o\left( -i\right) \prime }(%
\mathbb{I}_{N_{1}}\otimes \mathbb{M}_{\mathbf{\iota }_{T}})X^{o\left(
-i\right) }-\frac{1}{N_{1}T}X^{o\left( -i\right) \prime }(\mathbb{P}_{%
\mathbf{\iota }_{N_{1}}}\otimes \mathbb{M}_{\mathbf{\iota }_{T}})X^{o\left(
-i\right) }=Q^{\left( -i,d\right) }-C^{\left( -i\right) }C^{\left( -i\right)
\prime }.
\end{eqnarray*}%
Let $A=\left[ \mathbb{I}_{T}+C^{\prime }C\right] ^{-1}\ $and $A^{\left(
-i\right) }=\left[ \mathbb{I}_{T}+C^{\left( -i\right) \prime }C^{\left(
-i\right) }\right] ^{-1}.\ $By Sherman-Morrison Woodbury matrix identity
(e.g., Fact 6.4.31 in \cite{bernstein2005}), we have 
\begin{eqnarray*}
\hat{Q}^{-1} &=&\left[ Q^{\left( d\right) }-CC^{\prime }\right] ^{-1}=\left[
Q^{\left( d\right) }\right] ^{-1}\left[ Q^{\left( d\right) }+CAC^{\prime }%
\right] \left[ Q^{\left( d\right) }\right] ^{-1}\text{ and } \\
\left[ \hat{Q}^{(-i)}\right] ^{-1} &=&\left[ Q^{\left( -i,d\right)
}-C^{\left( -i\right) }C^{\left( -i\right) \prime }\right] ^{-1}=\left[
Q^{\left( -i,d\right) }\right] ^{-1}\left[ Q^{\left( -i,d\right) }+C^{\left(
-i\right) }A^{\left( -i\right) }C^{\left( -i\right) \prime }\right] \left[
Q^{\left( -i,d\right) }\right] ^{-1}.
\end{eqnarray*}%
For $\left[ \hat{Q}^{(-i)}\right] ^{-1},$ we make the following
decomposition: 
\begin{eqnarray}
\left[ \hat{Q}^{(-i)}\right] ^{-1} &=&[Q^{\left( -i,d\right) }]^{-1}\left[
Q^{\left( d\right) }+CAC^{\prime }\right] [Q^{\left( -i,d\right)
}]^{-1}+[Q^{\left( -i,d\right) }]^{-1}\left[ Q^{\left( -i,d\right)
}-Q^{\left( d\right) }\right] [Q^{\left( -i,d\right) }]^{-1}  \notag \\
&&+[Q^{\left( -i,d\right) }]^{-1}[C^{\left( -i\right) }A^{\left( -i\right)
}C^{\prime \left( -i\right) }-CAC^{\prime }][Q^{\left( -i,d\right) }]^{-1} 
\notag \\
&=&\hat{Q}^{-1}+[Q^{\left( d\right) }]^{-1}\left[ Q^{\left( d\right)
}+CAC^{\prime }\right] \left\{ [Q^{\left( -i,d\right) }]^{-1}-[Q^{\left(
d\right) }]^{-1}\right\}  \notag \\
&&+\left\{ [Q^{\left( -i,d\right) }]^{-1}-[Q^{\left( d\right)
}]^{-1}\right\} [Q^{\left( d\right) }+CAC^{\prime }]\left[ Q^{\left(
-i,d\right) }\right] ^{-1}  \notag \\
&&+[Q^{\left( -i,d\right) }]^{-1}[Q^{\left( -i,d\right) }-Q^{\left( d\right)
}][Q^{\left( -i,d\right) }]^{-1}+[Q^{\left( -i,d\right) }]^{-1}[C^{\left(
-i\right) }A^{\left( -i\right) }C^{\left( -i\right) \prime }-CAC^{\prime
}][Q^{\left( -i,d\right) }]^{-1}  \notag \\
&\equiv &\sum_{s=1}^{5}J_{s}^{P\left( -i\right) }.  \label{Qp_decomp}
\end{eqnarray}%
Let $\mathbf{\dot{x}}_{i}^{o}=\mathbb{M}_{\mathbf{\iota }_{T}}\mathbf{x}%
_{i}^{o}.$ To prove Lemma \ref{LemA.4}\textbf{, }we state and prove a
technical lemma.

\begin{lemma}
\label{LemS.2} Suppose that Assumption \ref{Assmp4.1} holds. Then

(i) $\left\Vert C\right\Vert =O_{p}\left( 1\right) $ and $\max_{i}\left\Vert
C^{\left( -i\right) }-C\right\Vert \lesssim \max_{i}N^{-1}\left( \left\Vert 
\mathbf{\dot{x}}_{i}^{o}\right\Vert +1\right) =O_{p}(N^{-1+1/(4+2\delta )});$

(ii) $\max_{i}\left\Vert A^{\left( -i\right) }-A\right\Vert \lesssim
\max_{i}N^{-1}\left( \left\Vert \mathbf{\dot{x}}_{i}^{o}\right\Vert
+1\right) =O_{p}(N^{-1+1/(4+2\delta )});$

(iii) $\max_{i}\left\Vert Q^{\left( -i,d\right) }-Q^{\left( d\right)
}\right\Vert \lesssim \max_{i}N_{1}^{-1}(\left\Vert \mathbf{\dot{x}}%
_{i}^{o}\right\Vert ^{2}+1)=O_{p}(N^{-1+1/(2+\delta )});$

(iv) $\max_{i}\left\Vert [Q^{\left( -i,d\right) }]^{-1}-[Q^{\left( d\right)
}]^{-1}\right\Vert \lesssim \max_{i}N_{1}^{-1}(\left\Vert \mathbf{\dot{x}}%
_{i}^{o}\right\Vert ^{2}+1)=O_{p}(N^{-1+1/(2+\delta )}).$
\end{lemma}

\noindent \textbf{Proof of Lemma \ref{LemS.2}}

(i)\ Noting that $C^{\prime }C=N^{-1}T^{-1}X^{o\prime }(P_{\mathbf{\iota }%
_{N}}\otimes \mathbb{M}_{\mathbf{\iota }_{T}})X^{o},$ we have 
\begin{equation*}
\left\Vert C\right\Vert ^{2}\leq \left( K_{x}\wedge T\right) \left\Vert
C\right\Vert _{\text{sp}}^{2}\lesssim \left\Vert CC^{\prime }\right\Vert _{%
\text{sp}}=\mu _{\max }(N^{-1}T^{-1}X^{o\prime }(P_{\mathbf{\iota }%
_{N}}\otimes \mathbb{M}_{\mathbf{\iota }_{T}})X^{o})=O_{p}\left( 1\right) .
\end{equation*}%
So $\left\Vert C\right\Vert =O_{p}\left( 1\right) .$ Noting that under
Assumption \ref{Assmp4.1}(ii) 
\begin{eqnarray*}
C-C^{\left( -i\right) } &=&N^{-1}T^{-1/2}X^{o\prime }(\mathbf{\iota }%
_{N}\otimes \mathbb{M}_{\mathbf{\iota }_{T}})-\left( N_{1}T\right)
^{-1/2}X^{o\left( -i\right) \prime }(\mathbf{\iota }_{N_{1}}\otimes \mathbb{M%
}_{\mathbf{\iota }_{T}}) \\
&=&N^{-1}T^{-1/2}\left[ X^{o\prime }(\mathbf{\iota }_{N}\otimes \mathbb{M}_{%
\mathbf{\iota }_{T}})-X^{o\left( -i\right) \prime }(\mathbf{\iota }%
_{N_{1}}\otimes \mathbb{M}_{\mathbf{\iota }_{T}})\right]
+(N^{-1}-N_{1}^{-1})T^{-1/2}X^{o\left( -i\right) \prime }(\mathbf{\iota }%
_{N_{1}}\otimes \mathbb{M}_{\mathbf{\iota }_{T}}) \\
&=&N^{-1}T^{-1/2}\mathbf{x}_{i}^{o\prime }\mathbb{M}_{\mathbf{\iota }%
_{T}}-N^{-1}N_{1}^{-1}T^{-1/2}\sum_{j=1}\mathbf{x}_{j}^{o\prime }\mathbb{M}_{%
\mathbf{\iota }_{T}}=N^{-1}T^{-1/2}\mathbf{\dot{x}}_{i}^{o\prime
}-N^{-1}N_{1}^{-1}T^{-1/2}\left( \sum_{j=1}^{N}\mathbf{\dot{x}}_{j}^{o\prime
}-\mathbf{\dot{x}}_{i}^{o\prime }\right) \\
&=&\left( NT\right) ^{-1/2}\mathbf{\dot{x}}_{i}^{o\prime
}-N^{-1}N_{1}^{-1}T^{-1/2}\sum_{j=1}^{N}\mathbf{\dot{x}}_{j}^{o\prime
}+o_{p}(N^{-1}),
\end{eqnarray*}%
we have $\max_{i}\left\Vert C^{\left( -i\right) }-C\right\Vert \lesssim
N^{-1}\left\Vert \mathbf{\dot{x}}_{i}^{o}\right\Vert +N^{-2}\left\Vert
\sum_{j}\mathbf{\dot{x}}_{j}^{o}\right\Vert \lesssim N^{-1/2}\left(
\left\Vert \mathbf{\dot{x}}_{i}^{o}\right\Vert +1\right)
=O_{p}(N^{-1+1/(4+2\delta )}).$

(ii) Noting that 
\begin{eqnarray*}
A^{\left( -i\right) }-A &=&\left[ \mathbb{I}_{T}+C^{\left( -i\right) \prime
}C^{\left( -i\right) }\right] ^{-1}-\left[ \mathbb{I}_{T}+C^{\prime }C\right]
^{-1} \\
&=&\left[ \mathbb{I}_{T}+C^{\left( -i\right) \prime }C^{\left( -i\right) }%
\right] ^{-1}\left( C^{\prime }C-C^{\left( -i\right) \prime }C^{\left(
-i\right) }\right) \left[ \mathbb{I}_{T}+C^{\prime }C\right] ^{-1}
\end{eqnarray*}%
and $C^{\prime }C-C^{\left( -i\right) \prime }C^{\left( -i\right) }=\left(
C-C^{\left( -i\right) }\right) ^{\prime }C+C^{\left( -i\right) \prime
}\left( C-C^{\left( -i\right) }\right) ,$ we have by the results in part
(i), 
\begin{eqnarray*}
\left\Vert A^{\left( -i\right) }-A\right\Vert &\lesssim &\left\Vert
C^{\prime }C-C^{\left( -i\right) \prime }C^{\left( -i\right) }\right\Vert
=\left\Vert \left( C-C^{\left( -i\right) }\right) ^{\prime }C+C^{\left(
-i\right) \prime }\left( C-C^{\left( -i\right) }\right) \right\Vert \\
&\lesssim &\left\Vert C-C^{\left( -i\right) }\right\Vert \left\Vert
C\right\Vert \left\{ 1+o_{p}(1)\right\} \lesssim N^{-1}\left( \left\Vert 
\mathbf{x}_{i}\right\Vert +1\right) =O_{p}(N^{-1+1/(4+2\delta )}).
\end{eqnarray*}

(iii)-(iv) Noting that 
\begin{eqnarray*}
Q^{\left( -i,d\right) }-Q^{\left( d\right) } &=&\frac{1}{N_{1}T}X^{o\left(
-i\right) \prime }(\mathbb{I}_{N_{1}}\otimes \mathbb{M}_{\mathbf{\iota }%
_{T}})X^{o\left( -i\right) }-\frac{1}{NT}X^{o\prime }(\mathbb{I}_{N}\otimes 
\mathbb{M}_{\mathbf{\iota }_{T}})X^{o} \\
&=&\frac{1}{N_{1}T}\left[ X^{o\left( -i\right) \prime }(\mathbb{I}%
_{N_{1}}\otimes \mathbb{M}_{\mathbf{\iota }_{T}})X^{o\left( -i\right)
}-X^{o\prime }(\mathbb{I}_{N}\otimes \mathbb{M}_{\mathbf{\iota }_{T}})X^{o}%
\right] +\frac{1}{N_{1}NT}X^{o\prime }(\mathbb{I}_{N}\otimes \mathbb{M}_{%
\mathbf{\iota }_{T}})X^{o} \\
&=&\frac{1}{N_{1}T}\mathbf{\dot{x}}_{i}^{o\prime }\mathbf{\dot{x}}_{i}^{o}-%
\frac{1}{N_{1}NT}\sum_{j=1}^{N}\mathbf{\dot{x}}_{j}^{o\prime }\mathbf{\dot{x}%
}_{j}^{o},
\end{eqnarray*}%
and $[Q^{\left( -i,d\right) }]^{-1}-[Q^{\left( d\right) }]^{-1}=-[Q^{\left(
-i,d\right) }]^{-1}(Q^{\left( -i,d\right) }-Q^{\left( d\right) })[Q^{\left(
d\right) }]^{-1},$ we have%
\begin{eqnarray*}
\left\Vert Q^{\left( -i,d\right) }-Q^{\left( d\right) }\right\Vert &\lesssim
&\frac{1}{N_{1}T}\left\Vert \mathbf{\dot{x}}_{i}^{o}\right\Vert ^{2}+\frac{1%
}{N_{1}NT}\left\Vert \sum_{j=1}^{N}\mathbf{\dot{x}}_{j}^{o\prime }\mathbf{%
\dot{x}}_{j}^{o}\right\Vert \lesssim \frac{1}{N_{1}}\left( \left\Vert 
\mathbf{\dot{x}}_{i}^{o}\right\Vert ^{2}+1\right) =O_{p}(N^{-1+1/(2+\delta
)})\text{ and} \\
\left\Vert \lbrack Q^{\left( -i,d\right) }]^{-1}-[Q^{\left( d\right)
}]^{-1}\right\Vert &\lesssim &\left\Vert Q^{\left( -i,d\right) }-Q^{\left(
d\right) }\right\Vert \lesssim \frac{1}{N_{1}}\left( \left\Vert \mathbf{\dot{%
x}}_{i}^{o}\right\Vert ^{2}+1\right) =O_{p}(N^{-1+1/(2+\delta )}).\text{ }%
{\tiny \blacksquare }
\end{eqnarray*}

\bigskip

\noindent \textbf{Proof of Lemma \ref{LemA.4}}

By (\ref{Lp_decomp}) and (\ref{Qp_decomp}), we can make the following
decomposition for $B_{h}^{P\left( -i\right) }:$ 
\begin{equation*}
B_{h}^{P\left( -i\right) }=\left[ \hat{Q}^{(-i)}\right] ^{-1}L_{h}^{P\left(
-i\right) }=\sum_{l=1}^{5}J_{l}^{P\left( -i\right) }L_{h}^{P\left( -i\right)
}\equiv \sum_{l=1}^{5}B_{h,l}^{P\left( -i\right) }\text{ for }h\in \left[ 4%
\right] ,\text{ }
\end{equation*}%
where $B_{h,l}^{P\left( -i\right) }=J_{l}^{P\left( -i\right) }L_{h}^{P\left(
-i\right) }.$ Below, we study $B_{h,l}^{\left( -i\right) }$ for $l\in \left[
5\right] $ in parts (i)--(iv) for $h\in \left[ 4\right] $ respectively. To
save space, let $\Delta _{1}^{\left( -i\right) }=Q^{\left( -i,d\right)
}-Q^{\left( d\right) }\ $and $\Delta _{2}^{\left( -i\right) }=\left[
Q^{\left( -i,d\right) }\right] ^{-1}-\left[ Q^{\left( d\right) }\right]
^{-1},$ and $\Delta _{3}^{\left( -i\right) }=C^{\left( -i\right) }A^{\left(
-i\right) }C^{\left( -i\right) \prime }-CAC^{\prime }.$

(i) We study $B_{1}^{P\left( -i\right) }=[\hat{Q}^{(-i)}]^{-1}L_{1}^{P\left(
-i\right) }=\sum_{l=1}^{5}J_{l}^{P\left( -i\right) }L_{1}^{P\left( -i\right)
}\equiv \sum_{l=1}^{5}B_{1,l}^{P\left( -i\right) }$ by studying $%
B_{1,l}^{P\left( -i\right) },$ $l\in \left[ 5\right] $, in turn. For $%
B_{1,1}^{P\left( -i\right) },$ we have%
\begin{equation*}
B_{1,1}^{P\left( -i\right) }=J_{1}^{P\left( -i\right) }L_{1}^{P\left(
-i\right) }=\hat{Q}^{-1}L_{1}^{P\left( -i\right) }=\frac{N}{N_{1}}\hat{Q}%
^{-1}\frac{1}{NT}X^{\prime }(\mathbb{M}_{\mathbf{\iota }_{N}}\otimes \mathbb{%
M}_{\mathbf{\iota }_{T}})\mathbf{U}^{\ddagger }=\frac{N}{N_{1}}\left( \hat{%
\beta}^{P}-\beta ^{0}\right) .
\end{equation*}%
By Assumption \ref{Assmp4.1}, we can readily show that 
\begin{equation*}
L_{1}^{P\left( -i\right) }=\frac{1}{N_{1}T}X^{\prime }(\mathbb{M}_{\mathbf{%
\iota }_{N}}\otimes \mathbb{M}_{\mathbf{\iota }_{T}})\mathbf{U}^{\ddagger }=%
\frac{1}{N_{1}T}\sum_{j}\mathbf{\ddot{x}}_{j}^{\prime }\mathbf{\dot{u}}%
_{j}^{\ddagger }=O_{p}(N^{-1/2}).
\end{equation*}%
For the other four terms, we have by Lemma \ref{LemS.2}(i)-(iv), 
\begin{eqnarray*}
\max_{i}\left\Vert B_{1,2}^{P\left( -i\right) }\right\Vert
&=&\max_{i}\left\Vert J_{2}^{P\left( -i\right) }L_{1}^{P\left( -i\right)
}\right\Vert =\max_{i}\left\Vert \left[ Q^{\left( d\right) }\right] ^{-1}%
\left[ Q^{\left( d\right) }+CAC^{\prime }\right] \Delta _{2}^{\left(
-i\right) }\frac{1}{N_{1}T}\sum_{j}\mathbf{\ddot{x}}_{j}^{\prime }\mathbf{u}%
_{j}^{\ddagger }\right\Vert \\
&\lesssim &\max_{i}\left\Vert \Delta _{2}^{\left( -i\right) }\right\Vert 
\frac{1}{N_{1}T}\left\Vert \sum_{j}\mathbf{\ddot{x}}_{j}^{\prime }\mathbf{%
\dot{u}}_{j}^{\ddagger }\right\Vert =O_{p}(N^{-1+1/(4+2\delta
)})O_{p}(N^{-1/2})=o_{p}(N^{-1}), \\
\max_{i}\left\Vert B_{1,3}^{P\left( -i\right) }\right\Vert
&=&\max_{i}\left\Vert J_{3}^{P\left( -i\right) }L_{1}^{P\left( -i\right)
}\right\Vert =\max_{i}\left\Vert \Delta _{2}^{\left( -i\right) }\left[
Q^{\left( d\right) }+CAC^{\prime }\right] \left[ Q^{\left( -i,d\right) }%
\right] ^{-1}\frac{1}{N_{1}T}\sum_{j}\mathbf{\ddot{x}}_{j}^{\prime }\mathbf{%
\dot{u}}_{j}^{\ddagger }\right\Vert \\
&\lesssim &\max_{i}\left\Vert \Delta _{2}^{\left( -i\right) }\right\Vert 
\frac{1}{N_{1}T}\left\Vert \sum_{j}\mathbf{\ddot{x}}_{j}^{\prime }\mathbf{%
\dot{u}}_{j}^{\ddagger }\right\Vert =o_{p}(N^{-1}),\ \text{and} \\
\max_{i}\left\Vert B_{1,4}^{P\left( -i\right) }\right\Vert
&=&\max_{i}\left\Vert J_{4}^{P\left( -i\right) }L_{1}^{P\left( -i\right)
}\right\Vert =\max_{i}\left\Vert \left[ Q^{\left( -i,d\right) }\right]
^{-1}\Delta _{1}^{\left( -i\right) }\left[ Q^{\left( -i,d\right) }\right]
^{-1}\frac{1}{N_{1}T}\sum_{j}\mathbf{\ddot{x}}_{j}^{\prime }\mathbf{u}%
_{j}^{\ddagger }\right\Vert \\
&\lesssim &\max_{i}\left\Vert \Delta _{1}^{\left( -i\right) }\right\Vert 
\frac{1}{N_{1}T}\left\Vert \sum_{j}\mathbf{\ddot{x}}_{j}^{\prime }\mathbf{u}%
_{j}^{\ddagger }\right\Vert =O_{p}(N^{-1+1/(2+\delta
)})O_{p}(N^{-1/2})=o_{p}(N^{-1}),\text{ and} \\
\max_{i}\left\Vert B_{1,5}^{P\left( -i\right) }\right\Vert
&=&\max_{i}\left\Vert J_{5}^{P\left( -i\right) }L_{1}^{P\left( -i\right)
}\right\Vert =\max_{i}\left\Vert \left[ Q^{\left( -i,d\right) }\right]
^{-1}\Delta _{3}^{\left( -i\right) }\left[ Q^{\left( -i,d\right) }\right]
^{-1}\frac{1}{N_{1}T}\sum_{j}\mathbf{\ddot{x}}_{j}^{\prime }\mathbf{u}%
_{j}^{\ddagger }\right\Vert \\
&\lesssim &\max_{i}\left\Vert \Delta _{3}^{\left( -i\right) }\right\Vert 
\frac{1}{N_{1}T}\left\Vert \sum_{j}\mathbf{\ddot{x}}_{j}^{\prime }\mathbf{u}%
_{j}^{\ddagger }\right\Vert =O_{p}(N^{-1+1/(4+2\delta
)})O_{p}(N^{-1/2})=o_{p}(N^{-1}).
\end{eqnarray*}%
In sum, $B_{1}^{P\left( -i\right) }=\frac{N}{N_{1}}(\hat{\beta}^{P}-\beta
^{0})+o_{p}(N^{-1})$ uniformly in\textbf{\ }$i\in \left[ N\right] .$ $%
\medskip $

(ii) We study $B_{2}^{P\left( -i\right) }=\left[ \hat{Q}^{(-i)}\right]
^{-1}L_{2}^{P\left( -i\right) }=\sum_{l=1}^{5}J_{l}^{P\left( -i\right)
}L_{2}^{P\left( -i\right) }\equiv \sum_{l=1}^{5}B_{2,l}^{P\left( -i\right) }$
by studying $B_{2,l}^{P\left( -i\right) },$ $l\in \left[ 5\right] $, in
turn. Noting that $m_{ij}=\mathbf{1}\left\{ i=j\right\} -\frac{1}{N},$ we
have%
\begin{eqnarray*}
L_{2}^{P\left( -i\right) } &=&\frac{-1}{N_{1}T}\breve{X}^{\left( i\right)
\prime }(\mathbb{M}_{\mathbf{\iota }_{N}}\otimes \mathbb{M}_{\mathbf{\iota }%
_{T}})\mathbf{U}^{\ddagger }=\frac{-1}{N_{1}T}\sum_{j=1}^{N}\mathbf{\dot{x}}%
_{i}^{\prime }\mathbf{u}_{j}^{\ddagger }m_{ij}=\frac{-1}{N_{1}T}\mathbf{\dot{%
x}}_{i}^{\prime }\mathbf{u}_{i}^{\ddagger }+\frac{1}{NN_{1}T}\sum_{j=1}^{N}%
\mathbf{\dot{x}}_{i}^{\prime }\mathbf{u}_{j}^{\ddagger } \\
&=&\frac{-1}{N_{1}T}\mathbf{\dot{x}}_{i}^{\prime }\mathbf{u}_{i}^{\ddagger
}+o_{p}(N^{-1})\text{ uniformly in }i\in \left[ N\right] .
\end{eqnarray*}%
Then we can readily show that $\max_{i}\left\Vert L_{2}^{P\left( -i\right)
}\right\Vert \lesssim \frac{1}{N_{1}T}\max_{i}\left\Vert \mathbf{\dot{x}}%
_{i}^{\prime }\mathbf{u}_{i}^{\ddagger }\right\Vert
+o_{p}(N^{-1})=O_{p}(N^{-1+1/(2+\delta )})$ under Assumption \ref{Assmp4.1}.
It follows that that uniformly in $i\in \left[ N\right] ,$ 
\begin{equation*}
B_{2,1}^{P\left( -i\right) }=J_{1}^{P\left( -i\right) }L_{2}^{P\left(
-i\right) }=\hat{Q}^{-1}\left[ \frac{-1}{N_{1}T}\mathbf{\dot{x}}_{i}^{\prime
}\mathbf{u}_{i}^{\ddagger }+o_{p}(N^{-1})\right] =-\hat{Q}^{-1}\frac{1}{%
N_{1}T}\mathbf{\dot{x}}_{i}^{\prime }\mathbf{u}_{i}^{\ddagger
}+o_{p}(N^{-1}).
\end{equation*}%
For the other four terms, we can use the above uniform bound for $%
L_{2}^{P\left( -i\right) }$ and Lemma \ref{LemS.2}(i)-(iv) to obtain the
rough probability bound:%
\begin{eqnarray*}
\max_{i}\left\Vert B_{2,2}^{P\left( -i\right) }\right\Vert
&=&\max_{i}\left\Vert J_{2}^{P\left( -i\right) }L_{2}^{P\left( -i\right)
}\right\Vert =\max_{i}\left\Vert -\left[ Q^{\left( d\right) }\right] ^{-1}%
\left[ Q^{\left( d\right) }+CAC^{\prime }\right] \Delta _{2}^{\left(
-i\right) }L_{2}^{P\left( -i\right) }\right\Vert , \\
&\lesssim &\max_{i}\left\Vert \Delta _{2}^{\left( -i\right) }\right\Vert
\max_{i}\left\Vert L_{2}^{P\left( -i\right) }\right\Vert
=O_{p}(N^{-1+1/(2+\delta )})O_{p}(N^{-1+1/(2+\delta )})=o_{p}(N^{-1}) \\
\max_{i}\left\Vert B_{2,3}^{P\left( -i\right) }\right\Vert
&=&\max_{i}\left\Vert J_{3}^{P\left( -i\right) }L_{2}^{P\left( -i\right)
}\right\Vert =\max_{i}\left\Vert -\Delta _{2}^{\left( -i\right) }\left[
Q^{\left( d\right) }+CAC^{\prime }\right] \left[ Q^{\left( -i,d\right) }%
\right] ^{-1}L_{2}^{P\left( -i\right) }\right\Vert \\
&\lesssim &\max_{i}\left\Vert \Delta _{2}^{\left( -i\right) }\right\Vert
\max_{i}\left\Vert L_{2}^{P\left( -i\right) }\right\Vert
=O_{p}(N^{-1+1/(2+\delta )})O_{p}(N^{-1+1/(2+\delta )})=o_{p}(N^{-1}) \\
\max_{i}\left\Vert B_{2,4}^{P\left( -i\right) }\right\Vert
&=&\max_{i}\left\Vert J_{4}^{P\left( -i\right) }L_{2}^{P\left( -i\right)
}\right\Vert =\max_{i}\left\Vert -\left[ Q^{\left( -i,d\right) }\right]
^{-1}\Delta _{1}^{\left( -i\right) }\left[ Q^{\left( -i,d\right) }\right]
^{-1}L_{2}^{P\left( -i\right) }\right\Vert \\
&\lesssim &\max_{i}\left\Vert \Delta _{1}^{\left( -i\right) }\right\Vert
\max_{i}\left\Vert L_{2}^{P\left( -i\right) }\right\Vert
=O_{p}(N^{-1+1/(2+\delta )})O_{p}(N^{-1+1/(2+\delta )})=o_{p}(N^{-1}), \\
\max_{i}\left\Vert B_{2,5}^{P\left( -i\right) }\right\Vert
&=&\max_{i}\left\Vert J_{5}^{P\left( -i\right) }L_{2}^{P\left( -i\right)
}\right\Vert =\max_{i}\left\Vert \left[ Q^{\left( -i,d\right) }\right]
^{-1}\Delta _{3}^{\left( -i\right) }\left[ Q^{\left( -i,d\right) }\right]
^{-1}L_{2}^{P\left( -i\right) }\right\Vert \\
&\lesssim &\max_{i}\left\Vert \Delta _{3}^{\left( -i\right) }\right\Vert
\max_{i}\left\Vert L_{2}^{P\left( -i\right) }\right\Vert
=O_{p}(N^{-1+1/(4+2\delta )})O_{p}(N^{-1+1/(2+\delta )})=o_{p}(N^{-1}).
\end{eqnarray*}%
It follows that $B_{2}^{P\left( -i\right) }=-\hat{Q}^{-1}\frac{1}{N_{1}T}%
\mathbf{\dot{x}}_{i}^{\prime }\mathbf{u}_{i}^{\ddagger
}+o_{p}(N^{-1}).\medskip $

(iii) We study $B_{3}^{P\left( -i\right) }=[\hat{Q}^{(-i)}]^{-1}L_{3}^{P%
\left( -i\right) }=\sum_{l=1}^{5}J_{l}^{P\left( -i\right) }L_{3}^{P\left(
-i\right) }\equiv \sum_{l=1}^{5}B_{3,l}^{P\left( -i\right) }$ by studying $%
B_{3,l}^{P\left( -i\right) },$ $l\in \left[ 5\right] $, in turn. Noting that 
$\breve{X}^{\left( -i\right) }=X-\breve{X}^{\left( i\right) },$ we have 
\begin{eqnarray*}
L_{3}^{P\left( -i\right) } &=&-\frac{1}{N_{1}T}\breve{X}^{\left( -i\right)
\prime }(\mathbb{M}_{\mathbf{\iota }_{N}}\otimes \mathbb{M}_{\mathbf{\iota }%
_{T}})\mathbf{\breve{U}}^{\ddagger \left( i\right) }=-\frac{1}{N_{1}T}%
X^{\prime }(\mathbb{M}_{\mathbf{\iota }_{N}}\otimes \mathbb{M}_{\mathbf{%
\iota }_{T}})\mathbf{\breve{U}}^{\ddagger \left( i\right) }+\frac{1}{N_{1}T}%
X^{\left( i\right) \prime }(\mathbb{M}_{\mathbf{\iota }_{N}}\otimes \mathbb{M%
}_{\mathbf{\iota }_{T}})\mathbf{\breve{U}}^{\ddagger \left( i\right) } \\
&=&-\frac{1}{N_{1}T}\mathbf{\ddot{x}}_{i}^{\prime }\mathbf{\dot{u}}%
_{i}^{\ddagger }+\frac{1}{N_{1}T}\mathbf{\dot{x}}_{i}^{\prime }\mathbf{\dot{u%
}}_{i}^{\ddagger }m_{ii},
\end{eqnarray*}%
and\ $\max_{i}\left\Vert L_{3}^{P\left( -i\right) }\right\Vert \lesssim 
\frac{1}{N_{1}T}\left( \left\Vert \mathbf{\ddot{x}}_{i}^{\prime }\mathbf{%
\dot{u}}_{i}^{\ddagger }\right\Vert +\left\Vert \mathbf{\dot{x}}_{i}^{\prime
}\mathbf{\dot{u}}_{i}^{\ddagger }\right\Vert \right)
=O_{p}(N^{-1+1/(2+\delta )}).$ For $B_{3,1}^{P\left( -i\right) },$ we use
the fact that $m_{ii}=1-\frac{1}{N}$ to obtain%
\begin{equation*}
B_{3,1}^{P\left( -i\right) }=J_{1}^{P\left( -i\right) }L_{3}^{P\left(
-i\right) }=\hat{Q}^{-1}\left( \frac{-1}{N_{1}T}\mathbf{\ddot{x}}%
_{i}^{\prime }\mathbf{\dot{u}}_{i}^{\ddagger }+\frac{1}{N_{1}T}\mathbf{\dot{x%
}}_{i}^{\prime }\mathbf{u}_{i}^{\ddagger }m_{ii}\right) =\frac{-1}{N_{1}T}%
\hat{Q}^{-1}\mathbf{\ddot{x}}_{i}^{\prime }\mathbf{\dot{u}}_{i}^{\ddagger }+%
\frac{1}{N_{1}T}\hat{Q}^{-1}\mathbf{\dot{x}}_{i}^{\prime }\mathbf{\dot{u}}%
_{i}^{\ddagger }+o_{p}(N^{-1})
\end{equation*}%
uniformly in $i\in \left[ N\right] .$ For the other four terms, we have by
Lemma \ref{LemS.2} 
\begin{eqnarray*}
\max_{i}\left\Vert B_{3,2}^{P\left( -i\right) }\right\Vert
&=&\max_{i}\left\Vert J_{2}^{P\left( -i\right) }L_{3}^{P\left( -i\right)
}\right\Vert =\max_{i}\left\Vert -\left[ Q^{\left( d\right) }\right] ^{-1}%
\left[ Q^{\left( d\right) }+CAC^{\prime }\right] \Delta _{2}^{\left(
-i\right) }L_{3}^{P\left( -i\right) }\right\Vert \\
&\lesssim &\max_{i}\left\Vert \Delta _{2}^{\left( -i\right) }\right\Vert
\max_{i}\left\Vert L_{3}^{P\left( -i\right) }\right\Vert
=O_{p}(N^{-1+1/(2+\delta )})O_{p}(N^{-1+1/(2+\delta )})=o_{p}\left(
N^{-1}\right) , \\
\max_{i}\left\Vert B_{3,3}^{P\left( -i\right) }\right\Vert
&=&\max_{i}\left\Vert J_{3}^{P\left( -i\right) }L_{3}^{P\left( -i\right)
}\right\Vert =\max_{i}\left\Vert -\Delta _{2}^{\left( -i\right) }\left[
Q^{\left( d\right) }+CAC^{\prime }\right] \left[ Q^{\left( -i,d\right) }%
\right] ^{-1}L_{3}^{P\left( -i\right) }\right\Vert \\
&\lesssim &\max_{i}\left\Vert \Delta _{2}^{\left( -i\right) }\right\Vert
\max_{i}\left\Vert L_{3}^{P\left( -i\right) }\right\Vert =o_{p}\left(
N^{-1}\right) , \\
\max_{i}\left\Vert B_{3,4}^{P\left( -i\right) }\right\Vert
&=&\max_{i}\left\Vert J_{4}^{P\left( -i\right) }L_{3}^{P\left( -i\right)
}\right\Vert =\max_{i}\left\Vert -\left[ Q^{\left( -i,d\right) }\right]
^{-1}\Delta _{1}^{\left( -i\right) }\left[ Q^{\left( -i,d\right) }\right]
^{-1}L_{3}^{P\left( -i\right) }\right\Vert \\
&\lesssim &\max_{i}\left\Vert \Delta _{1}^{\left( -i\right) }\right\Vert
\max_{i}\left\Vert L_{3}^{P\left( -i\right) }\right\Vert
=O_{p}(N^{-1+1/(2+\delta )})O_{p}(N^{-1+1/(2+\delta )})=o_{p}\left(
N^{-1}\right) ,\text{ and} \\
\max_{i}\left\Vert B_{3,5}^{P\left( -i\right) }\right\Vert
&=&\max_{i}\left\Vert J_{5}^{P\left( -i\right) }L_{3}^{P\left( -i\right)
}\right\Vert =\max_{i}\left\Vert -\left[ Q^{\left( -i,d\right) }\right]
^{-1}\Delta _{3}^{\left( -i\right) }\left[ Q^{\left( -i,d\right) }\right]
^{-1}L_{3}^{P\left( -i\right) }\right\Vert \\
&\lesssim &\max_{i}\left\Vert \Delta _{3}^{\left( -i\right) }\right\Vert
\max_{i}\left\Vert L_{3}^{P\left( -i\right) }\right\Vert
=O_{p}(N^{-1+1/(4+2\delta )})O_{p}(N^{-1+1/(2+\delta )})=o_{p}\left(
N^{-1}\right) .
\end{eqnarray*}%
Then $B_{3}^{P\left( -i\right) }=-\frac{1}{N_{1}T}\hat{Q}^{-1}\mathbf{\ddot{x%
}}_{i}^{\prime }\mathbf{\dot{u}}_{i}^{\ddagger }+\frac{1}{N_{1}T}\hat{Q}^{-1}%
\mathbf{\dot{x}}_{i}^{\prime }\mathbf{\dot{u}}_{i}^{\ddagger }+o_{p}\left(
N^{-1}\right) $ uniformly in $i\in \left[ N\right] .\medskip $

(iv) Now we study $B_{4}^{P\left( -i\right) }=\left[ \hat{Q}^{(-i)}\right]
^{-1}L_{4}^{P\left( -i\right) }$. By (\ref{PP1}), we have 
\begin{equation*}
L_{4}^{P\left( -i\right) }=\frac{-1}{N_{1}T}\breve{X}^{\left( -i\right) }%
\left[ (\mathbb{P}_{\mathbf{\iota }_{N}^{\left( -i\right) }}-\mathbb{P}_{%
\mathbf{\iota }_{N}})\otimes \mathbb{M}_{\mathbf{\iota }_{T}}\right] \mathbf{%
\breve{U}}^{\ddagger \left( -i\right) }=\frac{-1}{N_{1}T}\sum_{j\neq i,k\neq
i}\mathbf{\dot{x}}_{j}^{\prime }\mathbf{\dot{u}}_{k}^{\ddagger }\left[ 
\mathbb{P}_{\mathbf{\iota }_{N}^{\left( -i\right) }}-\mathbb{P}_{\mathbf{%
\iota }_{N}}\right] _{jk}=\frac{1}{NN_{1}^{2}T}\sum_{j\neq i,k\neq i}\mathbf{%
\dot{x}}_{j}^{\prime }\mathbf{\dot{u}}_{k}^{\ddagger }.
\end{equation*}%
It is easy to see that under Assumption \ref{Assmp4.1} 
\begin{eqnarray*}
\max_{i}\left\Vert L_{4}^{P\left( -i\right) }\right\Vert &=&\max_{i}\frac{1}{%
NN_{1}^{2}T}\left\Vert \sum_{j\neq i}\mathbf{\dot{x}}_{j}^{\prime }\left(
\sum_{k=1}^{N}\mathbf{\dot{u}}_{k}^{\ddagger }-\mathbf{\dot{u}}%
_{i}^{\ddagger }\right) \right\Vert \lesssim \frac{1}{N}\sum_{j\neq
i}\left\Vert \mathbf{\dot{x}}_{j}\right\Vert \left\{ \frac{1}{N_{1}^{2}}%
\left\Vert \sum_{k=1}^{N}\mathbf{\dot{u}}_{k}^{\ddagger }\right\Vert +\frac{1%
}{N_{1}^{2}}\max_{i}\left\Vert \mathbf{\dot{u}}_{i}^{\ddagger }\right\Vert
\right\} \\
&=&O_{p}(N^{-3/2}).
\end{eqnarray*}%
This, along with the fact that $\max_{i}\left\Vert \left[ \hat{Q}^{(-i)}%
\right] ^{-1}\right\Vert \leq \sum_{l=1}^{5}\max_{i}\left\Vert
J_{l}^{P\left( -i\right) }\right\Vert =O_{p}\left( 1\right) ,$ implies that 
\begin{equation*}
\max_{i}\left\Vert B_{4}^{P\left( -i\right) }\right\Vert =\max_{i}\left\Vert 
\left[ \hat{Q}^{(-i)}\right] ^{-1}L_{4}^{P\left( -i\right) }\right\Vert
\lesssim \max_{i}\left\Vert \left[ \hat{Q}^{(-i)}\right] ^{-1}\right\Vert
\max_{i}\left\Vert L_{4}^{P\left( -i\right) }\right\Vert =O_{p}(N^{-3/2}).
\end{equation*}%
This completes the proof of the lemma. ${\tiny \blacksquare }$

\section{Additional Simulation Results \label{SecS2}}

In this section, we consider two more DGPs (DGPs 5 and 6) with TWFEs only.
DGPs 5 and 6 are exactly the same as DGPs 3 and 4, respectively in Section %
\ref{Sim_DGP} of the main text except that the IFEs ($\lambda _{i}^{o\prime
}f_{t}^{o}$ and $\gamma _{i}^{o\prime }f_{t}^{o}$) are absent from both $x$-
and $y$-equations. We still consider the same six estimators as discussed in
Section \ref{Sim_DGP} and the implementations details are exactly the same.
Table \ref{estimationS1} reports the estimation results, which shows similar
conclusions to those presented in Section \ref{Sim_results} of the main
text. As shown in Table \ref{CoverageS1}, the proposed CI\ and testing
procedure also perform well for these two DGPs. We can conclude that our new
estimation and inference methods also work well for the DGPs with TWFEs only.

\renewcommand{\arraystretch}{1} \setlength{\tabcolsep}{2.5pt}

\begin{table}[tph]
\caption{Estimation Results}
\label{estimationS1}
\begin{center}
\begin{tabular}{ccccccccccccccccc}
\hline\hline
&  &  & \multicolumn{6}{c}{bias$\times 10$} &  &  & \multicolumn{6}{c}{MSE$%
\times 100$} \\ \cline{4-9}\cline{12-17}
&  &  & TW- & TW-MG & TW- & CCE- & CCE- &  & \ \ \ \  & \ \ \ \  & TW- & 
TW-MG & TW- & CCE- & CCE- &  \\ 
$N$ & $T$ &  & MG & ridge & pooled & MG & pooled & MG &  &  & MG & ridge & 
pooled & MG & pooled & MG \\ \cline{1-2}\cline{4-9}\cline{9-17}
\multicolumn{17}{c}{DGP 5} \\ \hline
50 & 3 &  & 0.18 & -0.21 & 5.08 & 0.37 & 11.26 & 2.84 &  &  & 3.63 & 2.59 & 
37.54 & 15.69 & 251.92 & 13.09 \\ 
100 & 3 &  & 0.03 & -0.16 & 5.11 & 0.05 & 14.88 & 2.79 &  &  & 2.09 & 1.70 & 
31.60 & 15.85 & 299.10 & 11.88 \\ 
200 & 3 &  & 0.01 & -0.13 & 5.30 & 0.11 & 16.86 & 2.81 &  &  & 1.33 & 0.87 & 
31.19 & 7.04 & 347.81 & 11.56 \\ 
50 & 5 &  & 0.12 & -0.12 & 5.12 & 0.37 & 5.94 & 2.86 &  &  & 2.25 & 2.13 & 
34.69 & 8.11 & 43.21 & 10.54 \\ 
100 & 5 &  & 0.00 & -0.11 & 4.99 & 0.35 & 6.24 & 2.77 &  &  & 1.23 & 1.21 & 
30.22 & 30.32 & 43.85 & 9.51 \\ 
200 & 5 &  & 0.03 & -0.03 & 5.13 & -0.03 & 6.60 & 2.76 &  &  & 0.59 & 0.57 & 
28.55 & 2.24 & 45.85 & 8.80 \\ 
50 & 10 &  & 0.17 & -0.03 & 5.03 & 0.23 & 6.20 & 2.91 &  &  & 1.91 & 1.80 & 
31.48 & 1.98 & 43.21 & 10.53 \\ 
100 & 10 &  & 0.00 & -0.09 & 4.99 & 0.02 & 6.27 & 2.78 &  &  & 1.15 & 1.14 & 
28.78 & 1.20 & 42.45 & 9.02 \\ 
200 & 10 &  & 0.02 & -0.03 & 5.13 & 0.03 & 6.52 & 2.65 &  &  & 0.52 & 0.52 & 
27.85 & 0.53 & 43.91 & 7.83 \\ \hline
\multicolumn{17}{c}{DGP 6} \\ \hline
50 & 4 &  & 0.11 & 0.02 & 6.59 & 0.36 & 11.85 & 2.22 &  &  & 13.59 & 6.39 & 
53.35 & 16.80 & 243.86 & 23.76 \\ 
100 & 4 &  & 0.01 & -0.08 & 6.51 & 0.25 & 12.76 & 2.09 &  &  & 6.02 & 3.05 & 
48.98 & 8.08 & 300.42 & 11.37 \\ 
200 & 4 &  & 0.18 & 0.07 & 6.90 & 0.12 & 14.68 & 2.22 &  &  & 2.96 & 1.54 & 
50.05 & 5.59 & 323.11 & 9.56 \\ 
50 & 5 &  & 0.17 & 0.01 & 6.73 & 0.11 & 6.31 & 2.23 &  &  & 3.98 & 3.46 & 
54.56 & 38.66 & 55.06 & 8.82 \\ 
100 & 5 &  & -0.05 & -0.12 & 6.50 & -0.42 & 6.02 & 2.02 &  &  & 2.52 & 2.22
& 47.49 & 13.39 & 44.71 & 6.52 \\ 
200 & 5 &  & 0.03 & 0.02 & 6.74 & 0.16 & 6.56 & 2.05 &  &  & 1.33 & 1.19 & 
47.70 & 8.61 & 47.86 & 5.88 \\ 
50 & 10 &  & 0.13 & 0.01 & 6.73 & 0.12 & 6.42 & 2.12 &  &  & 2.29 & 2.20 & 
52.36 & 2.78 & 47.31 & 6.56 \\ 
100 & 10 &  & -0.03 & -0.09 & 6.48 & -0.04 & 6.29 & 1.96 &  &  & 1.50 & 1.47
& 45.92 & 1.69 & 43.21 & 5.23 \\ 
200 & 10 &  & 0.01 & -0.02 & 6.63 & 0.02 & 6.51 & 1.91 &  &  & 0.69 & 0.68 & 
45.76 & 0.85 & 44.17 & 4.48 \\ \hline
\end{tabular}%
\end{center}
\par
Notes: \textquotedblleft TW-MG\textquotedblright\ refers to the
two-way-mean-group estimator proposed in (\ref{betahat1}), \textquotedblleft
TW-MG ridge\textquotedblright\ refers to the ridge-type estimator proposed
in (\ref{ridge}), \textquotedblleft TW-pooled\textquotedblright\ refers to
the pooled two-way fixed effects estimator in (\ref{B_tilde1}),
\textquotedblleft CCE-pooled\textquotedblright\ refers to \cite{pesaran2006}%
's CCE\ pooled estimator, and \textquotedblleft CCE-MG\textquotedblright\
refers to \cite{pesaran2006}'s CCE\ mean group estimator. and
\textquotedblleft MG\textquotedblright\ refers to \cite{pesaran_smith1995}'s
standard mean-group estimator.
\end{table}

\begin{table}[tph]
\caption{Inference Results}
\label{CoverageS1}
\begin{center}
\begin{tabular}{cccccccccccccc}
\hline\hline
&  &  & \multicolumn{5}{c}{Empirical coverage} &  & \multicolumn{5}{c}{
Rejection frequency of testing $H_{0}$} \\ \cline{10-14}
&  &  & \multicolumn{5}{c}{of 95\% CI for $\beta ^{0}$} &  & 
\multicolumn{5}{c}{power} \\ \cline{1-2}\cline{4-8}\cline{10-11}\cline{13-14}
&  &  & \multicolumn{2}{c}{TW-MG estimator} &  & \multicolumn{2}{c}{TW-MG
ridge estimator} &  & \multicolumn{2}{c}{TW-MG estimator} &  & 
\multicolumn{2}{c}{TW-MG ridge estimator} \\ 
\cline{1-2}\cline{4-5}\cline{7-8}\cline{10-11}\cline{13-14}
$N$ & $T$ &  & \ \ DGP\ 5\ \  & \ \ DGP\ 6\ \  &  & \ \ DGP\ 5\ \  & \ \
DGP\ 6\ \  &  & \ \ DGP\ 5\ \  & \ \ DGP6\ \  &  & \ \ DGP\ 5\ \  & \ \
DGP6\ \  \\ \cline{1-2}\cline{4-5}\cline{7-8}\cline{10-11}\cline{13-14}
50 & 3 &  & 0.95 & 0.96 &  & 0.96 & 0.96 &  & 0.46 & 0.44 &  & 0.56 & 0.52
\\ 
100 & 3 &  & 0.93 & 0.96 &  & 0.93 & 0.94 &  & 0.74 & 0.66 &  & 0.81 & 0.79
\\ 
200 & 3 &  & 0.96 & 0.97 &  & 0.96 & 0.97 &  & 0.92 & 0.92 &  & 0.94 & 0.97
\\ 
50 & 5 &  & 0.96 & 0.96 &  & 0.96 & 0.97 &  & 0.70 & 0.65 &  & 0.77 & 0.71
\\ 
100 & 5 &  & 0.94 & 0.94 &  & 0.94 & 0.94 &  & 0.85 & 0.89 &  & 0.87 & 0.90
\\ 
200 & 5 &  & 0.95 & 0.93 &  & 0.96 & 0.94 &  & 0.98 & 1.00 &  & 0.98 & 1.00
\\ 
50 & 10 &  & 0.97 & 0.96 &  & 0.97 & 0.95 &  & 0.84 & 0.93 &  & 0.86 & 0.96
\\ 
100 & 10 &  & 0.94 & 0.92 &  & 0.92 & 0.91 &  & 0.97 & 0.98 &  & 0.97 & 0.99
\\ 
200 & 10 &  & 0.95 & 0.94 &  & 0.95 & 0.94 &  & 1.00 & 1.00 &  & 1.00 & 1.00
\\ \hline
\end{tabular}%
\end{center}
\par
Notes: 95\% CI\ refers to 95\% confidence interval based on the proposed
jackknife method. The test of $\mathbb{H}_{0}$ (\ref{H0}) is based on the
test statistic (\ref{test}). The nominal size is $5\%$. \textquotedblleft
TW-MG\textquotedblright\ refers to the two-way-mean-group estimator proposed
in (\ref{betahat1}) and \textquotedblleft TW-MG ridge\textquotedblright\
refers to the ridge-type estimator proposed in (\ref{ridge}). For DGP\ 6
where $K_{x}=2$, the shortest $T$ is 4 instead of $3$.
\end{table}

\end{document}